\documentclass[a4paper,11pt]{article}
\pdfoutput=1
\usepackage{jheppub}
\usepackage[utf8]{inputenc}

\usepackage{pdflscape}
\usepackage{tikz}
\usetikzlibrary{positioning}
\usetikzlibrary{arrows}
\usetikzlibrary{arrows.meta}
\usetikzlibrary{shapes,arrows,fit,decorations.pathreplacing, shapes.misc,decorations.pathmorphing}

\usepackage{subcaption}
\usepackage{mathtools}
\usepackage{ragged2e}
\usepackage{booktabs}
\usepackage{paralist}
\usepackage[noabbrev,capitalise]{cleveref}
\usepackage{soul}
\usepackage{multirow}
\usepackage[colorinlistoftodos]{todonotes}
\usepackage{amsthm}
\usepackage{amsmath}
\usepackage{changepage}
\usepackage{longtable}
\usepackage{amssymb}
\usepackage{flafter}
\usepackage{adjustbox}
\usepackage[english]{babel}
\usepackage{tabularx}
\usepackage{ytableau}
\usepackage{setspace}
\usepackage{gensymb}
\usepackage{tcolorbox}
\usepackage{multicol}
\usepackage{placeins}
\definecolor{myGreen1}{RGB}{53, 157, 42}
\definecolor{myGreen}{RGB}{54, 150, 45}

\usepackage{amsthm}
\newtheorem{myrule}{Group}

\tikzset{Reddotted/.style={dashed,color=red}}
\tikzset{gauge1/.style={draw=none,minimum size=0.6cm,fill=white,circle, draw}}
\tikzset{gauge3/.style={draw=none,minimum size=0.35cm,fill=white,circle, draw}}
\tikzset{gauge5/.style={draw=none,minimum size=0.35cm,fill=white,circle, draw}}
\tikzset{crosses/.style={draw,circle,cross out}}
\tikzset{blank/.style={draw=none,minimum size=0.4cm,fill=none,circle, draw}}
\tikzset{flavor2/.style={draw=none,minimum size=0.4cm,fill=white,regular polygon sides=4,draw}}
\tikzset{flavour2/.style={draw=none,minimum size=0.4cm,fill=white,regular polygon sides=4,draw}}
\tikzset{flavorBlue/.style={draw=none,minimum size=0.4cm,fill=blue,regular polygon sides=4,draw}}
\tikzset{flavorRed/.style={draw=none,minimum size=0.4cm,fill=red,regular polygon sides=4,draw}}
\tikzset{none/.style={draw=none}}
\tikzset{flavourBlue/.style={draw=none,minimum size=0.4cm,fill=blue,regular polygon sides=4,draw}}
\tikzset{flavourRed/.style={draw=none,minimum size=0.4cm,fill=red,regular polygon sides=4,draw}}
\tikzset{none/.style={draw=none}}
\tikzset{redgauge/.style={draw=none,minimum size=0.4cm,fill=red,circle, draw}}
\tikzset{miniU/.style={draw=none,minimum size=0.1cm,fill=red,circle, draw}}
\tikzset{smallgauge1/.style={draw=none,minimum size=0.1cm,fill=white,circle, draw}}
\tikzset{blankflavor/.style={draw=none,minimum size=0.8mm,fill=none, regular polygon,regular polygon sides=4,draw}}
\tikzset{miniBlue/.style={draw=none,minimum size=0.1cm,fill=blue,circle, draw}}
\tikzset{gauge2/.style={draw=none,minimum size=0.35mm,fill=red,circle, draw}}
\tikzset{bluegauge/.style={draw=none,minimum size=0.4cm,fill=blue,circle, draw}}
\tikzset{flavor1/.style={draw=none,minimum size=0.35mm,fill=blue, regular polygon,regular polygon sides=4,draw}}
\tikzset{flavor0/.style={draw=none,minimum size=0.35mm,fill=white, regular polygon,regular polygon sides=4,draw}}
\tikzset{o5/.style={green,dotted}}
\tikzset{new edge style 0/.style={magenta}}
\tikzset{o3tildeplus/.style={dashed}}
\tikzset{o3plus/.style={dotted}}
\tikzset{smalldot/.style={draw=none,minimum size=0.1mm,fill=black, circle,draw}}
\tikzset{dotsize/.style={circle,fill,inner sep=1.5pt,draw}}
\tikzset{doubleguys/.style={double, double distance = 3pt}}
\tikzset{tripleguys/.style={triple}}
\tikzset{new edge style 1/.style={dashed}}
\tikzset{thickline/.style={line width=0.06cm}}
\tikzset{darke/.style={line width=0.3mm,black}}
\tikzset{brace/.style={decorate,decoration={brace,amplitude=10pt}}}
\tikzset{brace1/.style={decorate,decoration={brace,amplitude=10pt}}}
\pgfdeclarelayer{edgelayer}
\pgfdeclarelayer{nodelayer}
\pgfsetlayers{edgelayer,nodelayer,main}

\tikzset{wd/.style={circle, draw,inner sep=3.5pt}}
\tikzset{bd/.style={circle, draw,inner sep=3.5pt, fill=black}}

\tikzset{gauge/.style={circle, draw,inner sep=3pt}}
\tikzset{gaugeSO/.style={circle, draw,inner sep=3pt,fill=red}}
\tikzset{gaugeSp/.style={circle, draw,inner sep=3pt,fill=blue}}
\tikzset{flavour/.style={regular polygon,regular polygon sides=4,inner
sep=3pt, draw}}
\tikzset{flavourSO/.style={regular polygon,regular polygon sides=4,inner
sep=3pt, draw,fill=red}}
\tikzset{flavourSp/.style={regular polygon,regular polygon sides=4,inner
sep=3pt, draw,fill=blue}}

\usepackage{xargs}
\newlength{\myline}
\newcommandx*{\doublearrow}[4][1=0, 2=1]{
  \draw[line width=\myline,double distance=3\myline,#3] #4;
}

\newcommand{\Ncal}{\mathcal{N}}

\def\ns#1{
	\node[circle, draw, fill=white] at (#1){};
	\node[cross out, draw] at (#1){};
}
\def\onz#1{
	\node[circle, draw, fill=black] at (#1){};
}
\def\on#1{
	\node[circle, draw, fill=white] at (#1){};
}

\def\SevenB#1{
	\node[circle, draw, fill=white] at (#1){};
}

\def\monocut#1#2{
    \draw[dashdotted,red] (#1)--(#2);
}

\newcommand{\urm}{\mathrm{U}}
\newcommand{\surm}{\mathrm{SU}}
\newcommand{\sprm}{\mathrm{Sp}}
\newcommand{\usprm}{\mathrm{USp}}
\newcommand{\sorm}{\mathrm{SO}}

\newcommand{\Coulomb}{\mathcal{C}}
\newcommand{\Higgs}{\mathcal{H}}
\newcommand{\HWG}{\mathrm{HWG}}
\newcommand{\PE}{\mathrm{PE}}

\newcommand{\ra}[1]{\renewcommand{\arraystretch}{#1}}

\allowdisplaybreaks

\preprint{}

\title{Balanced $\boldsymbol{B}$ and $\boldsymbol{D}$-type orthosymplectic quivers\\ --- Magnetic quivers for product theories}

\author[a, b]{Marcus Sperling}
\author[c]{and Zhenghao Zhong}
\affiliation[a]{Shing-Tung Yau Center, Southeast University \\ Xuanwu District, Nanjing, Jiangsu, 210096, China}
\affiliation[b]{Yau Mathematical Sciences Center, Tsinghua University,\\ Haidian District, Beijing, 100084, China}
\affiliation[c]{Theoretical Physics Group, The Blackett Laboratory, Imperial College London,\\ Prince Consort Road
London, SW7 2AZ, UK}
\emailAdd{msperling@seu.edu.cn}
\emailAdd{zhenghao.zhong14@imperial.ac.uk}

\abstract{We investigate orthosymplectic quivers that take the shape of $D$-type and $B$-type Dynkin diagrams. The $D$-type orthosymplectic quivers explored here contain a balanced ``fork", i.e.\ a balanced subquiver with a $D$-type bifurcation, whereas the $B$-type orthosymplectic quivers are obtained by folding the $D$-type quivers. The Coulomb branches of these quivers are products of two moduli spaces. 
In the second part, the relevant orthosymplectic quivers are shown to emerge as magnetic quivers for brane configurations involving ON${}^0$ planes. Notably, the appearance of ON${}^0$ plane clarifies the product nature of the theories in question. The derivation leads to the analysis of magnetic quivers from branes systems with intersecting O$p$, O$(p+2)$, and ON${}^0$ planes.
}

\begin{document} 
\maketitle

\section{Introduction}
Coulomb branches of 3d $\mathcal{N}=4$ quiver gauge theories have rich geometric properties. In the last few decades, quivers composed of unitary gauge groups had been studied in great length. 
Prominent physics approaches include Hilbert series \cite{Cremonesi:2013lqa}, abelianisation \cite{Bullimore:2015lsa}, and Coulomb branch quantisation \cite{Dedushenko:2018icp}; while active mathematics research culminated in a rigorous definition \cite{Nakajima:2015txa,Braverman:2016wma}.
On the other hand, \emph{orthosymplectic quivers} --- quivers composed of alternating (special) orthogonal and symplectic gauge groups ---  are less frequented; both in physics and mathematics. One explanation is that orthosymplectic quivers are more difficult to study due to many subtle features. Recent developments in orthosymplectic quivers allow to expand the exploration. For instance, novel types of orthosymplectic quivers arise in the study of \emph{magnetic quivers} using brane configurations \cite{Cabrera:2019dob,Bourget:2020gzi,Akhond:2020vhc,Akhond:2021knl}. In particular, these include quivers without flavour nodes --- hereafter referred as \emph{unframed} quivers. These developments are largely based on an improved understanding of generating functions, like the Coulomb branch Hilbert series \cite{Cremonesi:2013lqa,Bourget:2020xdz,Bourget:2021xex}.
This greatly expands the set of orthosymplectic quivers one can probe and the time is ripe for a systematic exploration. 

One way to systematically explore orthosymplectic quivers is to follow the early developments of unitary quivers. 
For example, the $ADE$ Dynkin classification of unitary quiver theories 
describing the low-energy theories of D-branes probing an $ADE$ singularity $\mathbb{C}^2\slash \Gamma_{ADE}$ \cite{Douglas:1996sw,Johnson:1996py}, also known as McKay correspondence. The Dynkin type quivers are completed by unitary quivers \cite{Cremonesi:2014xha} in the shape of classical $BC$ type Dynkin diagrams and quivers in the shape of exceptional $G_2$, $F_4$ Dynkin diagrams. While the $ADE$-type quivers admit a Lagrangian description, the $BCFG$-type quivers do have a known Lagrangian. The Coulomb branch of a Dynkin quiver of the Lie algebra $\mathfrak{g}$ is the (reduced) one $G$-instanton moduli space, or equivalently, the minimal nilpotent orbit closure of $\mathfrak{g}$ \cite{Kronheimer:1990ay,1998math.....11032B,2000math......1025K}. This relations between  Coulomb branches and moduli spaces of instantons was pointed out for $ADE$ quivers in \cite{Intriligator:1996ex}, see also \cite{deBoer:1996mp,Porrati:1996xi}, and for $BCFG$ quivers in \cite{Cremonesi:2014xha}.
This classification yields a set of simple moduli spaces which are ubiquitous in quiver gauge theories.

The description of the 3d $\Ncal=4$ Coulomb branch has benefited tremendously from the realisation that monopole operators \cite{Aharony:1997bx,Borokhov:2002ib,Borokhov:2002cg} are suitable coordinates to describe the quantum-corrected moduli space. Moreover, monopole operators are instrumental for the phenomenon of the enhancement of the Coulomb branch global symmetry in the IR \cite{Borokhov:2002cg,Gaiotto:2008ak,Bashkirov:2010hj}. By studying which monopole operators sit inside the global symmetry current, it has been shown in \cite{Gaiotto:2008ak} that the enhanced Coulomb branch global symmetry of a unitary quiver can be read off by identifying the subset of \emph{balanced} nodes and comparing it to Dynkin diagrams. The ability to immediately read off the Coulomb branch global symmetry is a powerful tool, especially if the quiver is too complicated for other methods. However, it might be the case that the set of balanced nodes does not yield the full global symmetry in the IR, as further enhancement might occur, see for instance \cite{Mekareeya:2017jgc,Bourget:2020mez,Gledhill:2021cbe} as well as below.

The possibility to realise 3d $\Ncal=4$ theories by Type IIB brane configurations \cite{Hanany:1996ie} has undoubtedly advanced the understanding of intricate aspects: let it be dualities like 3d mirror symmetry or the geometry of moduli space. The set of possible low-energy theories is enriched by inclusion of orientifold and orbifold planes into brane the configurations \cite{Uranga:1998uj,Gimon:1996rq,Hanany:1997gh,Kapustin:1998fa,Hanany:1999sj,Feng:2000eq,Gaiotto:2008ak}. Recently, the \emph{magnetic phase} of D3-D5-NS5 branes systems has been generalised to D$p$-D$(p{+}2)$-NS5 configurations in order to characterise Higgs branches of $p$-dimensional theories with 8 supercharges. Such a magnetic phase gives rise to a \emph{magnetic quiver} \cite{Cabrera:2018jxt,Cabrera:2019izd,Cabrera:2019dob, Bourget:2020gzi,Akhond:2020vhc,vanBeest:2020kou,Giacomelli:2020gee, VanBeest:2020kxw,Closset:2020scj,Closset:2020afy, Akhond:2021knl,vanBeest:2021xyt}, i.e.\ an auxiliary 3d $\Ncal=4$ quiver theory whose Coulomb branch describes the Higgs branch of the higher-dimensional theory. Hence, 3d $\Ncal=4$ Coulomb branches are essential tools to study Higgs branches of strongly coupled theories, where (semi-)classical techniques break down.

\begin{table}[ht]
    \centering
    \renewcommand{\arraystretch}{2}
    \begin{tabular}{cc} \toprule
         & Dynkin type orthosymplectic quivers \\ \midrule
       $A$-type  & \raisebox{-.5\height}{\begin{tikzpicture}
	\begin{pgfonlayer}{nodelayer}
		\node [style=bluegauge] (212) at (13, -0.25) {};
		\node [style=redgauge] (213) at (14.25, -0.25) {};
		\node [style=bluegauge] (214) at (15.25, -0.25) {};
		\node [style=redgauge] (215) at (16.25, -0.25) {};
		\node [style=redgauge] (216) at (18.25, -0.25) {};
		\node [style=bluegauge] (217) at (14.25, -0.25) {};
		\node [style=bluegauge] (218) at (16.25, -0.25) {};
		\node [style=redgauge] (219) at (15.25, -0.25) {};
		\node [style=bluegauge] (220) at (18.25, -0.25) {};
		\node [style=redgauge] (221) at (19.25, -0.25) {};
		\node [style=redgauge] (222) at (13, -0.25) {};
		\node [style=none] (223) at (17.25, -0.25) {$\dots$};
		\node [style=none] (224) at (16.75, -0.25) {};
		\node [style=none] (225) at (17.75, -0.25) {};
	\end{pgfonlayer}
	\begin{pgfonlayer}{edgelayer}
		\draw (220) to (221);
		\draw (217) to (218);
		\draw (222) to (217);
		\draw (225.center) to (220);
		\draw (224.center) to (218);
	\end{pgfonlayer}
\end{tikzpicture}} \\ 
$B$-type & \raisebox{-.5\height}{\begin{tikzpicture}
	\begin{pgfonlayer}{nodelayer}
		\node [style=bluegauge] (152) at (13, -2.25) {};
		\node [style=redgauge] (153) at (14.25, -2.25) {};
		\node [style=bluegauge] (158) at (15.25, -2.25) {};
		\node [style=redgauge] (159) at (16.25, -2.25) {};
		\node [style=redgauge] (160) at (18.25, -2.25) {};
		\node [style=bluegauge] (163) at (14.25, -2.25) {};
		\node [style=bluegauge] (164) at (16.25, -2.25) {};
		\node [style=redgauge] (165) at (15.25, -2.25) {};
		\node [style=bluegauge] (166) at (18.25, -2.25) {};
		\node [style=redgauge] (167) at (19.25, -2.25) {};
		\node [style=redgauge] (170) at (13, -2.25) {};
		\node [style=none] (172) at (17.25, -2.25) {$\dots$};
		\node [style=none] (173) at (14.25, -2.125) {};
		\node [style=none] (174) at (14.25, -2.375) {};
		\node [style=none] (175) at (13, -2.125) {};
		\node [style=none] (176) at (13, -2.375) {};
		\node [style=none] (177) at (13.75, -1.875) {};
		\node [style=none] (178) at (13.75, -2.625) {};
		\node [style=none] (179) at (13.4, -2.25) {};
		\node [style=none] (180) at (16.75, -2.25) {};
		\node [style=none] (181) at (17.75, -2.25) {};
	\end{pgfonlayer}
	\begin{pgfonlayer}{edgelayer}
		\draw (166) to (167);
		\draw (163) to (164);
		\draw (173.center) to (175.center);
		\draw (174.center) to (176.center);
		\draw (179.center) to (177.center);
		\draw (179.center) to (178.center);
		\draw (164) to (180.center);
		\draw (181.center) to (166);
	\end{pgfonlayer}
\end{tikzpicture}}
 \\ 
 $C$-type & \raisebox{-.5\height}{\begin{tikzpicture}
	\begin{pgfonlayer}{nodelayer}
		\node [style=bluegauge] (180) at (13, -4.25) {};
		\node [style=redgauge] (181) at (14.25, -4.25) {};
		\node [style=bluegauge] (182) at (15.25, -4.25) {};
		\node [style=redgauge] (183) at (16.25, -4.25) {};
		\node [style=redgauge] (184) at (18.25, -4.25) {};
		\node [style=bluegauge] (185) at (14.25, -4.25) {};
		\node [style=bluegauge] (186) at (16.25, -4.25) {};
		\node [style=redgauge] (187) at (15.25, -4.25) {};
		\node [style=bluegauge] (188) at (18.25, -4.25) {};
		\node [style=redgauge] (189) at (19.25, -4.25) {};
		\node [style=redgauge] (190) at (13, -4.25) {};
		\node [style=none] (191) at (17.25, -4.25) {$\dots$};
		\node [style=none] (192) at (14.25, -4.125) {};
		\node [style=none] (193) at (14.25, -4.375) {};
		\node [style=none] (194) at (13, -4.125) {};
		\node [style=none] (195) at (13, -4.375) {};
		\node [style=none] (196) at (13.4, -3.875) {};
		\node [style=none] (197) at (13.4, -4.625) {};
		\node [style=none] (198) at (13.75, -4.25) {};
		\node [style=none] (199) at (16.75, -4.25) {};
		\node [style=none] (200) at (17.75, -4.25) {};
	\end{pgfonlayer}
	\begin{pgfonlayer}{edgelayer}
		\draw (188) to (189);
		\draw (185) to (186);
		\draw (192.center) to (194.center);
		\draw (193.center) to (195.center);
		\draw (198.center) to (196.center);
		\draw (198.center) to (197.center);
		\draw (186) to (199.center);
		\draw (200.center) to (188);
	\end{pgfonlayer}
\end{tikzpicture}}
\\ 
$D$-type
& \raisebox{-.5\height}{\begin{tikzpicture}
	\begin{pgfonlayer}{nodelayer}
		\node [style=redgauge] (200) at (14.25, -7.25) {};
		\node [style=bluegauge] (201) at (15.25, -7.25) {};
		\node [style=redgauge] (202) at (16.25, -7.25) {};
		\node [style=redgauge] (203) at (18.25, -7.25) {};
		\node [style=bluegauge] (204) at (14.25, -7.25) {};
		\node [style=bluegauge] (205) at (16.25, -7.25) {};
		\node [style=redgauge] (206) at (15.25, -7.25) {};
		\node [style=bluegauge] (207) at (18.25, -7.25) {};
		\node [style=redgauge] (208) at (19.25, -7.25) {};
		\node [style=redgauge] (209) at (13, -6.25) {};
		\node [style=none] (210) at (17.25, -7.25) {$\dots$};
		\node [style=redgauge] (211) at (13, -8.25) {};
		\node [style=none] (212) at (16.75, -7.25) {};
		\node [style=none] (213) at (17.75, -7.25) {};
	\end{pgfonlayer}
	\begin{pgfonlayer}{edgelayer}
		\draw (207) to (208);
		\draw (204) to (205);
		\draw (204) to (209);
		\draw (204) to (211);
		\draw (212.center) to (205);
		\draw (213.center) to (207);
	\end{pgfonlayer}
\end{tikzpicture}}
 \\ \bottomrule
    \end{tabular}
    \caption{The Dynkin types of orthosymplectic quivers relevant for this paper. Red nodes represent special orthogonal groups and blue represent symplectic groups, see \eqref{eq:conventions}.}
    \label{Dynkintable}
\end{table}

\paragraph{Dynkin-type classification.}
In this paper, orthosymplectic quivers are explored in a similar manner --- aiming to develop a Dynkin classification of balanced orthosymplectic quivers, see Tables \ref{Dynkintable} and \ref{tab:results}:
\begin{itemize}
    \item $A$-type orthosymplectic quivers, i.e.\ all edges are simply-laced and the gauge groups are arranged in a linear chain, are well known \cite{Feng:2000eq,Gaiotto:2008ak}.
    \item $C$-type orthosymplectic quivers, most notably those obtained by \emph{folding} $A$-type quivers, have been explored recently in \cite{Bourget:2021xex}.
    \item In this paper the exploration of classical Dynkin type orthosymplectic quivers is completed by studying $B$-type and $D$-type. This includes a systematic study of both \emph{framed} and \emph{unframed} quivers. It is also clarified how the balance of $\sorm$ and $\sprm$ gauge groups allows one to read off the global symmetry for such orthosymplectic quivers. One crucial feature, which all the examples in this paper displays, is that the Coulomb branch is the product of two moduli spaces, see Table \ref{tab:results}.
\end{itemize}

\begin{table}[ht]
    \centering
    \ra{2.5}
    \begin{tabular}{lll}
    \toprule
    Orthosymplectic quiver & Framed & Unframed \\\midrule
    Balanced $A$-type &  $\overline{\mathcal{O}}_{\min}^{D}$, e.g.\  \cite{Feng:2000eq}  &
      $\overline{\mathcal{O}}_{\min}^{E_n}$, e.g.\ \cite{Bourget:2020gzi} \\
    Balanced $B$-type & $\overline{\mathcal{O}}^{D}\times \overline{\mathcal{O}}^{B}$, Sec.\ \ref{sec:B-type_framed} & $\begin{aligned}
    \overline{\mathcal{O}}_{\min}^{E_6}\times \overline{\mathcal{O}}_{\min}^{F_4} ,\overline{\mathcal{O}}_{\min}^{E_6}\times \overline{\mathcal{O}}_{\min}^{B_4},
    \overline{\mathcal{O}}_{\min}^{E_8}\times \overline{\mathcal{O}}^{B_7}, \\
    \overline{\mathcal{O}}_{\min}^{F_4}\times \overline{\mathcal{O}}_{\min}^{B_4},
    \overline{\mathcal{O}}^{B}\times \overline{\mathcal{O}}^{D} , \; 
    \text{Sec.\ \ref{sec:unframed_B}}
    \end{aligned}$
    \\
    Balanced $C$-type &  $\overline{\mathcal{O}}_{\min}^{A}$, e.g.\ \cite{Bourget:2021xex} & $\overline{\mathcal{O}}_{\min}^{E_7,E_6,D_5,D_4,A_3}$, e.g.\ \cite{Bourget:2021xex}  \\
    Balanced $D$-type & $\overline{\mathcal{O}}^{D}\times \overline{\mathcal{O}}^{D}$, Sec.\ \ref{sec:framed_D-type} &
    $\begin{aligned}
    \overline{\mathcal{O}}_{\min}^{E_n}\times \overline{\mathcal{O}}_{\min}^{E_n} ,\overline{\mathcal{O}}_{\min}^{E_6}\times \overline{\mathcal{O}}_{\min}^{D_5},
    \overline{\mathcal{O}}_{\min}^{E_8}\times \overline{\mathcal{O}}^{D_8}, \\
    \overline{\mathcal{O}}_{\min}^{F_4}\times \overline{\mathcal{O}}_{\min}^{F_4},
    \overline{\mathcal{O}}^{B}\times \overline{\mathcal{O}}^{B} , \; 
    \text{Sec.\ \ref{forkingunframed}}
    \end{aligned}$
     \\ \bottomrule
    \end{tabular}
    \caption{Representative examples of $ABCD$-Dynkin type orthosymplectic quivers and their Coulomb branch moduli spaces. $\overline{\mathcal{O}}^{\mathfrak{g}}$ denotes the closure of a nilpotent orbit closure of the Lie algebra $\mathfrak{g}$. The subscript ${}_{\min}$ denotes the minimal orbit. }
    \label{tab:results}
\end{table}

The scope of this work is to complete the classification of balanced classical Dynkin-type orthosymplectic quivers. This does not equal a classification of orthosymplectic quivers, for the same reasons as the balanced Dynkin-type unitary quiver classification is not a classification of all unitary quivers. The set of balanced $B$- and $D$-type orthosymplectic quivers is understood as a distinguished set, for two reasons: firstly, there are no unnecessary unbalanced gauge nodes present. Secondly, the subdiagram of balanced nodes is a single connected diagram. Because if there exists an unbalanced gauge node, one could either connect it to more unbalanced nodes or connect it to another set of balanced gauge nodes. None of these operators would change the character of that single unbalanced node, but both operations would lead to more complicated moduli spaces. In other words, restricting to balanced quivers is equivalent to studying the fundamental features, that arise from balanced Dynkin-type orthosymplectic quivers. These features are
\begin{compactitem}
\item Framed quivers: Firstly, the linear tail begins with an $\sorm(2)$ gauge node and ends  on a $B$/$D$-type edge. Secondly, all gauge nodes are balanced.
\item Unframed quivers: Firstly, the allowed gauge nodes are either $\sorm(n)$, $\sprm(k)$, or $\urm(1)$ nodes. Secondly, the linear tail begins with an $\sorm(2)$ gauge node and ends with on a $B$/$D$-type edge. Thirdly, all gauge nodes outside the balanced $B$/$D$-type Dynkin diagram are included solely to balance the $B$/$D$-type Dynkin subgraph.
\end{compactitem}
The only exception to these features, known to the authors, is the so-called $E_7 \times E_7$ family detailed and explained below. The constraint that the linear tail of alternating orthosymplectic gauge nodes decreases in rank down to $\sorm(2)$ can be understood as vital condition for the product structure via explicit brane constructions. It is then promoted to a guiding principle for all balanced orthosymplectic $B$/$D$-type Dynkin quivers with a product structure. The absence of counterexamples is a reassuring sign.

\paragraph{Product theories.}
One overarching theme of this work are quivers whose moduli spaces are product spaces. In particular D-type quivers are interesting because they fall in the class of star-shaped orthosymplectic quivers. A rather large subset of these can be understood as mirrors of class $\mathcal{S}$ theories, with legs specified by partition data. Hence, the quiver theories detailed here are candidate mirrors of certain class $\mathcal{S}$ theories that should secretly be product theories. As class $\mathcal{S}$ theories are meant to be constructed using fundamental building blocks of three-punctured spheres, it is important to find which of these fundamental building blocks are not actually fundamental, but products of other SCFTs. Recalling from \cite{Distler:2017xba,Distler:2018gbc} that identifying product class $\mathcal{S}$ theories is based on a systematic search without a clear ``smoking gun" behaviour to look for. The results of this work suggest, that a balanced set of nodes in the shape of a D-type Dynkin diagram is such a pattern to look for.

\paragraph{Outline.} Section \ref{forkingit} explores $D$-type orthosymplectic quivers. 
 The \emph{framed} ($D$-type) orthosymplectic quivers  of Section \ref{sec:framed_D-type} exhibit Coulomb branch moduli spaces that are products of height 2 nilpotent orbit closures of $\mathfrak{so}(2n)$. Next, \emph{unframed} $D$-type orthosymplectic quivers are discussed in Section \ref{forkingunframed} and the Coulomb branches are products of minimal nilpotent orbits of $\mathfrak{e}_n$ and $\mathfrak{f}_4$ algebras. These quivers are then generalised to infinite sequences. In Section \ref{foldafterfork}, $B$-type orthosymplectic quivers, obtained from folding quivers in Section \ref{forkingit}, are studied. Thereafter, brane set-ups for theories in $d=3,5,6$ dimension are detailed in Section \ref{branes}; these brane configurations allow to derive framed and unframed  $D$-type as well as framed $B$-type orthosymplectic quivers. Lastly, Section \ref{sec:conclusion} contains conclusions. 
Several appendices complement the main text. Type II brane configuration are reviewed in Appendix \ref{app:branes}. Appendix \ref{u1guy} explores additional abelian Coulomb branch global symmetry factors beyond the balance condition. Appendix \ref{app:HS} provides computational details for the results of this paper.

\paragraph{Conventions and notation.}
Before starting the main part, some important conventions are outlined.
The main object are \emph{quivers} that encode supersymmetric theories with 8 supercharges. The quiver diagram is composed of nodes and edges. The nodes are either \emph{gauge nodes}, i.e.\ a circle $\bigcirc$ encodes a dynamical vector multiplet of some gauge group, or \emph{flavour nodes}, i.e.\ a box $\Box$ encodes a background vector multiplet of a global symmetry group. 
Throughout the paper, the adopted convention for gauge nodes is:
\begin{align}
\label{eq:conventions}
\raisebox{-.5\height}{
\begin{tikzpicture}
    \node (g1) [gauge,label=below:{\footnotesize{$k$}}] {};
\end{tikzpicture}
}
\quad \leftrightarrow \quad  \urm(k)
\qquad 
\raisebox{-.5\height}{
\begin{tikzpicture}
    \node (g1) [gaugeSO,label=below:{\footnotesize{$k$}}] {};
\end{tikzpicture}
}
\quad \leftrightarrow \quad  \sorm(k)
\qquad
\raisebox{-.5\height}{
\begin{tikzpicture}
    \node (g2) [gaugeSp,label=below:{\footnotesize{$2k$}}] {};
\end{tikzpicture}
}
\quad \leftrightarrow \quad \usprm(2k) 
\end{align}
and the same colouring scheme applies to flavour nodes. The edges in a quiver diagram, which are most relevant for this paper, are either \emph{simply laced} edges or \emph{non-simply laced} edges, where the naming is borrowed from Dynkin diagrams. A simply laced edge between two nodes encodes a hypermultiplet transforming in the bifundamental representation.

Recall from \cite{Gaiotto:2008ak} that a 3d $\Ncal=4$ $G= \urm(n), \sorm(n), \sprm(n)$ gauge theory with $N_f$ fundamental flavours is called \emph{balanced} if:
\begin{align}
\label{eq:def_balance}
    \text{for} \;\; \urm(n): \; N_f = 2n  
    \,, \qquad 
    \text{for} \;\; \sorm(n): \; N_f = n-1  
    \,, \qquad
    \text{for} \;\; \sprm(n):  \; N_f =  n+1 \,.
\end{align}
If the number of flavour is larger, then the gauge nodes is \emph{good} and referred to as \emph{overbalanced}. If the number of flavours is smaller, the gauge node becomes \emph{ugly} or \emph{bad}.

\section{\texorpdfstring{Forking orthosymplectic quivers ($\boldsymbol{D}$-type)}{Forking orthosymplectic quivers (D-type)}}
\label{forkingit}
To begin the exploration, the class of framed orthosymplectic quivers is considered. Here and throughout the paper, \emph{framed} refers to quivers that exhibit explicit flavour groups; which might also be known as \emph{flavoured} quivers in the physics literature.

\subsection{Global symmetry}
In a \emph{forked quiver}\footnote{This term is first used in \cite{Gulotta:2012yd}.}, the set of gauge nodes are arranged in the shape of a $D$-type Dynkin diagram. Here, a forked orthosymplectic quiver is defined as a quiver that has a subset of \emph{balanced} gauge groups that form a fork.
Forked quivers are non-linear quivers in the sense that the gauge groups are not arranged in a single line. This is in contrast to the $ABC$-type orthosymplectic quivers, which are a linear chain of gauge nodes. The Coulomb branch global symmetry of forked orthosymplectic quivers has been conjectured in \cite[Sec.\ 7.4]{Gaiotto:2008ak}: For a balanced fork composed of $n$ gauge nodes (i.e.\ \emph{all} gauge nodes are balanced)
\begin{equation}
\raisebox{-.5\height}{
\begin{tikzpicture}
	\begin{pgfonlayer}{nodelayer}
		\node [style=miniU] (0) at (5.25, 0) {};
		\node [style=miniBlue] (1) at (4.25, 0) {};
		\node [style=miniU] (2) at (3.25, 0) {};
		\node [style=none] (3) at (2.25, 0) {$\dots$};
		\node [style=miniU] (5) at (1.25, 0) {};
		\node [style=miniBlue] (6) at (0.25, 0.75) {};
		\node [style=miniBlue] (7) at (0.25, -0.75) {};
		\node [style=none] (8) at (5.25, -0.75) {};
		\node [style=none] (9) at (1.25, -0.75) {};
		\node [style=none] (10) at (3.5, -1.25) {$n-2$};
		\node [style=none] (11) at (-3.5, 0) {$G_{\mathrm{global}}= \sorm(n)\times \sorm(n)$};
		\node [style=none] (12) at (2.75, 0) {};
		\node [style=none] (13) at (1.75, 0) {};
	\end{pgfonlayer}
	\begin{pgfonlayer}{edgelayer}
		\draw (0) to (2);
		\draw (5) to (6);
		\draw (5) to (7);
		\draw [style=brace] (8.center) to (9.center);
		\draw (12.center) to (2);
		\draw (13.center) to (5);
	\end{pgfonlayer}
\end{tikzpicture}
}
\label{soxso}
\end{equation}
the global symmetry is a product. This is verified through explicit Hilbert series computations for multiple examples, and the rule \eqref{soxso} remains the same regardless of the rank of the gauge groups as long as they are balanced. Even if the tail begins with $\sprm$ gauge group or if the two bifurcated nodes are $\sorm$ gauge groups, the global symmetry remains the same.

\paragraph{Exceptions.}
An exception to \eqref{soxso} arises if the tail of the fork begins with an $\sorm(2)$ gauge group\footnote{This is perhaps more evident from a brane set-up with O3 planes, where one should include an $\usprm(0)$ gauge node connected to the $\sorm(2)$ node. By including the $\usprm(0)$ node, the global symmetry is consistent with \eqref{soxso}. In Section \ref{branes}, the $\usprm(0)$ nodes is always omitted.}. Here, the global symmetry is enhanced to
\begin{equation}
\raisebox{-.5\height}{
\begin{tikzpicture}
	\begin{pgfonlayer}{nodelayer}
		\node [style=miniU] (0) at (5.25, 0) {};
		\node [style=miniBlue] (1) at (4.25, 0) {};
		\node [style=miniU] (2) at (3.25, 0) {};
		\node [style=none] (3) at (2.25, 0) {$\dots$};
		\node [style=miniU] (5) at (1.25, 0) {};
		\node [style=miniBlue] (6) at (0.25, 0.75) {};
		\node [style=miniBlue] (7) at (0.25, -0.75) {};
		\node [style=none] (8) at (5.25, -0.75) {};
		\node [style=none] (9) at (1.25, -0.75) {};
		\node [style=none] (10) at (3.5, -1.25) {$n-2$};
		\node [style=none] (11) at (-3.5, 0) {$G_{\mathrm{global}}= \sorm(n{+}1)\times \sorm(n{+}1)$};
		\node [style=none] (12) at (5.25, -0.5) {2};
		\node [style=none] (13) at (2.75, 0) {};
		\node [style=none] (14) at (1.75, 0) {};
	\end{pgfonlayer}
	\begin{pgfonlayer}{edgelayer}
		\draw (0) to (2);
		\draw (5) to (6);
		\draw (5) to (7);
		\draw [style=brace] (8.center) to (9.center);
		\draw (13.center) to (2);
		\draw (14.center) to (5);
	\end{pgfonlayer}
\end{tikzpicture}
}
\end{equation}
and another exception arises if this $\sorm(2)$ is ungauged
\begin{equation}
\raisebox{-.5\height}{
\begin{tikzpicture}
	\begin{pgfonlayer}{nodelayer}
		\node [style=miniBlue] (35) at (17, 4) {};
		\node [style=miniU] (36) at (16, 4) {};
		\node [style=none] (37) at (15, 4) {$\dots$};
		\node [style=miniU] (38) at (14, 4) {};
		\node [style=miniBlue] (39) at (13, 4.75) {};
		\node [style=miniBlue] (40) at (13, 3.25) {};
		\node [style=none] (41) at (17, 3.25) {};
		\node [style=none] (42) at (14, 3.25) {};
		\node [style=none] (43) at (15.5, 2.625) {$n-3$};
		\node [style=none] (44) at (8.25, 4) {$G_{\mathrm{global}}= \sorm(n{-}1)\times \sorm(n{-}1)\times \sorm(2)$};
		\node [style=none] (45) at (18, 3.5) {2};
		\node [style=flavourRed] (46) at (18, 4) {};
		\node [style=none] (47) at (17, 3.5) {2};
		\node [style=none] (48) at (15.5, 4) {};
		\node [style=none] (49) at (14.5, 4) {};
	\end{pgfonlayer}
	\begin{pgfonlayer}{edgelayer}
		\draw (38) to (39);
		\draw (38) to (40);
		\draw [style=brace] (41.center) to (42.center);
		\draw (46) to (35);
		\draw (35) to (36);
		\draw (36) to (48.center);
		\draw (49.center) to (38);
	\end{pgfonlayer}
\end{tikzpicture}
}
\end{equation}
where an additional $\sorm(2)$ factor arises in the global symmetry group. This case is further explored in Appendix \ref{u1guy}. For framed orthosymplectic quivers, these are the only exceptions to \eqref{soxso}.
%
On the other hand, for an \textit{unframed} orthosymplectic quiver there could be additional symmetry enhancement, as discussed in Section \ref{forkingunframed}. 

In order for the Coulomb branch of a quiver to be a product of Coulomb branches of two theories, the global symmetry must be the product of (at least) two non-Abelian groups. The global symmetry of a framed balanced D-type quiver is always the product of two identical non-Abelian groups, see for instance \cite{Gaiotto:2008ak}. The next subsection provides examples of such quivers. This allows to argue that quivers with such a feature are natural candidates for product theories. As is explored in Section \ref{forkingunframed}, gauging the flavours in the framed quivers of Section \ref{sec:framed_D-type} also leads to product theories, although the non-abelian groups in the product may not be identical.

\subsection{\texorpdfstring{Framed $D$-type orthosymplectic quivers}{Framed D-type orthosymplectic quivers}}
\label{sec:framed_D-type}
In this paper, forked orthosymplectic quivers are restricted to quivers whose fork is balanced and the long (balanced) tail begins with $\sorm{(2)}$. As a consequence, such a tail results in an increasing sequence of the form
\begin{align}
    \raisebox{-.5\height}{
\scalebox{0.9}{\begin{tikzpicture}
	\begin{pgfonlayer}{nodelayer}
		\node [style=miniBlue] (10) at (9.75, 0) {};
		\node [style=miniU] (12) at (10.75, 0) {};
		\node [style=miniBlue] (13) at (11.75, 0) {};
		\node [style=none] (14) at (13.125, 0) {$\dots$};
		\node [style=miniU] (15) at (14.5, 0) {};
		\node [style=miniBlue] (16) at (15.5, 0) {};
		\node [style=miniU] (17) at (16.5, 0) {};
		\node [style=none] (29) at (10.75, -0.5) {$2k$};
		\node [style=none] (30) at (11.75, -0.5) {$2k{-}2$};
		\node [style=none] (31) at (14.5, -0.5) {4};
		\node [style=none] (32) at (15.5, -0.5) {2};
		\node [style=none] (33) at (16.5, -0.5) {2};
		\node [style=none] (37) at (9.75, -0.5) {$2k$};
		\node [style=none] (54) at (8.875, 0) {};
		\node [style=none] (55) at (12.375, 0) {};
		\node [style=none] (56) at (13.875, 0) {};
		\node [style=none] (57) at (8.25, 0) {$\dots$};
	\end{pgfonlayer}
	\begin{pgfonlayer}{edgelayer}
		\draw (10) to (13);
		\draw (15) to (17);
		\draw (56.center) to (15);
		\draw (55.center) to (13);
		\draw (54.center) to (10);
	\end{pgfonlayer}
\end{tikzpicture}}
    } 
\end{align}
up until the first flavour node.  
Restricting to such a class of quivers allows to construct the following three parameter family:
\begin{equation}
\raisebox{-.5\height}{
\scalebox{0.9}{\begin{tikzpicture}
	\begin{pgfonlayer}{nodelayer}
		\node [style=miniBlue] (2) at (0.1, 1) {};
		\node [style=miniBlue] (3) at (0.1, -1) {};
		\node [style=miniU] (4) at (1.1, 0) {};
		\node [style=miniBlue] (5) at (5.975, 0) {};
		\node [style=none] (7) at (7.875, 0) {$\dots$};
		\node [style=miniU] (8) at (9.75, 0) {};
		\node [style=miniBlue] (9) at (8.75, 0) {};
		\node [style=miniBlue] (10) at (10.75, 0) {};
		\node [style=flavourRed] (11) at (10.75, 1) {};
		\node [style=miniU] (12) at (11.75, 0) {};
		\node [style=miniBlue] (13) at (12.75, 0) {};
		\node [style=none] (14) at (13.625, 0) {$\dots$};
		\node [style=miniU] (15) at (14.5, 0) {};
		\node [style=miniBlue] (16) at (15.5, 0) {};
		\node [style=miniU] (17) at (16.5, 0) {};
		\node [style=none] (20) at (0.1, -1.5) {$2k$};
		\node [style=none] (21) at (0.1, 1.5) {$2k$};
		\node [style=none] (22) at (1.1, -0.5) {$4k{+}2$};
		\node [style=none] (25) at (8.75, -0.5) {$4k{+}2l$};
		\node [style=none] (28) at (10.75, 1.5) {1};
		\node [style=none] (29) at (11.75, 0.5) {$4k{+}2l$};
		\node [style=none] (30) at (12.75, -0.5) {$4k{+}2l{-}2$};
		\node [style=none] (31) at (14.5, -0.5) {4};
		\node [style=none] (32) at (15.5, -0.5) {2};
		\node [style=none] (33) at (16.5, -0.5) {2};
		\node [style=none] (34) at (5.725, -0.75) {};
		\node [style=none] (35) at (10.25, -0.75) {};
		\node [style=none] (36) at (8.25, -1.25) {$2n{-}4k{-}4l{-}2$};
		\node [style=none] (37) at (10.75, -0.5) {$4k{+}2l$};
		\node [style=none] (38) at (9.75, 0.5) {$4k{+}2l{+}1$};
		\node [style=miniU] (39) at (6.975, 0) {};
		\node [style=none] (40) at (6.975, 0.5) {$4k{+}2l{+}1$};
		\node [style=none] (41) at (5.975, -0.5) {$4k{+}2l$};
		\node [style=flavourRed] (42) at (5.975, 1) {};
		\node [style=none] (43) at (5.975, 1.5) {1};
		\node [style=miniU] (44) at (4.975, 0) {};
		\node [style=miniBlue] (45) at (3.975, 0) {};
		\node [style=none] (46) at (4.975, 0.5) {$4k{+}2l$};
		\node [style=none] (47) at (3.975, -0.5) {$4k{+}2l{-}2$};
		\node [style=none] (48) at (3.1, 0) {$\dots$};
		\node [style=miniBlue] (49) at (2.1, 0) {};
		\node [style=none] (50) at (2.1, -0.5) {$4k{+}2$};
		\node [style=none] (51) at (2.6, 0) {};
		\node [style=none] (52) at (3.475, 0) {};
		\node [style=none] (53) at (7.375, 0) {};
		\node [style=none] (54) at (8.375, 0) {};
		\node [style=none] (55) at (13.125, 0) {};
		\node [style=none] (56) at (14.125, 0) {};
	\end{pgfonlayer}
	\begin{pgfonlayer}{edgelayer}
		\draw (2) to (4);
		\draw (4) to (3);
		\draw (9) to (8);
		\draw (8) to (10);
		\draw (10) to (11);
		\draw (10) to (13);
		\draw (15) to (17);
		\draw [style=brace] (35.center) to (34.center);
		\draw (42) to (5);
		\draw (4) to (49);
		\draw (45) to (39);
		\draw (49) to (51.center);
		\draw (52.center) to (45);
		\draw (39) to (53.center);
		\draw (54.center) to (9);
		\draw (56.center) to (15);
		\draw (55.center) to (13);
	\end{pgfonlayer}
\end{tikzpicture}}
}
\label{weirdcase}
\end{equation}
such that  $n\geq 2k+2l \geq 4$ holds. In fact, one finds
\begin{tcolorbox}
\begin{myrule}\label{rule:framed_D}
Quiver \eqref{weirdcase} constitutes all \emph{framed} orthosymplectic quivers which satisfy: 
\begin{compactitem}
    \item Tail begins with $\sorm(2)$ and end with a fork.  
    \item All nodes are balanced. 
\end{compactitem}
\end{myrule}
\end{tcolorbox}
\noindent
The long tail starting with $\sorm(2)$ seems to be characteristic of forked orthosymplectic quivers whose Coulomb branch is a product of two moduli spaces\footnote{If the tail does not begin with $\sorm(2)$, one can still construct a balanced fork with an $\mathrm{SO}\times \mathrm{SO}$ Coulomb branch global symmetry. However, so far, it is not clear how to construct any other forked orthosymplectic quivers that are framed and whose Coulomb branches are also products of moduli spaces.  In Section \ref{branes}, the relevant brane configurations demonstrate that it is quite a coincidence which leads to the fact that \eqref{weirdcase} is mirror dual to a product of two SQCD theories.}.

The significance of \eqref{weirdcase} comes from the following:
 Explicit Hilbert series computations, see Tables \ref{weirdcase1}, suggest that the Coulomb branch is of the form
\begin{equation}
    \mathcal{C}\eqref{weirdcase} = \overline{\mathcal{O}}^{\mathfrak{so}(2n)}_{(2^{2k},1^{2n-4k})}\times \overline{\mathcal{O}}^{\mathfrak{so}(2n)}_{(2^{2k+2l},1^{2n-4k-4l})} \,,
    \label{modulispaceprodeqn}
\end{equation}
where $\overline{\mathcal{O}}^{\mathfrak{so}(2n)}_{(\dots)}$ refers to the closure of the $\mathfrak{so}(2n)$ nilpotent orbit labelled by the partition $(\dots)$ \cite{Collingwood:1993fk}.  The Coulomb branch is the product of two \emph{non-identical} height 2 nilpotent orbits for $l\geq 1$. The highest weight generating function (HWG) are known to be \cite{Hanany:2016gbz}:
\begin{equation}
     \HWG\eqref{weirdcase} 
    =\PE\left[\sum_{i=1}^k \mu_{2i}t^{2i} \right]\cdot
    \PE\left[\sum_{i=1}^{k+l} m_{2i} t^{2i}\right]
    \label{firstHWG}
\end{equation}
 where $\mu_i, m_i$ are the highest weight fugacities for $\sorm(2n)\times \sorm(2n)$. For orthosymplectic quivers, it is not known how to compute the refined Coulomb branch Hilbert series via the monopole formula due to lack of topological symmetries in the quiver description. Nevertheless, the unrefined Hilbert series computed with the monopole formula is consistent with the unrefined HWG. Therefore, it is strictly speaking a conjecture that the refined Coulomb branch Hilbert series of \eqref{weirdcase} is given by \eqref{firstHWG}. The same conjecture is made whenever a HWG is given in this paper.

It is illuminating to consider special parameter regions of \eqref{weirdcase}.
\begin{itemize}
    \item  For $l=0$ and $n\geq 2k+2\geq 4$, the family reduces to a two parameter family:
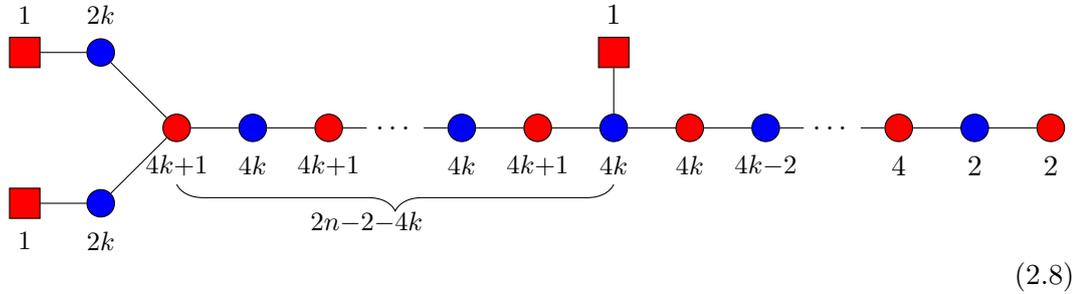
\begin{equation}
\raisebox{-.5\height}{
\begin{tikzpicture}
	\begin{pgfonlayer}{nodelayer}
		\node [style=flavourRed] (0) at (-5.25, 1) {};
		\node [style=flavourRed] (1) at (-5.25, -1) {};
		\node [style=miniBlue] (2) at (-4.25, 1) {};
		\node [style=miniBlue] (3) at (-4.25, -1) {};
		\node [style=miniU] (4) at (-3.25, 0) {};
		\node [style=miniBlue] (5) at (-2.25, 0) {};
		\node [style=miniU] (6) at (-1.25, 0) {};
		\node [style=none] (7) at (-0.375, 0) {$\dots$};
		\node [style=miniU] (8) at (1.5, 0) {};
		\node [style=miniBlue] (9) at (0.5, 0) {};
		\node [style=miniBlue] (10) at (2.5, 0) {};
		\node [style=flavourRed] (11) at (2.5, 1) {};
		\node [style=miniU] (12) at (3.5, 0) {};
		\node [style=miniBlue] (13) at (4.5, 0) {};
		\node [style=none] (14) at (5.375, 0) {$\dots$};
		\node [style=miniU] (15) at (6.25, 0) {};
		\node [style=miniBlue] (16) at (7.25, 0) {};
		\node [style=miniU] (17) at (8.25, 0) {};
		\node [style=none] (18) at (-5.25, 1.5) {\small{1}};
		\node [style=none] (19) at (-5.25, -1.5) {\small{1}};
		\node [style=none] (20) at (-4.25, -1.5) {\small{$2k$}};
		\node [style=none] (21) at (-4.25, 1.5) {\small{$2k$}};
		\node [style=none] (22) at (-3.25, -0.5) {\small{$4k{+}1$}};
		\node [style=none] (23) at (-2.25, -0.5) {\small{$4k$}};
		\node [style=none] (24) at (-1.25, -0.5) {\small{$4k{+}1$}};
		\node [style=none] (25) at (0.5, -0.5) {\small{$4k$}};
		\node [style=none] (26) at (1.5, -0.5) {\small{$4k{+}1$}};
		\node [style=none] (27) at (2.5, -0.5) {\small{$4k$}};
		\node [style=none] (28) at (2.5, 1.5) {1};
		\node [style=none] (29) at (3.5, -0.5) {\small{$4k$}};
		\node [style=none] (30) at (4.5, -0.5) {\small{$4k{-}2$}};
		\node [style=none] (31) at (6.25, -0.5) {4};
		\node [style=none] (32) at (7.25, -0.5) {2};
		\node [style=none] (33) at (8.25, -0.5) {2};
		\node [style=none] (34) at (-3.25, -0.75) {};
		\node [style=none] (35) at (2.5, -0.75) {};
		\node [style=none] (36) at (-0.75, -1.25) {\small{$2n{-}2{-}4k$}};
		\node [style=none] (37) at (-0.75, 0) {};
		\node [style=none] (38) at (0, 0) {};
		\node [style=none] (39) at (5.75, 0) {};
		\node [style=none] (40) at (5, 0) {};
	\end{pgfonlayer}
	\begin{pgfonlayer}{edgelayer}
		\draw (0) to (2);
		\draw (1) to (3);
		\draw (2) to (4);
		\draw (4) to (3);
		\draw (4) to (5);
		\draw (5) to (6);
		\draw (9) to (8);
		\draw (8) to (10);
		\draw (10) to (11);
		\draw (10) to (13);
		\draw (15) to (17);
		\draw [style=brace] (35.center) to (34.center);
		\draw (6) to (37.center);
		\draw (38.center) to (9);
		\draw (13) to (40.center);
		\draw (39.center) to (15);
	\end{pgfonlayer}
\end{tikzpicture}
}
\label{prodDn}
\end{equation}
and the Coulomb branch is the product of two identical moduli spaces.
\item For $l\geq 1$ and $n = 2k+2l+1$, the family reduced to another two parameter family:
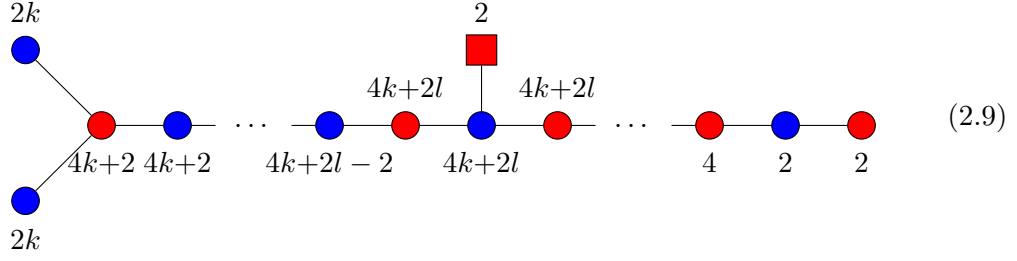
\begin{equation}
\raisebox{-.5\height}{
    \begin{tikzpicture}
	\begin{pgfonlayer}{nodelayer}
		\node [style=miniBlue] (2) at (0.5, 1) {};
		\node [style=miniBlue] (3) at (0.5, -1) {};
		\node [style=miniU] (4) at (1.5, 0) {};
		\node [style=miniBlue] (5) at (6.5, 0) {};
		\node [style=none] (14) at (8.5, 0) {$\dots$};
		\node [style=miniU] (15) at (9.5, 0) {};
		\node [style=miniBlue] (16) at (10.5, 0) {};
		\node [style=miniU] (17) at (11.5, 0) {};
		\node [style=none] (20) at (0.5, -1.5) {$2k$};
		\node [style=none] (21) at (0.5, 1.5) {$2k$};
		\node [style=none] (22) at (1.5, -0.5) {$4k{+}2$};
		\node [style=none] (31) at (9.5, -0.5) {4};
		\node [style=none] (32) at (10.5, -0.5) {2};
		\node [style=none] (33) at (11.5, -0.5) {2};
		\node [style=none] (34) at (6.25, -0.75) {};
		\node [style=none] (41) at (6.5, -0.5) {$4k{+}2l$};
		\node [style=none] (43) at (6.5, 1.5) {2};
		\node [style=miniU] (44) at (5.5, 0) {};
		\node [style=miniBlue] (45) at (4.5, 0) {};
		\node [style=none] (46) at (5.5, 0.5) {$4k{+}2l$};
		\node [style=none] (47) at (4.5, -0.5) {$4k{+}2l-{2}$};
		\node [style=none] (48) at (3.5, 0) {$\dots$};
		\node [style=miniBlue] (49) at (2.5, 0) {};
		\node [style=none] (50) at (2.5, -0.5) {$4k{+}2$};
		\node [style=none] (51) at (7.5, 0.5) {$4k{+}2l$};
		\node [style=miniU] (52) at (7.5, 0) {};
		\node [style=none] (54) at (8, 0) {};
		\node [style=none] (55) at (9, 0) {};
		\node [style=none] (56) at (3, 0) {};
		\node [style=none] (57) at (4, 0) {};
		\node [style=flavorRed] (58) at (6.5, 1) {};
	\end{pgfonlayer}
	\begin{pgfonlayer}{edgelayer}
		\draw (2) to (4);
		\draw (4) to (3);
		\draw (15) to (17);
		\draw (4) to (49);
		\draw (45) to (44);
		\draw (44) to (52);
		\draw (54.center) to (52);
		\draw (55.center) to (15);
		\draw (56.center) to (49);
		\draw (57.center) to (45);
		\draw (58) to (5);
	\end{pgfonlayer}
\end{tikzpicture}
}
\label{onelasttime}
\end{equation}
Furthermore, for $l=0$ and $n=2k+1$, a one parameter family arises
        \begin{equation}
\raisebox{-.5\height}
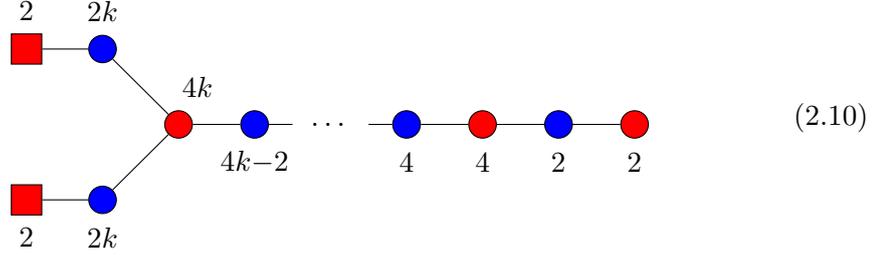
{
\begin{tikzpicture}
	\begin{pgfonlayer}{nodelayer}
		\node [style=flavourRed] (0) at (0.5, 1) {};
		\node [style=flavourRed] (1) at (0.5, -1) {};
		\node [style=miniBlue] (2) at (1.5, 1) {};
		\node [style=miniBlue] (3) at (1.5, -1) {};
		\node [style=miniU] (4) at (2.5, 0) {};
		\node [style=miniBlue] (5) at (3.5, 0) {};
		\node [style=none] (6) at (4.5, 0) {$\dots$};
		\node [style=miniBlue] (7) at (5.5, 0) {};
		\node [style=miniU] (8) at (6.5, 0) {};
		\node [style=miniBlue] (9) at (7.5, 0) {};
		\node [style=miniU] (10) at (8.5, 0) {};
		\node [style=none] (11) at (0.5, 1.5) {2};
		\node [style=none] (12) at (0.5, -1.5) {2};
		\node [style=none] (13) at (1.5, -1.5) {$2k$};
		\node [style=none] (14) at (1.5, 1.5) {$2k$};
		\node [style=none] (15) at (2.75, 0.5) {$4k$};
		\node [style=none] (16) at (3.5, -0.5) {$4k{-}2$};
		\node [style=none] (17) at (5.5, -0.5) {4};
		\node [style=none] (18) at (6.5, -0.5) {4};
		\node [style=none] (19) at (7.5, -0.5) {2};
		\node [style=none] (20) at (8.5, -0.5) {2};
		\node [style=none] (21) at (4, 0) {};
		\node [style=none] (22) at (5, 0) {};
	\end{pgfonlayer}
	\begin{pgfonlayer}{edgelayer}
		\draw (0) to (2);
		\draw (1) to (3);
		\draw (2) to (4);
		\draw (4) to (3);
		\draw (4) to (5);
		\draw (8) to (10);
		\draw (8) to (7);
		\draw (22.center) to (7);
		\draw (21.center) to (5);
	\end{pgfonlayer}
\end{tikzpicture}
}
\label{firstfork}
\end{equation}
which was studied in \cite{Gaiotto:2008ak}.   
    \item  For $l\geq 1$ and $n=2k+2l$, the larger of the two moduli spaces in the product $\overline{\mathcal{O}}^{\mathfrak{so}(4k+4l)}_{(2^{2k+2l})}$ is the union of two identical cones $\overline{\mathcal{O}}^{\mathfrak{so}(4k+4l),I}_{(2^{2k+2l})}$ and $ \overline{\mathcal{O}}^{\mathfrak{so}(4k+4l),II}_{(2^{2k+2l})}$ \cite{Kraft1982}.
The product only includes one of the two cones. The quiver then takes the following form:
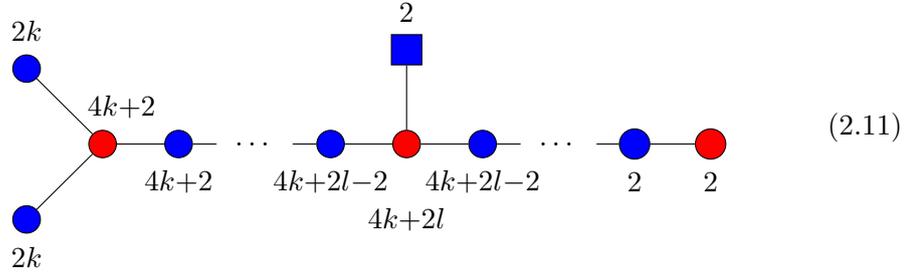
\begin{equation}
\raisebox{-.5\height}{
    \begin{tikzpicture}
	\begin{pgfonlayer}{nodelayer}
		\node [style=miniBlue] (2) at (1.5, 1) {};
		\node [style=miniBlue] (3) at (1.5, -1) {};
		\node [style=miniU] (4) at (2.5, 0) {};
		\node [style=miniBlue] (5) at (3.5, 0) {};
		\node [style=none] (6) at (4.5, 0) {$\dots$};
		\node [style=miniBlue] (7) at (5.5, 0) {};
		\node [style=miniU] (8) at (6.5, 0) {};
		\node [style=miniBlue] (9) at (7.5, 0) {};
		\node [style=none] (13) at (1.5, -1.5) {$2k$};
		\node [style=none] (14) at (1.5, 1.5) {$2k$};
		\node [style=none] (15) at (2.75, 0.5) {$4k{+}2$};
		\node [style=none] (16) at (3.5, -0.5) {$4k{+}2$};
		\node [style=none] (17) at (5.5, -0.5) {$4k{+}2l{-}2$};
		\node [style=none] (21) at (4, 0) {};
		\node [style=none] (22) at (5, 0) {};
		\node [style=none] (23) at (6.5, -1) {$4k{+}2l$};
		\node [style=none] (24) at (7.5, -0.5) {$4k{+}2l{-}2$};
		\node [style=flavourBlue] (25) at (6.5, 1.25) {};
		\node [style=none] (26) at (6.5, 1.75) {2};
		\node [style=none] (27) at (8, 0) {};
		\node [style=none] (28) at (8.5, 0) {$\dots$};
		\node [style=none] (29) at (9, 0) {};
		\node [style=bluegauge] (30) at (9.5, 0) {};
		\node [style=redgauge] (31) at (10.5, 0) {};
		\node [style=none] (32) at (10.5, -0.5) {2};
		\node [style=none] (33) at (9.5, -0.5) {2};
	\end{pgfonlayer}
	\begin{pgfonlayer}{edgelayer}
		\draw (2) to (4);
		\draw (4) to (3);
		\draw (4) to (5);
		\draw (8) to (7);
		\draw (5) to (21.center);
		\draw (22.center) to (7);
		\draw (25) to (8);
		\draw (8) to (9);
		\draw (31) to (30);
		\draw (27.center) to (9);
		\draw (29.center) to (30);
	\end{pgfonlayer}
\end{tikzpicture}
}
\label{secondlasttime}
\end{equation}
Taking further $l=0$, the bifurcated nodes have different ranks:
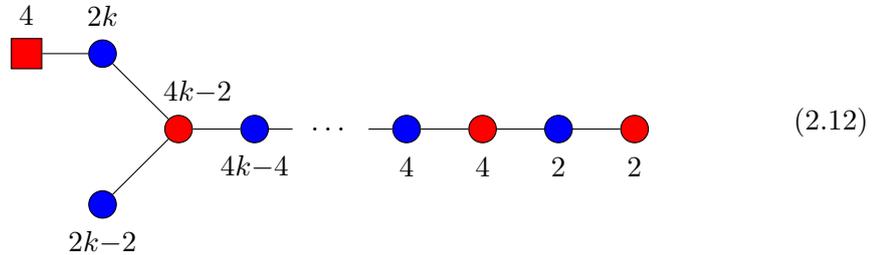
\begin{equation}
\raisebox{-.5\height}{
\begin{tikzpicture}
	\begin{pgfonlayer}{nodelayer}
		\node [style=flavourRed] (0) at (0.5, 1) {};
		\node [style=miniBlue] (2) at (1.5, 1) {};
		\node [style=miniBlue] (3) at (1.5, -1) {};
		\node [style=miniU] (4) at (2.5, 0) {};
		\node [style=miniBlue] (5) at (3.5, 0) {};
		\node [style=none] (6) at (4.5, 0) {$\dots$};
		\node [style=miniBlue] (7) at (5.5, 0) {};
		\node [style=miniU] (8) at (6.5, 0) {};
		\node [style=miniBlue] (9) at (7.5, 0) {};
		\node [style=miniU] (10) at (8.5, 0) {};
		\node [style=none] (11) at (0.5, 1.5) {4};
		\node [style=none] (13) at (1.5, -1.5) {$2k{-}2$};
		\node [style=none] (14) at (1.5, 1.5) {$2k$};
		\node [style=none] (15) at (2.75, 0.5) {$4k{-}2$};
		\node [style=none] (16) at (3.5, -0.5) {$4k{-}4$};
		\node [style=none] (17) at (5.5, -0.5) {4};
		\node [style=none] (18) at (6.5, -0.5) {4};
		\node [style=none] (19) at (7.5, -0.5) {2};
		\node [style=none] (20) at (8.5, -0.5) {2};
		\node [style=none] (21) at (4, 0) {};
		\node [style=none] (22) at (5, 0) {};
	\end{pgfonlayer}
	\begin{pgfonlayer}{edgelayer}
		\draw (0) to (2);
		\draw (2) to (4);
		\draw (4) to (3);
		\draw (4) to (5);
		\draw (8) to (10);
		\draw (8) to (7);
		\draw (5) to (21.center);
		\draw (22.center) to (7);
	\end{pgfonlayer}
\end{tikzpicture}
}
\label{finalfinitequiver}
\end{equation}
\end{itemize}

By virtue of 3d mirror symmetry, the Coulomb branches of \eqref{weirdcase} are given by the Higgs branches of the following SQCD quivers:
\begin{equation}
\raisebox{-.5\height}{
 \begin{tikzpicture}
	\begin{pgfonlayer}{nodelayer}
		\node [style=miniBlue] (0) at (0, 0) {};
		\node [style=flavourRed] (1) at (0, 1.5) {};
		\node [style=none] (2) at (0, -0.5) {$\usprm(2k)$};
		\node [style=none] (3) at (0, 2) {$\sorm(2n)$};
		\node [style=none] (4) at (1, 0.75) {$\times$};
		\node [style=miniBlue] (5) at (2, 0) {};
		\node [style=flavourRed] (6) at (2, 1.5) {};
		\node [style=none] (7) at (2.05, -0.5) {$\usprm(2k{+}2l)$};
		\node [style=none] (8) at (2, 2) {$\sorm(2n)$};
	\end{pgfonlayer}
	\begin{pgfonlayer}{edgelayer}
		\draw (1) to (0);
		\draw (6) to (5);
	\end{pgfonlayer}
\end{tikzpicture}
}
\label{productfiniteX}
\end{equation}
which is discussed in Section \ref{branes} by using brane systems. This underpins the claim \eqref{modulispaceprodeqn}.
\subsection{\texorpdfstring{Unframed $D$-type orthosymplectic quivers}{Unframed D-type orthosymplectic quivers}}
\label{forkingunframed}
So far, framed orthosymplectic quivers, whose Coulomb branches are products of nilpotent orbit closures of $\mathfrak{so}(2n)$ algebra, have been considered. In this section, the analysis is extended to unframed\footnote{In this section, the term ``unframed'' is reserved for quivers without flavour groups connected with charge 1 hypermultiplets.} orthosymplectic quivers with a balanced fork.

The gauge group $G$ of the quiver is a product of various $\sorm(2k)$, $\usprm(2m)$ and $\urm(1)$ gauge groups. It has been shown in \cite{Bourget:2020xdz} that  such a combination of gauge groups, in the absence of flavour groups connected with charge 1 hypermultiplets, contains a subgroup $H=\mathbb{Z}_2\subset G$ that acts trivially on the matter content of the quiver gauge theory. This group $H$ is chosen to be ungauged for all unframed quiver in this section. By ungauging $H$, the computation of the Coulomb branch Hilbert series requires summing over monopole operators with both integer and integer-plus-half magnetic charges in the monopole formula.  The integer-plus-half magnetic charges may enhance the global symmetry group that is predicted by \eqref{soxso}. As a result, many of the Coulomb branches here are products of nilpotent orbit closures of exceptional algebras. 

\begin{table}[t]
    \centering
    \scalebox{0.86}{
    \begin{tabular}{cc} \toprule
    Orthosymplectic quiver & Coulomb branch \\ \midrule 
      
         \scalebox{0.8}{ \begin{tikzpicture}
	\begin{pgfonlayer}{nodelayer}
		\node [style=redgauge] (0) at (-0.25, -1.25) {};
		\node [style=bluegauge] (1) at (0.75, -1.25) {};
		\node [style=redgauge] (2) at (2, 0) {};
		\node [style=bluegauge] (3) at (3, 0) {};
		\node [style=redgauge] (4) at (4, 0) {};
		\node [style=bluegauge] (5) at (5, 0) {};
		\node [style=redgauge] (6) at (6, 0) {};
		\node [style=bluegauge] (7) at (7, 0) {};
		\node [style=redgauge] (8) at (8, 0) {};
		\node [style=bluegauge] (9) at (9, 0) {};
		\node [style=redgauge] (10) at (10, 0) {};
		\node [style=none] (11) at (2, -0.5) {14};
		\node [style=none] (12) at (3, -0.5) {12};
		\node [style=none] (13) at (4, -0.5) {12};
		\node [style=none] (14) at (5, -0.5) {10};
		\node [style=none] (15) at (6, -0.5) {10};
		\node [style=none] (16) at (7, -0.5) {8};
		\node [style=none] (17) at (8, -0.5) {8};
		\node [style=none] (18) at (9, -0.5) {6};
		\node [style=none] (19) at (10, -0.5) {6};
		\node [style=none] (20) at (11, -0.5) {4};
		\node [style=none] (21) at (12, -0.5) {4};
		\node [style=none] (22) at (13, -0.5) {2};
		\node [style=none] (23) at (14, -0.5) {2};
		\node [style=bluegauge] (25) at (11, 0) {};
		\node [style=redgauge] (26) at (12, 0) {};
		\node [style=bluegauge] (27) at (13, 0) {};
		\node [style=redgauge] (28) at (14, 0) {};
		\node [style=bluegauge] (30) at (0.75, 1.25) {};
		\node [style=none] (31) at (0.75, 1.75) {6};
		\node [style=none] (32) at (0.75, -1.75) {8};
		\node [style=none] (33) at (-0.25, -1.75) {\color{red}{4}};
	\end{pgfonlayer}
	\begin{pgfonlayer}{edgelayer}
		\draw (30) to (2);
		\draw (0) to (1);
		\draw (1) to (2);
		\draw (7) to (2);
		\draw (28) to (7);
	\end{pgfonlayer}
\end{tikzpicture}
} &\raisebox{1.3cm}{\Large{ $\overline{\mathcal{O}}^{\mathfrak{e}_8}_{\text{min}}\times \overline{\mathcal{O}}^{\mathfrak{e}_8}_{\text{min}}$ }}   \\
\scalebox{0.8}{\begin{tikzpicture}
	\begin{pgfonlayer}{nodelayer}
		\node [style=redgauge] (0) at (3.5, -1.25) {};
		\node [style=bluegauge] (1) at (4.5, -1.25) {};
		\node [style=redgauge] (6) at (6, 0) {};
		\node [style=bluegauge] (7) at (7, 0) {};
		\node [style=redgauge] (8) at (8, 0) {};
		\node [style=bluegauge] (9) at (9, 0) {};
		\node [style=redgauge] (10) at (10, 0) {};
		\node [style=none] (15) at (6, -0.5) {10};
		\node [style=none] (16) at (7, -0.5) {8};
		\node [style=none] (17) at (8, -0.5) {8};
		\node [style=none] (18) at (9, -0.5) {6};
		\node [style=none] (19) at (10, -0.5) {6};
		\node [style=none] (20) at (11, -0.5) {4};
		\node [style=none] (21) at (12, -0.5) {4};
		\node [style=none] (22) at (13, -0.5) {2};
		\node [style=none] (23) at (14, -0.5) {2};
		\node [style=bluegauge] (25) at (11, 0) {};
		\node [style=redgauge] (26) at (12, 0) {};
		\node [style=bluegauge] (27) at (13, 0) {};
		\node [style=redgauge] (28) at (14, 0) {};
		\node [style=bluegauge] (30) at (4.5, 1.25) {};
		\node [style=none] (31) at (4.5, 1.75) {4};
		\node [style=none] (32) at (4.5, -1.75) {6};
		\node [style=none] (33) at (3.5, -1.75) {\color{red}{4}};
		\node [style=bluegauge] (34) at (2.5, -1.25) {};
		\node [style=redgauge] (35) at (1.5, -1.25) {};
		\node [style=none] (36) at (2.5, -1.75) {2};
		\node [style=none] (37) at (1.5, -1.75) {2};
	\end{pgfonlayer}
	\begin{pgfonlayer}{edgelayer}
		\draw (0) to (1);
		\draw (30) to (6);
		\draw (6) to (1);
		\draw (7) to (6);
		\draw (0) to (35);
		\draw (6) to (28);
	\end{pgfonlayer}
\end{tikzpicture}
}
&\raisebox{1.3cm}{\Large{ $\overline{\mathcal{O}}^{\mathfrak{e}_7}_{\text{min}}\times \overline{\mathcal{O}}^{\mathfrak{e}_7}_{\text{min}}$}}     \\
\scalebox{0.8}{\begin{tikzpicture}
	\begin{pgfonlayer}{nodelayer}
		\node [style=redgauge] (0) at (5.5, -1.25) {};
		\node [style=bluegauge] (1) at (6.5, -1.25) {};
		\node [style=redgauge] (8) at (8, 0) {};
		\node [style=bluegauge] (9) at (9, 0) {};
		\node [style=redgauge] (10) at (10, 0) {};
		\node [style=none] (17) at (8, -0.5) {8};
		\node [style=none] (18) at (9, -0.5) {6};
		\node [style=none] (19) at (10, -0.5) {6};
		\node [style=none] (20) at (11, -0.5) {4};
		\node [style=none] (21) at (12, -0.5) {4};
		\node [style=none] (22) at (13, -0.5) {2};
		\node [style=none] (23) at (14, -0.5) {2};
		\node [style=bluegauge] (25) at (11, 0) {};
		\node [style=redgauge] (26) at (12, 0) {};
		\node [style=bluegauge] (27) at (13, 0) {};
		\node [style=redgauge] (28) at (14, 0) {};
		\node [style=bluegauge] (30) at (6.5, 1.25) {};
		\node [style=none] (31) at (6.5, 1.75) {4};
		\node [style=none] (32) at (6.5, -1.75) {4};
		\node [style=none] (33) at (5.5, -1.75) {\color{red}{1}};
		\node [style=redgauge] (34) at (5.5, 1.25) {};
		\node [style=none] (35) at (5.5, 1.75) {\color{red}{1}};
		\node [style=gauge3] (36) at (5.5, 1.25) {};
		\node [style=gauge3] (37) at (5.5, -1.25) {};
	\end{pgfonlayer}
	\begin{pgfonlayer}{edgelayer}
		\draw (0) to (1);
		\draw (8) to (1);
		\draw (8) to (30);
		\draw (30) to (34);
		\draw (8) to (28);
	\end{pgfonlayer}
\end{tikzpicture}} &\raisebox{1.3cm}{\Large{  $\overline{\mathcal{O}}^{\mathfrak{e}_6}_{\text{min}}\times \overline{\mathcal{O}}^{\mathfrak{e}_6}_{\text{min}}$}  }
\\
\scalebox{0.8}{\begin{tikzpicture}
	\begin{pgfonlayer}{nodelayer}
		\node [style=redgauge] (0) at (7.5, -0.5) {};
		\node [style=bluegauge] (1) at (8.5, -1.25) {};
		\node [style=redgauge] (10) at (10, 0) {};
		\node [style=none] (19) at (10, -0.5) {6};
		\node [style=none] (20) at (11, -0.5) {4};
		\node [style=none] (21) at (12, -0.5) {4};
		\node [style=none] (22) at (13, -0.5) {2};
		\node [style=none] (23) at (14, -0.5) {2};
		\node [style=bluegauge] (25) at (11, 0) {};
		\node [style=redgauge] (26) at (12, 0) {};
		\node [style=bluegauge] (27) at (13, 0) {};
		\node [style=redgauge] (28) at (14, 0) {};
		\node [style=bluegauge] (30) at (8.5, 1.25) {};
		\node [style=none] (31) at (8.5, 1.75) {2};
		\node [style=none] (32) at (8.5, -1.75) {4};
		\node [style=none] (33) at (7.5, 0) {\color{red}{1}};
		\node [style=redgauge] (34) at (7.5, -2) {};
		\node [style=none] (35) at (7.5, -2.5) {\color{red}{1}};
		\node [style=gauge3] (36) at (7.5, -0.5) {};
		\node [style=gauge3] (37) at (7.5, -2) {};
	\end{pgfonlayer}
	\begin{pgfonlayer}{edgelayer}
		\draw (0) to (1);
		\draw (30) to (10);
		\draw (10) to (1);
		\draw (1) to (34);
		\draw (10) to (28);
	\end{pgfonlayer}
\end{tikzpicture}
} &\raisebox{1.8cm}{ \Large{ $ \overline{\mathcal{O}}^{\mathfrak{so}(10)}_{\text{min}}\times  \overline{\mathcal{O}}^{\mathfrak{so}(10)}_{\text{min}}$} } \\
 \scalebox{0.8}{
 \raisebox{-.5\height}{
\begin{tikzpicture}
	\begin{pgfonlayer}{nodelayer}
		\node [style=bluegauge] (1) at (10.5, -1.25) {};
		\node [style=none] (21) at (12, -0.5) {4};
		\node [style=none] (23) at (13, -0.5) {2};
		\node [style=none] (24) at (14, -0.5) {2};
		\node [style=redgauge] (26) at (12, 0) {};
		\node [style=redgauge] (28) at (13, 0) {};
		\node [style=bluegauge] (29) at (14, 0) {};
		\node [style=bluegauge] (30) at (10.5, 1.25) {};
		\node [style=none] (31) at (10.5, 1.75) {2};
		\node [style=none] (32) at (10.5, -1.75) {2};
		\node [style=gauge3] (33) at (9.5, 1.25) {};
		\node [style=gauge3] (34) at (9.5, -1.25) {};
		\node [style=none] (35) at (9.5, 1.75) {\color{red}{1}};
		\node [style=none] (36) at (9.5, -1.75) {\color{red}{1}};
		\node [style=none] (37) at (8.75, 0) {};
		\node [style=bluegauge] (38) at (13, 0) {};
		\node [style=redgauge] (39) at (14, 0) {};
	\end{pgfonlayer}
	\begin{pgfonlayer}{edgelayer}
		\draw (33) to (30);
		\draw (30) to (26);
		\draw (26) to (1);
		\draw (1) to (34);
		\draw (33) to (34);
		\draw [style=new edge style 1, bend right, looseness=0.75] (33) to (37.center);
		\draw [style=new edge style 1, bend right=45, looseness=0.75] (37.center) to (34);
		\draw (29) to (26);
	\end{pgfonlayer}
\end{tikzpicture}
}}  &\Large{ $\overline{\mathcal{O}}^{\mathfrak{su}(5)}_{\text{min}}\times \overline{\mathcal{O}}^{\mathfrak{su}(5)}_{\text{min}}$} \\
\bottomrule
    \end{tabular}}
    \caption{The Coulomb branches of orthosymplectic quivers are products of two copies of the minimal nilpotent orbits closures of exceptional algebras  $\mathfrak{e}_{n}$ for $n=4,\ldots,8$. The numbers coloured in red represent gauge nodes that are overbalanced.}
    \label{EnProdtable}
\end{table}

\subsubsection{\texorpdfstring{$E_{n+1}\times E_{n+1}$ family}{E n+1 x E n+1 family}}
\label{sec:En_EN_5d}
The first set of unframed orthosymplectic quivers is related to 5d $\Ncal=1$ theories \cite{Bourget:2020gzi,Akhond:2020vhc}. To be exact, the Coulomb branches of the orthosymplectic quivers are the same as the Higgs branch of the product of two identical rank 1 5d $\Ncal=1$ SCFTs.  These are the UV completions of $\surm(2)$ gauge theories with $n\leq 7$ fundamental flavours which exhibit $E_{n+1}$ Higgs branch global symmetries \cite{Seiberg:1996bd,Morrison:1996xf}. 

\paragraph{Product of rank 1 theories.}
The $E_{n+1}\times E_{n+1}$  orthosymplectic quivers whose Coulomb branches are the product of two minimal nilpotent orbit closures of $\mathfrak{e}_{n+1}$ are displayed in Table \ref{EnProdtable}. For these quivers, integer-plus-half contributions enhances the Coulomb branch global symmetry. When this enhancement happens, \eqref{soxso} no longer predicts the correct global symmetry and explicit Hilbert series computations are required. 
For $3\leq n \leq 7$, the Coulomb branch Hilbert series are computed and, upon taking the square root, are compared with the known Hilbert series of $\overline{\mathcal{O}}^{\mathfrak{e}_{n+1}}_{\mathrm{min}}$. Details are provided in Table \ref{EnProdtableT}.

For the $n=5,6,7$ cases of Table \ref{EnProdtable}, the orthosymplectic quivers can be understood as class $\mathcal{S}$ theories with untwisted $D_4$, $D_5$, and $D_7$ punctures, respectively. For $n=5$, the $E_6\times E_6$  orthosymplectic quiver already appeared in \cite{Chacaltana:2011ze}. For $n=6$, the orthosymplectic quiver is derived in \cite{Akhond:2021knl} using brane webs and O5 planes. It has been noted before that certain class $\mathcal{S}$ theories can be factorised into the product of two theories  
\cite{Distler:2017xba,Distler:2018gbc,Ergun:2020fnm}.

\paragraph{General product families.}
In \cite{Bourget:2020xdz}, the $E_{n+1}$ orthosymplectic quivers have been extended to infinite families for each $n$.  The orthosymplectic quivers in Table \ref{EnProdtable} can be extended in a similar fashion to the infinite families shown in Table \ref{EnProdtable2}. 
These magnetic quivers describe the Higgs branches of two copies of 5d $\Ncal=1$ $\sprm(k)$ SQCD theories at the UV fix point, see Section \ref{sec:branes_5d}.
For $n<7$ and small $k$, Coulomb branch Hilbert series have 
\newpage
\begin{table}[t]
    \centering
    \scalebox{0.86}{
    \hspace{-1.25cm}
\begin{tabular}{ccc} \toprule
        Family &  Orthosymplectic quiver & Products \\
\bottomrule
    \end{tabular}}
    \caption{The extended infinite families of the orthosymplectic quivers in Table \ref{EnProdtable}. The Coulomb branch of the forked orthosymplectic quivers on the left are the same as the Coulomb branch of product theories on the right. The numbers coloured in red represent gauge nodes that are overbalanced. }
    \label{EnProdtable2}
\end{table}
\clearpage
\noindent
been computed, see Table \ref{EnProdtableT}, and compared against the results (after taking their products) in \cite{Bourget:2020xdz}. Also, the Coulomb branch dimension, computed directly from the quiver, is compared and agreement is found for all the infinite sequences.

One observes that the quivers in Table \ref{EnProdtable2} can be obtained through gauging flavour symmetries of the framed quivers in Section \ref{forkingit}. 
This can occur only when the rank of the flavour nodes is at least one. For instance, one may gauge the flavour symmetries in the cases \eqref{firstfork} and \eqref{finalfinitequiver}. Gauging the two $\sorm(2)$ flavour symmetries in \eqref{firstfork} reproduces the $E_6\times E_6 $ family. The $\sorm(4)$ flavour symmetry in \eqref{finalfinitequiver} can be gauged to a $\sorm(4)$ gauge symmetry to give the $E_8\times E_8$ family or gauged to a $\urm(1)\times \urm(1)$ symmetry to give the $E_5 \times E_5$ family. Connecting the $\sorm(4)$ gauge group in the $E_8\times E_8$ family with addition $\sorm(2)-\usprm(2)-$ gauge groups reproduces the $E_7\times E_7$ family. Table \ref{EnProdtable2} provides the $E_{4-2l}\times E_{4-2l}$ and $E_{3-2l} \times E_{3-2l}$ families with $l \geq 0$. These families take the same form as the $E_6\times E_6$ and $E_5 \times E_5$ families, respectively, but with additional hypermultiplets between the $\urm(1)$ gauge groups. The dashed line here represents fundamental-fundamental hypermultiplets which have been introduced in \cite{Akhond:2020vhc}. This shows that, up to additional hypermultiplets between $\urm(1)$s, the $E_n\times E_n$ families are obtained through gauging flavour symmetries of the framed orthosymplectic quivers classified in Section \ref{forkingit}. The appearing 5d $\Ncal=1$ SQCD theories exhaust \emph{all} $\sprm(m)$ gauge theories with the allowed range $0 \leq N_f \leq 2m+5$ of fundamental flavours, with $m\geq 1$. 

\subsubsection{\texorpdfstring{$E_6\times \sorm(10)$ family}{E6 x SO(10) family}}
For the case \eqref{weirdcase} with $n=2k+2l+1$ and $l\geq 1$, the two flavour nodes collide to form an $\sorm(2)$ flavour in \eqref{onelasttime}. This flavour node can then be gauged to give a family of unframed forked orthosymplectic quivers:
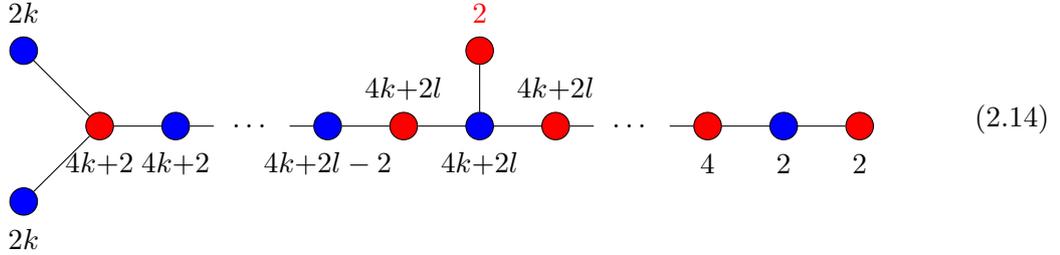
\begin{equation}
\raisebox{-.5\height}{
\begin{tikzpicture}
	\begin{pgfonlayer}{nodelayer}
		\node [style=miniBlue] (2) at (0.5, 1) {};
		\node [style=miniBlue] (3) at (0.5, -1) {};
		\node [style=miniU] (4) at (1.5, 0) {};
		\node [style=miniBlue] (5) at (6.5, 0) {};
		\node [style=none] (14) at (8.5, 0) {$\dots$};
		\node [style=miniU] (15) at (9.5, 0) {};
		\node [style=miniBlue] (16) at (10.5, 0) {};
		\node [style=miniU] (17) at (11.5, 0) {};
		\node [style=none] (20) at (0.5, -1.5) {$2k$};
		\node [style=none] (21) at (0.5, 1.5) {$2k$};
		\node [style=none] (22) at (1.5, -0.5) {$4k{+}2$};
		\node [style=none] (31) at (9.5, -0.5) {4};
		\node [style=none] (32) at (10.5, -0.5) {2};
		\node [style=none] (33) at (11.5, -0.5) {2};
		\node [style=none] (34) at (6.25, -0.75) {};
		\node [style=none] (41) at (6.5, -0.5) {$4k{+}2l$};
		\node [style=none] (43) at (6.5, 1.5) {$\textcolor{red}{2}$};
		\node [style=miniU] (44) at (5.5, 0) {};
		\node [style=miniBlue] (45) at (4.5, 0) {};
		\node [style=none] (46) at (5.5, 0.5) {$4k{+}2l$};
		\node [style=none] (47) at (4.5, -0.5) {$4k{+}2l-{2}$};
		\node [style=none] (48) at (3.5, 0) {$\dots$};
		\node [style=miniBlue] (49) at (2.5, 0) {};
		\node [style=none] (50) at (2.5, -0.5) {$4k{+}2$};
		\node [style=none] (51) at (7.5, 0.5) {$4k{+}2l$};
		\node [style=miniU] (52) at (7.5, 0) {};
		\node [style=miniU] (53) at (6.5, 1) {};
		\node [style=none] (54) at (8, 0) {};
		\node [style=none] (55) at (9, 0) {};
		\node [style=none] (56) at (3, 0) {};
		\node [style=none] (57) at (4, 0) {};
	\end{pgfonlayer}
	\begin{pgfonlayer}{edgelayer}
		\draw (2) to (4);
		\draw (4) to (3);
		\draw (15) to (17);
		\draw (4) to (49);
		\draw (45) to (44);
		\draw (44) to (52);
		\draw (53) to (5);
		\draw (54.center) to (52);
		\draw (55.center) to (15);
		\draw (56.center) to (49);
		\draw (57.center) to (45);
	\end{pgfonlayer}
\end{tikzpicture}
}
\label{weirder}
\end{equation}
where the red label indicates the node is overbalanced. Explicit Hilbert series computations, see Table \ref{weirder1}, indicate that the Coulomb branch is compatible with the product of the following two Coulomb branch moduli spaces:
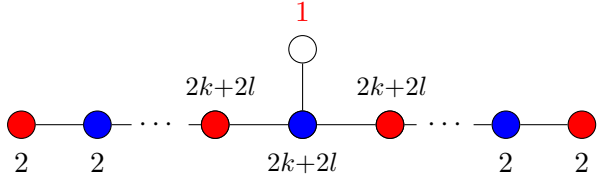
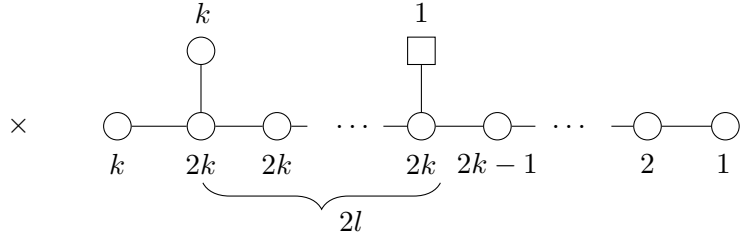
\begin{equation}
\raisebox{-.5\height}{
\begin{tikzpicture}
	\begin{pgfonlayer}{nodelayer}
		\node [style=miniU] (34) at (13.75, 0) {};
		\node [style=miniBlue] (35) at (14.75, 0) {};
		\node [style=miniBlue] (36) at (20.125, 0) {};
		\node [style=miniU] (37) at (21.125, 0) {};
		\node [style=none] (38) at (13.75, -0.5) {2};
		\node [style=none] (39) at (14.75, -0.5) {2};
		\node [style=none] (40) at (17.45, -0.5) {\small$2k{+}2l$};
		\node [style=none] (41) at (20.125, -0.5) {2};
		\node [style=none] (42) at (21.125, -0.5) {2};
		\node [style=miniU] (43) at (17.45, 0) {};
		\node [style=miniBlue] (44) at (16.3, 0) {};
		\node [style=miniBlue] (45) at (18.6, 0) {};
		\node [style=none] (46) at (15.55, 0) {$\cdots$};
		\node [style=none] (47) at (19.35, 0) {$\cdots$};
		\node [style=none] (48) at (15.2, 0) {};
		\node [style=none] (49) at (15.95, 0) {};
		\node [style=none] (50) at (18.95, 0) {};
		\node [style=none] (51) at (19.7, 0) {};
		\node [style=gauge3] (52) at (17.45, 1) {};
		\node [style=none] (53) at (17.45, 1.5) {\textcolor{red}{1}};
		\node [style=miniBlue] (54) at (17.45, 0) {};
		\node [style=miniU] (55) at (18.6, 0) {};
		\node [style=miniU] (56) at (16.3, 0) {};
		\node [style=none] (57) at (11.95, -4.25) {$\times$};
		\node [style=none] (58) at (18.625, 0.5) {\small$2k{+}2l$};
		\node [style=none] (59) at (16.375, 0.5) {\small$2k{+}2l$};
		\node [style=gauge3] (60) at (13.25, -4.25) {};
		\node [style=gauge3] (61) at (14.35, -4.25) {};
		\node [style=gauge3] (62) at (15.35, -4.25) {};
		\node [style=none] (63) at (14.35, -4.75) {$2k$};
		\node [style=none] (64) at (15.35, -4.75) {$2k$};
		\node [style=none] (65) at (19.225, -4.25) {$\dots$};
		\node [style=gauge3] (66) at (20.225, -4.25) {};
		\node [style=none] (67) at (21.225, -4.75) {$1$};
		\node [style=none] (68) at (20.225, -4.75) {$2$};
		\node [style=none] (69) at (13.25, -4.75) {$k$};
		\node [style=none] (75) at (14, -4.625) {};
		\node [style=gauge3] (77) at (21.25, -4.25) {};
		\node [style=gauge3] (78) at (17.25, -4.25) {};
		\node [style=gauge3] (79) at (18.25, -4.25) {};
		\node [style=none] (80) at (16.375, -4.25) {$\dots$};
		\node [style=none] (81) at (17.25, -4.75) {$2k$};
		\node [style=none] (82) at (18.25, -4.75) {$2k-1$};
		\node [style=blankflavor] (83) at (17.25, -3.25) {};
		\node [style=none] (84) at (17.25, -2.75) {1};
		\node [style=none] (85) at (14.375, -5) {};
		\node [style=none] (86) at (17.5, -5) {};
		\node [style=none] (87) at (16.325, -5.5) {$2l$};
		\node [style=gauge3] (88) at (14.35, -3.25) {};
		\node [style=none] (89) at (14.375, -2.75) {$k$};
		\node [style=none] (90) at (19.75, -4.25) {};
		\node [style=none] (91) at (18.75, -4.25) {};
		\node [style=none] (92) at (16.75, -4.25) {};
		\node [style=none] (93) at (15.75, -4.25) {};
	\end{pgfonlayer}
	\begin{pgfonlayer}{edgelayer}
		\draw (34) to (35);
		\draw (36) to (37);
		\draw (44) to (43);
		\draw (43) to (45);
		\draw (35) to (48.center);
		\draw (49.center) to (44);
		\draw (45) to (50.center);
		\draw (51.center) to (36);
		\draw (52) to (43);
		\draw (61) to (62);
		\draw (83) to (78);
		\draw [style=brace] (86.center) to (85.center);
		\draw (77) to (66);
		\draw (79) to (78);
		\draw (61) to (60);
		\draw (88) to (61);
		\draw (62) to (93.center);
		\draw (92.center) to (78);
		\draw (79) to (91.center);
		\draw (90.center) to (66);
	\end{pgfonlayer}
\end{tikzpicture}
}
\label{weirder_product}
\end{equation}
where the top is the $E_6$ family \cite{Bourget:2020gzi} and the bottom is the unitary quiver whose Coulomb branch is $\overline{\mathcal{O}}^{\mathfrak{so}(4k+4l+2)}_{(2^{2k},1^{4l+2})}$ \cite{Hanany:2016gbz,Ferlito:2016grh}. For $k>1$ the Coulomb branch global symmetry is $\sorm(4k+4l+2)\times \sorm(4k+4l+2)\times U(1)$. For $l=1$, this family coincides with the $E_6 \times E_5'$ family of \cite[Sec.\ 3.9]{Akhond:2021knl}. 
\subsubsection{\texorpdfstring{$E_8\times\sorm(16)$ family}{SO(16)xE8 family}}\label{finalfamily}
For $n=2k+2l$ and $l \geq 1$, the quiver \eqref{weirdcase} reduces to \eqref{secondlasttime},  which contains a $\usprm(2)$ flavour node that can be gauged to produce the following quiver:
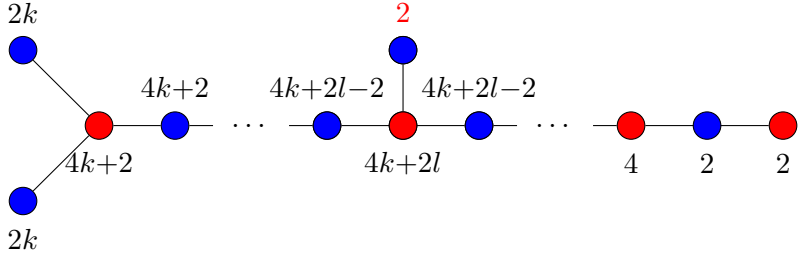
\begin{equation}
\raisebox{-.5\height}{
\begin{tikzpicture}
	\begin{pgfonlayer}{nodelayer}
		\node [style=miniBlue] (2) at (1, 1) {};
		\node [style=miniBlue] (3) at (1, -1) {};
		\node [style=miniU] (4) at (2, 0) {};
		\node [style=miniBlue] (5) at (6, 0) {};
		\node [style=none] (14) at (8, 0) {$\dots$};
		\node [style=miniU] (15) at (9, 0) {};
		\node [style=miniBlue] (16) at (10, 0) {};
		\node [style=miniU] (17) at (11, 0) {};
		\node [style=none] (20) at (1, -1.5) {$2k$};
		\node [style=none] (21) at (1, 1.5) {$2k$};
		\node [style=none] (22) at (2, -0.5) {$4k{+}2$};
		\node [style=none] (31) at (9, -0.5) {4};
		\node [style=none] (32) at (10, -0.5) {2};
		\node [style=none] (33) at (11, -0.5) {2};
		\node [style=none] (41) at (6, -0.5) {$4k{+}2l$};
		\node [style=none] (43) at (6, 1.5) {\textcolor{red}{2}};
		\node [style=miniU] (44) at (5, 0) {};
		\node [style=none] (46) at (5, 0.5) {$4k{+}2l{-}2$};
		\node [style=none] (48) at (4, 0) {$\dots$};
		\node [style=miniBlue] (49) at (3, 0) {};
		\node [style=none] (50) at (3, 0.5) {$4k{+}2$};
		\node [style=none] (51) at (7, 0.5) {$4k{+}2l{-}2$};
		\node [style=miniU] (52) at (7, 0) {};
		\node [style=miniU] (53) at (6, 1) {};
		\node [style=none] (54) at (4.5, 0) {};
		\node [style=none] (55) at (3.5, 0) {};
		\node [style=none] (56) at (8.5, 0) {};
		\node [style=none] (57) at (7.5, 0) {};
		\node [style=miniBlue] (58) at (7, 0) {};
		\node [style=miniU] (59) at (6, 0) {};
		\node [style=miniBlue] (60) at (5, 0) {};
		\node [style=miniBlue] (61) at (6, 1) {};
	\end{pgfonlayer}
	\begin{pgfonlayer}{edgelayer}
		\draw (2) to (4);
		\draw (4) to (3);
		\draw (15) to (17);
		\draw (4) to (49);
		\draw (44) to (52);
		\draw (53) to (5);
		\draw (54.center) to (44);
		\draw (55.center) to (49);
		\draw (56.center) to (15);
		\draw (57.center) to (52);
	\end{pgfonlayer}
\end{tikzpicture}
}
\label{betterbelastguy}
\end{equation}
and Hilbert series computations, see Table \ref{betterbelastguy1}, suggest that the Coulomb branch for $k >1$ is the product of the following two Coulomb branches:
\begin{equation}
\raisebox{-.5\height}{
\begin{tikzpicture}
	\begin{pgfonlayer}{nodelayer}
		\node [style=miniU] (34) at (13.75, 0) {};
		\node [style=miniBlue] (35) at (14.75, 0) {};
		\node [style=miniBlue] (36) at (20.125, 0) {};
		\node [style=miniU] (37) at (21.125, 0) {};
		\node [style=none] (38) at (13.75, -0.5) {2};
		\node [style=none] (39) at (14.75, -0.5) {2};
		\node [style=none] (40) at (17.45, -0.5) {\small$2k{+}2l$};
		\node [style=none] (41) at (20.125, -0.5) {2};
		\node [style=none] (42) at (21.125, -0.5) {2};
		\node [style=miniU] (43) at (17.45, 0) {};
		\node [style=miniBlue] (44) at (16.3, 0) {};
		\node [style=miniBlue] (45) at (18.6, 0) {};
		\node [style=none] (46) at (15.55, 0) {$\cdots$};
		\node [style=none] (47) at (19.35, 0) {$\cdots$};
		\node [style=none] (48) at (15.2, 0) {};
		\node [style=none] (49) at (15.95, 0) {};
		\node [style=none] (50) at (18.95, 0) {};
		\node [style=none] (51) at (19.7, 0) {};
		\node [style=gauge3] (52) at (17.45, 1) {};
		\node [style=none] (53) at (17.45, 1.5) {\textcolor{red}{2}};
		\node [style=miniBlue] (54) at (17.45, 0) {};
		\node [style=miniU] (55) at (18.6, 0) {};
		\node [style=miniU] (56) at (16.3, 0) {};
		\node [style=none] (57) at (12.2, -3.5) {$\times$};
		\node [style=none] (58) at (18.625, 0.5) {\small$2k{+}2l{-}2$};
		\node [style=none] (59) at (16.375, 0.5) {\small$2k{+}2l{-}2$};
		\node [style=gauge3] (60) at (13.375, -3.5) {};
		\node [style=gauge3] (61) at (14.475, -3.5) {};
		\node [style=gauge3] (62) at (15.475, -3.5) {};
		\node [style=none] (63) at (14.475, -4) {$2k$};
		\node [style=none] (64) at (15.475, -4) {$2k$};
		\node [style=none] (65) at (19.475, -3.5) {$\dots$};
		\node [style=gauge3] (66) at (20.475, -3.5) {};
		\node [style=none] (67) at (21.475, -4) {$1$};
		\node [style=none] (68) at (20.475, -4) {$2$};
		\node [style=none] (69) at (13.375, -4) {$k$};
		\node [style=none] (75) at (14.125, -3.875) {};
		\node [style=gauge3] (77) at (21.5, -3.5) {};
		\node [style=gauge3] (78) at (17.5, -3.5) {};
		\node [style=gauge3] (79) at (18.5, -3.5) {};
		\node [style=none] (80) at (16.5, -3.5) {$\dots$};
		\node [style=none] (81) at (17.5, -4) {$2k$};
		\node [style=none] (82) at (18.5, -4) {$2k-1$};
		\node [style=blankflavor] (83) at (17.5, -2.5) {};
		\node [style=none] (84) at (17.5, -2) {1};
		\node [style=none] (85) at (14.5, -4.25) {};
		\node [style=none] (86) at (17.75, -4.25) {};
		\node [style=none] (87) at (16.325, -5) {$2l{-}1$};
		\node [style=gauge3] (88) at (14.475, -2.5) {};
		\node [style=none] (89) at (14.5, -2) {$k$};
		\node [style=miniU] (90) at (17.45, 0) {};
		\node [style=miniBlue] (91) at (16.3, 0) {};
		\node [style=miniBlue] (92) at (18.6, 0) {};
		\node [style=miniBlue] (93) at (17.45, 1) {};
		\node [style=none] (94) at (16, -3.5) {};
		\node [style=none] (95) at (17, -3.5) {};
		\node [style=none] (96) at (20, -3.5) {};
		\node [style=none] (97) at (19, -3.5) {};
	\end{pgfonlayer}
	\begin{pgfonlayer}{edgelayer}
		\draw (34) to (35);
		\draw (36) to (37);
		\draw (44) to (43);
		\draw (43) to (45);
		\draw (35) to (48.center);
		\draw (49.center) to (44);
		\draw (45) to (50.center);
		\draw (51.center) to (36);
		\draw (52) to (43);
		\draw (61) to (62);
		\draw (83) to (78);
		\draw [style=brace] (86.center) to (85.center);
		\draw (77) to (66);
		\draw (79) to (78);
		\draw (61) to (60);
		\draw (88) to (61);
		\draw (95.center) to (78);
		\draw (94.center) to (62);
		\draw (97.center) to (79);
		\draw (96.center) to (66);
	\end{pgfonlayer}
\end{tikzpicture}

}
\label{finallygotitright}
\end{equation}
The Coulomb branch of the top quiver is the $E_8$ family \cite{Bourget:2020gzi} of Table \ref{EnProdtable2} and the bottom quiver is the closure of the nilpotent orbit $\overline{\mathcal{O}}^{\mathfrak{so}(4k+4l)}_{(2^{2k},1^{4l})}$ \cite{Hanany:2016gbz,Ferlito:2016grh}. 

For $k=1$, $l=1$ one observes that the Coulomb branch Hilbert series of \eqref{betterbelastguy} diverges. This is already known in \cite{Bourget:2021zyc} and might be attributed to the fact that all three nodes in the trifucation are balanced. The resulting quiver is not a forked orthosymplectic quiver, but something that does not take the shape of a finite Dynkin diagram. However, the pattern in \eqref{finallygotitright} shows that, the top quiver for $k=l=1$ has the central $\sorm(4)$ connected to an $\usprm(2)$. The latter node is \emph{bad}; hence, leading to the divergent Hilbert series in \eqref{betterbelastguy}. For $k=2$, $l=1$, the Coulomb branch of the orthosymplectic quiver is known to be the flat space $\mathbb{H}^{16}$ \cite{Bourget:2020xdz} and the Coulomb branch of the unitary quiver is  $\overline{\mathcal{O}}^{\mathfrak{so}(12)}_{(2^{4},1^{4})}$. For $k=3$, $l=1$, the product is formed by the minimal nilpotent orbit closure of $E_8$ and $\overline{\mathcal{O}}^{\mathfrak{so}(16)}_{(2^{6},1^{4})}$. For $k>3$, the Coulomb branch global symmetry is $\sorm(4k+4l)\times \sorm(4k+4l)$.

The results of this section can be condensed into the following:
\begin{tcolorbox}
\begin{myrule} \label{rule:En}
The $E_n \times E_n$, $E_6\times \sorm(10)$, and $ E_8 \times \sorm(16) $ families include all \emph{unframed} orthosymplectic quivers which satisfy the following properties:
\begin{compactitem}
    \item Composed of only $\sorm(\mathrm{even})$, $\usprm(\mathrm{even})$, $\urm(1)$ gauge groups.
    \item A long tail beginning with $\sorm(2)$ and ending with fork that is balanced.
    \item $\sorm(\mathrm{even})$ gauge group in the centre of the fork.
    \item All gauge groups that are not part of the fork are necessary to balance it.
\end{compactitem}
\end{myrule}
\end{tcolorbox}
\noindent
The last rule is necessary because one can, in principle, connect an arbitrary number of gauge nodes to existing overbalanced gauge nodes whilst preserving the balanced fork. So far, the only case where including gauge nodes not needed to balance the quiver, but still give a Coulomb branch that is the product of two moduli spaces, is the $E_7\times E_7$ family. This is just the $E_8\times E_8$ family, but with an additional $\sorm(2) - \usprm(2)-$ connected to the $\sorm(4)$ node. Another exception to this rule is the number of bifundamental hypermultiplets (solid lines) and fundamental-fundamental hypermultiplets (dashed lines) between the unbalanced $\urm(1)$ nodes in Table \ref{EnProdtable2} can in principle take any value. However, other values have yet to provide Coulomb branches that are products of moduli spaces.

As a comment, a set of unframed unitary-orthosymplectic quivers $\widetilde{\mathsf{Q}}_{\mathrm{UOSp}}$ that is also denoted as $E_{n+1}\times E_{n+1}$ family has been proposed in \cite{Akhond:2021knl}. However, the underlying 5d $\Ncal=1$ $\sorm(4)$ theories with hypermultiplets in the spinor and conjugate spinor representation are fundamentally different to the ones considered in Section \ref{sec:En_EN_5d} up to accidental dualities. Comparing the two sets of magnetic quivers $\widetilde{\mathsf{Q}}_{\mathrm{UOSp}}$ of \cite{Akhond:2021knl} and $\mathsf{Q}_{\mathrm{OSp}}$ discussed here, the magnetic quivers only coincide for $n=5,6$ cases of the $E_{n+1} \times E_{n+1}$ families and $l=1$ case of \eqref{weirder}. The expectation is that the underlying 5d theories share the same UV fixed point. See Section \ref{sec:ON_5d} for an example of a duality between a brane web for $\sorm(4)$ theories and a brane web for a product of $\surm(k)$ theories.

\subsubsection{\texorpdfstring{$F_4 \times F_4$ family}{F4 x F4 family}}
So far, the forked orthosymplectic quivers discussed had an $\sorm(\mathrm{even})$ central node; instead, the focus is now placed on quivers with a $\usprm$ central node. For example:
\begin{equation}
\raisebox{-.5\height}{
\begin{tikzpicture}
	\begin{pgfonlayer}{nodelayer}
		\node [style=bluegauge] (1) at (8.5, -1.25) {};
		\node [style=redgauge] (10) at (10, 0) {};
		\node [style=none] (19) at (10, -0.5) {\footnotesize{$6$}};
		\node [style=none] (20) at (11, -0.5) {\footnotesize{$6$}};
		\node [style=none] (21) at (12, -0.5) {\footnotesize{$4$}};
		\node [style=none] (22) at (13, -0.5) {\footnotesize{$4$}};
		\node [style=none] (23) at (14, -0.5) {\footnotesize{$2$}};
		\node [style=bluegauge] (25) at (11, 0) {};
		\node [style=redgauge] (26) at (12, 0) {};
		\node [style=bluegauge] (27) at (13, 0) {};
		\node [style=redgauge] (28) at (14, 0) {};
		\node [style=bluegauge] (30) at (8.5, 1.25) {};
		\node [style=none] (32) at (8.5, -1.75) {4};
		\node [style=bluegauge] (38) at (10, 0) {};
		\node [style=bluegauge] (39) at (12, 0) {};
		\node [style=redgauge] (40) at (11, 0) {};
		\node [style=redgauge] (41) at (13, 0) {};
		\node [style=bluegauge] (42) at (14, 0) {};
		\node [style=redgauge] (43) at (15, 0) {};
		\node [style=none] (44) at (15, -0.5) {2};
		\node [style=redgauge] (45) at (8.5, 1.25) {};
		\node [style=redgauge] (46) at (8.5, -1.25) {};
		\node [style=none] (47) at (8.5, 1.75) {4};
		\node [style=none] (48) at (6, 0) {$ \overline{\mathcal{O}}^{\mathfrak{f}_4}_{\text{min}}\times \overline{\mathcal{O}}^{\mathfrak{f}_4}_{\text{min}}$};
	\end{pgfonlayer}
	\begin{pgfonlayer}{edgelayer}
		\draw (30) to (10);
		\draw (10) to (1);
		\draw (10) to (28);
		\draw (42) to (43);
	\end{pgfonlayer}
\end{tikzpicture}
}
\label{f4}
\end{equation}
Notice that the fork is balanced without the need of additional flavour groups or overbalanced gauge nodes. 
Using the rules from \eqref{soxso}, one would expect a $\sorm(9)\times \sorm(9)$ global symmetry. As explained above, the symmetry is enhanced due to monopole operators with half-plus-integer magnetic charges in the monopole formula. As a result, the spinor representations of $\sorm(9) \times \sorm(9)$ enhance the global symmetry to $F_4 \times F_4$. 

The quiver in \eqref{f4} can be extended to the $F_4 \times F_4$ family:
\begin{equation}
\raisebox{-.5\height}{
\scalebox{0.8}{\begin{tikzpicture}
	\begin{pgfonlayer}{nodelayer}
		\node [style=bluegauge] (1) at (8.5, -1.25) {};
		\node [style=redgauge] (10) at (10, 0) {};
		\node [style=none] (19) at (10, -0.5) {$4k{+}2$};
		\node [style=none] (20) at (11, 0.5) {$4k{+}2$};
		\node [style=none] (21) at (12, -0.5) {$4k$};
		\node [style=none] (23) at (14, -0.5) {2};
		\node [style=bluegauge] (25) at (11, 0) {};
		\node [style=redgauge] (26) at (12, 0) {};
		\node [style=redgauge] (28) at (14, 0) {};
		\node [style=bluegauge] (30) at (8.5, 1.25) {};
		\node [style=none] (32) at (8.5, -1.75) {$2k{+}2$};
		\node [style=bluegauge] (38) at (10, 0) {};
		\node [style=bluegauge] (39) at (12, 0) {};
		\node [style=redgauge] (40) at (11, 0) {};
		\node [style=bluegauge] (42) at (14, 0) {};
		\node [style=redgauge] (43) at (15, 0) {};
		\node [style=none] (44) at (15, -0.5) {2};
		\node [style=redgauge] (45) at (8.5, 1.25) {};
		\node [style=redgauge] (46) at (8.5, -1.25) {};
		\node [style=none] (47) at (8.5, 1.75) {$2k{+}2$};
		\node [style=none] (48) at (13, 0) {$\dots$};
		\node [style=none] (49) at (16.25, 0) {$\leftrightarrow$};
		\node [style=none] (50) at (17.75, 2) {};
		\node [style=none] (51) at (17.75, -2) {};
		\node [style=gauge3] (52) at (17.75, 0) {};
		\node [style=gauge3] (53) at (18.75, 0) {};
		\node [style=gauge3] (54) at (19.85, 0) {};
		\node [style=gauge3] (55) at (20.85, 0) {};
		\node [style=none] (57) at (19.85, -0.5) {$2k{+}1$};
		\node [style=none] (58) at (20.85, 0.5) {$2k$};
		\node [style=none] (59) at (21.85, 0) {$\dots$};
		\node [style=gauge3] (60) at (22.85, 0) {};
		\node [style=none] (62) at (23.85, -0.5) {$1$};
		\node [style=none] (63) at (22.85, -0.5) {$2$};
		\node [style=none] (64) at (18.75, 0.5) {$k{+}1$};
		\node [style=none] (65) at (17.75, -0.5) {$1$};
		\node [style=none] (66) at (20, 0.125) {};
		\node [style=none] (67) at (20, -0.125) {};
		\node [style=none] (68) at (18.75, 0.125) {};
		\node [style=none] (69) at (18.75, -0.125) {};
		\node [style=none] (70) at (19.5, 0.375) {};
		\node [style=none] (71) at (19.5, -0.375) {};
		\node [style=none] (72) at (19.15, 0) {};
		\node [style=none] (73) at (23.85, 2) {};
		\node [style=none] (74) at (23.85, -2) {};
		\node [style=none] (75) at (24.5, 2.25) {\Large{2}};
		\node [style=flavor2] (76) at (23.85, 0) {};
		\node [style=none] (77) at (13.5, 0) {};
		\node [style=none] (78) at (12.5, 0) {};
		\node [style=none] (79) at (22.5, 0) {};
		\node [style=none] (80) at (21.25, 0) {};
	\end{pgfonlayer}
	\begin{pgfonlayer}{edgelayer}
		\draw (30) to (10);
		\draw (10) to (1);
		\draw (42) to (43);
		\draw (38) to (39);
		\draw [bend right=60, looseness=0.50] (50.center) to (51.center);
		\draw (52) to (53);
		\draw (54) to (55);
		\draw (66.center) to (68.center);
		\draw (67.center) to (69.center);
		\draw (72.center) to (70.center);
		\draw (72.center) to (71.center);
		\draw [bend left=60, looseness=0.50] (73.center) to (74.center);
		\draw (76) to (60);
		\draw (39) to (78.center);
		\draw (77.center) to (42);
		\draw (55) to (80.center);
		\draw (79.center) to (60);
	\end{pgfonlayer}
\end{tikzpicture}}
}
\label{f4f4}
\end{equation}
which gives back $F_4\times F_4$ for $k=1$. For $k>1$, the Coulomb branch global symmetry is $\sorm(4k+5)\times \sorm(4k+5)$. 
The Coulomb branches of \eqref{f4f4} are not products of the Coulomb branch of orthosymplectic quivers, but rather of non-simply laced unitary quivers on the right hand side. As mentioned above, the proposed equivalence \eqref{f4f4} here is only established at the level of the Coulomb branch Hilbert series\footnote{In contrast to the $E_n\times E_n$ family, which is further validated by brane constructions in Section \ref{branes}.}, see for instance Table \ref{f4f41}. The HWG of 
\eqref{f4f4} is the product of two copies of the unitary quiver of the right hand side:
\begin{equation}
    \HWG\eqref{f4f4}(t;\mu_i,\rho_i)=\PE\left[\sum_{i=1}^k\mu_{2i}t^{2i}+\mu_{2k+2}t^{k+1}\right] \cdot \PE\left[\sum_{i=1}^k\rho_{2i}t^{2i}+\rho_{2k+2}t^{k+1}\right]
\end{equation}
where $\mu_i$ and $\rho_i$ are the highest weight fugacities of $\sorm(4k+5)\times\sorm(4k+5) $ which is the Coulomb branch global symmetry group of the orthosymplectic quiver \eqref{f4f4}. 
\subsubsection{\texorpdfstring{$\sorm(\mathrm{odd})\times \sorm(\mathrm{odd})$ family}{SO odd x SO odd family}}
The last family of unframed orthosymplectic quivers with a balanced fork has a $\usprm$ node in the centre and $\sorm(\mathrm{odd})$ nodes at the bifurcated nodes. The quiver is composed of $\sorm(\mathrm{odd})$, $\sorm(\mathrm{even})$ and $\usprm(\mathrm{even})$ gauge nodes. According to \cite{Bourget:2020xdz}, the subgroup $H$ is now trivial and the monopole formula only contains integer valued magnetic charges for the monopole operators. 
The resulting quiver takes the following form:
\begin{equation}
\raisebox{-.5\height}{
\begin{tikzpicture}
	\begin{pgfonlayer}{nodelayer}
		\node [style=bluegauge] (1) at (8.5, -1.25) {};
		\node [style=redgauge] (10) at (10, 0) {};
		\node [style=none] (19) at (10, -0.5) {$4k$};
		\node [style=none] (20) at (11, 0.5) {$4k$};
		\node [style=none] (21) at (12, -0.5) {$4k{-}2$};
		\node [style=none] (23) at (14, -0.5) {$2$};
		\node [style=bluegauge] (25) at (11, 0) {};
		\node [style=redgauge] (26) at (12, 0) {};
		\node [style=redgauge] (28) at (14, 0) {};
		\node [style=bluegauge] (30) at (8.5, 1.25) {};
		\node [style=none] (32) at (8.5, -1.75) {$2k{+}1$};
		\node [style=bluegauge] (38) at (10, 0) {};
		\node [style=bluegauge] (39) at (12, 0) {};
		\node [style=redgauge] (40) at (11, 0) {};
		\node [style=bluegauge] (42) at (14, 0) {};
		\node [style=redgauge] (43) at (15, 0) {};
		\node [style=none] (44) at (15, -0.5) {$2$};
		\node [style=redgauge] (45) at (8.5, 1.25) {};
		\node [style=redgauge] (46) at (8.5, -1.25) {};
		\node [style=none] (47) at (8.5, 1.75) {$2k{+}1$};
		\node [style=none] (48) at (13, 0) {$\dots$};
		\node [style=none] (49) at (12.5, 0) {};
		\node [style=none] (50) at (13.5, 0) {};
	\end{pgfonlayer}
	\begin{pgfonlayer}{edgelayer}
		\draw (30) to (10);
		\draw (10) to (1);
		\draw (42) to (43);
		\draw (38) to (39);
		\draw (39) to (49.center);
		\draw (50.center) to (42);
	\end{pgfonlayer}
\end{tikzpicture}
}
\label{soddxsodd}
\end{equation}
whose Coulomb branch is conjectured to be
\begin{align}
    \Coulomb \eqref{soddxsodd} = \overline{\mathcal{O}}^{\mathfrak{so}_{4k+3}}_{(2^{2k},1^3)}\times \overline{\mathcal{O}}^{\mathfrak{so}_{4k+3}}_{(2^{2k},1^3)} \,,
\end{align}
 based on the explicit Hilbert series computation given in Table \ref{soddxsodd1}. 
Equivalently, this is the magnetic quiver whose Coulomb branch is the same as the Higgs branch of the following two copies of SQCD theories\footnote{\label{fn:Sp_odd}The 3d $\Ncal=4$ $\sprm(k)$ theory with odd number of half-hypermultiplets suffers from a parity anomaly, which can be cured by a suitable CS-level. This is natural in the brane setup of Section \ref{sec:branes_3d} and can be neglected here, as the Higgs branch is not affected by a CS-level.}:
\begin{equation}
\raisebox{-.5\height}{
\begin{tikzpicture}
	\begin{pgfonlayer}{nodelayer}
		\node [style=miniBlue] (0) at (0, 0) {};
		\node [style=flavourRed] (1) at (0, 1.5) {};
		\node [style=none] (2) at (0, -0.5) {$\usprm(2k)$};
		\node [style=none] (3) at (0, 2) {$\sorm(4k{+}3)$};
		\node [style=none] (4) at (1, 0.75) {$\times$};
		\node [style=miniBlue] (5) at (2, 0) {};
		\node [style=flavourRed] (6) at (2, 1.5) {};
		\node [style=none] (7) at (2, -0.5) {$\usprm(2k)$};
		\node [style=none] (8) at (2, 2) {$\sorm(4k{+}3)$};
	\end{pgfonlayer}
	\begin{pgfonlayer}{edgelayer}
		\draw (1) to (0);
		\draw (6) to (5);
	\end{pgfonlayer}
\end{tikzpicture}
}
\end{equation}
This case has an interesting feature as there is no known orthosymplectic quiver whose Coulomb branch is a single $\overline{\mathcal{O}}^{\mathfrak{so}_{4k+3}}_{(2^{2k},1^3)}$ orbit, because these are non-special nilpotent orbits. 

The results of the above two sections can be condensed into
\begin{tcolorbox}
\begin{myrule} \label{rule:unframed_D} The $F_4 \times F_4$ and $\mathrm{SO(odd)}\times \mathrm{SO(odd})$
families include all \emph{unframed} orthosymplectic quivers which satisfy the following properties:
\begin{compactitem}
    \item Composed of only $\sorm(\mathrm{odd})$, $\sorm(\mathrm{even})$, $\usprm(\mathrm{even})$, $\urm(1)$ gauge groups.
    \item A long tail beginning with $\sorm(2)$ and ending with fork that is balanced.
    \item $\usprm(\mathrm{even})$ gauge group in the centre of the fork.
\end{compactitem}
\end{myrule}
\end{tcolorbox}
\noindent
For these quivers the forks are balanced without the need  of additional flavour groups or overbalanced gauge groups. 
\FloatBarrier
\section{\texorpdfstring{Folding after forking ($B$-type)}{Folding after Forking (B-type)}}
\label{foldafterfork}
In this section, the remaining class of classical Dynkin-type orthosymplectic quivers is explored: $B$-type orthosymplectic quivers, which contain a non-simply laced edge. 

The simplest way to construct $B$-type orthosymplectic quivers is to fold forked/$D$-type orthosymplectic quivers. As long as the two bifurcated nodes are identical, this can be achieved by folding them into a non-simply laced edge. The $B$-type quivers in this section are all obtain through folding of the forked quivers of Section \ref{forkingit}. The resulting Coulomb branches are products of two \emph{non-identical} moduli spaces. 
\subsection{Global symmetry}
 Through explicit Hilbert series computations, one finds the Coulomb branch global symmetry for a balanced $B_n$ orthosymplectic quiver to be the product:
\begin{equation}
\raisebox{-.5\height}{
\begin{tikzpicture}
	\begin{pgfonlayer}{nodelayer}
		\node [style=gauge3] (53) at (18.75, 0) {};
		\node [style=gauge3] (54) at (19.85, 0) {};
		\node [style=gauge3] (55) at (20.85, 0) {};
		\node [style=none] (59) at (21.775, 0) {$\dots$};
		\node [style=gauge3] (60) at (22.85, 0) {};
		\node [style=gauge3] (61) at (23.85, 0) {};
		\node [style=none] (66) at (20, 0.125) {};
		\node [style=none] (67) at (20, -0.125) {};
		\node [style=none] (68) at (18.75, 0.125) {};
		\node [style=none] (69) at (18.75, -0.125) {};
		\node [style=none] (70) at (19.5, 0.375) {};
		\node [style=none] (71) at (19.5, -0.375) {};
		\node [style=none] (72) at (19.15, 0) {};
		\node [style=none] (73) at (18.75, -0.75) {};
		\node [style=none] (74) at (24, -0.75) {};
		\node [style=none] (75) at (21.25, -1.25) {$n$ balanced nodes};
		\node [style=redgauge] (76) at (18.75, 0) {};
		\node [style=bluegauge] (77) at (19.85, 0) {};
		\node [style=redgauge] (78) at (20.85, 0) {};
		\node [style=redgauge] (79) at (23.85, 0) {};
		\node [style=bluegauge] (80) at (22.85, 0) {};
		\node [style=none] (81) at (14, 0) {$G_{\mathrm{global}}=\sorm(n+1)\times \sorm(n)$};
		\node [style=none] (82) at (21.25, 0) {};
		\node [style=none] (83) at (22.25, 0) {};
	\end{pgfonlayer}
	\begin{pgfonlayer}{edgelayer}
		\draw (54) to (55);
		\draw (60) to (61);
		\draw (66.center) to (68.center);
		\draw (67.center) to (69.center);
		\draw (72.center) to (70.center);
		\draw (72.center) to (71.center);
		\draw [style=brace] (74.center) to (73.center);
		\draw (83.center) to (80);
		\draw (78) to (82.center);
	\end{pgfonlayer}
\end{tikzpicture}
}
\end{equation}
where, as usual, the first node from the left or the right need not be special orthogonal so long as it is a balanced orthosymplectic gauge node. 
\noindent
In the event that the first node from the right is an $\sorm(2)$ node, the global symmetry exhibits a, by now, familiar enhancement:
\begin{equation}
\raisebox{-.5\height}{
\begin{tikzpicture}
	\begin{pgfonlayer}{nodelayer}
		\node [style=gauge3] (53) at (18.75, 0) {};
		\node [style=gauge3] (54) at (19.85, 0) {};
		\node [style=gauge3] (55) at (20.85, 0) {};
		\node [style=none] (59) at (21.85, 0) {$\dots$};
		\node [style=gauge3] (60) at (22.85, 0) {};
		\node [style=gauge3] (61) at (23.85, 0) {};
		\node [style=none] (66) at (20, 0.125) {};
		\node [style=none] (67) at (20, -0.125) {};
		\node [style=none] (68) at (18.75, 0.125) {};
		\node [style=none] (69) at (18.75, -0.125) {};
		\node [style=none] (70) at (19.5, 0.375) {};
		\node [style=none] (71) at (19.5, -0.375) {};
		\node [style=none] (72) at (19.15, 0) {};
		\node [style=none] (73) at (18.75, -0.75) {};
		\node [style=none] (74) at (24, -0.75) {};
		\node [style=none] (75) at (21.25, -1.25) {$n$ balanced nodes};
		\node [style=redgauge] (76) at (18.75, 0) {};
		\node [style=bluegauge] (77) at (19.85, 0) {};
		\node [style=redgauge] (78) at (20.85, 0) {};
		\node [style=redgauge] (79) at (23.85, 0) {};
		\node [style=bluegauge] (80) at (22.85, 0) {};
		\node [style=none] (81) at (14, 0) {$G_{\mathrm{global}}=\sorm(n+2)\times \sorm(n+1)$};
		\node [style=none] (82) at (23.875, 0.5) {2};
		\node [style=none] (83) at (21.25, 0) {};
		\node [style=none] (84) at (22.35, 0) {};
	\end{pgfonlayer}
	\begin{pgfonlayer}{edgelayer}
		\draw (54) to (55);
		\draw (60) to (61);
		\draw (66.center) to (68.center);
		\draw (67.center) to (69.center);
		\draw (72.center) to (70.center);
		\draw (72.center) to (71.center);
		\draw [style=brace] (74.center) to (73.center);
		\draw (83.center) to (78);
		\draw (84.center) to (80);
	\end{pgfonlayer}
\end{tikzpicture}
}
\end{equation}
Similar to the case of $D$-type quivers, if the $\sorm(2)$ is ungauged, the global symmetry carries an additional $\sorm(2)$ factor, i.e.
\begin{equation}
\raisebox{-.5\height}{
\begin{tikzpicture}
	\begin{pgfonlayer}{nodelayer}
		\node [style=gauge3] (53) at (18.75, 0) {};
		\node [style=gauge3] (54) at (19.85, 0) {};
		\node [style=gauge3] (55) at (20.85, 0) {};
		\node [style=none] (59) at (21.85, 0) {$\dots$};
		\node [style=gauge3] (60) at (22.8, 0) {};
		\node [style=gauge3] (61) at (23.8, 0) {};
		\node [style=none] (66) at (20, 0.125) {};
		\node [style=none] (67) at (20, -0.125) {};
		\node [style=none] (68) at (18.75, 0.125) {};
		\node [style=none] (69) at (18.75, -0.125) {};
		\node [style=none] (70) at (19.5, 0.375) {};
		\node [style=none] (71) at (19.5, -0.375) {};
		\node [style=none] (72) at (19.15, 0) {};
		\node [style=none] (73) at (18.75, -0.75) {};
		\node [style=none] (74) at (22.7, -0.75) {};
		\node [style=none] (75) at (21, -1.25) {$(n-1)$ balanced nodes};
		\node [style=redgauge] (76) at (18.75, 0) {};
		\node [style=bluegauge] (77) at (19.85, 0) {};
		\node [style=redgauge] (78) at (20.85, 0) {};
		\node [style=flavorRed] (79) at (23.8, 0) {};
		\node [style=bluegauge] (80) at (22.8, 0) {};
		\node [style=none] (81) at (14, 0) {$G_{\mathrm{global}}=\sorm(n)\times \sorm(n-1)\times \sorm(2)$};
		\node [style=none] (82) at (23.825, 0.5) {2};
		\node [style=none] (83) at (21.325, 0) {};
		\node [style=none] (84) at (22.35, 0) {};
	\end{pgfonlayer}
	\begin{pgfonlayer}{edgelayer}
		\draw (54) to (55);
		\draw (60) to (61);
		\draw (66.center) to (68.center);
		\draw (67.center) to (69.center);
		\draw (72.center) to (70.center);
		\draw (72.center) to (71.center);
		\draw [style=brace] (74.center) to (73.center);
		\draw (84.center) to (80);
		\draw (83.center) to (78);
	\end{pgfonlayer}
\end{tikzpicture}
}
\label{anomalousfold}
\end{equation}
For unframed $B$-type orthosymplectic quivers, there could be additional symmetry enhancement due to monopole operators with half-plus-integer magnetic charges in the monopole formula. For instance, the examples in Section \ref{sec:unframed_B} display such enhancement.

\subsection{\texorpdfstring{Framed $B$-type orthosymplectic quivers}{Framed B-type orthosymplectic quivers}}
\label{sec:B-type_framed}
The simplest example of a framed $B$-type orthosymplectic quiver is obtained by folding the quiver \eqref{firstfork}; in detail
\begin{equation}
\raisebox{-.5\height}{
\begin{tikzpicture}
	\begin{pgfonlayer}{nodelayer}
		\node [style=flavourRed] (0) at (0.5, 1) {};
		\node [style=flavourRed] (1) at (0.5, -1) {};
		\node [style=miniBlue] (2) at (1.5, 1) {};
		\node [style=miniBlue] (3) at (1.5, -1) {};
		\node [style=miniU] (4) at (2.5, 0) {};
		\node [style=miniBlue] (5) at (3.5, 0) {};
		\node [style=none] (7) at (4.5, 0) {$\dots$};
		\node [style=miniBlue] (9) at (5.5, 0) {};
		\node [style=miniU] (15) at (6.5, 0) {};
		\node [style=miniBlue] (16) at (7.5, 0) {};
		\node [style=miniU] (17) at (8.5, 0) {};
		\node [style=none] (18) at (0.5, 1.5) {2};
		\node [style=none] (19) at (0.5, -1.5) {2};
		\node [style=none] (20) at (1.5, -1.5) {$2k$};
		\node [style=none] (21) at (1.5, 1.5) {$2k$};
		\node [style=none] (22) at (2.75, 0.5) {$4k$};
		\node [style=none] (23) at (3.5, -0.5) {$4k{-}2$};
		\node [style=none] (25) at (5.5, -0.5) {4};
		\node [style=none] (31) at (6.5, -0.5) {4};
		\node [style=none] (32) at (7.5, -0.5) {2};
		\node [style=none] (33) at (8.5, -0.5) {2};
		\node [style=flavourRed] (34) at (0, -4.5) {};
		\node [style=miniBlue] (36) at (1, -4.5) {};
		\node [style=miniU] (38) at (2.5, -4.5) {};
		\node [style=miniBlue] (39) at (3.5, -4.5) {};
		\node [style=none] (40) at (4.5, -4.5) {$\dots$};
		\node [style=miniBlue] (41) at (5.5, -4.5) {};
		\node [style=miniU] (42) at (6.5, -4.5) {};
		\node [style=miniBlue] (43) at (7.5, -4.5) {};
		\node [style=miniU] (44) at (8.5, -4.5) {};
		\node [style=none] (45) at (0, -5) {2};
		\node [style=none] (48) at (1, -5) {$2k$};
		\node [style=none] (51) at (5.5, -5) {4};
		\node [style=none] (52) at (6.5, -5) {4};
		\node [style=none] (53) at (7.5, -5) {2};
		\node [style=none] (54) at (8.5, -5) {2};
		\node [style=miniBlue] (95) at (1, -4.5) {};
		\node [style=miniU] (96) at (2.5, -4.5) {};
		\node [style=none] (97) at (1, -4.375) {};
		\node [style=none] (98) at (1, -4.625) {};
		\node [style=none] (99) at (2.5, -4.375) {};
		\node [style=none] (100) at (2.5, -4.625) {};
		\node [style=none] (101) at (2, -4.125) {};
		\node [style=none] (102) at (2, -4.875) {};
		\node [style=none] (103) at (1.625, -4.5) {};
		\node [style=none] (104) at (3.75, -1.25) {};
		\node [style=none] (105) at (3.75, -3.5) {};
		\node [style=none] (106) at (4.25, -2.5) {Fold};
		\node [style=none] (107) at (4, 0) {};
		\node [style=none] (108) at (5, 0) {};
		\node [style=none] (109) at (5, -4.5) {};
		\node [style=none] (110) at (4, -4.5) {};
		\node [style=none] (111) at (2.5, -4) {$4k$};
		\node [style=none] (112) at (3.5, -5) {$4k{-}2$};
	\end{pgfonlayer}
	\begin{pgfonlayer}{edgelayer}
		\draw (0) to (2);
		\draw (1) to (3);
		\draw (2) to (4);
		\draw (4) to (3);
		\draw (4) to (5);
		\draw (15) to (17);
		\draw (15) to (9);
		\draw (34) to (36);
		\draw (38) to (39);
		\draw (42) to (44);
		\draw (42) to (41);
		\draw (97.center) to (99.center);
		\draw (100.center) to (98.center);
		\draw (101.center) to (103.center);
		\draw (103.center) to (102.center);
		\draw [style=->] (104.center) to (105.center);
		\draw (107.center) to (5);
		\draw (108.center) to (9);
		\draw (109.center) to (41);
		\draw (110.center) to (39);
	\end{pgfonlayer}
\end{tikzpicture}}
\label{foldedBtypefirst}
\end{equation}
where all the gauge nodes are balanced. The explicit Hilbert series results of Table \ref{foldedBtypefirst1} provide evidence that the Coulomb branch of the resulting theory  in \eqref{foldedBtypefirst} is the product of two nilpotent orbit closures
\begin{equation}
    \mathcal{C}\eqref{foldedBtypefirst} = \overline{\mathcal{O}}^{\mathfrak{so}(4k+1)}_{(2^{2k},1)} \times \overline{\mathcal{O}}^{\mathfrak{so}(4k+2)}_{(2^{2k},1^{2})}
\end{equation}
Compared with the Coulomb branch of the orthosymplectic quiver \eqref{firstfork} before folding, one of the two factors of $\overline{\mathcal{O}}^{\mathfrak{so}(4k+2)}_{(2^{2k},1^{2})}$ remains the same, whereas the other factor reduces to $\overline{\mathcal{O}}^{\mathfrak{so}(4k+1)}_{(2^{2k},1)}$. In view of the results presented so far, this seems to be the general pattern of folding forked orthosymplectic quivers: \emph{one of the two spaces remains the same, whereas the other becomes smaller.} 

One can then fold the most general family \eqref{weirdcase}, which yields:
\begin{equation}
\raisebox{-.5\height}{
   \scalebox{0.82}{\begin{tikzpicture}
	\begin{pgfonlayer}{nodelayer}
		\node [style=miniU] (2) at (1.25, 0) {};
		\node [style=miniBlue] (3) at (6.25, 0) {};
		\node [style=none] (4) at (8.25, 0) {$\dots$};
		\node [style=miniU] (5) at (10.25, 0) {};
		\node [style=miniBlue] (6) at (9.25, 0) {};
		\node [style=miniBlue] (7) at (11.25, 0) {};
		\node [style=flavourRed] (8) at (11.25, 1) {};
		\node [style=miniU] (9) at (12.25, 0) {};
		\node [style=miniBlue] (10) at (13.25, 0) {};
		\node [style=none] (11) at (14.25, 0) {$\dots$};
		\node [style=miniU] (12) at (15.25, 0) {};
		\node [style=miniBlue] (13) at (16.25, 0) {};
		\node [style=miniU] (14) at (17.25, 0) {};
		\node [style=none] (17) at (1.25, -0.5) {$4k{+}2$};
		\node [style=none] (18) at (9.25, -0.5) {$4k{+}2l$};
		\node [style=none] (19) at (11.25, 1.5) {1};
		\node [style=none] (20) at (12.25, 0.5) {$4k{+}2l$};
		\node [style=none] (21) at (13.25, -0.5) {$4k{+}2l{-}2$};
		\node [style=none] (22) at (15.25, -0.5) {4};
		\node [style=none] (23) at (16.25, -0.5) {2};
		\node [style=none] (24) at (17.25, -0.5) {2};
		\node [style=none] (25) at (6, -0.75) {};
		\node [style=none] (26) at (10.75, -0.75) {};
		\node [style=none] (27) at (8.5, -1.25) {$2n{-}4k{-}4l{-}2$};
		\node [style=none] (28) at (11.25, -0.5) {$4k{+}2l$};
		\node [style=none] (29) at (10.25, 0.5) {$4k{+}2l{+}1$};
		\node [style=miniU] (30) at (7.25, 0) {};
		\node [style=none] (31) at (7.25, 0.5) {$4k{+}2l{+}1$};
		\node [style=none] (32) at (6.25, -0.5) {$4k{+}2l$};
		\node [style=flavourRed] (33) at (6.25, 1) {};
		\node [style=none] (34) at (6.25, 1.5) {1};
		\node [style=miniU] (35) at (5.25, 0) {};
		\node [style=miniBlue] (36) at (4.25, 0) {};
		\node [style=none] (37) at (5.25, 0.5) {$4k{+}2l$};
		\node [style=none] (38) at (4.25, -0.5) {$4k{+}2l{-}1$};
		\node [style=none] (39) at (3.25, 0) {$\dots$};
		\node [style=miniBlue] (40) at (2.25, 0) {};
		\node [style=none] (41) at (2.25, -0.5) {$4k{+}2$};
		\node [style=miniBlue] (42) at (-0.25, 0) {};
		\node [style=miniU] (43) at (1.25, 0) {};
		\node [style=none] (44) at (-0.25, -0.5) {$2k$};
		\node [style=none] (46) at (-0.25, 0.125) {};
		\node [style=none] (47) at (-0.25, -0.125) {};
		\node [style=none] (48) at (1.25, 0.125) {};
		\node [style=none] (49) at (1.25, -0.125) {};
		\node [style=none] (50) at (0.75, 0.375) {};
		\node [style=none] (51) at (0.75, -0.375) {};
		\node [style=none] (52) at (0.375, 0) {};
		\node [style=none] (53) at (3.75, 0) {};
		\node [style=none] (54) at (2.75, 0) {};
		\node [style=none] (55) at (7.75, 0) {};
		\node [style=none] (56) at (8.75, 0) {};
		\node [style=none] (57) at (13.75, 0) {};
		\node [style=none] (58) at (14.75, 0) {};
	\end{pgfonlayer}
	\begin{pgfonlayer}{edgelayer}
		\draw (6) to (5);
		\draw (5) to (7);
		\draw (7) to (8);
		\draw (7) to (10);
		\draw (12) to (14);
		\draw [style=brace] (26.center) to (25.center);
		\draw (33) to (3);
		\draw (2) to (40);
		\draw (36) to (30);
		\draw (46.center) to (48.center);
		\draw (49.center) to (47.center);
		\draw (50.center) to (52.center);
		\draw (52.center) to (51.center);
		\draw (40) to (54.center);
		\draw (53.center) to (36);
		\draw (30) to (55.center);
		\draw (56.center) to (6);
		\draw (10) to (57.center);
		\draw (58.center) to (12);
	\end{pgfonlayer}
\end{tikzpicture}}
}
\label{missingguy}
\end{equation}
where $n\geq 2k+2l$ if $l\geq 1$ and $n \geq 2k +1 $ if $l=0$ (because \eqref{finalfinitequiver} cannot be folded). 

\begin{tcolorbox}
\begin{myrule} \label{rule:framed_B}
Quiver \eqref{missingguy} includes all \emph{framed} orthosymplectic quivers which satisfy:
\begin{compactitem}
    \item Tail begins with $\sorm(2)$ and end with a $B$-type non-simply laced edge.
    \item All nodes are balanced.
\end{compactitem}
\end{myrule}
\end{tcolorbox}

Based on the Hilbert series presented in Table \ref{missingguy1}, it is suggestive to identify the Coulomb branch of \eqref{missingguy} with the product of nilpotent orbit closures of $\sorm(2n-1)$ and $\sorm(2n)$:
\begin{equation}
    \mathcal{C}\eqref{weirdcase} =\overline{\mathcal{O}}^{\mathfrak{so}(2n-1)}_{(2^{2k},1^{2n-4k-1})}\times   \overline{\mathcal{O}}^{\mathfrak{so}(2n)}_{(2^{2k+2l},1^{2n-4k-4l})}\,.
\end{equation}
Again, a 3d mirror symmetry argument \eqref{eq:ex_product_OSp_different_diff_ranks} shows that the Coulomb branch of \eqref{missingguy} is also equivalent to the Higgs branch of the product of two SQCD theories\footnote{The same comment as in footnote \ref{fn:Sp_odd} applies to $\sprm(k)$ with $2n-1$ half-hypermultiplets.}:
\begin{equation}
\raisebox{-.5\height}{
\begin{tikzpicture}
	\begin{pgfonlayer}{nodelayer}
		\node [style=miniBlue] (0) at (0, 0) {};
		\node [style=flavourRed] (1) at (0, 1.5) {};
		\node [style=none] (2) at (0, -0.5) {$\usprm(2k)$};
		\node [style=none] (3) at (0, 2) {$\sorm(2n-1)$};
		\node [style=none] (4) at (1, 0.75) {$\times$};
		\node [style=miniBlue] (5) at (2, 0) {};
		\node [style=flavourRed] (6) at (2, 1.5) {};
		\node [style=none] (7) at (2, -0.5) {$\usprm(2k{+}2l)$};
		\node [style=none] (8) at (2, 2) {$\sorm(2n)$};
	\end{pgfonlayer}
	\begin{pgfonlayer}{edgelayer}
		\draw (1) to (0);
		\draw (6) to (5);
	\end{pgfonlayer}
\end{tikzpicture}}
\end{equation}
which, again, are two theories with gauge groups of different ranks. 

\subsection{\texorpdfstring{Unframed $B$-type orthosymplectic quivers}{Unframed B-type orthosymplectic quivers}}
\label{sec:unframed_B}
Now, the attention is turned to $B$-type unframed orthosymplectic quivers, which are obtained by folding the forked unframed quivers of Section \ref{forkingunframed}. Only five families of unframed forked quivers have identical bifurcated nodes that allow to be folded. In detail, these are the $E_6 \times E_6$ family in Table \ref{EnProdtable2},  $E_6 \times \sorm(10)$ family in \eqref{weirder}, $E_8 \times \sorm(16)$ family in \eqref{betterbelastguy}, $F_4 \times F_4$ family in \eqref{f4f4} and $\sorm(\mathrm{odd})\times \sorm(\mathrm{odd})$ family in \eqref{soddxsodd}. 
\subsubsection{\texorpdfstring{$E_6 \times F_4$ family}{E6 x F4 family}}
Taking the $E_6 \times E_6$ family in Table \ref{EnProdtable2}, the Coulomb branch after folding seems consistent with a product space: one factor is the $E_6$ family and the second factor is the $F_4$ family introduced in \eqref{f4f4}. Hence, there is correspondence between 
\begin{equation}
\scalebox{0.7}{
\raisebox{-.5\height}{
\begin{tikzpicture}
	\begin{pgfonlayer}{nodelayer}
		\node [style=redgauge] (0) at (-0.25, 0) {};
		\node [style=bluegauge] (1) at (0.75, 0) {};
		\node [style=redgauge] (2) at (2, 0) {};
		\node [style=bluegauge] (3) at (3.5, 0) {};
		\node [style=none] (11) at (2, -0.5) {$4k+4$};
		\node [style=none] (12) at (3.5, -0.5) {$4k+2$};
		\node [style=none] (22) at (7, -0.5) {2};
		\node [style=none] (23) at (8, -0.5) {2};
		\node [style=bluegauge] (27) at (7, 0) {};
		\node [style=redgauge] (28) at (8, 0) {};
		\node [style=none] (32) at (0.75, -0.5) {$2k+2$};
		\node [style=none] (33) at (-0.25, -0.5) {\color{red}{1}};
		\node [style=none] (34) at (6, 0) {$\dots$};
		\node [style=redgauge] (35) at (5, 0) {};
		\node [style=none] (36) at (5, -0.5) {$4k+2$};
		\node [style=gauge3] (40) at (-0.25, 0) {};
		\node [style=none] (41) at (2, 0.125) {};
		\node [style=none] (42) at (2, -0.125) {};
		\node [style=none] (43) at (0.75, 0.125) {};
		\node [style=none] (44) at (0.75, -0.125) {};
		\node [style=none] (45) at (1.5, 0.375) {};
		\node [style=none] (46) at (1.5, -0.375) {};
		\node [style=none] (47) at (1.15, 0) {};
		\node [style=miniU] (48) at (11, 0) {};
		\node [style=miniBlue] (49) at (12, 0) {};
		\node [style=miniBlue] (50) at (17.375, 0) {};
		\node [style=miniU] (51) at (18.375, 0) {};
		\node [style=none] (52) at (11, -0.5) {2};
		\node [style=none] (53) at (12, -0.5) {2};
		\node [style=none] (54) at (14.7, -0.5) {\small$2k+2$};
		\node [style=none] (55) at (15.85, -0.5) {\small$2k+2$};
		\node [style=none] (56) at (17.375, -0.5) {2};
		\node [style=none] (57) at (18.375, -0.5) {2};
		\node [style=miniU] (58) at (14.7, 0) {};
		\node [style=miniBlue] (59) at (13.55, 0) {};
		\node [style=miniBlue] (60) at (15.85, 0) {};
		\node [style=none] (61) at (12.8, 0) {$\cdots$};
		\node [style=none] (62) at (16.6, 0) {$\cdots$};
		\node [style=none] (63) at (12.45, 0) {};
		\node [style=none] (64) at (13.2, 0) {};
		\node [style=none] (65) at (16.2, 0) {};
		\node [style=none] (66) at (16.95, 0) {};
		\node [style=none] (67) at (13.55, -0.5) {\small$2k+2$};
		\node [style=gauge3] (68) at (14.7, 1) {};
		\node [style=none] (69) at (14.7, 1.5) {\textcolor{red}{1}};
		\node [style=miniBlue] (70) at (14.7, 0) {};
		\node [style=miniU] (71) at (15.85, 0) {};
		\node [style=miniU] (72) at (13.55, 0) {};
		\node [style=none] (73) at (10.25, -3) {$\times$};
		\node [style=gauge3] (74) at (11.625, -3) {};
		\node [style=gauge3] (75) at (12.625, -3) {};
		\node [style=gauge3] (76) at (13.725, -3) {};
		\node [style=gauge3] (77) at (14.725, -3) {};
		\node [style=none] (78) at (13.725, -3.5) {$2k+1$};
		\node [style=none] (79) at (14.725, -2.5) {$2k$};
		\node [style=none] (80) at (15.725, -3) {$\dots$};
		\node [style=gauge3] (81) at (16.725, -3) {};
		\node [style=flavor2] (82) at (17.725, -3) {};
		\node [style=none] (83) at (17.725, -3.5) {1};
		\node [style=none] (84) at (16.725, -3.5) {2};
		\node [style=none] (85) at (12.625, -2.5) {$k+1$};
		\node [style=none] (86) at (11.625, -3.5) {\textcolor{red}{1}};
		\node [style=none] (87) at (13.875, -2.875) {};
		\node [style=none] (88) at (13.875, -3.125) {};
		\node [style=none] (89) at (12.625, -2.875) {};
		\node [style=none] (90) at (12.625, -3.125) {};
		\node [style=none] (91) at (13.375, -2.625) {};
		\node [style=none] (92) at (13.375, -3.375) {};
		\node [style=none] (93) at (13.025, -3) {};
		\node [style=none] (94) at (9.5, 0) {$\leftrightarrow$};
		\node [style=none] (95) at (6.5, 0) {};
		\node [style=none] (96) at (5.5, 0) {};
		\node [style=none] (97) at (15.25, -3) {};
		\node [style=none] (98) at (16.25, -3) {};
	\end{pgfonlayer}
	\begin{pgfonlayer}{edgelayer}
		\draw (0) to (1);
		\draw (2) to (3);
		\draw (27) to (28);
		\draw (3) to (35);
		\draw (41.center) to (43.center);
		\draw (42.center) to (44.center);
		\draw (47.center) to (45.center);
		\draw (47.center) to (46.center);
		\draw (48) to (49);
		\draw (50) to (51);
		\draw (59) to (58);
		\draw (58) to (60);
		\draw (49) to (63.center);
		\draw (64.center) to (59);
		\draw (60) to (65.center);
		\draw (66.center) to (50);
		\draw (68) to (58);
		\draw (74) to (75);
		\draw (76) to (77);
		\draw (81) to (82);
		\draw (87.center) to (89.center);
		\draw (88.center) to (90.center);
		\draw (93.center) to (91.center);
		\draw (93.center) to (92.center);
		\draw (35) to (96.center);
		\draw (95.center) to (27);
		\draw (98.center) to (81);
		\draw (77) to (97.center);
	\end{pgfonlayer}
\end{tikzpicture}}
}
\label{E6F4family}
\end{equation}
due to their Coulomb branches. This proposal receives further validation from the explicit Hilbert series results of Table \ref{E6F4family1}.
For $k=1$, the Coulomb branch is  $\overline{\mathcal{O}}^{\mathfrak{e}_6}_{\text{min}}\times \overline{\mathcal{O}}^{\mathfrak{f}_4}_{\text{min}}$ and, hence, the namesake for the family. For $k=0$, the quiver is a free theory whose Coulomb branch is $\mathbb{H}^4\times \mathbb{H}^2 \cong \mathbb{H}^6$. For $k>1$, the Coulomb branch global symmetry is $\sorm(4k+6)\times SO(4k+5)\times \urm(1)$. 
\subsubsection{\texorpdfstring{$E_6\times \sorm(9)$ family}{E6 x  SO9 family}}
Considering the $E_6\times \sorm(10) $ family \eqref{weirder} and folding it yields the following equivalences of Coulomb branches:
\begin{equation}
\scalebox{0.7}{\raisebox{-.5\height}{\begin{tikzpicture}
	\begin{pgfonlayer}{nodelayer}
		\node [style=miniBlue] (0) at (6.5, 0) {};
		\node [style=none] (1) at (8.5, 0) {$\dots$};
		\node [style=miniU] (2) at (9.5, 0) {};
		\node [style=miniBlue] (3) at (10.5, 0) {};
		\node [style=miniU] (4) at (11.5, 0) {};
		\node [style=none] (5) at (9.5, -0.5) {4};
		\node [style=none] (6) at (10.5, -0.5) {2};
		\node [style=none] (7) at (11.5, -0.5) {2};
		\node [style=none] (8) at (6.25, -0.75) {};
		\node [style=none] (9) at (6.5, -0.5) {$4k{+}2l$};
		\node [style=none] (10) at (6.5, 1.5) {\textcolor{red}{2}};
		\node [style=miniU] (11) at (5.5, 0) {};
		\node [style=miniBlue] (12) at (4.5, 0) {};
		\node [style=none] (13) at (5.5, 0.5) {$4k{+}2l$};
		\node [style=none] (14) at (4.5, -0.5) {$4k{+}2l-{2}$};
		\node [style=none] (15) at (3.5, 0) {$\dots$};
		\node [style=miniBlue] (16) at (2.5, 0) {};
		\node [style=none] (17) at (2.5, -0.5) {$4k{+}2$};
		\node [style=none] (18) at (7.5, 0.5) {$4k{+}2l$};
		\node [style=miniU] (19) at (7.5, 0) {};
		\node [style=miniU] (20) at (6.5, 1) {};
		\node [style=miniU] (21) at (1.25, 0) {};
		\node [style=none] (22) at (1.25, -0.5) {$4k{+}2$};
		\node [style=miniBlue] (23) at (-0.25, 0) {};
		\node [style=miniU] (24) at (1.25, 0) {};
		\node [style=none] (25) at (-0.25, -0.5) {$2k$};
		\node [style=none] (26) at (-0.25, 0.125) {};
		\node [style=none] (27) at (-0.25, -0.125) {};
		\node [style=none] (28) at (1.25, 0.125) {};
		\node [style=none] (29) at (1.25, -0.125) {};
		\node [style=none] (30) at (0.75, 0.375) {};
		\node [style=none] (31) at (0.75, -0.375) {};
		\node [style=none] (32) at (0.375, 0) {};
		\node [style=none] (33) at (12.5, 0) {$\leftrightarrow$};
		\node [style=miniU] (34) at (13.75, 0) {};
		\node [style=miniBlue] (35) at (14.75, 0) {};
		\node [style=miniBlue] (36) at (20.125, 0) {};
		\node [style=miniU] (37) at (21.125, 0) {};
		\node [style=none] (38) at (13.75, -0.5) {2};
		\node [style=none] (39) at (14.75, -0.5) {2};
		\node [style=none] (40) at (17.45, -0.5) {\small$2k{+}2l$};
		\node [style=none] (41) at (20.125, -0.5) {2};
		\node [style=none] (42) at (21.125, -0.5) {2};
		\node [style=miniU] (43) at (17.45, 0) {};
		\node [style=miniBlue] (44) at (16.3, 0) {};
		\node [style=miniBlue] (45) at (18.6, 0) {};
		\node [style=none] (46) at (15.55, 0) {$\cdots$};
		\node [style=none] (47) at (19.35, 0) {$\cdots$};
		\node [style=none] (48) at (15.2, 0) {};
		\node [style=none] (49) at (15.95, 0) {};
		\node [style=none] (50) at (18.95, 0) {};
		\node [style=none] (51) at (19.7, 0) {};
		\node [style=gauge3] (52) at (17.45, 1) {};
		\node [style=none] (53) at (17.45, 1.5) {\textcolor{red}{1}};
		\node [style=miniBlue] (54) at (17.45, 0) {};
		\node [style=miniU] (55) at (18.6, 0) {};
		\node [style=miniU] (56) at (16.3, 0) {};
		\node [style=none] (57) at (12.2, -4.25) {$\times$};
		\node [style=none] (58) at (18.625, 0.5) {\small$2k{+}2l$};
		\node [style=none] (59) at (16.375, 0.5) {\small$2k{+}2l$};
		\node [style=gauge3] (60) at (13.375, -4.25) {};
		\node [style=gauge3] (61) at (14.475, -4.25) {};
		\node [style=gauge3] (62) at (15.475, -4.25) {};
		\node [style=none] (63) at (14.475, -4.75) {$2k$};
		\node [style=none] (64) at (15.475, -4.75) {$2k$};
		\node [style=none] (65) at (19.225, -4.25) {$\dots$};
		\node [style=gauge3] (66) at (20.225, -4.25) {};
		\node [style=none] (67) at (21.225, -4.75) {$1$};
		\node [style=none] (68) at (20.225, -4.75) {$2$};
		\node [style=none] (69) at (13.375, -4.75) {$k$};
		\node [style=none] (70) at (14.625, -4.125) {};
		\node [style=none] (71) at (14.625, -4.375) {};
		\node [style=none] (72) at (13.375, -4.125) {};
		\node [style=none] (73) at (13.375, -4.375) {};
		\node [style=none] (74) at (14.125, -3.875) {};
		\node [style=none] (75) at (14.125, -4.625) {};
		\node [style=none] (76) at (13.775, -4.25) {};
		\node [style=gauge3] (77) at (21.25, -4.25) {};
		\node [style=gauge3] (78) at (17.25, -4.25) {};
		\node [style=gauge3] (79) at (18.25, -4.25) {};
		\node [style=none] (80) at (16.375, -4.25) {$\dots$};
		\node [style=none] (81) at (17.25, -4.75) {$2k$};
		\node [style=none] (82) at (18.25, -4.75) {$2k-1$};
		\node [style=blankflavor] (83) at (17.25, -3.25) {};
		\node [style=none] (84) at (17.25, -2.75) {1};
		\node [style=none] (85) at (14.5, -5) {};
		\node [style=none] (86) at (17.5, -5) {};
		\node [style=none] (87) at (16.575, -5.5) {$2l$};
		\node [style=none] (88) at (3, 0) {};
		\node [style=none] (89) at (4, 0) {};
		\node [style=none] (90) at (9, 0) {};
		\node [style=none] (91) at (8, 0) {};
		\node [style=none] (92) at (16.75, -4.25) {};
		\node [style=none] (93) at (16, -4.25) {};
		\node [style=none] (94) at (19.75, -4.25) {};
		\node [style=none] (95) at (18.75, -4.25) {};
	\end{pgfonlayer}
	\begin{pgfonlayer}{edgelayer}
		\draw (2) to (4);
		\draw (12) to (11);
		\draw (11) to (19);
		\draw (20) to (0);
		\draw (26.center) to (28.center);
		\draw (29.center) to (27.center);
		\draw (30.center) to (32.center);
		\draw (32.center) to (31.center);
		\draw (16) to (24);
		\draw (34) to (35);
		\draw (36) to (37);
		\draw (44) to (43);
		\draw (43) to (45);
		\draw (35) to (48.center);
		\draw (49.center) to (44);
		\draw (45) to (50.center);
		\draw (51.center) to (36);
		\draw (52) to (43);
		\draw (61) to (62);
		\draw (70.center) to (72.center);
		\draw (71.center) to (73.center);
		\draw (76.center) to (74.center);
		\draw (76.center) to (75.center);
		\draw (83) to (78);
		\draw [style=brace] (86.center) to (85.center);
		\draw (77) to (66);
		\draw (79) to (78);
		\draw (89.center) to (12);
		\draw (88.center) to (16);
		\draw (90.center) to (2);
		\draw (19) to (91.center);
		\draw (62) to (93.center);
		\draw (92.center) to (78);
		\draw (79) to (95.center);
		\draw (94.center) to (66);
	\end{pgfonlayer}
\end{tikzpicture}}
\label{foldedannoyingquiver}
}
\end{equation}
where the quiver on the left-hand side results from folding \eqref{weirder}, and the right-hand side quivers are the $E_6$ family (top) and the unitary non-simply laced quiver (bottom) whose Coulomb branch is $\overline{\mathcal{O}}^{\mathfrak{so}(4k+4l+1)}_{(2^{2k},1^{4l+1})}$.
The explicit Hilbert series computations summarised in Table \ref{foldedannoyingquiver1} provide further evidence for this proposal.
For $k=1$, $l=1$, the Coulomb branch is  $\overline{\mathcal{O}}^{\mathfrak{e}_6}_{\text{min}}\times \overline{\mathcal{O}}^{\mathfrak{so}(9)}_{\text{min}}$ and hence the namesake for the family. For $k>1$, the Coulomb branch global symmetry is $\sorm(4k+4l+2)\times \sorm(4k+4l+1)\times \urm(1)$. 
\subsubsection{\texorpdfstring{$E_8 \times \sorm(15)$ family}{E8 x SO15 family}}
Similarly, folding \eqref{betterbelastguy} gives the following correspondence:
\begin{equation}
\scalebox{0.7}{\raisebox{-.5\height}{\begin{tikzpicture}
	\begin{pgfonlayer}{nodelayer}
		\node [style=none] (0) at (10.45, -4) {$\times$};
		\node [style=none] (1) at (11.5, 0) {$\leftrightarrow$};
		\node [style=miniU] (31) at (2.375, 0) {};
		\node [style=miniBlue] (32) at (3.375, 0) {};
		\node [style=none] (33) at (4.375, 0) {$\dots$};
		\node [style=miniBlue] (34) at (5.375, 0) {};
		\node [style=miniU] (35) at (6.375, 0) {};
		\node [style=miniBlue] (36) at (7.375, 0) {};
		\node [style=none] (37) at (2.375, 0.5) {$4k{+}2$};
		\node [style=none] (38) at (3.375, -0.5) {$4k{+}2$};
		\node [style=none] (39) at (5.375, -0.5) {$4k{+}2l{-}2$};
		\node [style=none] (40) at (3.875, 0) {};
		\node [style=none] (41) at (4.875, 0) {};
		\node [style=none] (42) at (6.375, -1) {$4k{+}2l$};
		\node [style=none] (43) at (7.375, -0.5) {$4k{+}2l{-}2$};
		\node [style=none] (44) at (6.375, 1.5) {\textcolor{red}{2}};
		\node [style=none] (45) at (7.875, 0) {};
		\node [style=none] (46) at (8.375, 0) {$\dots$};
		\node [style=none] (47) at (8.875, 0) {};
		\node [style=bluegauge] (48) at (9.375, 0) {};
		\node [style=redgauge] (49) at (10.375, 0) {};
		\node [style=none] (50) at (10.375, -0.5) {2};
		\node [style=none] (51) at (9.375, -0.5) {2};
		\node [style=miniBlue] (52) at (1.125, 0) {};
		\node [style=miniU] (53) at (2.375, 0) {};
		\node [style=none] (54) at (2.375, 0.125) {};
		\node [style=none] (55) at (2.375, -0.125) {};
		\node [style=none] (56) at (1.125, 0.125) {};
		\node [style=none] (57) at (1.125, -0.125) {};
		\node [style=none] (58) at (1.875, 0.375) {};
		\node [style=none] (59) at (1.875, -0.375) {};
		\node [style=none] (60) at (1.525, 0) {};
		\node [style=none] (61) at (1.125, -0.5) {2k};
		\node [style=gauge3] (62) at (13.225, -4) {};
		\node [style=gauge3] (63) at (14.225, -4) {};
		\node [style=none] (64) at (13.225, -4.5) {$2k$};
		\node [style=none] (65) at (14.225, -4.5) {$2k$};
		\node [style=none] (66) at (18.225, -4) {$\dots$};
		\node [style=gauge3] (67) at (19.225, -4) {};
		\node [style=none] (68) at (20.225, -4.5) {$1$};
		\node [style=none] (69) at (19.225, -4.5) {$2$};
		\node [style=gauge3] (70) at (20.25, -4) {};
		\node [style=gauge3] (71) at (16.25, -4) {};
		\node [style=gauge3] (72) at (17.25, -4) {};
		\node [style=none] (73) at (15.25, -4) {$\dots$};
		\node [style=none] (74) at (16.25, -4.5) {$2k$};
		\node [style=none] (75) at (17.25, -4.5) {$2k-1$};
		\node [style=blankflavor] (76) at (16.25, -3) {};
		\node [style=none] (77) at (16.25, -2.5) {1};
		\node [style=none] (78) at (13.25, -4.75) {};
		\node [style=none] (79) at (16.5, -4.75) {};
		\node [style=none] (80) at (15.075, -5.5) {$2l{-}1$};
		\node [style=none] (81) at (14.75, -4) {};
		\node [style=none] (82) at (15.75, -4) {};
		\node [style=none] (83) at (18.75, -4) {};
		\node [style=none] (84) at (17.75, -4) {};
		\node [style=gauge3] (85) at (11.85, -4) {};
		\node [style=none] (86) at (11.85, -4.5) {$k$};
		\node [style=none] (87) at (13.1, -3.875) {};
		\node [style=none] (88) at (13.1, -4.125) {};
		\node [style=none] (89) at (11.85, -3.875) {};
		\node [style=none] (90) at (11.85, -4.125) {};
		\node [style=none] (91) at (12.6, -3.625) {};
		\node [style=none] (92) at (12.6, -4.375) {};
		\node [style=none] (93) at (12.25, -4) {};
		\node [style=gauge3] (94) at (13.225, -4) {};
		\node [style=bluegauge] (95) at (6.375, 1) {};
		\node [style=miniU] (96) at (12.5, 0) {};
		\node [style=miniBlue] (97) at (13.5, 0) {};
		\node [style=miniBlue] (98) at (18.875, 0) {};
		\node [style=miniU] (99) at (19.875, 0) {};
		\node [style=none] (100) at (12.5, -0.5) {2};
		\node [style=none] (101) at (13.5, -0.5) {2};
		\node [style=none] (102) at (16.2, -0.5) {\small$2k{+}2l$};
		\node [style=none] (103) at (18.875, -0.5) {2};
		\node [style=none] (104) at (19.875, -0.5) {2};
		\node [style=miniU] (105) at (16.2, 0) {};
		\node [style=miniBlue] (106) at (15.05, 0) {};
		\node [style=miniBlue] (107) at (17.35, 0) {};
		\node [style=none] (108) at (14.3, 0) {$\cdots$};
		\node [style=none] (109) at (18.1, 0) {$\cdots$};
		\node [style=none] (110) at (13.95, 0) {};
		\node [style=none] (111) at (14.7, 0) {};
		\node [style=none] (112) at (17.7, 0) {};
		\node [style=none] (113) at (18.45, 0) {};
		\node [style=gauge3] (114) at (16.2, 1) {};
		\node [style=none] (115) at (16.2, 1.5) {\textcolor{red}{2}};
		\node [style=miniBlue] (116) at (16.2, 0) {};
		\node [style=miniU] (117) at (17.35, 0) {};
		\node [style=miniU] (118) at (15.05, 0) {};
		\node [style=none] (119) at (17.375, 0.5) {\small$2k{+}2l{-}2$};
		\node [style=none] (120) at (15.125, 0.5) {\small$2k{+}2l{-}2$};
		\node [style=miniU] (121) at (16.2, 0) {};
		\node [style=miniBlue] (122) at (15.05, 0) {};
		\node [style=miniBlue] (123) at (17.35, 0) {};
		\node [style=miniBlue] (124) at (16.2, 1) {};
	\end{pgfonlayer}
	\begin{pgfonlayer}{edgelayer}
		\draw (31) to (32);
		\draw (35) to (34);
		\draw (32) to (40.center);
		\draw (41.center) to (34);
		\draw (35) to (36);
		\draw (49) to (48);
		\draw (45.center) to (36);
		\draw (47.center) to (48);
		\draw (54.center) to (56.center);
		\draw (55.center) to (57.center);
		\draw (60.center) to (58.center);
		\draw (60.center) to (59.center);
		\draw (62) to (63);
		\draw (76) to (71);
		\draw [style=brace] (79.center) to (78.center);
		\draw (70) to (67);
		\draw (72) to (71);
		\draw (82.center) to (71);
		\draw (81.center) to (63);
		\draw (84.center) to (72);
		\draw (83.center) to (67);
		\draw (87.center) to (89.center);
		\draw (88.center) to (90.center);
		\draw (93.center) to (91.center);
		\draw (93.center) to (92.center);
		\draw (95) to (35);
		\draw (96) to (97);
		\draw (98) to (99);
		\draw (106) to (105);
		\draw (105) to (107);
		\draw (97) to (110.center);
		\draw (111.center) to (106);
		\draw (107) to (112.center);
		\draw (113.center) to (98);
		\draw (114) to (105);
	\end{pgfonlayer}
\end{tikzpicture}
}}
\label{foldedfinaleguy}
\end{equation}
which has been tested by Hilbert series methods, see Table \ref{foldedfinaleguy1} for details. The top quiver on the right-hand side is the $E_8$ family and the bottom quiver has Coulomb branch $\overline{\mathcal{O}}^{\mathfrak{so}(4k+4l-1)}_{(2^{2k},1^{4l-1})}$. 
For $k=1,l=1$, the orthosymplectic quiver on the left-hand side has divergent Coulomb branch Hilbert series and is \textit{not} a $B$-type orthosymplectic quiver since there is an additional $\usprm(2)$ node that is balanced. The resulting balanced set of nodes does not give a finite Dynkin diagram. For $k=2$, $l=1$, the Coulomb branch of the top quiver on the right-hand side is $\mathbb{H}^{16}$ and the bottom is $\overline{\mathcal{O}}^{\mathfrak{so}(11)}_{(2^{4},1^{3})}$. For $k=3$, $l=1$, the Coulomb branch is  $\overline{\mathcal{O}}^{\mathfrak{e}_8}_{\text{min}}\times \overline{\mathcal{O}}^{\mathfrak{so}(15)}_{(2^{6},1^{3})}$ and hence the namesake of the family. For $k>3$, the Coulomb branch global symmetry is $\sorm(4k+4l)\times \sorm(4k+4l-1)$.

Note that for the $E_8 \times \sorm(16)$ and $E_8 \times \sorm(15)$ families, the difference before and after folding is that the spinor nodes of the unitary quivers whose Coulomb branch is $\overline{\mathcal{O}}^{\mathfrak{so}(4k+4l)}_{(2^{2k},1^{4l})}$ are folded as well to become $\overline{\mathcal{O}}^{\mathfrak{so}(4k+4l-1)}_{(2^{2k},1^{4l-1})}$.
\subsubsection{\texorpdfstring{$F_4 \times \sorm(9)$ family}{F4 x SO9 family}}
Folding the $F_4 \times F_4$ family \eqref{f4f4} results again in a product moduli space, which is realised by the following quivers:
\begin{equation}
\scalebox{0.8}{
\raisebox{-.5\height}{
\begin{tikzpicture}
	\begin{pgfonlayer}{nodelayer}
		\node [style=bluegauge] (1) at (8.5, 0) {};
		\node [style=redgauge] (10) at (10, 0) {};
		\node [style=none] (19) at (10, -0.5) {$4k+2$};
		\node [style=none] (20) at (11, 0.5) {$4k+2$};
		\node [style=none] (21) at (12, -0.5) {$4k$};
		\node [style=none] (23) at (14, -0.5) {2};
		\node [style=bluegauge] (25) at (11, 0) {};
		\node [style=redgauge] (26) at (12, 0) {};
		\node [style=redgauge] (28) at (14, 0) {};
		\node [style=none] (32) at (8.5, -0.5) {$2k+2$};
		\node [style=bluegauge] (38) at (10, 0) {};
		\node [style=bluegauge] (39) at (12, 0) {};
		\node [style=redgauge] (40) at (11, 0) {};
		\node [style=bluegauge] (42) at (14, 0) {};
		\node [style=redgauge] (43) at (15, 0) {};
		\node [style=none] (44) at (15, -0.5) {2};
		\node [style=redgauge] (46) at (8.5, 0) {};
		\node [style=none] (48) at (13, 0) {$\dots$};
		\node [style=none] (49) at (9.75, 0.125) {};
		\node [style=none] (50) at (9.75, -0.125) {};
		\node [style=none] (51) at (8.5, 0.125) {};
		\node [style=none] (52) at (8.5, -0.125) {};
		\node [style=none] (53) at (9.25, 0.375) {};
		\node [style=none] (54) at (9.25, -0.375) {};
		\node [style=none] (55) at (8.9, 0) {};
		\node [style=none] (104) at (15.75, -3.75) {$\times$};
		\node [style=gauge3] (105) at (18.125, 0) {};
		\node [style=gauge3] (106) at (19.125, 0) {};
		\node [style=gauge3] (107) at (20.225, 0) {};
		\node [style=gauge3] (108) at (21.225, 0) {};
		\node [style=none] (109) at (20.225, -0.5) {$2k+1$};
		\node [style=none] (110) at (21.225, 0.5) {$2k$};
		\node [style=none] (111) at (22.225, 0) {$\dots$};
		\node [style=gauge3] (112) at (23.225, 0) {};
		\node [style=none] (114) at (24.225, -0.5) {1};
		\node [style=none] (115) at (23.225, -0.5) {2};
		\node [style=none] (116) at (19.125, 0.5) {$k+1$};
		\node [style=none] (117) at (18.125, -0.5) {\textcolor{red}{1}};
		\node [style=none] (118) at (20.375, 0.125) {};
		\node [style=none] (119) at (20.375, -0.125) {};
		\node [style=none] (120) at (19.125, 0.125) {};
		\node [style=none] (121) at (19.125, -0.125) {};
		\node [style=none] (122) at (19.875, 0.375) {};
		\node [style=none] (123) at (19.875, -0.375) {};
		\node [style=none] (124) at (19.525, 0) {};
		\node [style=none] (125) at (16.75, 0) {$\leftrightarrow$};
		\node [style=gauge3] (127) at (17.65, -3.75) {};
		\node [style=gauge3] (128) at (18.75, -3.75) {};
		\node [style=gauge3] (129) at (19.75, -3.75) {};
		\node [style=none] (130) at (18.75, -3.25) {$k+1$};
		\node [style=none] (131) at (19.75, -4.25) {$2k$};
		\node [style=none] (132) at (21.75, -3.75) {$\dots$};
		\node [style=gauge3] (133) at (22.75, -3.75) {};
		\node [style=none] (135) at (23.75, -4.25) {1};
		\node [style=none] (136) at (22.75, -4.25) {2};
		\node [style=none] (137) at (17.65, -4.25) {\textcolor{red}{1}};
		\node [style=none] (139) at (18.9, -3.625) {};
		\node [style=none] (140) at (18.9, -3.875) {};
		\node [style=none] (141) at (17.65, -3.625) {};
		\node [style=none] (142) at (17.65, -3.875) {};
		\node [style=none] (143) at (18.4, -3.375) {};
		\node [style=none] (144) at (18.4, -4.125) {};
		\node [style=none] (145) at (18.05, -3.75) {};
		\node [style=gauge3] (146) at (19.75, -2.75) {};
		\node [style=none] (147) at (20.225, -2.75) {$k$};
		\node [style=gauge3] (148) at (20.75, -3.75) {};
		\node [style=none] (149) at (20.725, -4.25) {$2k-1$};
		\node [style=flavor2] (150) at (24.25, 0) {};
		\node [style=flavor2] (151) at (23.75, -3.75) {};
		\node [style=none] (152) at (12.5, 0) {};
		\node [style=none] (153) at (13.5, 0) {};
		\node [style=none] (154) at (21.75, 0) {};
		\node [style=none] (155) at (22.75, 0) {};
		\node [style=none] (156) at (22.25, -3.75) {};
		\node [style=none] (157) at (21.25, -3.75) {};
	\end{pgfonlayer}
	\begin{pgfonlayer}{edgelayer}
		\draw (42) to (43);
		\draw (38) to (39);
		\draw (49.center) to (51.center);
		\draw (50.center) to (52.center);
		\draw (55.center) to (53.center);
		\draw (55.center) to (54.center);
		\draw (105) to (106);
		\draw (107) to (108);
		\draw (118.center) to (120.center);
		\draw (119.center) to (121.center);
		\draw (124.center) to (122.center);
		\draw (124.center) to (123.center);
		\draw (128) to (129);
		\draw (139.center) to (141.center);
		\draw (140.center) to (142.center);
		\draw (145.center) to (143.center);
		\draw (145.center) to (144.center);
		\draw (146) to (129);
		\draw (129) to (148);
		\draw (150) to (112);
		\draw (133) to (151);
		\draw (39) to (152.center);
		\draw (153.center) to (42);
		\draw (108) to (154.center);
		\draw (155.center) to (112);
		\draw (148) to (157.center);
		\draw (156.center) to (133);
	\end{pgfonlayer}
\end{tikzpicture}}}
\label{F4SO9family}
\end{equation}
where the quiver on the left-hand side is the result of folding \eqref{f4f4}; while the quivers on the right-hand side have suitable Coulomb branches that agree with the product. This has been verified by Hilbert series computations, see Table \ref{F4SO9family1}. 
For $k=0$, the quiver is a free theory whose Coulomb branch is $\mathbb{H}^3$.
For $k=1$, the Coulomb branch is $\overline{\mathcal{O}}^{\mathfrak{f}_4}_{\text{min}}\times \overline{\mathcal{O}}^{\mathfrak{so}_9}_{\text{min}}$ and, hence, the namesake. For $k>1$, the Coulomb branch global symmetry is $\sorm(4k+5)\times \sorm(4k+4)$.  The HWG of 
\eqref{F4SO9family} is then conjectured to be the product of the HWGs for the two unitary quivers on the right-hand side:
\begin{equation}
    \HWG\eqref{F4SO9family}(t;\mu_i,\rho_i)=
    \PE\left[\sum_{i=1}^k\mu_{2i}t^{2i}+\mu_{2k+2}t^{k+1}\right]
    \cdot
    \PE\left[\sum_{i=1}^k\rho_{2i}t^{2i}+\rho_{2k+2}t^{k+1}\right]
\end{equation}
where $\mu_i$ and $\rho_i$ are the highest weight fugacities of $\sorm(4k+5)\times\sorm(4k+4) $ which is the Coulomb branch global symmetry group of the orthosymplectic quiver. Inside the product, the term of the left is the HWG for the top unitary quiver whereas the term on the right is the HWG for the bottom unitary quiver.  

Here, there is an interesting observation: the top unitary quiver can be obtained by folding the quiver in  \cite[Tab.\ 17]{Cabrera:2018jxt} which is the magnetic quiver corresponding to the 5d $\Ncal=1$ $\surm(N_c)$ theory  with $N_f$ even flavours and CS-level $|k|=2=N_c-N_f/2 +2$ at infinite gauge coupling. The bottom unitary quiver can be obtained by folding the quiver in \cite[Tab.\ 17]{Cabrera:2018jxt} which is the magnetic quiver corresponding to the 5d $\Ncal=1$ $\surm(N_c)$ theory with $N_f$ odd flavours and CS-level $|k|=3/2=N_c-N_f/2 +2$ at infinite gauge coupling. 
\subsubsection{\texorpdfstring{$\sorm(\mathrm{even})\times \sorm(\mathrm{odd})$ family}{SO(odd)xSO(odd)}}
Folding the $\sorm(\mathrm{odd})\times \sorm(\mathrm{odd})$ family \eqref{soddxsodd} results in:
\begin{equation}
\raisebox{-.5\height}{
\begin{tikzpicture}
	\begin{pgfonlayer}{nodelayer}
		\node [style=bluegauge] (152) at (13, -2.25) {};
		\node [style=redgauge] (153) at (14.25, -2.25) {};
		\node [style=none] (154) at (14.25, -2.75) {$4k$};
		\node [style=none] (155) at (15.25, -1.75) {$4k$};
		\node [style=none] (156) at (16.25, -2.75) {$4k-2$};
		\node [style=none] (157) at (18.25, -2.75) {2};
		\node [style=bluegauge] (158) at (15.25, -2.25) {};
		\node [style=redgauge] (159) at (16.25, -2.25) {};
		\node [style=redgauge] (160) at (18.25, -2.25) {};
		\node [style=none] (162) at (13, -2.75) {$2k+1$};
		\node [style=bluegauge] (163) at (14.25, -2.25) {};
		\node [style=bluegauge] (164) at (16.25, -2.25) {};
		\node [style=redgauge] (165) at (15.25, -2.25) {};
		\node [style=bluegauge] (166) at (18.25, -2.25) {};
		\node [style=redgauge] (167) at (19.25, -2.25) {};
		\node [style=none] (168) at (19.25, -2.75) {2};
		\node [style=redgauge] (170) at (13, -2.25) {};
		\node [style=none] (172) at (17.25, -2.25) {$\dots$};
		\node [style=none] (173) at (14.25, -2.125) {};
		\node [style=none] (174) at (14.25, -2.375) {};
		\node [style=none] (175) at (13, -2.125) {};
		\node [style=none] (176) at (13, -2.375) {};
		\node [style=none] (177) at (13.75, -1.875) {};
		\node [style=none] (178) at (13.75, -2.625) {};
		\node [style=none] (179) at (13.4, -2.25) {};
		\node [style=none] (180) at (16.75, -2.25) {};
		\node [style=none] (181) at (17.75, -2.25) {};
	\end{pgfonlayer}
	\begin{pgfonlayer}{edgelayer}
		\draw (166) to (167);
		\draw (163) to (164);
		\draw (173.center) to (175.center);
		\draw (174.center) to (176.center);
		\draw (179.center) to (177.center);
		\draw (179.center) to (178.center);
		\draw (164) to (180.center);
		\draw (181.center) to (166);
	\end{pgfonlayer}
\end{tikzpicture}
}
\label{soddfold}
\end{equation}
and the associated Coulomb branches are again products of nilpotent orbit closures:
\begin{equation}
  \mathcal{C}\eqref{soddfold} = \overline{\mathcal{O}}^{\mathfrak{so}_{4k+2}}_{(2^{2k},1^2)}\times \overline{\mathcal{O}}^{\mathfrak{so}_{4k+3}}_{(2^{2k},1^3)} \,,
\end{equation}
which can be verified by the Hilbert series results given in Table \ref{soddfold1}. 
One can understand \eqref{soddfold} as magnetic quiver for the product of two classical Higgs branches of SQCD theories:
\begin{equation}
\raisebox{-.5\height}{
   \begin{tikzpicture}
	\begin{pgfonlayer}{nodelayer}
		\node [style=miniBlue] (0) at (0, 0) {};
		\node [style=flavourRed] (1) at (0, 1.5) {};
		\node [style=none] (2) at (0, -0.5) {$\usprm(2k)$};
		\node [style=none] (3) at (0, 2) {$\sorm(4k+2)$};
		\node [style=none] (4) at (1, 0.75) {$\times$};
		\node [style=miniBlue] (5) at (2, 0) {};
		\node [style=flavourRed] (6) at (2, 1.5) {};
		\node [style=none] (7) at (2, -0.5) {$\usprm(2k)$};
		\node [style=none] (8) at (2, 2) {$\sorm(4k+3)$};
	\end{pgfonlayer}
	\begin{pgfonlayer}{edgelayer}
		\draw (1) to (0);
		\draw (6) to (5);
	\end{pgfonlayer}
\end{tikzpicture} 
}
\end{equation}
where one theory has a half-hypermultiplet more than the other. 

The findings of this section can be condensed into 
\begin{tcolorbox}
\begin{myrule} \label{rule:unframed_B}
The $E_6 \times F_4$, $E_{6} \times \sorm(9)$, $E_8 \times \sorm(15)$, $F_4 \times \sorm(9)$, and $\sorm(\mathrm{even}) \times \sorm(\mathrm{odd})$ families include all $B$-type \emph{unframed} orthosymplectic quivers which satisfy:
\begin{compactitem}
    \item Composed of only $\sorm(\mathrm{odd})$, $\sorm(\mathrm{even})$, $\usprm(\mathrm{even})$, $\urm(1)$ gauge groups.
    \item There is a $B$-type non-simply laced edge connected by two balanced nodes. 
    \item The long end (tail) of the quiver (in the sense of the simple roots of $B$-type Lie algebras) begins with $\sorm(2)$. 
    \item All overbalanced gauge groups that are not part of the long tail or the non-simply laced edge are necessary in order to balance B-type quiver.
\end{compactitem}
\end{myrule}
\end{tcolorbox}
\noindent
The last point prevents the possibility of connecting an arbitrary number of gauge nodes to the overbalanced nodes in the quiver. 
\section{Derivation from Type II brane configurations}
\label{branes}
3d $\Ncal=4$ quiver gauge theories and brane configurations enjoy an intricate relationship. In this section, the orthosymplectic quivers derived above are constructed either as 3d theories via the Type IIB configurations of \cite{Hanany:1996ie} or as magnetic quivers for 5d or 6d brane configurations, along the lines of \cite{Cabrera:2018jxt,Cabrera:2019izd,Cabrera:2019dob,Bourget:2020gzi,Akhond:2020vhc,Akhond:2021knl}. Appendix \ref{app:branes} provides background material.

\begin{table}[t]
\centering
\begin{tabular}{c|cccccccccc}
\toprule
     IIB & $x^0$ & $x^1$& $x^2$& $x^3$& $x^4$& $x^5$& $x^6$& $x^7$& $x^8$& $x^9$ \\ \midrule
    NS5/ON & $\times$ & $\times$& $\times$& $\times$& $\times$& $\times$ & & & & \\
    D3/O3 & $\times$ & $\times$& $\times$& & &  & $\times$ & & & \\
    D5/O5 & $\times$ & $\times$& $\times$ & & & & & $\times$& $\times$& $\times$ \\  \bottomrule 
\end{tabular}
\caption{The D3-D5-NS5 brane configuration for 3d $\Ncal=4$ theories. The $2+1$ dimensional world-volume theory is realised on the D3 branes. The various orientifolds O3, O5, NS5 are introduced parallel to the respective D3, D5, NS5 brane, i.e.\ the ON may loosely speaking be considered as orientifold for NS branes.}
\label{tab:branes}
\end{table}

\subsection{3d brane systems}
\label{sec:branes_3d}
Starting from the Type IIB brane setup introduced in \cite{Hanany:1996ie}, the resulting classes of quiver gauge theories can be enriched by inclusion of orientifolds and orbifold planes. There are three types to consider: O3, O5, and ON planes, and the reader is referred to \cite{Uranga:1998uj,Gimon:1996rq,Hanany:1997gh,Kapustin:1998fa,Hanany:1999sj,Feng:2000eq,Gaiotto:2008ak} for details.
\subsubsection{\texorpdfstring{A first example: ON${}^0$ planes}{A first example: ON planes}}
It is well-appreciated that $D$-type Dynkin quivers can be realised by inclusion of  ON${}^0$ planes \cite{Kapustin:1998fa}. Having in mind product theories, one can consider a degenerate example of a $D$-type Dynkin quiver as follows:
\begin{align}
\raisebox{-.5\height}{
    \begin{tikzpicture}
    \draw (-1,0.1)--(2,0.1) (-1,-0.1)--(2,-0.1);
    \draw (1,-1)--(1,1) (1.1,-1)--(1.1,1) (1.7,-1)--(1.7,1);
    \node at (1.4,0.5) {$\cdots$};
    \node at (1.4,-0.5) {$\cdots$};
        \ns{-1,0}
        \onz{2,0}
        \draw[decoration={brace,mirror,raise=10pt},decorate,thick]
  (0.9,-0.8) -- node[below=15pt] {\footnotesize{$N_f$ D5}} (1.8,-0.8);
  \node at (0,0.3) {\footnotesize{$2N$ D3}};
  \node at (-0.75,-0.35) {\footnotesize{NS5}};
  \node at (2.25,-0.35) {\footnotesize{ON${}^0$}};
    \end{tikzpicture}
    }
    \qquad 
        \raisebox{-.5\height}{
    \begin{tikzpicture}
	\node (g1) [gauge,label=below:{\footnotesize{$N$}}] {};
	\node (f1) [flavour,above of=g1, label=above:{\footnotesize{$N_f$}}] {};
    \node at (1,0.5) {$\times$};
	\node (g2) [gauge,label=below:{\footnotesize{$N$}}] at (2,0) {};
	\node (f2) [flavour,above of=g2, label=above:{\footnotesize{$N_f$}}] {};
	\draw (g1)--(f1) (g2)--(f2);
	\end{tikzpicture}
    }
    \label{eq:ex_product}
\end{align}
for which the electric theory is given by two copies of $\urm(N)$ SQCD with $N_f$ fundamental flavours. For the SQCD theories to be \emph{good}, the number of flavours and colours are constraint by $N_f\geq 2N$. Next, one can derive the 3d mirror from the brane configuration. After  moving a sufficient number of D5 branes through the left NS5 brane and taking care of D3 brane creation, the brane system becomes
\begin{align}
    \raisebox{-.5\height}{
    \begin{tikzpicture}
    \draw (0,-1)--(0,1) (1,-1)--(1,1) (2,-1)--(2,1) (3,-1)--(3,1) 
    (4,-1)--(4,1) (5,-1)--(5,1) (6,-1)--(6,1) (7,-1)--(7,1)
    (9,-1)--(9,1) (10,-1)--(10,1) (11,-1)--(11,1);
    \node at (3.5,0) {$\cdots$};
    \node at (9.5,0) {$\cdots$};
    \draw (0,0)--(3,0) (4,0)--(9,0) (10,0)--(13,0);
        \ns{8,0}
        \ns{12,0}
        \on{13,0}
        \draw[decoration={brace,mirror,raise=10pt},decorate,thick]
  (-0.1,-0.8) -- node[below=15pt] {\footnotesize{$2N$ D5}} (7.1,-0.8);
  \draw[decoration={brace,mirror,raise=10pt},decorate,thick]
  (9-0.1,-0.8) -- node[below=15pt] {\footnotesize{$N_f-2N$ D5}} (11.1,-0.8);
  \draw[decoration={brace,mirror,raise=10pt},decorate,thick]
  (12-0.1,-0.2) -- node[below=15pt] {\footnotesize{$\cong$ ON${}^0$}} (13.1,-0.2);
  \node at (0.5,0.25) {\footnotesize{$1$}};
  \node at (1.5,0.25) {\footnotesize{$2$}};
  \node at (2.5,0.25) {\footnotesize{$3$}};
  \node at (4.5,0.25) {\footnotesize{$2N{-}3$}};
  \node at (5.5,0.25) {\footnotesize{$2N{-}2$}};
  \node at (6.5,0.25) {\footnotesize{$2N{-}1$}};
  \node at (7.5,0.25) {\footnotesize{$2N$}};
  \node at (8.5,0.25) {\footnotesize{$2N$}};
  \node at (10.5,0.25) {\footnotesize{$2N$}};
  \node at (11.5,0.25) {\footnotesize{$2N$}};
  \node at (13,0.5) {\footnotesize{ON${}^-$}};
    \end{tikzpicture}
    }
\end{align}
where the ON${}^0$ is understood as ON${}^-$ plus an NS5. Upon S-duality, the configuration reads
\begin{align}
    \raisebox{-.5\height}{
    \begin{tikzpicture}
    \draw (8,-1)--(8,1) (12,-1)--(12,1);
    \node at (3.5,0) {$\cdots$};
    \node at (9.5,0) {$\cdots$};
    \draw[dashed] (13,-1)--(13,0.7);
    \draw (0,0)--(3,0) (4,0)--(9,0) (10,0)--(13,0);
        \ns{0,0}
        \ns{1,0}
        \ns{2,0}
        \ns{3,0}
        \ns{4,0}
        \ns{5,0}
        \ns{6,0}
        \ns{7,0}
        \ns{9,0}
        \ns{10,0}
        \ns{11,0}
        \draw[decoration={brace,mirror,raise=10pt},decorate,thick]
  (-0.1,-0.2) -- node[below=15pt] {\footnotesize{$2N$ NS5}} (7.1,-0.2);
  \draw[decoration={brace,mirror,raise=10pt},decorate,thick]
  (9-0.1,-0.2) -- node[below=15pt] {\footnotesize{$N_f-2N$ NS5}} (11.1,-0.2);
  \node at (0.5,0.35) {\footnotesize{$1$}};
  \node at (1.5,0.35) {\footnotesize{$2$}};
  \node at (2.5,0.35) {\footnotesize{$3$}};
  \node at (4.5,0.35) {\footnotesize{$2N{-}3$}};
  \node at (5.5,0.35) {\footnotesize{$2N{-}2$}};
  \node at (6.5,0.35) {\footnotesize{$2N{-}1$}};
  \node at (7.5,0.35) {\footnotesize{$2N$}};
  \node at (8.5,0.35) {\footnotesize{$2N$}};
  \node at (10.5,0.35) {\footnotesize{$2N$}};
  \node at (11.5,0.35) {\footnotesize{$2N$}};
  \node at (13,0.85) {\footnotesize{O5${}^-$}};
    \end{tikzpicture}
    }
\end{align}
and the corresponding quiver gauge theory is given by
\begin{align}
        \raisebox{-.5\height}{
    \begin{tikzpicture}
	\node (g1) [gauge,label=below:{\footnotesize{$1$}}] {};
	\node (g2) [gauge,right of =g1,label=below:{\footnotesize{$2$}}] {};
	\node (g3) [gauge,right of =g2,label=below:{\footnotesize{$3$}}] {};
	\node (g4) [right of =g3] {$\ldots$};
	\node (g5) [gauge,right of =g4,label=below:{\footnotesize{$2N{-}3$}}] {};
	\node (g6) [gauge,right of =g5,label=below:{\footnotesize{$2N{-}2$}}] {};
	\node (g7) [gauge,right of =g6,label=below:{\footnotesize{$2N{-}1$}}] {};
	\node (g8) [gauge,right of =g7,label=below:{\footnotesize{$2N$}}] {};
	\node (g9) [right of =g8] {$\ldots$};
	\node (g10) [gauge,right of =g9,label=below:{\footnotesize{$2N$}}] {};
	\node (g11) [gaugeSp,right of =g10,label=below:{\footnotesize{$2N$}}] {};
	\node (f1) [flavour,above of=g8, label=above:{\footnotesize{$1$}}] {};
    \node (f2) [flavourSO,above of=g11, label=above:{\footnotesize{$2$}}] {};
	\draw  (g1)--(g2) (g2)--(g3) (g3)--(g4) (g4)--(g5) (g5)--(g6) (g6)--(g7) (g7)--(g8) (g8)--(g9) (g9)--(g10) (g10)--(g11) (g8)--(f1) (g11)--(f2);
  \draw[decoration={brace,mirror,raise=10pt},decorate,thick]
  (7-0.1,-0.2) -- node[below=15pt] {\footnotesize{$N_f{-}2N$ nodes}} (9.1,-0.2);
	\end{tikzpicture}
    }
    \label{eq:ex_product_mirror}
\end{align}
and the Coulomb and the Higgs branch of \eqref{eq:ex_product_mirror} are, consequently, product spaces
\begin{align}
    \Coulomb\eqref{eq:ex_product_mirror} = 
    \Higgs\left(  
    \scalebox{0.6}{
    \raisebox{-.4\height}{
    \begin{tikzpicture}
	\node (g1) [gauge,label=right:{$N$}] {};
	\node (f1) [flavour,above of=g1, label=right:{$N_f$}] {}; 	
	\draw (g1)--(f1);
	\end{tikzpicture}
    }}
    \right)^2
    \; , \qquad 
    \Higgs\eqref{eq:ex_product_mirror} = 
    \Coulomb\left(  
    \scalebox{0.6}{
    \raisebox{-.4\height}{
    \begin{tikzpicture}
	\node (g1) [gauge,label=right:{$N$}] {};
	\node (f1) [flavour,above of=g1, label=right:{$N_f$}] {}; 	
	\draw (g1)--(f1);
	\end{tikzpicture}
    }}
    \right)^2
\end{align}
and arise naturally in brane configurations.
Note that the balanced case $N_f=2N$ yields 
\begin{align}
        \raisebox{-.5\height}{
    \begin{tikzpicture}
	\node (g1) [gauge,label=below:{\footnotesize{$1$}}] {};
	\node (g2) [gauge,right of =g1,label=below:{\footnotesize{$2$}}] {};
	\node (g3) [gauge,right of =g2,label=below:{\footnotesize{$3$}}] {};
	\node (g4) [right of =g3] {$\ldots$};
	\node (g5) [gauge,right of =g4,label=below:{\footnotesize{$2N{-}3$}}] {};
	\node (g6) [gauge,right of =g5,label=below:{\footnotesize{$2N{-}2$}}] {};
	\node (g7) [gauge,right of =g6,label=below:{\footnotesize{$2N{-}1$}}] {};
	\node (g8) [gaugeSp,right of =g7,label=below:{\footnotesize{$2N$}}] {};
    \node (f2) [flavourSO,above of=g8, label=above:{\footnotesize{$4$}}] {};
	\draw  (g1)--(g2) (g2)--(g3) (g3)--(g4) (g4)--(g5) (g5)--(g6) (g6)--(g7) (g7)--(g8) (g8)--(f2);
	\end{tikzpicture}
    }
    \label{eq:ex_product_mirror_Nf=2N}
\end{align}
which had already been considered in \cite{Gaiotto:2008ak}.

Next, the setup can be generalised by choosing a non-symmetric splitting at the ON plane, see for instance \cite{Hanany:1999sj,Gaiotto:2008ak}. To be specific, choose a splitting such that $n$ D3s split as $p+q$ near the ON${}^-$. Here, without loss of generality, one may assume $p\geq q$ such that the setup becomes
\begin{align}
\raisebox{-.5\height}{
    \begin{tikzpicture}
    \draw (-1,0.1)--(3,0.1) (-1,-0.1)--(3,-0.1);
    \draw (1,-1)--(1,1) (1.1,-1)--(1.1,1) (1.7,-1)--(1.7,1);
    \node at (1.4,0.5) {$\cdots$};
    \node at (1.4,-0.5) {$\cdots$};
        \ns{-1,0}
        \ns{2,0}
        \on{3,0}
        \draw[decoration={brace,mirror,raise=10pt},decorate,thick]
  (0.9,-0.8) -- node[below=15pt] {\footnotesize{$n{+}k$ D5}} (1.8,-0.8);
  \node at (0,0.3) {\footnotesize{$(p{+}q)$ D3}};
  \node at (2.5,0.3) {\footnotesize{$2q$}};
  \node at (-0.75,-0.35) {\footnotesize{NS5}};
  \node at (3.25,-0.35) {\footnotesize{ON${}^-$}};
    \end{tikzpicture}
    }
    \qquad 
        \raisebox{-.5\height}{
    \begin{tikzpicture}
	\node (g1) [gauge,label=below:{\footnotesize{$p$}}] {};
	\node (f1) [flavour,above of=g1, label=above:{\footnotesize{$n+k$}}] {};
    \node at (1,0.5) {$\times$};
	\node (g2) [gauge,label=below:{\footnotesize{$q$}}] at (2,0) {};
	\node (f2) [flavour,above of=g2, label=above:{\footnotesize{$n+k$}}] {};
	\draw (g1)--(f1) (g2)--(f2);
	\end{tikzpicture}
    }
    \label{eq:ex_product_different}
\end{align}
with $n=p+q$, $k\geq p-q$, and $p\geq q$. The parametrisation implies that $n+k\geq 2p \geq 2q$, i.e.\ both gauge nodes are always \emph{good}. For the special case $k=p-q$, the $\urm(p)$ gauge node is \emph{balanced}. By analogous arguments, one finds that the mirror quiver is given by 
\begin{align}
        \raisebox{-.5\height}{
    \begin{tikzpicture}
	\node (g1) [gauge,label=below:{\footnotesize{$1$}}] {};
	\node (g2) [gauge,right of =g1,label=below:{\footnotesize{$2$}}] {};
	\node (g3) [right of =g2] {$\ldots$};
	\node (g4) [gauge,right of =g3,label=below:{\footnotesize{$n{-}1$}}] {};
	\node (g5) [gauge,right of =g4,label=below:{\footnotesize{$n$}}] {};
	\node (g6) [gauge,right of =g5,label=below:{\footnotesize{$n$}}] {};
	\node (g7) [right of =g6] {$\ldots$};
	\node (g8) [gauge,right of =g7,label=below:{\footnotesize{$n$}}] {};
	\node (g9) [gauge,right of =g8,label=below:{\footnotesize{$n$}}] {};
	\node (g10) [gauge,right of =g9,label=below:{\footnotesize{$n{-}1$}}] {};
	\node (g11) [right of =g10] {$\ldots$};
	\node (g13) [gauge,right of =g11,label=below:{\footnotesize{$2q{+}1$}}] {};
	\node (g14) [gaugeSp,right of =g13, label=below:{\footnotesize{$2q$}}] {};
	\node (f1) [flavour,above of=g5, label=above:{\footnotesize{$1$}}] {};
    \node (f2) [flavour,above of=g9, label=above:{\footnotesize{$1$}}] {};
	\draw  (g1)--(g2) (g2)--(g3) (g3)--(g4) (g4)--(g5) (g5)--(g6) (g6)--(g7) (g7)--(g8) (g8)--(g9) (g9)--(g10) (g10)--(g11)  (g11)--(g13) (g13)--(g14) (g5)--(f1) (g9)--(f2);
  \draw[decoration={brace,mirror,raise=10pt},decorate,thick]
  (4-0.1,-0.2) -- node[below=15pt] {\footnotesize{$k{-}(p{-}q){+}1$ nodes}} (8.1,-0.2);
	\end{tikzpicture}
    }
    \label{eq:ex_product_different_mirror}
\end{align}
and its Coulomb branch equals the product of the Higgs branches of the two different SQCD theories in \eqref{eq:ex_product_different}. Likewise, the Higgs branch of \eqref{eq:ex_product_different_mirror} is a product space too.
For the special case $k=p-q$, the two flavour nodes collide and form an non-abelian flavour symmetry. This is a manifestation of the enhanced Coulomb branch global symmetry in the mirror $\urm(p)$ gauge node, which is balanced in this case. The Coulomb branch Hilbert series for some of the cases of \eqref{eq:ex_product_different_mirror} are given in Table \ref{marcussuperquiver}.
\subsubsection{\texorpdfstring{A second example: ON${}^0$ with O3 and O5 planes}{A second example: ON with O3 and O5 planes}}
\label{sec:ON_O3_O5}
One can repeat the same logic in the presence of O5 or O3 orientifolds. Since the presence of any two of the O3, O5, ON planes imply the presence of the third orientifold, the setup necessarily has all planes present. To be specific, consider an O$5^-$, an O$3^+$, and an ON${}^0$ together with one NS5, $2k$ full D3, and $n$ full D5 arranged as follows:
\begin{align}
\raisebox{-.5\height}{
    \begin{tikzpicture}
    \draw (-1,0.1)--(2,0.1) (-1,-0.1)--(2,-0.1);
    \draw[dashed] (-1,0)--(2,0);
    \draw (1,-1)--(1,1) (1.1,-1)--(1.1,1) (1.7,-1)--(1.7,1);
    \node at (1.4,0.5) {$\cdots$};
    \node at (1.4,-0.5) {$\cdots$};
    \draw[dashed] (2,1)--(2,-1);
        \ns{-1,0}
        \onz{2,0}
        \draw[decoration={brace,mirror,raise=10pt},decorate,thick]
  (0.9,-0.8) -- node[below=15pt] {\footnotesize{$n$ D5}} (1.8,-0.8);
  \node at (0,0.3) {\footnotesize{$2k$ D3}};
  \node at (-0.75,-0.35) {\footnotesize{NS5}};
  \node at (2.5,-0.35) {\footnotesize{ON${}^0$}};
  \node at (2.5,0.85) {\footnotesize{O5${}^-$}};
  \node at (0.25,-0.5) {\footnotesize{O3${}^+$}};
    \end{tikzpicture}
    }
    \qquad 
        \raisebox{-.5\height}{
    \begin{tikzpicture}
	\node (g1) [gaugeSp,label=below:{\footnotesize{$2k$}}] {};
	\node (f1) [flavourSO,above of=g1, label=above:{\footnotesize{$2n$}}] {};
    \node at (1,0.5) {$\times$};
	\node (g2) [gaugeSp,label=below:{\footnotesize{$2k$}}] at (2,0) {};
	\node (f2) [flavourSO,above of=g2, label=above:{\footnotesize{$2n$}}] {};
	\draw (g1)--(f1) (g2)--(f2);
	\end{tikzpicture}
    }
    \label{eq:ex_product_OSp}
\end{align}
such that the electric theory is two copies of $\sprm(k)$ SQCD with $n$ fundamental flavours. The number of colours and flavours is constraint by $n \geq 2k+1$ such that the SQCD theories are good. After S-duality and suitable brane creation, the brane configuration can be brought into the form
\begin{align}
    \raisebox{-.5\height}{
    \begin{tikzpicture}
    \draw (8,-1)--(8,1) (12,-1)--(12,1);
    \node at (3.5,0) {$\cdots$};
    \node at (9.5,0) {$\cdots$};
    \draw[dashed] (13,-1)--(13,0.7);
    \draw (1,0.1)--(3,0.1)  (1,-0.1)--(3,-0.1) 
    (4,0.1)--(9,0.1) (4,-0.1)--(9,-0.1)
    (10,0.1)--(13,0.1) (10,-0.1)--(13,-0.1);
    \draw[dashed] (0,0)--(3,0) (4,0)--(9,0) (10,0)--(13,0);
        \ns{0,0}
        \ns{1,0}
        \ns{2,0}
        \ns{3,0}
        \ns{4,0}
        \ns{5,0}
        \ns{6,0}
        \ns{7,0}
        \ns{9,0}
        \ns{10,0}
        \ns{11,0}
        \onz{13,0}
        \draw[decoration={brace,mirror,raise=10pt},decorate,thick]
  (-0.1,-0.2) -- node[below=15pt] {\footnotesize{$4k+1$ half NS5}} (7.1,-0.2);
  \draw[decoration={brace,mirror,raise=10pt},decorate,thick]
  (9-0.1,-0.2) -- node[below=15pt] {\footnotesize{$2(n-2k-1)$ half NS5}} (11.1,-0.2);
  \node at (1.5,0.35) {\footnotesize{$1$}};
  \node at (2.5,0.35) {\footnotesize{$1$}};
  \node at (4.5,0.35) {\footnotesize{$2k{-}1$}};
  \node at (5.5,0.35) {\footnotesize{$2k{-}1$}};
  \node at (6.5,0.35) {\footnotesize{$2k$}};
  \node at (7.5,0.35) {\footnotesize{$2k$}};
  \node at (8.5,0.35) {\footnotesize{$2k$}};
  \node at (10.5,0.35) {\footnotesize{$2k$}};
  \node at (11.5,0.35) {\footnotesize{$2k$}};
  \node at (13,0.85) {\footnotesize{O5${}^-$}};
    \node at (13.5,-0.35) {\footnotesize{ON${}^0$}};
  \node at (0.5,-0.35) {\footnotesize{$+$}};
  \node at (1.5,-0.35) {\footnotesize{$-$}};
  \node at (2.5,-0.35) {\footnotesize{$+$}};
  \node at (4.5,-0.35) {\footnotesize{$-$}};
  \node at (5.5,-0.35) {\footnotesize{$+$}};
  \node at (6.5,-0.35) {\footnotesize{$-$}};
  \node at (7.5,-0.35) {\footnotesize{$+$}};
  \node at (8.5,-0.35) {\footnotesize{$\widetilde{+}$}};
  \node at (10.5,-0.35) {\footnotesize{$\widetilde{-}$}};
  \node at (11.5,-0.35) {\footnotesize{$\widetilde{+}$}};
  \node at (12.5,-0.35) {\footnotesize{$+$}};
    \end{tikzpicture}
    }
\end{align}
and the corresponding quiver gauge theory is given by
\begin{align}
        \raisebox{-.5\height}{
    \begin{tikzpicture}
	\node (g1) [gaugeSO,label=below:{\footnotesize{$2$}}] {};
	\node (g2) [gaugeSp,right of =g1,label=below:{\footnotesize{$2$}}] {};
	\node (g3) [gaugeSO,right of =g2,label=below:{\footnotesize{$4$}}] {};
	\node (g4) [right of =g3] {$\ldots$};
	\node (g5) [gaugeSO,right of =g4,label=below:{\footnotesize{$4k{-}2$}}] {};
	\node (g6) [gaugeSp,right of =g5,label=below:{\footnotesize{$4k{-}2$}}] {};
	\node (g7) [gaugeSO,right of =g6,label=below:{\footnotesize{$4k$}}] {};
	\node (g8) [gaugeSp,right of =g7,label=below:{\footnotesize{$4k$}}] {};
	\node (g9) [gaugeSO,right of =g8,label=below:{\footnotesize{$4k{+}1$}}] {};
	\node (g10) [gaugeSp,right of =g9,label=below:{\footnotesize{$4k$}}] {};
	\node (g11) [right of =g10] {$\ldots$};
	\node (g12) [gaugeSO,right of =g11,label=below:{\footnotesize{$4k{+}1$}}] {};
	\node (g13) [gaugeSp,below right of =g12,label=below:{\footnotesize{$2k$}}] {};
	\node (g14) [gaugeSp,above right of =g12,label=below:{\footnotesize{$2k$}}] {};
	\node (f1) [flavourSO,above of=g8, label=above:{\footnotesize{$1$}}] {};
    \node (f2) [flavourSO,right of=g13, label=below:{\footnotesize{$1$}}] {};
    \node (f3) [flavourSO,right of=g14, label=below:{\footnotesize{$1$}}] {};
	\draw  (g1)--(g2) (g2)--(g3) (g3)--(g4) (g4)--(g5) (g5)--(g6) (g6)--(g7) (g7)--(g8) (g8)--(g9) (g9)--(g10) (g10)--(g11) (g11)--(g12) (g12)--(g13) (g12)--(g14) (g8)--(f1) (g13)--(f2) (g14)--(f3);
\draw[decoration={brace,mirror,raise=10pt},decorate,thick]
  (7-0.1,-0.3) -- node[below=15pt] {\footnotesize{$2(n-2k-1)$ nodes}} (11.1,-0.3);
	\end{tikzpicture}
    }
    \label{eq:ex_product_mirror_OSp}
\end{align}
 Again, the Higgs and Coulomb branch of \eqref{eq:ex_product_mirror_OSp} are products of the Coulomb and Higgs branch of $\sprm(k)$ SQCD with $n$ fundamental flavours, respectively. This example shows that the quiver \eqref{prodDn} and its non-obvious relation to product theories \eqref{eq:ex_product_OSp} appear naturally in string theory.
As a comment, the balanced case $n=2k+1$ has a slightly simpler mirror quiver given by 
\begin{align}
        \raisebox{-.5\height}{
    \begin{tikzpicture}
	\node (g1) [gaugeSO,label=below:{\footnotesize{$2$}}] {};
	\node (g2) [gaugeSp,right of =g1,label=below:{\footnotesize{$2$}}] {};
	\node (g3) [gaugeSO,right of =g2,label=below:{\footnotesize{$4$}}] {};
	\node (g4) [right of =g3] {$\ldots$};
	\node (g5) [gaugeSO,right of =g4,label=below:{\footnotesize{$4k{-}2$}}] {};
	\node (g6) [gaugeSp,right of =g5,label=below:{\footnotesize{$4k{-}2$}}] {};
	\node (g7) [gaugeSO,right of =g6,label=below:{\footnotesize{$4k$}}] {};
	\node (g13) [gaugeSp,below right of =g7,label=below:{\footnotesize{$2k$}}] {};
	\node (g14) [gaugeSp,above right of =g7,label=below:{\footnotesize{$2k$}}] {};
    \node (f2) [flavourSO,right of=g13, label=below:{\footnotesize{$2$}}] {};
    \node (f3) [flavourSO,right of=g14, label=below:{\footnotesize{$2$}}] {};
	\draw  (g1)--(g2) (g2)--(g3) (g3)--(g4) (g4)--(g5) (g5)--(g6) (g6)--(g7) (g7)--(g13) (g7)--(g14) (g13)--(f2) (g14)--(f3); 	
	\end{tikzpicture}
    }
    \label{eq:ex_product_mirror_OSp_balanced}
\end{align}
which reproduced the example \eqref{firstfork} from Section \ref{forkingit}. Note that this particular case \eqref{eq:ex_product_mirror_OSp_balanced} had also appeared in \cite{Gaiotto:2008ak}.

Analogous to the brane systems without orientifolds, one can generalise the setup by ON planes with a non-symmetric splitting. To begin with, consider a stack of $2k+l$ full D3 branes which split as $k$ and $k+l$ near an ON${}^-$. This leads to
\begin{align}
\raisebox{-.5\height}{
    \begin{tikzpicture}
    \draw (-1,0.1)--(3,0.1) (-1,-0.1)--(3,-0.1);
    \draw[dashed] (-1,0)--(3,0);
    \draw (1,-1)--(1,1) (1.1,-1)--(1.1,1) (1.7,-1)--(1.7,1);
    \node at (1.4,0.5) {$\cdots$};
    \node at (1.4,-0.5) {$\cdots$};
    \draw[dashed] (3,1)--(3,-1);
        \ns{-1,0}
        \ns{2,0}
        \on{3,0}
        \draw[decoration={brace,mirror,raise=10pt},decorate,thick]
  (0.9,-0.8) -- node[below=15pt] {\footnotesize{$n$ D5}} (1.8,-0.8);
  \node at (0,0.3) {\footnotesize{$2k{+}l$ D3}};
  \node at (2.5,0.3) {\footnotesize{$k$ D3}};
  \node at (-0.75,-0.35) {\footnotesize{NS5}};
  \node at (3.5,-0.35) {\footnotesize{ON${}^-$}};
  \node at (3.5,0.85) {\footnotesize{O5${}^-$}};
  \node at (0.25,-0.5) {\footnotesize{O3${}^+$}};
    \end{tikzpicture}
    }
    \qquad 
        \raisebox{-.5\height}{
    \begin{tikzpicture}
	\node (g1) [gaugeSp,label=below:{\footnotesize{$2k$}}] {};
	\node (f1) [flavourSO,above of=g1, label=above:{\footnotesize{$2n$}}] {};
    \node at (1,0.5) {$\times$};
	\node (g2) [gaugeSp,label=below:{\footnotesize{$2k+2l$}}] at (2,0) {};
	\node (f2) [flavourSO,above of=g2, label=above:{\footnotesize{$2n$}}] {};
	\draw (g1)--(f1) (g2)--(f2);
	\end{tikzpicture}
    }
    \label{eq:ex_product_OSp_different}
\end{align}
with $2n=4k+2l+x$ and $x\geq 2l+2$, $x\in 2\mathbb{N}$. The logic of the parametrisation is as follows: keeping both $\sprm$ gauge nodes \emph{good} requires $n  \geq 2k +1$,  $n \geq 2k +2l+1$ such that $x\geq 2(l+1)$. I.e. $l$ controls the difference in gauge group ranks. $x=0$ keeps $\sprm(k+l)$ \emph{balanced} and $\sprm(k)$ \emph{good}; while any $x>0$ renders both $\sprm$ nodes \emph{good}.  From the brane configuration \eqref{eq:ex_product_OSp_different}, the mirror theory is derived to be
\begin{align}
        \raisebox{-.5\height}{
    \begin{tikzpicture}
	\node (g1) [gaugeSO,label=below:{\footnotesize{$2$}}] {};
	\node (g2) [gaugeSp,right of =g1,label=below:{\footnotesize{$2$}}] {};
	\node (g3) [right of =g2] {$\ldots$};
	\node (g4) [gaugeSO,right of =g3,label=above:{\footnotesize{$4k{+}2l$}}] {};
	\node (g5) [gaugeSp,right of =g4,label=below:{\footnotesize{$4k{+}2l$}}] {};
	\node (g6) [gaugeSO,right of =g5,label=above:{\footnotesize{$4k{+}2l{+}1$}}] {};
	\node (g7) [right of =g6] {$\ldots$};
	\node (g8) [gaugeSO,right of =g7,label=above:{\footnotesize{$4k{+}2l{+}1$}}] {};
	\node (g9) [gaugeSp,right of =g8,label=below:{\footnotesize{$4k{+}2l$}}] {};
	\node (g10) [gaugeSO,right of =g9,label=above:{\footnotesize{$4k{+}2l$}}] {};
	\node (g11) [right of =g10] {$\ldots$};
	\node (g12) [gaugeSp,right of =g11,label=below:{\footnotesize{$4k{+}2$}}] {};
	\node (g13) [gaugeSO,right of =g12,label=below:{\footnotesize{$4k{+}2$}}] {};
	\node (g14) [gaugeSp,below right of =g13,label=below:{\footnotesize{$2k$}}] {};
	\node (g15) [gaugeSp,above right of =g13,label=below:{\footnotesize{$2k$}}] {};
	\node (f1) [flavourSO,above of=g5, label=above:{\footnotesize{$1$}}] {};
    \node (f2) [flavourSO,above of=g9, label=above:{\footnotesize{$1$}}] {};
	\draw  (g1)--(g2) (g2)--(g3) (g3)--(g4) (g4)--(g5) (g5)--(g6) (g6)--(g7) (g7)--(g8) (g8)--(g9) (g9)--(g10) (g10)--(g11) (g11)--(g12) (g12)--(g13) (g13)--(g14) (g13)--(g15) (g5)--(f1) (g9)--(f2);
\draw[decoration={brace,mirror,raise=10pt},decorate,thick]
  (4-0.1,-0.3) -- node[below=15pt] {\footnotesize{$x-(2l+2)$ nodes}} (7.1,-0.3);
	\end{tikzpicture}
    }
    \label{eq:ex_product_mirror_OSp_different}
\end{align}
which reproduces \eqref{weirdcase} from Section \ref{forkingit}. The origin \eqref{eq:ex_product_OSp_different}, implies immediately that the Coulomb branch of \eqref{eq:ex_product_mirror_OSp_different} is the product \eqref{modulispaceprodeqn} of the Higgs branches of two different $\sprm$ SQCD theories. Analogous statements hold for the Higgs branch of \eqref{eq:ex_product_mirror_OSp_different}. 
For the special case $x=2l+2$, the two flavour nodes collide and form a $\sorm(2)$ flavour. The resulting mass parameter reflects the fact that the mirror $\sprm(2k+2l)$ gauge node is balance, i.e. has enhance Coulomb branch global symmetry.
\subsubsection{\texorpdfstring{A last example: ON${}^0$ with O3 and $\widetilde{\mathrm{O5}}$ planes}{A last example: ON with O3 and tilde-O5 planes}}
So far, the discussion focused on ON${}^-$ planes intersecting an O5${}^-$ orientifold. Instead, one could also consider a $\widetilde{\mathrm{O5}}^-$. If a stack of $2k$ D3 branes intersects a stack of $n$ D5 on top of an $\widetilde{\mathrm{O5}}^-$ plane, the results 3d theory is $\sprm(k)$ gauge theory with a $\sorm(2n+1)$ flavour symmetry and some suitable Chern-Simons level. The CS level renders the parity-anomalous $\sprm(k)$ theory with $2n+1$ half-hypermultiplets consistent. The (classical) Higgs branch, which is insensitive to the CS-level, is known to be a $B$-type nilpotent orbit closure \cite{Hanany:2016gbz}. 
Applying S-duality, yields a brane system with an $\widetilde{\mathrm{ON}}^-$ plane, which gives rise to a $B$-type Dynkin quiver theory composed of unitary nodes \cite{Cremonesi:2014xha}. However, the potential mirror symmetry between these two theories is blurred by the fact that the Coulomb branch of the $\sprm$ theory is rather poorly understood or, equivalently, the Higgs branch a $B$-type Dynkin quivers suffers from a lack of Lagrangian description. Hence, it is not clear what the low-energy description for the brane system is.

Following this train of thought, another possibility is to consider a $\widetilde{\mathrm{O5}}^-$ plane instead of an O5$^-$ plane in the setups of Section \ref{sec:ON_O3_O5}. For instance, a suitable brane configuration is as follows:
\begin{align}
\raisebox{-.5\height}{
    \begin{tikzpicture}
    \draw (-1,0.1)--(2,0.1) (-1,-0.1)--(2,-0.1);
    \draw[dashed] (-1,0)--(2,0);
    \draw (1,-1)--(1,1) (1.1,-1)--(1.1,1) (1.7,-1)--(1.7,1);
    \node at (1.4,0.5) {$\cdots$};
    \node at (1.4,-0.5) {$\cdots$};
    \draw[dashed] (2,1)--(2,-1);
        \ns{-1,0}
        \onz{2,0}
        \draw[decoration={brace,mirror,raise=10pt},decorate,thick]
  (0.9,-0.8) -- node[below=15pt] {\footnotesize{$2n{-}1$  half D5}} (1.8,-0.8);
  \node at (0,0.3) {\footnotesize{$2k$ D3}};
  \node at (-0.75,-0.35) {\footnotesize{NS5}};
  \node at (2.5,-0.35) {\footnotesize{ON${}^0$}};
  \node at (2.5,0.85) {\footnotesize{$\widetilde{\mathrm{O5}}^-$}};
  \node at (0.25,-0.5) {\footnotesize{O3${}^+$}};
    \end{tikzpicture}
    }
    \qquad 
        \raisebox{-.5\height}{
    \begin{tikzpicture}
	\node (g1) [gaugeSp,label=below:{\footnotesize{$2k$}}] {};
	\node (f1) [flavourSO,above of=g1, label=above:{\footnotesize{$2n$}}] {};
    \node at (1,0.5) {$\times$};
	\node (g2) [gaugeSp,label=below:{\footnotesize{$2k$}}] at (2,0) {};
	\node (f2) [flavourSO,above of=g2, label=above:{\footnotesize{$2n{-}1$}}] {};
	\draw (g1)--(f1) (g2)--(f2);
	\end{tikzpicture}
    }
    \label{eq:ex_product_OSp_diff_ranks}
\end{align}
such that the electric theory is a product of a $\sprm(k)$ SQCD with $\sorm(2n)$ flavour symmetry and $\sorm(2n-1)$ flavour symmetry, respectively. Chern-Simons levels are neglected here, as these are not relevant for the Higgs branch moduli space.
After an S-duality and suitable brane creation, the brane configuration can be brought into the form
\begin{align}
    \raisebox{-.5\height}{
    \begin{tikzpicture}
    \draw (8,-1)--(8,1) (12,-1)--(12,1);
    \node at (3.5,0) {$\cdots$};
    \node at (9.5,0) {$\cdots$};
    \draw[dashed] (13,-1)--(13,0.7);
    \draw (1,0.1)--(3,0.1)  (1,-0.1)--(3,-0.1) 
    (4,0.1)--(9,0.1) (4,-0.1)--(9,-0.1)
    (10,0.1)--(13,0.1) (10,-0.1)--(13,-0.1);
    \draw[dashed] (0,0)--(3,0) (4,0)--(9,0) (10,0)--(13,0);
        \ns{0,0}
        \ns{1,0}
        \ns{2,0}
        \ns{3,0}
        \ns{4,0}
        \ns{5,0}
        \ns{6,0}
        \ns{7,0}
        \ns{9,0}
        \ns{10,0}
        \ns{11,0}
        \onz{13,0}
        \draw[decoration={brace,mirror,raise=10pt},decorate,thick]
  (-0.1,-0.2) -- node[below=15pt] {\footnotesize{$4k+1$ half NS5}} (7.1,-0.2);
  \draw[decoration={brace,mirror,raise=10pt},decorate,thick]
  (9-0.1,-0.2) -- node[below=15pt] {\footnotesize{$2(n-2k-1)$ half NS5}} (11.1,-0.2);
  \node at (1.5,0.35) {\footnotesize{$1$}};
  \node at (2.5,0.35) {\footnotesize{$1$}};
  \node at (4.5,0.35) {\footnotesize{$2k{-}1$}};
  \node at (5.5,0.35) {\footnotesize{$2k{-}1$}};
  \node at (6.5,0.35) {\footnotesize{$2k$}};
  \node at (7.5,0.35) {\footnotesize{$2k$}};
  \node at (8.5,0.35) {\footnotesize{$2k$}};
  \node at (10.5,0.35) {\footnotesize{$2k$}};
  \node at (11.5,0.35) {\footnotesize{$2k$}};
  \node at (13.5,-0.35) {\footnotesize{$\widetilde{\mathrm{ON}}^-$}};
  \node at (13,0.85) {\footnotesize{O5${}^-$}};
  \node at (0.5,-0.35) {\footnotesize{$+$}};
  \node at (1.5,-0.35) {\footnotesize{$-$}};
  \node at (2.5,-0.35) {\footnotesize{$+$}};
  \node at (4.5,-0.35) {\footnotesize{$-$}};
  \node at (5.5,-0.35) {\footnotesize{$+$}};
  \node at (6.5,-0.35) {\footnotesize{$-$}};
  \node at (7.5,-0.35) {\footnotesize{$+$}};
  \node at (8.5,-0.35) {\footnotesize{$\widetilde{+}$}};
  \node at (10.5,-0.35) {\footnotesize{$\widetilde{-}$}};
  \node at (11.5,-0.35) {\footnotesize{$\widetilde{+}$}};
  \node at (12.5,-0.35) {\footnotesize{$+$}};
    \end{tikzpicture}
    }
\end{align}
and the S-dual displays an $\widetilde{\mathrm{ON}}^-$ plane due to the original $\widetilde{\mathrm{O5}}^-$. As depicted in Figure \ref{fig:D1_and_F1_on_O3}, the presence of the $\widetilde{\mathrm{ON}}^-$ plane introduces a non-simply laced link between the $\sorm(4k+1)$ gauge node and the $\sprm(k)$ node. The resulting quiver gauge theory is given by
\begin{align}
        \raisebox{-.5\height}{
    \begin{tikzpicture}
	\node (g1) [gaugeSO,label=below:{\footnotesize{$2$}}] {};
	\node (g2) [gaugeSp,right of =g1,label=below:{\footnotesize{$2$}}] {};
	\node (g3) [gaugeSO,right of =g2,label=below:{\footnotesize{$4$}}] {};
	\node (g4) [right of =g3] {$\ldots$};
	\node (g5) [gaugeSO,right of =g4,label=below:{\footnotesize{$4k{-}2$}}] {};
	\node (g6) [gaugeSp,right of =g5,label=below:{\footnotesize{$4k{-}2$}}] {};
	\node (g7) [gaugeSO,right of =g6,label=below:{\footnotesize{$4k$}}] {};
	\node (g8) [gaugeSp,right of =g7,label=below:{\footnotesize{$4k$}}] {};
	\node (g9) [gaugeSO,right of =g8,label=below:{\footnotesize{$4k{+}1$}}] {};
	\node (g10) [gaugeSp,right of =g9,label=below:{\footnotesize{$4k$}}] {};
	\node (g11) [right of =g10] {$\ldots$};
	\node (g12) [gaugeSO,right of =g11,label=below:{\footnotesize{$4k{+}1$}}] {};
	\node (gaux) [right of =g12,xshift=-0.05cm] {};
	\node (g13) [gaugeSp, right of =gaux,xshift=-0.45cm,label=below:{\footnotesize{$2k$}}] {};
	\node (f1) [flavourSO,above of=g8, label=above:{\footnotesize{$1$}}] {};
    \node (f2) [flavourSO,above of=g13, label=above:{\footnotesize{$1$}}] {};
	\draw  (g1)--(g2) (g2)--(g3) (g3)--(g4) (g4)--(g5) (g5)--(g6) (g6)--(g7) (g7)--(g8) (g8)--(g9) (g9)--(g10) (g10)--(g11) (g11)--(g12)  (g8)--(f1) (g13)--(f2);
	\doublearrow{arrows={}}{(g12) -- (g13)};
	\doublearrow{arrows={-Implies}}{(g12) -- (gaux)};
\draw[decoration={brace,mirror,raise=10pt},decorate,thick]
  (7-0.1,-0.3) -- node[below=15pt] {\footnotesize{$2(n-2k-1)$ nodes}} (11.1,-0.3);
	\end{tikzpicture}
    }
    \label{eq:ex_product_mirror_OSp_diff_ranks}
\end{align}
which reproduces the $l=0$ limit of \eqref{missingguy}, i.e. the folded version of \eqref{prodDn}.

The setting \eqref{eq:ex_product_OSp_diff_ranks} can be further generalised by an ON plane with a non-symmetric splitting. To begin with, consider a stack of $2k+l$ full D3 branes which split as $k$ and $k+l$ near an  $\widetilde{\mathrm{O5}}^-$ and $\mathrm{ON}^-$. This leads to
\begin{align}
\raisebox{-.5\height}{
    \begin{tikzpicture}
    \draw (-1,0.1)--(3,0.1) (-1,-0.1)--(3,-0.1);
    \draw[dashed] (-1,0)--(3,0);
    \draw (1,-1)--(1,1) (1.1,-1)--(1.1,1) (1.7,-1)--(1.7,1);
    \node at (1.4,0.5) {$\cdots$};
    \node at (1.4,-0.5) {$\cdots$};
    \draw[dashed] (3,1)--(3,-1);
        \ns{-1,0}
        \ns{2,0}
        \on{3,0}
        \draw[decoration={brace,mirror,raise=10pt},decorate,thick]
  (0.9,-0.8) -- node[below=15pt] {\footnotesize{$2n{-}1$ half D5}} (1.8,-0.8);
  \node at (0,0.3) {\footnotesize{$2k{+}l$ D3}};
  \node at (2.5,0.3) {\footnotesize{$k$ D3}};
  \node at (-0.75,-0.35) {\footnotesize{NS5}};
  \node at (3.5,-0.35) {\footnotesize{ON${}^-$}};
  \node at (3.5,0.85) {\footnotesize{$\widetilde{\mathrm{O5}}^-$}};
  \node at (0.25,-0.5) {\footnotesize{O3${}^+$}};
    \end{tikzpicture}
    }
    \qquad 
        \raisebox{-.5\height}{
    \begin{tikzpicture}
	\node (g1) [gaugeSp,label=below:{\footnotesize{$2k$}}] {};
	\node (f1) [flavourSO,above of=g1, label=above:{\footnotesize{$2n-1$}}] {};
    \node at (1,0.5) {$\times$};
	\node (g2) [gaugeSp,label=below:{\footnotesize{$2k+2l$}}] at (2,0) {};
	\node (f2) [flavourSO,above of=g2, label=above:{\footnotesize{$2n$}}] {};
	\draw (g1)--(f1) (g2)--(f2);
	\end{tikzpicture}
    }
    \label{eq:ex_product_OSp_different_diff_ranks}
\end{align}
with $2n=4k+2l+x$ and $x\geq 2l+2$, $x\in 2\mathbb{N}$. Again, CS-levels are neglected.
From the brane configuration \eqref{eq:ex_product_OSp_different_diff_ranks}, the mirror theory is derived to be
\begin{align}
        \raisebox{-.5\height}{
    \begin{tikzpicture}
	\node (g1) [gaugeSO,label=below:{\footnotesize{$2$}}] {};
	\node (g2) [gaugeSp,right of =g1,label=below:{\footnotesize{$2$}}] {};
	\node (g3) [right of =g2] {$\ldots$};
	\node (g4) [gaugeSO,right of =g3,label=above:{\footnotesize{$4k{+}2l$}}] {};
	\node (g5) [gaugeSp,right of =g4,label=below:{\footnotesize{$4k{+}2l$}}] {};
	\node (g6) [gaugeSO,right of =g5,label=above:{\footnotesize{$4k{+}2l{+}1$}}] {};
	\node (g7) [right of =g6] {$\ldots$};
	\node (g8) [gaugeSO,right of =g7,label=above:{\footnotesize{$4k{+}2l{+}1$}}] {};
	\node (g9) [gaugeSp,right of =g8,label=below:{\footnotesize{$4k{+}2l$}}] {};
	\node (g10) [gaugeSO,right of =g9,label=above:{\footnotesize{$4k{+}2l$}}] {};
	\node (g11) [right of =g10] {$\ldots$};
	\node (g12) [gaugeSp,right of =g11,label=below:{\footnotesize{$4k{+}2$}}] {};
	\node (g13) [gaugeSO,right of =g12,label=below:{\footnotesize{$4k{+}2$}}] {};
	\node (gaux) [right of =g13,xshift=-0.05cm] {};
	\node (g14) [gaugeSp,right of =gaux,xshift=-0.45cm,label=below:{\footnotesize{$2k$}}] {};
	\node (f1) [flavourSO,above of=g5, label=above:{\footnotesize{$1$}}] {};
    \node (f2) [flavourSO,above of=g9, label=above:{\footnotesize{$1$}}] {};
	\draw  (g1)--(g2) (g2)--(g3) (g3)--(g4) (g4)--(g5) (g5)--(g6) (g6)--(g7) (g7)--(g8) (g8)--(g9) (g9)--(g10) (g10)--(g11) (g11)--(g12) (g12)--(g13) (g13)--(g14) (g5)--(f1) (g9)--(f2);
	\doublearrow{arrows={}}{(g13) -- (g14)};
	\doublearrow{arrows={-Implies}}{(g13) -- (gaux)};
\draw[decoration={brace,mirror,raise=10pt},decorate,thick]
  (4-0.1,-0.3) -- node[below=15pt] {\footnotesize{$x-(2l+2)$ nodes}} (7.1,-0.3);
	\end{tikzpicture}
    }
    \label{eq:ex_product_mirror_OSp_different_diff_ranks}
\end{align}
 which reproduces \eqref{missingguy} from Section \ref{foldafterfork}.
 \paragraph{Comment.}
The general form of \eqref{eq:ex_product_mirror_OSp_different} and \eqref{eq:ex_product_mirror_OSp_different_diff_ranks} allows an intuitive understanding of the requirement that the long balanced tail has to terminate in an $\sorm(2)$ node. Suppose the orthosymplectic quiver has a set of balanced gauge nodes in the shape of a $B$/$D$-type Dynkin diagram, but the long tail of rank-decreasing gauge nodes terminates early on a orthogonal/symplectic node with rank larger than one. In order to balance this node, it needs to be connected to a suitably chosen flavour node. This flavour node, with rank larger or equal than one, in turn implies that the mirror configuration contains an additional NS5 brane. As a result, the mirror theory is not a simple product theory anymore, but is itself a D-type Dynkin quiver. Consequently, the product structure is lost as soon as the long tail of rank-decreasing nodes does not terminate on an $\sorm(2)$.
\subsection{6d brane systems}
\label{sec:branes_6d}
In view of the magnetic quiver constructions for 6d theories \cite{Cabrera:2019izd,Cabrera:2019dob}, a natural next step is to consider D6-D8-NS5 brane configurations \cite{Hanany:1997gh,Brunner:1997gk,Brunner:1997gf} in the presence of ON planes \cite{Hanany:1999sj}. 
\subsubsection{\texorpdfstring{Inclusion of ON${}^0$ plane}{Inclusion of ON plane}}
Consider the brane configuration, see also \cite{Hanany:1999sj},
\begin{align}
\raisebox{-.5\height}{
\begin{tikzpicture}
    \draw (-1-7,0)--(1-7,0) 
    (1-7,0.1)--(3-7,0.1) (1-7,-0.1)--(3-7,-0.1);
    \ns{1-7,0}
    \onz{3-7,0}
    \node at (1-7,0.375) {\footnotesize{NS5}};
    \node at (3-7,0.375) {\footnotesize{ON${}^0$}};
    \node at (-7,0.25) {\footnotesize{$l$}};
    \node at (-5,0.25) {\footnotesize{$k$}};
    \node at (-5,-0.25) {\footnotesize{$h$}};
 \node at (-2.75,0) {$\cong$};
    \draw (-2,0)--(0,0) 
    (0,0.1)--(2,0.1) (0,-0.1)--(2,-0.1)
    (2,0.1)--(4,0.1) (2,-0.1)--(4,-0.1);
    \ns{0,0}
    \ns{2,0}
    \on{4,0}
    \node at (0,0.375) {\footnotesize{NS5}};
    \node at (2,0.375) {\footnotesize{NS5}};
    \node at (2,0.675) {\footnotesize{virtual}};
    \node at (4,0.375) {\footnotesize{ON${}^-$}};
    \node at (-1,0.25) {\footnotesize{$l$}};
    \node at (1,0.25) {\footnotesize{$k$}};
    \node at (1,-0.25) {\footnotesize{$h$}};
    \node at (3,0.25) {\footnotesize{$k$}};
    \node at (3,-0.25) {\footnotesize{$k$}};
\end{tikzpicture}
}
\label{eq:6d_branes_ON}
\end{align}
for which charge conservation or, equivalently, anomaly cancellation implies $k=h$ and $l=2k$ such that the 6d $\Ncal=(1,0)$ world-volume theory is given by
\begin{align}
        \raisebox{-.5\height}{
    \begin{tikzpicture}
	\node (g1) [gauge,label=right:{\footnotesize{$\surm(k)$}}] {};
\node (f1) [flavour,above of=g1, label=right:{\footnotesize{$2k$}}] {};
\node at (2,0.5) {$\times$};
	\node (g2) [gauge,label=right:{\footnotesize{$\surm(k)$}}] at (3,0) {};
\node (f2) [flavour,above of=g2, label=right:{\footnotesize{$2k$}}] {};
	\draw  (g1)--(f1) (g2)--(f2);
	\end{tikzpicture}
    } \,.
\end{align}
The finite and infinite Higgs branches of $\surm(k)$ with $N_f=2k$ flavours are known to be related by discrete gauging \cite{Hanany:2018vph,Hanany:2018cgo,Hanany:2018dvd,Cabrera:2019izd}. In brief,
\begin{align}
        \Higgs \left(
        \raisebox{-.5\height}{
    \begin{tikzpicture}
	\node (g1) [gauge,label=right:{\footnotesize{$\surm(k)$}}] {};
    \node (f1) [flavour,above of=g1, label=right:{\footnotesize{$2k$}}] {};
	\draw  (g1)--(f1);
	\end{tikzpicture}
    } 
    \bigg|_{g^2<\infty}\right)
    &= 
    \Coulomb \left(
    \mathsf{M}_{k,(1^2)}       
    \right) 
    \qquad 
    \Higgs \left(
        \raisebox{-.5\height}{
    \begin{tikzpicture}
	\node (g1) [gauge,label=right:{\footnotesize{$\surm(k)$}}] {};
    \node (f1) [flavour,above of=g1, label=right:{\footnotesize{$2k$}}] {};
	\draw  (g1)--(f1);
	\end{tikzpicture}
    } 
    \bigg|_{g^2=\infty}\right)
    = 
    \Coulomb \left(
       \mathsf{M}_{k,(2)} 
    \right)
    \end{align}
with the following magnetic quivers:
\begin{subequations}
    \begin{align}
    \mathsf{M}_{p,(1^2)} &= 
    \raisebox{-.5\height}{
    \begin{tikzpicture}
	\node (g1) [gauge,label=below:{\footnotesize{$1$}}] {};
	\node (g2) [gauge,right of=g1 ,label=below:{\footnotesize{$2$}}] {};
	\node (g3) [right of =g2] {$\ldots$};
	\node (g4) [gauge,right of =g3,label=below:{\footnotesize{$k{-}1$}}] {};
	\node (g5) [gauge,right of =g4,label=below:{\footnotesize{$k$}}] {};
	\node (g6) [gauge,right of =g5,label=below:{\footnotesize{$k{-}1$}}] {};
	\node (g7) [right of =g6] {$\ldots$};
	\node (g8) [gauge,right of=g7 ,label=below:{\footnotesize{$2$}}] {};
	\node (g9) [gauge,right of=g8 ,label=below:{\footnotesize{$1$}}] {};
    \node (b1) [gauge,above left of=g5, label=left:{\footnotesize{$1$}}] {};
    \node (b2) [gauge,above right of=g5, label=right:{\footnotesize{$1$}}] {};
	\draw  (g1)--(g2) (g2)--(g3) (g3)--(g4) (g4)--(g5) (g5)--(g6) (g6)--(g7) (g7)--(g8) (g8)--(g9)
	(g5)--(b1) (g5)--(b2);
	\end{tikzpicture}
    }
    \\
        \mathsf{M}_{p,(2)} &= 
        \raisebox{-.5\height}{
    \begin{tikzpicture}
	\node (g1) [gauge,label=below:{\footnotesize{$1$}}] {};
	\node (g2) [gauge,right of=g1 ,label=below:{\footnotesize{$2$}}] {};
	\node (g3) [right of =g2] {$\ldots$};
	\node (g4) [gauge,right of =g3,label=below:{\footnotesize{$k{-}1$}}] {};
	\node (g5) [gauge,right of =g4,label=below:{\footnotesize{$k$}}] {};
	\node (g6) [gauge,right of =g5,label=below:{\footnotesize{$k{-}1$}}] {};
	\node (g7) [right of =g6] {$\ldots$};
	\node (g8) [gauge,right of=g7 ,label=below:{\footnotesize{$2$}}] {};
	\node (g9) [gauge,right of=g8 ,label=below:{\footnotesize{$1$}}] {};
    \node (b1) [gauge,above of=g5, label=right:{\footnotesize{$2$}}] {};
    \draw (b1) to [out=45,in=135,looseness=10] (b1);
	\draw  (g1)--(g2) (g2)--(g3) (g3)--(g4) (g4)--(g5) (g5)--(g6) (g6)--(g7) (g7)--(g8) (g8)--(g9)
	(g5)--(b1);
	\end{tikzpicture}
    }
\end{align}
\end{subequations}
Now, the aim is to derive the magnetic quiver for finite and infinite coupling from the brane configuration \eqref{eq:6d_branes_ON}.
Moving in $2k$ D8 branes from infinity, the brane system can be written as 
\begin{align}
\raisebox{-.5\height}{
\begin{tikzpicture}
\draw (0,-1)--(0,1) (1,-1)--(1,1)  (2,-1)--(2,1)
(3,-1)--(3,1) (4,-1)--(4,1) (5,-1)--(5,1); 
    \draw (0,-0.25)--(1,-0.25) 
    (1,0.25)--(2,0.25) 
    (3,-0.25)--(4,-0.25)
    (4,0.25)--(5,0.25) 
    (5,0)--(8,0);
    \ns{6,0.5}
    \ns{7,0.5}
    \on{8,0}
    \node at (2.5,0) {$\ldots$};
    \node at (8,0.375) {\footnotesize{ON${}^-$}};
    \node at (0.5,0) {\footnotesize{$1$}};
    \node at (1.5,0.5) {\footnotesize{$2$}};
    \node at (3.5,0) {\footnotesize{$2k{-}2$}};
    \node at (4.5,0.5) {\footnotesize{$2k{-}1$}};
    \node at (5.5,0.25) {\footnotesize{$2k$}};
    \draw[decoration={brace,mirror,raise=10pt},decorate,thick]
  (-0.1,-0.8) -- node[below=15pt] {\footnotesize{$2k$ D8}} (5.1,-0.8);
\end{tikzpicture}
}
\label{eq:6d_branes_ON_Higgs}
\end{align}
for which the proposed magnetic quiver reads (see also Appendix \ref{app:branes})
\begin{align}
        \raisebox{-.5\height}{
    \begin{tikzpicture}
	\node (g1) [gauge,label=below:{\footnotesize{$1$}}] {};
	\node (g2) [gauge,right of =g1,label=below:{\footnotesize{$2$}}] {};
	\node (g3) [gauge,right of =g2,label=below:{\footnotesize{$3$}}] {};
	\node (g4) [right of =g3] {$\ldots$};
	\node (g5) [gauge,right of =g4,label=below:{\footnotesize{$2k{-}3$}}] {};
	\node (g6) [gauge,right of =g5,label=below:{\footnotesize{$2k{-}2$}}] {};
	\node (g7) [gauge,right of =g6,label=below:{\footnotesize{$2k{-}1$}}] {};
	\node (g8) [gaugeSp,right of =g7,label=below:{\footnotesize{$2k$}}] {};
    \node (b1) [gaugeSO,above left of=g8, label=above:{\footnotesize{$2$}}] {};
    \node (b2) [gaugeSO,above right of=g8, label=above:{\footnotesize{$2$}}] {};
	\draw  (g1)--(g2) (g2)--(g3) (g3)--(g4) (g4)--(g5) (g5)--(g6) (g6)--(g7) (g7)--(g8) (g8)--(b1) (g8)--(b2);
	\end{tikzpicture}
    }
    \label{eq:6d_magQuiv_Unitary_finite}
\end{align}
and the Coulomb branch satisfies
\begin{align}
\Coulomb  \eqref{eq:6d_magQuiv_Unitary_finite} 
= \left[ \Higgs \left( \text{$\surm(k)$ SQCD, $2k$ flavours}\right) \right]^2
= \left[ \Coulomb \left( \mathsf{M}_{k,(1^2)}\right) \right]^2 \,.
\end{align}
This expectation receives further validation by the explicit Hilbert series computations presented in Table \ref{eq:6d_magQuiv_Unitary_finite1}.
As a remark, one recognises that the 6d finite coupling magnetic quiver \eqref{eq:6d_magQuiv_Unitary_finite} results from the 3d magnetic quiver \eqref{eq:ex_product_mirror_Nf=2N} via explosion \cite{Bourget:2021zyc} of the $\sorm(4)$ flavour node into a bouquet of two $\sorm(2)$ magnetic gauge nodes.

Next, the transition to infinite coupling is realised by moving the NS5s onto the locus of the ON. Recall that there are two NS5 branes on each side of the ON plane; hence, the total number of NS5 is four and the brane configuration may be written as 
\begin{align}
\raisebox{-.5\height}{
\begin{tikzpicture}
\draw (0,-1)--(0,1) (1,-1)--(1,1)  (2,-1)--(2,1)
(3,-1)--(3,1) (4,-1)--(4,1) (5,-1)--(5,1); 
    \draw (0,-0.25)--(1,-0.25) 
    (1,0.25)--(2,0.25) 
    (3,-0.25)--(4,-0.25)
    (4,0.25)--(5,0.25) 
    (5,0)--(8,0);
    \ns{8,0.5}
    \ns{8,1}
    \ns{8,-0.5}
    \ns{8,-1}
    \on{8,0}
    \node at (2.5,0) {$\ldots$};
    \node at (8.75,0) {\footnotesize{ON${}^-$}};
    \node at (0.5,0) {\footnotesize{$1$}};
    \node at (1.5,0.5) {\footnotesize{$2$}};
    \node at (3.5,0) {\footnotesize{$2k{-}2$}};
    \node at (4.5,0.5) {\footnotesize{$2k{-}1$}};
    \node at (5.5,0.25) {\footnotesize{$2k$}};
    \draw[decoration={brace,mirror,raise=10pt},decorate,thick]
  (-0.1,-0.8) -- node[below=15pt] {\footnotesize{$2k$ D8}} (5.1,-0.8);
\end{tikzpicture}
}
\label{eq:6d_branes_ON_Higgs_infinite}
\end{align}
for which the proposed magnetic quiver reads
\begin{align}
        \raisebox{-.5\height}{
    \begin{tikzpicture}
	\node (g1) [gauge,label=below:{\footnotesize{$1$}}] {};
	\node (g2) [gauge,right of =g1,label=below:{\footnotesize{$2$}}] {};
	\node (g3) [gauge,right of =g2,label=below:{\footnotesize{$3$}}] {};
	\node (g4) [right of =g3] {$\ldots$};
	\node (g5) [gauge,right of =g4,label=below:{\footnotesize{$2k{-}3$}}] {};
	\node (g6) [gauge,right of =g5,label=below:{\footnotesize{$2k{-}2$}}] {};
	\node (g7) [gauge,right of =g6,label=below:{\footnotesize{$2k{-}1$}}] {};
	\node (g8) [gaugeSp,right of =g7,label=below:{\footnotesize{$2k$}}] {};
    \node (b1) [gaugeSO,above of=g8, label=right:{\footnotesize{$4$}}] {};
    \draw (b1) to [out=45,in=135,looseness=10] (b1);
	\draw  (g1)--(g2) (g2)--(g3) (g3)--(g4) (g4)--(g5) (g5)--(g6) (g6)--(g7) (g7)--(g8) (g8)--(b1);
	\end{tikzpicture}
    }
    \label{eq:6d_magQuiv_Unitary_infinite}
\end{align}
and the corresponding Coulomb branch satisfies
\begin{align}
\Coulomb  \eqref{eq:6d_magQuiv_Unitary_infinite} 
= \left[ \Higgs \left( \text{$\surm(k)$ SQCD, $2k$ flavours}\big|_{g^2=\infty}\right) \right]^2
= \left[ \Coulomb \left( \mathsf{M}_{k,(2)}\right) \right]^2 
\end{align}
which can be verified by Hilbert series techniques, see Table \ref{eq:6d_magQuiv_Unitary_infinite1} for details.

Before moving to orientifolds, a brief comment on non-symmetric splittings along the ON is in order. Following \cite{Hanany:1999sj}, a brane configuration of the form
\begin{align}
\raisebox{-.5\height}{
\begin{tikzpicture}
    \draw (-1-7,0)--(1-7,0) 
    (1-7,0.1)--(3-7,0.1) (1-7,-0.1)--(3-7,-0.1);
    \draw (3-7,1)--(3-7,-1);
    \ns{1-7,0}
    \onz{3-7,0}
    \node at (1-7,0.375) {\footnotesize{NS5}};
    \node at (3.5-7,-0.25) {\footnotesize{ON${}^0$}};
    \node at (-7,0.25) {\footnotesize{$l$}};
    \node at (-5,0.25) {\footnotesize{$k$}};
    \node at (-5,-0.25) {\footnotesize{$h$}};
    \node at (3.5-7,0.75) {\footnotesize{$w$ D8}};
 \node at (-2.75,0) {$\cong$};
    \draw (-2,0)--(0,0) 
    (0,0.1)--(2,0.1) (0,-0.1)--(2,-0.1)
    (2,0.1)--(4,0.1) (2,-0.1)--(4,-0.1);
    \draw (3.25,1)--(3.25,-1);
    \ns{0,0}
    \ns{2,0}
    \on{4,0}
    \node at (0,0.375) {\footnotesize{NS5}};
    \node at (2,0.375) {\footnotesize{NS5}};
    \node at (2,0.675) {\footnotesize{virtual}};
    \node at (4.15,0.375) {\footnotesize{ON${}^-$}};
    \node at (-1,0.25) {\footnotesize{$l$}};
    \node at (1,0.25) {\footnotesize{$k$}};
    \node at (1,-0.25) {\footnotesize{$h$}};
    \node at (2.75,0.25) {\footnotesize{$k$}};
    \node at (2.75,-0.25) {\footnotesize{$k$}};
    \node at (3.75,0.75) {\footnotesize{$w$ D8}};
\end{tikzpicture}
}
\label{eq:6d_branes_ON_different}
\end{align}
is constraint by charge conservation. The mismatch in number of D6 branes (for $k\neq h$) can be compensate by additional $w$ D8 branes, such that $w=k-h\geq 0$. The 6d $\Ncal=(1,0)$ world-volume theory is given by
\begin{align}
        \raisebox{-.5\height}{
    \begin{tikzpicture}
	\node (g1) [gauge,label=right:{\footnotesize{$\surm(k)$}}] {};
\node (f1) [flavour,above of=g1, label=right:{\footnotesize{$2k$}}] {};
\node at (2,0.5) {$\times$};
	\node (g2) [gauge,label=right:{\footnotesize{$\surm(h)$}}] at (3,0) {};
\node (f2) [flavour,above of=g2, label=right:{\footnotesize{$2h$}}] {};
	\draw  (g1)--(f1) (g2)--(f2);
	\end{tikzpicture}
    }
\end{align}
which are anomaly-free since the charges are conserved in the brane configuration. The potential finite coupling magnetic quiver is given by
\begin{align}
        \raisebox{-.5\height}{
    \begin{tikzpicture}
	\node (g1) [gauge,label=below:{\footnotesize{$1$}}] {};
	\node (g2) [gauge,right of =g1,label=below:{\footnotesize{$2$}}] {};
	\node (g3) [right of =g2] {$\ldots$};
	\node (g4) [gauge,right of =g3,label=below:{\footnotesize{$2h{-}1$}}] {};
	\node (g5) [gauge,right of =g4,label=below:{\footnotesize{$2h$}}] {};
	\node (g6) [gauge,right of =g5,label=below:{\footnotesize{$2h{+}2$}}] {};
	\node (g7) [gauge,right of =g6,label=below:{\footnotesize{$2h{+}4$}}] {};
	\node (g8) [right of =g7] {$\ldots$};
	\node (g9) [gauge,right of =g8,label=below:{\footnotesize{$2k{-}4$}}] {};
	\node (g10) [gauge,right of =g9,label=below:{\footnotesize{$2k{-}2$}}] {};
	\node (g11) [gaugeSp,right of =g10,label=below:{\footnotesize{$2k$}}] {};
    \node (b1) [gaugeSO,above left of=g11, label=above:{\footnotesize{$2$}}] {};
    \node (b2) [gaugeSO,above right of=g11, label=above:{\footnotesize{$2$}}] {};
	\draw  (g1)--(g2) (g2)--(g3) (g3)--(g4) (g4)--(g5) (g5)--(g6) (g6)--(g7) (g7)--(g8) (g8)--(g9) (g9)--(g10) (g10)--(g11) (g11)--(b1) (g11)--(b2);
	\end{tikzpicture}
    }
    \label{eq:6d_magQuiv_Unitary_different_finite}
\end{align}
for which the $\urm(2h)$ node is \emph{good}, but not balanced. As the $\sprm(k)$ node is \emph{bad}, the Coulomb branch is more intricate and cannot be analysed using the monopole formula.
\subsubsection{\texorpdfstring{ON${}^0$ with O6 and O8 planes}{ON with O6 and O8 planes}}
\label{sec:6d_ON_and_Op}
Besides the ON plane, O6 and O8 planes can be introduced alongside. Analogously to \cite{Hanany:1999sj}, consider the brane configuration 
\begin{align}
\raisebox{-.5\height}{
\begin{tikzpicture}
    \draw[dashed] (-8,0)--(-4,0);
    \draw[dashed] (-4,-1)--(-4,1);
    \draw (-1-7,0.1)--(1-7,0.1) (-1-7,-0.1)--(1-7,-0.1)
    (1-7,0.1)--(3-7,0.1) (1-7,-0.1)--(3-7,-0.1);
    \ns{1-7,0}
    \onz{3-7,0}
    \node at (-6,0.375) {\footnotesize{NS5}};
    \node at (-3.5,0) {\footnotesize{ON${}^0$}};
    \node at (-7,0.25) {\footnotesize{$l$}};
    \node at (-5,0.25) {\footnotesize{$k+h$}};
    \node at (-3.5,0.75) {\footnotesize{O$8^-$}};
    \node at (-7,-0.5) {\footnotesize{O$6^-$}};
    \node at (-5,-0.5) {\footnotesize{O$6^+$}};
 \node at (-2.75,0) {$\cong$};
    \draw[dashed] (-2,0)--(4,0);
    \draw[dashed] (4,-1)--(4,1);
    \draw (-2,0.1)--(0,0.1) (-2,-0.1)--(0,-0.1)
    (0,0.1)--(2,0.1) (0,-0.1)--(2,-0.1)
    (2,0.1)--(4,0.1) (2,-0.1)--(4,-0.1);
    \ns{0,0}
    \ns{2,0}
    \on{4,0}
    \node at (0,0.375) {\footnotesize{NS5}};
    \node at (2,0.375) {\footnotesize{NS5}};
    \node at (2,0.675) {\footnotesize{virtual}};
    \node at (4.55,0) {\footnotesize{ON${}^-$}};
    \node at (-1,0.25) {\footnotesize{$l$}};
    \node at (1,0.25) {\footnotesize{$k+h$}};
    \node at (3,0.25) {\footnotesize{$2k$}};
    \node at (-1,-0.5) {\footnotesize{O$6^-$}};
    \node at (1,-0.5) {\footnotesize{O$6^+$}};
    \node at (3,-0.5) {\footnotesize{O$6^-$}};
    \node at (4.55,0.75) {\footnotesize{O${8}^-$}};
\end{tikzpicture}
}
\label{eq:6d_branes_ON_and_Op}
\end{align}
where the numbers on top of the D6 branes denote full brane.
Charge conservation or, equivalently, anomaly cancellation dictate that $k=h$ and $l=2k+8$ such that the 6d $\Ncal=(1,0)$ world-volume theory is given by
\begin{align}
        \raisebox{-.5\height}{
    \begin{tikzpicture}
	\tikzstyle{gaugeSO} = [circle, draw,inner sep=3pt,fill=red];
	\tikzstyle{gaugeSp} = [circle, draw,inner sep=3pt,fill=blue];
	\tikzstyle{flavour} = [regular polygon,regular polygon sides=4,inner
sep=3pt, draw];
\tikzstyle{flavourSO} = [regular polygon,regular polygon sides=4,inner
sep=3pt, draw, fill=red];
\tikzstyle{flavourSp} = [regular polygon,regular polygon sides=4,inner
sep=3pt, draw, fill=blue];
	\node (g1) [gaugeSp,label=right:{\footnotesize{$\usprm(2k)$}}] {};
    \node (f1) [flavourSO,above of=g1, label=right:{\footnotesize{$\sorm(4k+16)$}}] {};
\node at (2,0.5) {$\times$};
	\node (g2) [gaugeSp,label=right:{\footnotesize{$\usprm(2k)$}}] at (3,0) {};
\node (f2) [flavourSO,above of=g2, label=right:{\footnotesize{$\sorm(4k+16)$}}] {};
	\draw  (g1)--(f1) (g2)--(f2);
	\end{tikzpicture}
    } 
\end{align}
The finite and infinite Higgs branches of $\usprm(2k)$ with $N_f=2k+8$ fundamental flavours are known to be related by the small $E_8$ instanton transition \cite{Ganor:1996mu,Hanany:2018uhm,Cabrera:2019izd,Cabrera:2019dob}. As above, the first challenge is to derive the magnetic quiver for the brane configuration \eqref{eq:6d_branes_ON_and_Op}. For this, one moves in $4k+16$ half D8 branes such that all of the semi-infinite D6 flavour branes can be terminated on one such half D8. After suitable brane transitions, the configuration \eqref{eq:6d_branes_ON_and_Op} is equivalently described by
\begin{align}
\raisebox{-.5\height}{
\begin{tikzpicture}
    \draw[dashed] (-6.5,0)--(4,0);
    \draw (-5.5,0.1)--(4,0.1) (-5.5,-0.1)--(4,-0.1);
    \draw (0,1)--(0,-1) (0.2,1)--(0.2,-1) 
    (0.8,1)--(0.8,-1) (1,1)--(1,-1);
    \node at (0.5,0.5) {$\cdots$};
    \node at (0.5,-0.5) {$\cdots$};
    \draw (-2,1)--(-2,-1) (-2.5,1)--(-2.5,-1) 
    (-3,1)--(-3,-1) (-3.5,1)--(-3.5,-1)
    (-4.5,1)--(-4.5,-1) (-5,1)--(-5,-1) 
    (-5.5,1)--(-5.5,-1) (-6,1)--(-6,-1);
    \node at (-4,0.5) {$\cdots$};
    \node at (-4,-0.5) {$\cdots$};
    \draw[dashed] (4,-1)--(4,1);
    \ns{-1,0}
    \ns{2,0}
    \on{4,0}
    \node at (4.55,0) {\footnotesize{ON${}^-$}};
    \node at (-5.25,0.25) {\footnotesize{$1$}};
    \node at (-4.75,0.25) {\footnotesize{$1$}};
    \node at (-1.5,0.25) {\footnotesize{$2k$}};
    \node at (-0.5,0.25) {\footnotesize{$2k$}};
    \node at (1.5,0.25) {\footnotesize{$2k$}};
    \node at (3,0.25) {\footnotesize{$2k$}};
    \node at (-6.25,-0.8) {\tiny{$-$}};
    \node at (-5.75,-0.8) {\tiny{$\widetilde{-}$}};
    \node at (-5.25,-0.8) {\tiny{$-$}};
    \node at (-4.75,-0.8) {\tiny{$\widetilde{-}$}};
    \node at (-3.25,-0.8) {\tiny{$\widetilde{-}$}};
    \node at (-2.75,-0.8) {\tiny{$\widetilde{-}$}};
    \node at (-2.25,-0.8) {\tiny{$-$}};
    \node at (-1.5,-0.8) {\tiny{$\widetilde{-}$}};
    \node at (-0.5,-0.8) {\tiny{$\widetilde{+}$}};
    \node at (1.5,-0.8) {\tiny{$+$}};
    \node at (3,-0.8) {\tiny{$-$}};
    \node at (4.55,0.75) {\footnotesize{O${8}^-$}};
    \draw[decoration={brace,mirror,raise=10pt},decorate,thick]
  (-0.1,-0.8) -- node[below=15pt] {\footnotesize{$15$ half D8}} (1.1,-0.8);
  \draw[decoration={brace,mirror,raise=10pt},decorate,thick]
  (-6.1,-0.8) -- node[below=15pt] {\footnotesize{$4k{+}1$ half D8}} (-1.9,-0.8);
\end{tikzpicture}
}
\label{eq:6d_branes_ON_and_Op_Higgs}
\end{align}
and the task is to derive the magnetic quiver. To begin with, all D6 branes need to be suspended between D8 branes and one observes that the branes away from the intersection of O6, O8 and ON only perceive the presence of the O6 orientifold. Hence, the major part of the brane system can be treated by the rules derived in \cite{Cabrera:2019dob}.  Therefore, to focus is place at the branes near the intersection of the O-planes. Concretely, the brane configuration becomes
\begin{align}
\raisebox{-.5\height}{
\begin{tikzpicture}
    \draw[dashed] (-1.5,0)--(3,0);
    \draw (-1.5,0.25)--(-1,0.25) (-1.5,-0.25)--(-1,-0.25);
    \draw[thick,myGreen] (-1,0.5)--(0,0.5) (-1,-0.5)--(0,-0.5);
    \draw[thick,blue] (0,0.35)--(2,0.35) (0,-0.35)--(2,-0.35);
    \draw[thick,red] (0,0.75)--(2.8,0.75)--(3,0.65)--(2.8,0.55)--(2,0.55) (0,-0.75)--(2.8,-0.75)--(3,-0.65)--(2.8,-0.55)--(2,-0.55);
    \draw (-1,1)--(-1,-1) (0,1)--(0,-1) (2,1)--(2,-1);
    \draw[dashed] (3,-1)--(3,1);
    \ns{1,0}
    \on{3,0}
    \node at (3.55,0) {\footnotesize{ON${}^-$}};
    \node at (-1.5,0.45) {\footnotesize{$2k$}};
    \node[myGreen] at (-0.5,0.75) {\footnotesize{$2k$}};
    \node[blue] at (0.5,0.55) {\footnotesize{$k$}};
    \node[red] at (1.5,0.95) {\footnotesize{$k$}};
    \node at (-1.5,-1) {\tiny{$+$}};
    \node at (-0.5,-1) {\tiny{$\widetilde{+}$}};
    \node at (0.5,-1) {\tiny{$\widetilde{+}$}};
    \node at (1.5,-1) {\tiny{$\widetilde{-}$}};
    \node at (2.5,-1) {\tiny{$-$}};
    \node at (3.55,0.75) {\footnotesize{O${8}^-$}};
\end{tikzpicture}
}
\quad \rightarrow \quad 
        \raisebox{-.5\height}{
    \begin{tikzpicture}
	\node (g11)  {$\ldots$};
	\node[thick,myGreen] (g12) [gauge,right of =g11,label=below:{\footnotesize{$B_{2k}$}}] {};
	\node[thick,blue] (g13) [gauge,below right of =g12,label=below:{\footnotesize{$C_k$}}] {};
	\node[thick,red] (g14) [gauge,above right of =g12,label=below:{\footnotesize{$C_k$}}] {};
    \node (f2) [flavour,right of=g13, label=below:{\footnotesize{$B_0$}}] {};
    \node (f3) [flavour,right of=g14, label=below:{\footnotesize{$B_0$}}] {};
	\draw (g11)--(g12) (g12)--(g13) (g12)--(g14) (g13)--(f2) (g14)--(f3);
	\end{tikzpicture}
    }
\end{align}
where the stuck half NS5 brane has been moved passed one half D8 brane. This brane transition is not accompanied by creation or creation of any physical brane and, moreover, is convenient for reading off the magnetic quiver \cite{Cabrera:2019dob}. Next, the branes are split as indicated, where the O8 plane has the same bifurcating effect as in \cite{Cabrera:2019izd}; however, the bifurcation is into two $C$-type magnetic gauge algebras, which originate from the red and blue stack of D6 branes. The stuck NS5 brane then becomes a flavour half-hypermultiplet, denoted by $B_0$.
In summary, the magnetic quiver is proposed to be (see also Appendix \ref{app:branes})
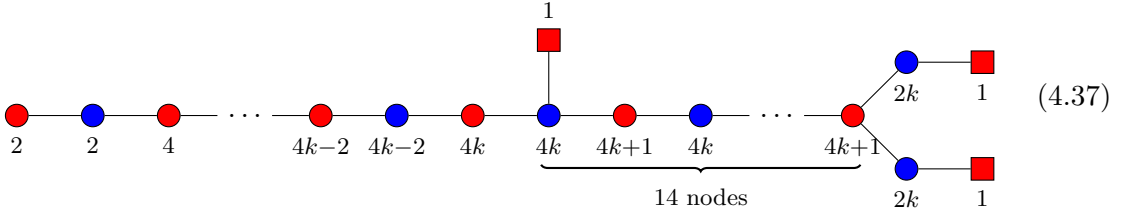
\begin{align}
        \raisebox{-.5\height}{
    \begin{tikzpicture}
	\node (g1) [gaugeSO,label=below:{\footnotesize{$2$}}] {};
	\node (g2) [gaugeSp,right of =g1,label=below:{\footnotesize{$2$}}] {};
	\node (g3) [gaugeSO,right of =g2,label=below:{\footnotesize{$4$}}] {};
	\node (g4) [right of =g3] {$\ldots$};
	\node (g5) [gaugeSO,right of =g4,label=below:{\footnotesize{$4k{-}2$}}] {};
	\node (g6) [gaugeSp,right of =g5,label=below:{\footnotesize{$4k{-}2$}}] {};
	\node (g7) [gaugeSO,right of =g6,label=below:{\footnotesize{$4k$}}] {};
	\node (g8) [gaugeSp,right of =g7,label=below:{\footnotesize{$4k$}}] {};
	\node (g9) [gaugeSO,right of =g8,label=below:{\footnotesize{$4k{+}1$}}] {};
	\node (g10) [gaugeSp,right of =g9,label=below:{\footnotesize{$4k$}}] {};
	\node (g11) [right of =g10] {$\ldots$};
	\node (g12) [gaugeSO,right of =g11,label=below:{\footnotesize{$4k{+}1$}}] {};
	\node (g13) [gaugeSp,below right of =g12,label=below:{\footnotesize{$2k$}}] {};
	\node (g14) [gaugeSp,above right of =g12,label=below:{\footnotesize{$2k$}}] {};
	\node (f1) [flavourSO,above of=g8, label=above:{\footnotesize{$1$}}] {};
    \node (f2) [flavourSO,right of=g13, label=below:{\footnotesize{$1$}}] {};
    \node (f3) [flavourSO,right of=g14, label=below:{\footnotesize{$1$}}] {};
	\draw  (g1)--(g2) (g2)--(g3) (g3)--(g4) (g4)--(g5) (g5)--(g6) (g6)--(g7) (g7)--(g8) (g8)--(g9) (g9)--(g10) (g10)--(g11) (g11)--(g12) (g12)--(g13) (g12)--(g14) (g8)--(f1) (g13)--(f2) (g14)--(f3);
\draw[decoration={brace,mirror,raise=10pt},decorate,thick]
  (7-0.1,-0.3) -- node[below=15pt] {\footnotesize{$14$ nodes}} (11.1,-0.3);
	\end{tikzpicture}
    }
    \label{eq:6d_sp_magQuiv_finite}
\end{align}
which coincides with the pure 3d result \eqref{eq:ex_product_mirror_OSp} for the $n=2k+8$. Hence, the associated moduli space satisfies the product space property.

As a next step, the brane system is transitioned into the infinite coupling regime by moving the NS5s onto the orientifold. Taking care of brane creation, the system becomes
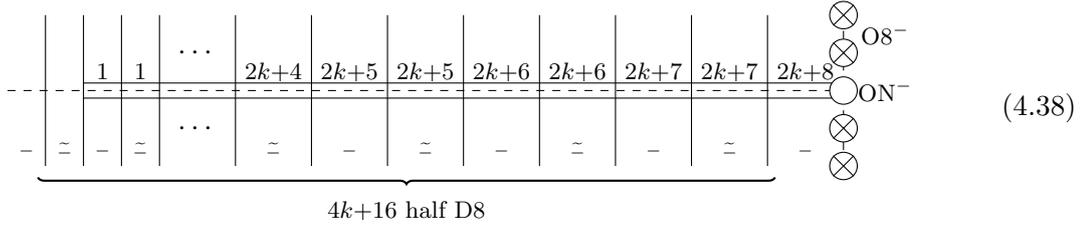
\begin{align}
\raisebox{-.5\height}{
\begin{tikzpicture}
    \draw[dashed] (-6.5,0)--(4.5,0);
    \draw (-5.5,0.1)--(4.5,0.1) (-5.5,-0.1)--(4.5,-0.1);
    \draw (3.5,1)--(3.5,-1) (2.5,1)--(2.5,-1)
    (1.5,1)--(1.5,-1) (0.5,1)--(0.5,-1)
    (-0.5,1)--(-0.5,-1) (-1.5,1)--(-1.5,-1) 
    (-2.5,1)--(-2.5,-1) (-3.5,1)--(-3.5,-1)
    (-4.5,1)--(-4.5,-1) (-5,1)--(-5,-1) 
    (-5.5,1)--(-5.5,-1) (-6,1)--(-6,-1);
    \node at (-4,0.5) {$\cdots$};
    \node at (-4,-0.5) {$\cdots$};
    \draw[dashed] (4.5,-1)--(4.5,1);
    \ns{4.5,1}
    \ns{4.5,0.5}
    \on{4.5,0}
    \ns{4.5,-1}
    \ns{4.5,-0.5}
    \node at (5.05,0) {\footnotesize{ON${}^-$}};
    \node at (-5.25,0.25) {\footnotesize{$1$}};
    \node at (-4.75,0.25) {\footnotesize{$1$}};
    \node at (-3,0.25) {\footnotesize{$2k{+}4$}};
    \node at (-2,0.25) {\footnotesize{$2k{+}5$}};
    \node at (-1,0.25) {\footnotesize{$2k{+}5$}};
    \node at (0,0.25) {\footnotesize{$2k{+}6$}};
    \node at (1,0.25) {\footnotesize{$2k{+}6$}};
    \node at (2,0.25) {\footnotesize{$2k{+}7$}};
    \node at (3,0.25) {\footnotesize{$2k{+}7$}};
    \node at (4,0.25) {\footnotesize{$2k{+}8$}};
    \node at (-6.25,-0.8) {\tiny{$-$}};
    \node at (-5.75,-0.8) {\tiny{$\widetilde{-}$}};
    \node at (-5.25,-0.8) {\tiny{$-$}};
    \node at (-4.75,-0.8) {\tiny{$\widetilde{-}$}};
    \node at (-3,-0.8) {\tiny{$\widetilde{-}$}};
    \node at (-2,-0.8) {\tiny{$-$}};
    \node at (-1,-0.8) {\tiny{$\widetilde{-}$}};
    \node at (0,-0.8) {\tiny{$-$}};
    \node at (1,-0.8) {\tiny{$\widetilde{-}$}};
    \node at (2,-0.8) {\tiny{$-$}};
    \node at (3,-0.8) {\tiny{$\widetilde{-}$}};
    \node at (4,-0.8) {\tiny{$-$}};
    \node at (5.05,0.75) {\footnotesize{O${8}^-$}};
  \draw[decoration={brace,mirror,raise=10pt},decorate,thick]
  (-6.1,-0.8) -- node[below=15pt] {\footnotesize{$4k{+}16$ half D8}} (3.6,-0.8);
\end{tikzpicture}
}
\label{eq:6d_branes_ON_and_Op_infinite}
\end{align}
where passing the half NS5 through the 15 half D8 branes has lead to D6 brane creation. As above, the brane away from the O8 and ON only perceive the O6 plane such that the magnetic quiver construction for this part follows from \cite{Cabrera:2019dob}. The novel part lies in the behaviour of the branes close to the intersection of the O-planes. Concretely, the brane configuration becomes
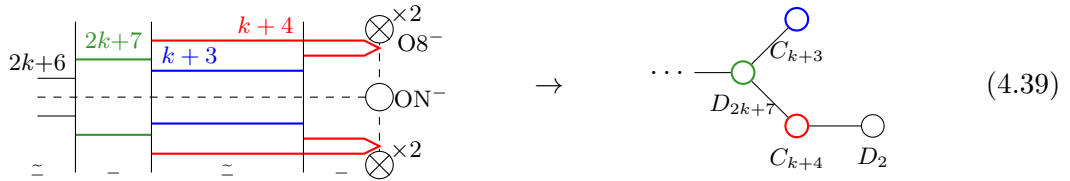
\begin{align}
\raisebox{-.5\height}{
\begin{tikzpicture}
    \draw[dashed] (-1.5,0)--(3,0);
    \draw (-1.5,0.25)--(-1,0.25) (-1.5,-0.25)--(-1,-0.25);
    \draw[thick,myGreen] (-1,0.5)--(0,0.5) (-1,-0.5)--(0,-0.5);
    \draw[thick,blue] (0,0.35)--(2,0.35) (0,-0.35)--(2,-0.35);
    \draw[thick,red] (0,0.75)--(2.8,0.75)--(3,0.65)--(2.8,0.55)--(2,0.55) (0,-0.75)--(2.8,-0.75)--(3,-0.65)--(2.8,-0.55)--(2,-0.55);
    \draw (-1,1)--(-1,-1) (0,1)--(0,-1) (2,1)--(2,-1);
    \draw[dashed] (3,-1)--(3,1);
    \ns{3,0.9}
    \ns{3,-0.9}
    \on{3,0}
    \node at (3.55,0) {\footnotesize{ON${}^-$}};
    \node at (3.35,1.1) {\footnotesize{$\times 2$}};
    \node at (3.35,-0.7) {\footnotesize{$\times 2$}};
    \node at (-1.5,0.45) {\footnotesize{$2k{+}6$}};
    \node[myGreen] at (-0.5,0.75) {\footnotesize{$2k{+}7$}};
    \node[blue] at (0.5,0.55) {\footnotesize{$k+3$}};
    \node[red] at (1.5,0.95) {\footnotesize{$k+4$}};
    \node at (-1.5,-1) {\tiny{$\widetilde{-}$}};
    \node at (-0.5,-1) {\tiny{$-$}};
    \node at (1,-1) {\tiny{$\widetilde{-}$}};
    \node at (2.5,-1) {\tiny{$-$}};
    \node at (3.55,0.75) {\footnotesize{O${8}^-$}};
\end{tikzpicture}
}
\qquad \rightarrow \qquad 
        \raisebox{-.5\height}{
    \begin{tikzpicture}
	\node (g11)  {$\ldots$};
	\node[thick,myGreen] (g12) [gauge,right of =g11,label=below:{\footnotesize{$D_{2k{+}7}$}}] {};
	\node[thick,blue] (g13) [gauge,above right of =g12,label=below:{\footnotesize{$C_{k{+}3}$}}] {};
	\node[thick,red] (g14) [gauge,below right of =g12,label=below:{\footnotesize{$C_{k{+}4}$}}] {};
\node (f3) [gauge,right of=g14, label=below:{\footnotesize{$D_2$}}] {};
	\draw (g11)--(g12) (g12)--(g13) (g12)--(g14) (g14)--(f3);
	\end{tikzpicture}
    }
\end{align}
in which the presence of the O8 plane led to the splitting into the red and blue stack of D6 branes. The type of magnetic gauge algebra is determined from the magnetic orientifolds associated to the O6 planes. Compared to the finite coupling brane configuration, the NS5 branes are now confined to the O8 plane. The associated magnetic gauge algebra is $D_2$. In summary, the magnetic quiver is proposed to be
\begin{align}
        \raisebox{-.5\height}{
    \begin{tikzpicture}
	\node (g1) [gaugeSO,label=below:{\footnotesize{$2$}}] {};
	\node (g2) [gaugeSp,right of =g1,label=below:{\footnotesize{$2$}}] {};
	\node (g3) [gaugeSO,right of =g2,label=below:{\footnotesize{$4$}}] {};
	\node (g4) [right of =g3] {$\ldots$};
	\node (g5) [gaugeSp,right of =g4,label=above:{\footnotesize{$4k{+}10$}}] {};
	\node (g6) [gaugeSO,right of =g5,label=below:{\footnotesize{$4k{+}12$}}] {};
	\node (g7) [gaugeSp,right of =g6,label=above:{\footnotesize{$4k{+}12$}}] {};
	\node (g8) [gaugeSO,right of =g7,label=below:{\footnotesize{$4k{+}14$}}] {};
	\node (g9) [gaugeSp, above right of =g8,label=below:{\footnotesize{$2k{+}6$}}] {};
	\node (g10) [gaugeSp,below right of =g8,label=below:{\footnotesize{$2k{+}8$}}] {};
	\node (g11) [gaugeSO,right of =g10,label=below:{\footnotesize{$4$}}] {};
	\draw  (g1)--(g2) (g2)--(g3) (g3)--(g4) (g4)--(g5) (g5)--(g6) (g6)--(g7) (g7)--(g8) (g8)--(g9) (g8)--(g10) (g10)--(g11);
	\end{tikzpicture}
    }
    \label{eq:6d_sp_magQuiv_infinite}
\end{align}
which coincides with the result of Table \ref{EnProdtable2} after shifting indices $k\to k+1$.

The $k=0$ case is of importance as the finite coupling theory is two copies of $\sprm(0)$ with 8 fundamental flavours which in the UV becomes two copies of the E-string theory. The infinite coupling magnetic quiver simplifies to quiver in the first row of Table \ref{EnProdtable}, with associated moduli space being the product of the minimal nilpotent orbit closure of $E_8$.

Contrary to the 3d brane configuration of Section \ref{sec:branes_3d}, one cannot replace the O8${}^-$ plane with a hypothetical $\widetilde{\text{O8}}^-$, because such orientifolds are inconsistent \cite{Hyakutake:2000mr}.

\subsection{5d brane systems}
\label{sec:branes_5d}
Lastly, one may consider 5-branes with ON planes. Magnetic quivers for 5-brane webs have been developed in \cite{Cabrera:2018jxt,Bourget:2020gzi,Akhond:2020vhc,Akhond:2021knl}. Here, the approach is generalised to accommodate for ON planes.
\subsubsection{\texorpdfstring{Inclusion of ON${}^0$ plane}{Inclusion of ON plane}}
\label{sec:ON_5d}
The simplest 5-brane which gives rise to a  5d $\Ncal=1$ product theory is given by
\begin{align}
            \raisebox{-.5\height}{
    \begin{tikzpicture}
    \draw[dotted] (4,-1)--(4,1);
    \draw (3.8,1)--(3.8,-1) (2,1)--(2,-1) 
    (0,0.5)--(4,0.5) (0,0.7)--(3.8,0.7) (3.8,0.8)--(4,0.8)
    (0,0.0)--(4,0.0) (0,0.2)--(3.8,0.2) (3.8,0.3)--(4,0.3)
    (0,-0.6)--(3.8,-0.6) (0,-0.8)--(4,-0.8) (3.8,-0.5)--(4,-0.5);
    \node at (1,-0.2) {$\vdots$};
    \node at (3,-0.2) {$\vdots$};
    \node at (4.3,1){\footnotesize{ON${}^-$}};
    \node at (2.3,1){\footnotesize{NS5}}; 
    \node at (1,1){\footnotesize{D5}}; 
    \draw[decoration={brace,mirror,raise=10pt},decorate,thick]
  (0,0.8) -- node[left=15pt] {\footnotesize{$2k$ D5}} (0,-0.9);
    \end{tikzpicture}
    }
    \qquad 
    \leftrightarrow
    \qquad
            \raisebox{-.5\height}{
    \begin{tikzpicture}
	\node (g1) [gauge,label=right:{\footnotesize{$\surm(k)$}}] {};
    \node (f1) [flavour,above of=g1, label=right:{\footnotesize{$2k$}}] {};
    \node at (2,0.5) {$\times$};
	\node (g2) [gauge,label=right:{\footnotesize{$\surm(k)$}}] at (3,0) {};
    \node (f2) [flavour,above of=g2, label=right:{\footnotesize{$2k$}}] {};
	\draw  (g1)--(f1) (g2)--(f2);
	\end{tikzpicture}
    }
    \label{eq:5-brane_ON}
\end{align}
which describes a product theory due to the ON${}^0$. In order to read the magnetic quiver, the 5-brane web is transferred into the Higgs branch phase. For this, the $(p,q)$ 5-brane are terminated on $[p,q]$ 7-branes and the system becomes
\begin{align}
    \raisebox{-.5\height}{
    \begin{tikzpicture}
    \draw[dotted] (4,-1)--(4,1);
    \draw (-2,0)--(0,0) (1,0)--(4,0);
    \draw (3,-1)--(3,1) (3.5,-1)--(3.5,1);
    \SevenB{2,0}
    \SevenB{1,0}
    \node at (0.5,0) {$\cdots$};
    \SevenB{0,0}
    \SevenB{-1,0}
    \SevenB{-2,0}
    \SevenB{3,-1}
    \SevenB{3,1}
    \SevenB{3.5,-1}
    \SevenB{3.5,1}
    \node at (4.35,1){\footnotesize{ON${}^-$}};
    \node at (-2.1,0.35){\footnotesize{D7}}; 
    \node at (-1.5,0.25) {\footnotesize{$1$}};
    \node at (-0.5,0.25) {\footnotesize{$2$}};
    \node at (1.5,0.25) {\footnotesize{$2k{-}1$}};
    \node at (2.5,0.25) {\footnotesize{$2k$}};
    \draw[decoration={brace,mirror,raise=10pt},decorate,thick]
  (-2.1,-0.2) -- node[below=15pt] {\footnotesize{$2k$ D7}} (2.1,-0.2);
    \end{tikzpicture}
    }
\end{align}
from which one derives that finite coupling magnetic quiver is given by \eqref{eq:6d_magQuiv_Unitary_finite}.

Moving on to the infinite coupling case, one noticed that the S-dual 5-brane web of \eqref{eq:5-brane_ON} is given by \cite{Hayashi:2015vhy}
\begin{align}
    \raisebox{-.5\height}{
    \begin{tikzpicture}
    \draw[dashed] (0,0)--(9,0);
    \foreach \i in {1,...,8}  \draw (\i,0)--(\i,1.5); 
    \foreach \i in {0,1,3,4}
    \foreach \j in {0,1}
    \draw (0+2*\i,0.25+0.25*\j)--(1+2*\i,0.25+0.25*\j);
    \node at (4.5,1) {$\cdots$};
    \draw[decoration={brace,mirror,raise=10pt},decorate,thick]
  (8.1,1.5) -- node[above=15pt] {\footnotesize{$2k$ half NS5}} (0.9,1.5);
 \foreach \i in {0,1,3,4} 
 \node at (0.5+2*\i,-0.2) {\tiny{O5${}^-$}};
 \foreach \i in {0,...,3} 
 \node at (1.5+2*\i,-0.2) {\tiny{O5${}^+$}};
    \end{tikzpicture}
    }
\end{align}
such that the 5d low-energy gauge theory description becomes
\begin{align}
        \raisebox{-.5\height}{
    \begin{tikzpicture}
	\node (g1) [gaugeSp,label=below:{\footnotesize{$0$}}] {};
	\node (g2) [gaugeSO,right of =g1,label=below:{\footnotesize{$2$}}] {};
	\node (g3) [gaugeSp,right of =g2,label=below:{\footnotesize{$0$}}] {};
	\node (g4) [right of =g3] {$\ldots$};
	\node (g5) [gaugeSp,right of =g4,label=below:{\footnotesize{$0$}}] {};
	\node (g6) [gaugeSO,right of =g5,label=below:{\footnotesize{$2$}}] {};
	\node (g7) [gaugeSp,right of =g6,label=below:{\footnotesize{$0$}}] {};
	\node (f1) [flavourSO,left of=g1,label=below:{\footnotesize{$2$}}] {};
	\node (f2) [flavourSO,right of=g7,label=below:{\footnotesize{$2$}}] {};
	\draw  (g1)--(g2) (g2)--(g3) (g3)--(g4) (g4)--(g5) (g5)--(g6) (g6)--(g7) (g7)--(f2) (g1)--(f1);
	\end{tikzpicture}
    }
    \label{eq:5d_OSp_lin_quiver}
\end{align}
However, the point is that S-duality preserves the UV fix point; hence, the theories \eqref{eq:5-brane_ON} and \eqref{eq:5d_OSp_lin_quiver} have the same infinite coupling Higgs branch.

For this class of 5d $\Ncal=1$ linear orthosymplectic gauge theories \eqref{eq:5d_OSp_lin_quiver}, the magnetic quiver construction has been derived in \cite{Bourget:2020gzi}. Specialising to the case of trivial $\sprm(0)$ nodes yields the following magnetic quiver \cite[eq.\ (6.22)]{Bourget:2020gzi}
\begin{align}
        \raisebox{-.5\height}{
    \begin{tikzpicture}
	\node (g1) [gaugeSO,label=below:{\footnotesize{$2$}}] {};
	\node (g2) [gaugeSp,right of =g1,label=below:{\footnotesize{$2$}}] {};
	\node (g3) [gaugeSO,right of =g2,label=below:{\footnotesize{$4$}}] {};
	\node (g4) [gaugeSp,right of =g3,label=below:{\footnotesize{$2$}}] {};
	\node (g5) [gaugeSO,right of =g4,label=below:{\footnotesize{$2$}}] {};
	\node (b1) [gaugeSp,above of =g3,label=right:{\footnotesize{$2k$}}] {};
	\node (b2) [gauge,above of=b1,label=right:{\footnotesize{$2k{-}1$}}] {};
	\node (b3) [above of =b2] {$\vdots$};
	\node (b4) [gauge,above of=b3,label=right:{\footnotesize{$2$}}] {};
	\node (b5) [gauge,above of=b4,label=right:{\footnotesize{$1$}}] {};
	\draw  (g1)--(g2) (g2)--(g3) (g3)--(g4) (g4)--(g5) (g3)--(b1) (b1)--(b2) (b2)--(b3) (b3)--(b4) (b4)--(b5);
	\end{tikzpicture}
    }
    \label{eq:5d_maqQuiv_ON_infinite}
\end{align}
which is the infinite coupling magnetic quiver for \eqref{eq:5-brane_ON}.
The associated Coulomb branch is a product space; for example, for $k=2$ the moduli space is $\Coulomb \left( \eqref{eq:5d_maqQuiv_ON_infinite} \right) = \overline{\mathcal{O}}^{\min}_{E_5}  \times \overline{\mathcal{O}}^{\min}_{E_5}$.

Equivalently, the magnetic quiver can be derived from the brane web without resorting to S-duality. At the fixed point, the 5-brane web can be explicitly written as 
\begin{align}
    \raisebox{-.5\height}{
    \begin{tikzpicture}
    \draw[dotted] (4,-3)--(4,3);
    \draw (-2,0)--(0,0) (1,0)--(4,0);
    \draw (3.9,-2.25)--(3.9,2.25) (4.1,-2.25)--(4.1,2.25);
    \SevenB{2,0}
    \SevenB{1,0}
    \node at (0.5,0) {$\cdots$};
    \SevenB{0,0}
    \SevenB{-1,0}
    \SevenB{-2,0}
    \SevenB{4,0.75}
    \SevenB{4,1.5}
    \SevenB{4,2.25}
    \SevenB{4,3}
    \SevenB{4,-0.75}
    \SevenB{4,-1.5}
    \SevenB{4,-2.25}
    \SevenB{4,-3}
    \node at (4.75,0.5){\footnotesize{ON${}^-$}};
    \node at (4.75,0.5+0.75){\footnotesize{$\widetilde{\mathrm{ON}}^-$}};
    \node at (4.75,0.5+1.5){\footnotesize{ON${}^-$}};
    \node at (4.75,0.5+2.25){\footnotesize{$\widetilde{\mathrm{ON}}^-$}};
    \node at (-2.15,0.4){\footnotesize{$[1,0]$}}; 
    \node at (3.5,3){\footnotesize{$[0,1]$}};
    \node at (-1.5,0.25) {\footnotesize{$1$}};
    \node at (-0.5,0.25) {\footnotesize{$2$}};
    \node at (1.5,0.25) {\footnotesize{$2k{-}1$}};
    \node at (2.5,0.25) {\footnotesize{$2k$}};
    \node at (3.75,0.5) {\footnotesize{$2$}};
    \node at (3.75,0.5+0.75) {\footnotesize{$1$}};
    \node at (3.75,0.5+1.5) {\footnotesize{$1$}};
    \node at (3.75,-0.5) {\footnotesize{$2$}};
    \node at (3.75,-0.5-0.75) {\footnotesize{$1$}};
    \node at (3.75,-0.5-1.5) {\footnotesize{$1$}};
    \draw[decoration={brace,mirror,raise=10pt},decorate,thick]
  (-2.1,-0.2) -- node[below=15pt] {\footnotesize{$2k$ D7}} (2.1,-0.2);
    \end{tikzpicture}
    }
\end{align}
such that a stack of NS5 branes on a ON${}^-$ yields a $D$-type magnetic gauge algebra; while a stack of NS5 branes on top of a $\widetilde{\mathrm{ON}}^-$ results in a $C$-type magnetic gauge algebra. Hence, the rules for a stack $(1,0)$ 5-branes on a O5 plane are the same as for a stack of $(0,1)$ 5-branes on a ON plane.
The generalised intersection numbers are straightforwardly computed \cite{Bourget:2020gzi}, as there are only $(1,0)$ and $(0,1)$ 5-branes and their $[p,q]$ 7-branes involved.
\subsubsection{\texorpdfstring{ON${}^0$ with O$5$ and O$7$ planes}{ON with O5 and O7 planes}}
\label{sec:5d_ON_and_Op}
As a last setup, consider a 5-brane web in the presence of an ON${}^0$, an O7$^-$, and an O5${}^+$. For reference, brane webs in the presence of either O5 or O7 planes are discussed in \cite{Brunner:1997gk,DeWolfe:1999hj,Zafrir:2015ftn,Bergman:2015dpa}; while ON planes are consider in \cite{Sen:1998rg,Sen:1998ii,Kapustin:1998fa,Hanany:1999sj,Hayashi:2015vhy}.
In order to compensate the brane bending of the NS5 near the ON${}^-$ due to the O5 plane, some additional brane are necessary.
Locally, the intersection of ON${}^0$, O5${}^+$ and O7${}^-$ can be thought of as 
\begin{subequations}
    \label{eq:intersection_O5s}
\begin{align}
    \raisebox{-.5\height}{
    \begin{tikzpicture}
    \draw[dashed] (-3,0)--(0,0);
    \node at (-3,-0.2) {\tiny{O5${}^+$}};
    \node at (-1,-0.2) {\tiny{O5${}^-$}};
    \draw[dotted] (0,-2)--(0,2);
    \node at (0.4,1.5) {\tiny{ON${}^-$}};
    \draw (-1,2)--(-1,1)--(-1.5,0.5)--(-2.5,0);
    \draw (-1,1)--(-0.5,1) (-1.5,0.5)--(-0.5,0.5);
    \SevenB{-1,2}
    \monocut{-0.5,1}{0,1}
    \monocut{-0.5,0.5}{0,0.5}
    \SevenB{-0.5,1}
    \SevenB{-0.5,0.5}
    \node at (0,0) {$\times$};
    \node at (0.4,0) {\tiny{O7${}^-$}};
    \begin{scope}[yscale=-1,xscale=1]
    \draw (-1,2)--(-1,1)--(-1.5,0.5)--(-2.5,0);
    \draw (-1,1)--(-0.5,1) (-1.5,0.5)--(-0.5,0.5);
    \SevenB{-1,2}
    \monocut{-0.5,1}{0,1}
    \monocut{-0.5,0.5}{0,0.5}
    \SevenB{-0.5,1}
    \SevenB{-0.5,0.5}
  \end{scope}
    \end{tikzpicture}
    }
    \quad \leftrightarrow \quad 
    \raisebox{-.5\height}{
    \begin{tikzpicture}
    \draw[dashed] (-5.5,0)--(0,0);
    \node at (-5.25,-0.3) {\tiny{O5${}^+$}};
    \node at (-4.5,-0.3) {\tiny{O5${}^-$}};
    \node at (-3.5,-0.3) {\tiny{$\widetilde{\text{O5}}^-$}};
    \node at (-2.5,-0.3) {\tiny{O5${}^-$}};
    \node at (-1.5,-0.3) {\tiny{$\widetilde{\text{O5}}^-$}};
    \node at (-0.5,-0.3) {\tiny{O5${}^-$}};
    \node at (-4.5,0.25) {\tiny{$2$}};
    \node at (-3.5,0.25) {\tiny{$1$}};
    \node at (-2.5,0.25) {\tiny{$1$}};
    \draw[dotted] (0,-2)--(0,2);
    \node at (0.4,1.5) {\tiny{ON${}^-$}};
    \draw (-5,-2)--(-5,2);
    \draw (-5,0.075)--(-2,0.075) (-5,-0.075)--(-2,-0.075);
    \SevenB{-5,2}
    \SevenB{-5,-2}
    \monocut{-4,-0.15}{0,-0.15}
    \monocut{-3,-0.075}{0,-0.075}
    \monocut{-2,0.075}{0,0.075}
    \monocut{-1,0.15}{0,0.15}
    \SevenB{-1,0}
    \SevenB{-2,0}
    \SevenB{-3,0}
    \SevenB{-4,0}
    \node at (0,0) {$\times$};
    \node at (0.4,0) {\tiny{O7${}^-$}};
    \end{tikzpicture}
    }
        \label{eq:intersection_O5s_1}
\end{align}
and the half D7 branes can be moved through the $(0,1)$ 5-brane 
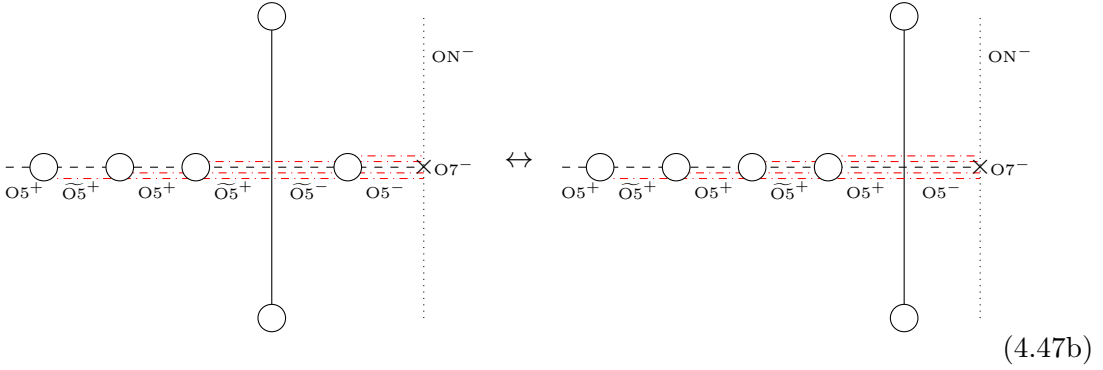
\begin{align}
    \raisebox{-.5\height}{
    \begin{tikzpicture}
    \draw[dashed] (-5.5,0)--(0,0);
    \node at (-5.25,-0.3) {\tiny{O5${}^+$}};
    \node at (-4.5,-0.3) {\tiny{$\widetilde{\text{O5}}^+$}};
    \node at (-3.5,-0.3) {\tiny{O5${}^+$}};
    \node at (-2.5,-0.3) {\tiny{$\widetilde{\text{O5}}^+$}};
    \node at (-1.5,-0.3) {\tiny{$\widetilde{\text{O5}}^-$}};
    \node at (-0.5,-0.3) {\tiny{O5${}^-$}};
    \draw[dotted] (0,-2)--(0,2);
    \node at (0.4,1.5) {\tiny{ON${}^-$}};
    \draw (-2,-2)--(-2,2);
    \SevenB{-2,2}
    \SevenB{-2,-2}
    \monocut{-5,-0.15}{0,-0.15}
    \monocut{-4,-0.075}{0,-0.075}
    \monocut{-3,0.075}{0,0.075}
    \monocut{-1,0.15}{0,0.15}
    \SevenB{-1,0}
    \SevenB{-3,0}
    \SevenB{-4,0}
    \SevenB{-5,0}
    \node at (0,0) {$\times$};
    \node at (0.4,0) {\tiny{O7${}^-$}};
    \end{tikzpicture}
    }
        \leftrightarrow 
       \raisebox{-.5\height}{
    \begin{tikzpicture}
    \draw[dashed] (-5.5,0)--(0,0);
    \node at (-5.25,-0.3) {\tiny{O5${}^+$}};
    \node at (-4.5,-0.3) {\tiny{$\widetilde{\text{O5}}^+$}};
    \node at (-3.5,-0.3) {\tiny{O5${}^+$}};
    \node at (-2.5,-0.3) {\tiny{$\widetilde{\text{O5}}^+$}};
    \node at (-1.5,-0.3) {\tiny{O5${}^+$}};
    \node at (-0.5,-0.3) {\tiny{O5${}^-$}};
    \draw[dotted] (0,-2)--(0,2);
    \node at (0.4,1.5) {\tiny{ON${}^-$}};
    \draw (-1,-2)--(-1,2);
    \SevenB{-1,2}
    \SevenB{-1,-2}
    \monocut{-5,-0.15}{0,-0.15}
    \monocut{-4,-0.075}{0,-0.075}
    \monocut{-3,0.075}{0,0.075}
    \monocut{-2,0.15}{0,0.15}
    \SevenB{-2,0}
    \SevenB{-3,0}
    \SevenB{-4,0}
    \SevenB{-5,0}
    \node at (0,0) {$\times$};
    \node at (0.4,0) {\tiny{O7${}^-$}};
    \end{tikzpicture}
    }
    \label{eq:intersection_O5s_2}
\end{align}
\end{subequations}
where brane annihilation has been taken into account. Note that the last two configurations do not have D5 branes, but the four half monodromy cuts compensate the brane bending effects due to the O5 brane, see also \cite{Zafrir:2015ftn,Bourget:2020gzi}. The left-hand side configuration in \eqref{eq:intersection_O5s_2} is preferred for reading the finite coupling magnetic quivers, since the magnetic orientifolds for $\widetilde{\mathrm{O5}}^\pm$ both yield C-type magnetic gauge groups \cite{Cabrera:2019dob}.

With this preliminaries, one can consider the following brane web:
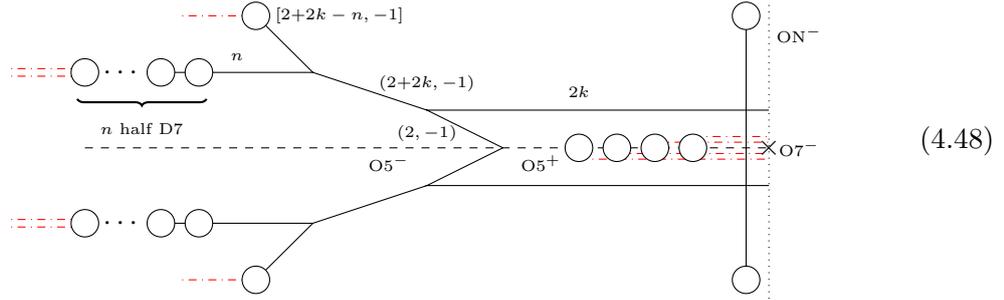
\begin{align}
    \raisebox{-.5\height}{
    \begin{tikzpicture}
    \draw[dashed] (-3,0)--(6,0);
    \node at (3,-0.2) {\tiny{O5${}^+$}};
    \node at (1,-0.2) {\tiny{O5${}^-$}};
    \draw[dotted] (6,-2)--(6,2);
    \node at (6.4,1.5) {\tiny{ON${}^-$}};
    \monocut{3.5,-0.15}{6,-0.15}
    \monocut{4,-0.075}{6,-0.075}
    \monocut{4.5,0.075}{6,0.075}
    \monocut{5,0.15}{6,0.15}
    \SevenB{5,0}
    \SevenB{4.5,0}
    \SevenB{4,0}
    \SevenB{3.5,0}
    \node at (6,0) {$\times$};
    \node at (6.4,0) {\tiny{O7${}^-$}};
    \draw (5.7,-1.75)--(5.7,1.75);
    \SevenB{5.7,-1.75}
    \SevenB{5.7,1.75}
    \draw (6,0.5)--(1.5,0.5)--(2.5,0)
    (1.5,0.5)--(0,1)--(-2,1) (0,1)--(-0.75,1.75);
    \monocut{-3,1.05}{-4,1.05}
    \monocut{-3,0.95}{-4,0.95}
    \monocut{-0.75,1.75}{-1.75,1.75}
    \SevenB{-1.5,1}
    \SevenB{-2,1}
    \node at (-2.5,1) {$\cdots$};
    \SevenB{-3,1}
    \SevenB{-0.75,1.75}
    \node at (3.5,0.75) {\tiny{$2k$}};
    \node at (1.5,0.2) {\tiny{$(2,-1)$}};
    \node at (1.5,0.85) {\tiny{$(2{+}2k,-1)$}};
    \node at (0.35,1.75) {\tiny{$[2{+}2k-n,-1]$}};
    \node at (-1,1.2) {\tiny{$n$}};
    \draw[decoration={brace,mirror,raise=10pt},decorate,thick]
  (-3.1,1) -- node[below=15pt] {\tiny{$n$ half D7}} (-1.4,1);
      \begin{scope}[yscale=-1,xscale=1]
    \draw (6,0.5)--(1.5,0.5)--(2.5,0)
    (1.5,0.5)--(0,1)--(-2,1) (0,1)--(-0.75,1.75);
    \monocut{-3,1.05}{-4,1.05}
    \monocut{-3,0.95}{-4,0.95}
    \monocut{-0.75,1.75}{-1.75,1.75}
    \SevenB{-1.5,1}
    \SevenB{-2,1}
    \node at (-2.5,1) {$\cdots$};
    \SevenB{-3,1}
    \SevenB{-0.75,1.75}
      \end{scope}
    \end{tikzpicture}
    }
    \label{eq:5-brane_ON_and_Op}
\end{align}
and the 5d $\Ncal=1$ effective gauge theory description is a product of the form
\begin{align}
        \raisebox{-.5\height}{
    \begin{tikzpicture}
	\tikzstyle{gaugeSO} = [circle, draw,inner sep=3pt,fill=red];
	\tikzstyle{gaugeSp} = [circle, draw,inner sep=3pt,fill=blue];
	\tikzstyle{flavour} = [regular polygon,regular polygon sides=4,inner
sep=3pt, draw];
\tikzstyle{flavourSO} = [regular polygon,regular polygon sides=4,inner
sep=3pt, draw, fill=red];
\tikzstyle{flavourSp} = [regular polygon,regular polygon sides=4,inner
sep=3pt, draw, fill=blue];
	\node (g1) [gaugeSp,label=right:{\footnotesize{$\usprm(2k)$}}] {};
    \node (f1) [flavourSO,above of=g1, label=right:{\footnotesize{$\sorm(2n+4)$}}] {};
\node at (2,0.5) {$\times$};
	\node (g2) [gaugeSp,label=right:{\footnotesize{$\usprm(2k)$}}] at (3,0) {};
\node (f2) [flavourSO,above of=g2, label=right:{\footnotesize{$\sorm(2n+4)$}}] {};
	\draw  (g1)--(f1) (g2)--(f2);
	\end{tikzpicture}
    }
\end{align}
with the assumption $2+2k-n\geq 0$ such that the two branes with NS charge do not intersect each other. The total number of fundamental flavours is $N_f =n +2\leq 2k+4$, due to the $n$ flavour 7-branes on the left-hand side and the 4 half 7-branes near the O7${}^-$.

For finite coupling, the brane web \eqref{eq:5-brane_ON_and_Op} is transferred into the Higgs branch phase 
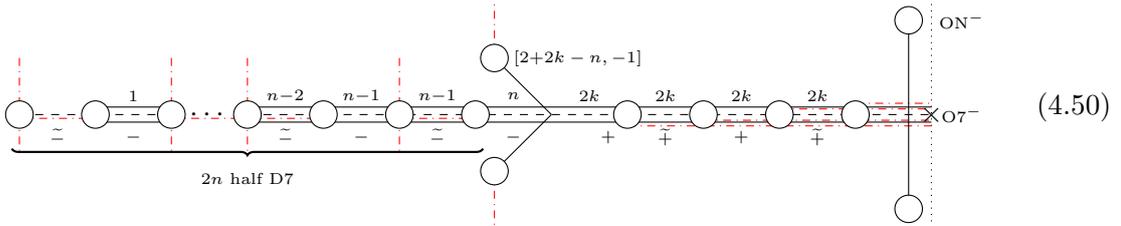
\begin{align}
    \raisebox{-.5\height}{
    \begin{tikzpicture}
    \draw[dashed] (-3,0)--(6,0) (-4,0)--(-6,0);
    \node at (4.5,-0.3) {\tiny{$\widetilde{+}$}};
    \node at (3.5,-0.3) {\tiny{$+$}};
    \node at (2.5,-0.3) {\tiny{$\widetilde{+}$}};
    \node at (1.75,-0.3) {\tiny{$+$}};
    \node at (0.5,-0.3) {\tiny{$-$}};
    \node at (-0.5,-0.3) {\tiny{$\widetilde{-}$}};
    \node at (-1.5,-0.3) {\tiny{$-$}};
    \node at (-2.5,-0.3) {\tiny{$\widetilde{-}$}};
    \node at (-4.5,-0.3) {\tiny{$-$}};
    \node at (-5.5,-0.3) {\tiny{$\widetilde{-}$}};
    \draw[dotted] (6,-1.5)--(6,1.5);
    \node at (6.4,1.25) {\tiny{ON${}^-$}};
    \foreach \i in {-1,1}
    \draw (6,0.1*\i)--(-3,0.1*\i) (-4,0.1*\i)--(-5,0.1*\i);
    \draw (1,0)--(0.25,0.75);
    \foreach \i in {-6,-4,-3,-1}
    \monocut{\i,0.75}{\i,-0.5} 
    \monocut{\i,-0.05}{\i+1,-0.05};
    \monocut{0.25,0.75}{0.25,1.5};
    \monocut{0.25,-0.75}{0.25,-1.5};
    \monocut{2,-0.15}{6,-0.15};
    \monocut{3,-0.075}{6,-0.075};
    \monocut{4,0.075}{6,0.075};
    \monocut{5,0.15}{6,0.15};
    \SevenB{5,0}
    \SevenB{4,0}
    \SevenB{3,0}
    \SevenB{2,0}
    \node at (6,0) {$\times$};
    \node at (6.4,0) {\tiny{O7${}^-$}};
    \draw (5.7,-1.25)--(5.7,1.25);
    \SevenB{5.7,-1.25}
    \SevenB{5.7,1.25}
    \SevenB{0,0}
    \SevenB{-1,0}
    \SevenB{-2,0}
    \SevenB{-3,0}
    \node at (-3.5,0) {$\cdots$};
    \SevenB{-4,0}
    \SevenB{-5,0}
    \SevenB{-6,0}
    \SevenB{0.25,0.75}
    \foreach \i in {1,2,3,4}
    \node at (0.5+\i,0.25) {\tiny{$2k$}};
    \node at (1.35,0.75) {\tiny{$[2{+}2k-n,-1]$}};
    \node at (0.5,0.25) {\tiny{$n$}};
    \node at (-0.5,0.25) {\tiny{$n{-}1$}};
    \node at (-1.5,0.25) {\tiny{$n{-}1$}};
    \node at (-2.5,0.25) {\tiny{$n{-}2$}};
    \node at (-4.5,0.25) {\tiny{$1$}};
    \draw[decoration={brace,mirror,raise=10pt},decorate,thick]
  (-6.1,-0.1) -- node[below=15pt] {\tiny{$2n$ half D7}} (0.1,-0.1);
      \begin{scope}[yscale=-1,xscale=1]
      \draw (1,0)--(0.25,0.75);
      \SevenB{0.25,0.75}
      \end{scope}
    \end{tikzpicture}
    }
    \label{eq:5-brane_ON_and_Op_Higgs}
\end{align}
but the derivation of the finite coupling magnetic quivers splits into three distinct cases.
\begin{itemize}
\item \ul{Case $2k+2\geq n\geq  2k$.}
In this region, the electric theory exhibits complete Higgsing. Transferring the brane web into the Higgs branch phase yields
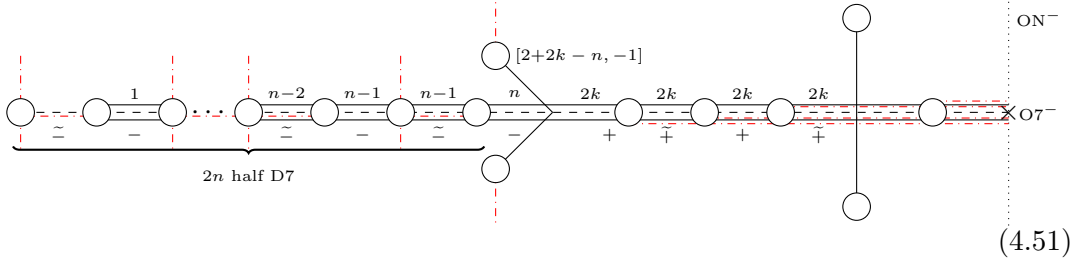
\begin{align}
    \raisebox{-.5\height}{
    \begin{tikzpicture}
    \draw[dashed] (-3,0)--(7,0) (-4,0)--(-6,0);
    \node at (4.5,-0.3) {\tiny{$\widetilde{+}$}};
    \node at (3.5,-0.3) {\tiny{$+$}};
    \node at (2.5,-0.3) {\tiny{$\widetilde{+}$}};
    \node at (1.75,-0.3) {\tiny{$+$}};
    \node at (0.5,-0.3) {\tiny{$-$}};
    \node at (-0.5,-0.3) {\tiny{$\widetilde{-}$}};
    \node at (-1.5,-0.3) {\tiny{$-$}};
    \node at (-2.5,-0.3) {\tiny{$\widetilde{-}$}};
    \node at (-4.5,-0.3) {\tiny{$-$}};
    \node at (-5.5,-0.3) {\tiny{$\widetilde{-}$}};
    \draw[dotted] (7,-1.5)--(7,1.5);
    \node at (7.4,1.25) {\tiny{ON${}^-$}};
    \foreach \i in {-1,1}
    \draw (7,0.1*\i)--(-3,0.1*\i) (-4,0.1*\i)--(-5,0.1*\i);
    \draw (1,0)--(0.25,0.75);
    \foreach \i in {-6,-4,-3,-1}
    \monocut{\i,0.75}{\i,-0.5} 
    \monocut{\i,-0.05}{\i+1,-0.05};
    \monocut{0.25,0.75}{0.25,1.5};
    \monocut{0.25,-0.75}{0.25,-1.5};
    \monocut{2,-0.15}{7,-0.15};
    \monocut{3,-0.075}{7,-0.075};
    \monocut{4,0.075}{7,0.075};
    \monocut{6,0.15}{7,0.15};
    \SevenB{6,0}
    \SevenB{4,0}
    \SevenB{3,0}
    \SevenB{2,0}
    \node at (7,0) {$\times$};
    \node at (7.4,0) {\tiny{O7${}^-$}};
    \draw (5,1.25)--(5,-1.25);
    \SevenB{5,-1.25}
    \SevenB{5,1.25}
    \SevenB{0,0}
    \SevenB{-1,0}
    \SevenB{-2,0}
    \SevenB{-3,0}
    \node at (-3.5,0) {$\cdots$};
    \SevenB{-4,0}
    \SevenB{-5,0}
    \SevenB{-6,0}
    \SevenB{0.25,0.75}
    \foreach \i in {1,2,3,4}
    \node at (0.5+\i,0.25) {\tiny{$2k$}};
    \node at (1.35,0.75) {\tiny{$[2{+}2k-n,-1]$}};
    \node at (0.5,0.25) {\tiny{$n$}};
    \node at (-0.5,0.25) {\tiny{$n{-}1$}};
    \node at (-1.5,0.25) {\tiny{$n{-}1$}};
    \node at (-2.5,0.25) {\tiny{$n{-}2$}};
    \node at (-4.5,0.25) {\tiny{$1$}};
    \draw[decoration={brace,mirror,raise=10pt},decorate,thick]
  (-6.1,-0.1) -- node[below=15pt] {\tiny{$2n$ half D7}} (0.1,-0.1);
      \begin{scope}[yscale=-1,xscale=1]
      \draw (1,0)--(0.25,0.75);
      \SevenB{0.25,0.75}
      \end{scope}
    \end{tikzpicture}
    }
    \label{eq:5-brane_ON_and_Op_Higgs_E765}
\end{align}
and since the $n\geq 2k$, the $(2+2k-n,-1)$ 5-brane is connected to $n-2k$ frozen D5 branes on the left-hand side. Moving this 5-brane through a sufficient number of half D7 branes and accounting for brane annihilation, the magnetic quiver is given by \eqref{eq:ex_product_mirror_OSp} for $n\to N_f = n+2$. Note in particular, that the 5-brane close to the ON plane gives rise to a flavour to both C-type gauge nodes at the bifurcation. This yields the finite coupling magnetic quivers for the families $E_7\times E_7$, $E_6\times E_6$, and $E_5\times E_5$.
\item \ul{Case $n= 2k-2l-1$ for $l\geq 0$.}
In this region, all cases expect $l=0$ are affected from incomplete Higgsing. To see this, the brane web \eqref{eq:5-brane_ON_and_Op} becomes
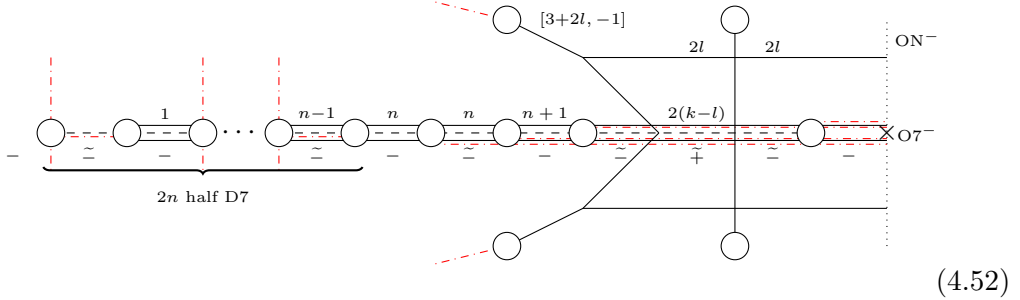
\begin{align}
    \raisebox{-.5\height}{
    \begin{tikzpicture}
    \draw[dashed] (-1,0)--(6,0) (-2,0)--(-4,0);
    \node at (6.5,-0.3) {\tiny{$-$}};
    \node at (5.5,-0.3) {\tiny{$\widetilde{-}$}};
    \node at (4.5,-0.3) {\tiny{$\widetilde{+}$}};
    \node at (3.5,-0.3) {\tiny{$\widetilde{-}$}};
    \node at (2.5,-0.3) {\tiny{$-$}};
    \node at (1.5,-0.3) {\tiny{$\widetilde{-}$}};
    \node at (0.5,-0.3) {\tiny{$-$}};
    \node at (-0.5,-0.3) {\tiny{$\widetilde{-}$}};
    \node at (-2.5,-0.3) {\tiny{$-$}};
    \node at (-3.5,-0.3) {\tiny{$\widetilde{-}$}};
    \node at (-4.5,-0.3) {\tiny{$-$}};
    \draw[dotted] (7,-1.5)--(7,1.5);
    \node at (7.4,1.25) {\tiny{ON${}^-$}};
    \foreach \i in {-1,1}
    \draw (7,0.1*\i)--(-1,0.1*\i) (-2,0.1*\i)--(-3,0.1*\i);
    \draw (4,0)--(3,1)--(2,1.5);
    \draw (3,1)--(7,1);
    \foreach \i in {-4,-1}
    \monocut{\i,1}{\i,-0.5} 
    \monocut{\i,-0.05}{\i+1,-0.05};
    \monocut{-2,1}{-2,-0.5} 
    \monocut{2,1.5}{1,1.75};
    \monocut{2,-1.5}{1,-1.75};
    \monocut{1,-0.15}{7,-0.15};
    \monocut{2,-0.075}{7,-0.075};
    \monocut{3,0.075}{7,0.075};
    \monocut{6,0.15}{7,0.15};
    \SevenB{6,0}
    \SevenB{3,0}
    \SevenB{2,0}
    \SevenB{1,0}
    \node at (7,0) {$\times$};
    \node at (7.4,0) {\tiny{O7${}^-$}};
    \draw (5,1.5)--(5,-1.5);
    \SevenB{5,1.5}
    \SevenB{5,-1.5}
    \SevenB{0,0}
    \SevenB{-1,0}
    \SevenB{-2,0}
    \SevenB{-3,0}
    \node at (-1.5,0) {$\cdots$};
    \SevenB{-4,0}
    \SevenB{2,1.5}
    \node at (3,1.5) {\tiny{$[3{+}2l,-1]$}};
    \node at (4.5,1.15) {\tiny{$2l$}};
    \node at (5.5,1.15) {\tiny{$2l$}};
    \node at (4.5,0.25) {\tiny{$2(k{-}l)$}};
    \node at (2.5,0.25) {\tiny{$n+1$}};
    \node at (1.5,0.25) {\tiny{$n$}};
    \node at (0.5,0.25) {\tiny{$n$}};
    \node at (-0.5,0.25) {\tiny{$n{-}1$}};
    \node at (-2.5,0.25) {\tiny{$1$}};
    \draw[decoration={brace,mirror,raise=10pt},decorate,thick]
  (-4.1,-0.1) -- node[below=15pt] {\tiny{$2n$ half D7}} (0.1,-0.1);
      \begin{scope}[yscale=-1,xscale=1]
    \draw (4,0)--(3,1)--(2,1.5);
    \draw (3,1)--(7,1);
      \SevenB{2,1.5}
      \end{scope}
    \end{tikzpicture}
    }
    \label{eq:5-brane_ON_and_Op_Higgs_E4}
\end{align}
where the $[2{+}2k-n,-1]=[3{+}2l,-1]$ 7-brane leads to a $(3,-1)$ 5-brane close to the O5 plane. Due to the presence of three half monodromy lines, this effectively looks like a $(2,-1)$ 5-brane. Note that $n+1=2(k-l)$. The brane web clearly displays a stack of $2l$ D5 branes in the Coulomb branch phase such that there is a residual $\sprm(l)\times \sprm(l)$ gauge symmetry. The stack of $2(k-l)$ D5 branes near the intersection of the O-planes splits evenly and gives rise to two $\sprm(k-l)$ nodes at the bifurcation. Moreover, the subweb that is still in the electric phase contributes a $\sorm(2)$ flavour to each of these nodes. The magnetic quiver is read off to be
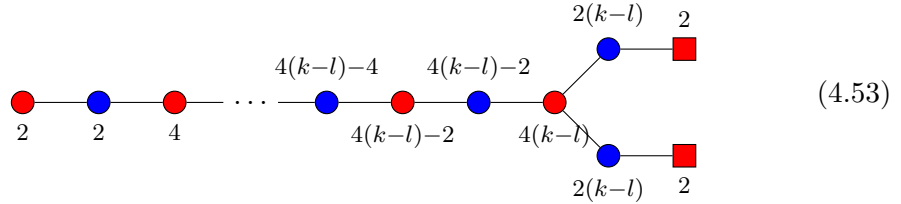
\begin{align}
        \raisebox{-.5\height}{
    \begin{tikzpicture}
	\node (g1) [gaugeSO,label=below:{\footnotesize{$2$}}] {};
	\node (g2) [gaugeSp,right of =g1,label=below:{\footnotesize{$2$}}] {};
	\node (g3) [gaugeSO,right of =g2,label=below:{\footnotesize{$4$}}] {};
	\node (g4) [right of =g3] {$\ldots$};
	\node (g5) [gaugeSp,right of =g4,label=above:{\footnotesize{$4(k{-}l){-}4$}}] {};
	\node (g6) [gaugeSO,right of =g5,label=below:{\footnotesize{$4(k{-}l){-}2$}}] {};
	\node (g7) [gaugeSp,right of =g6,label=above:{\footnotesize{$4(k{-}l){-}2$}}] {};
	\node (g8) [gaugeSO,right of =g7,label=below:{\footnotesize{$4(k{-}l)$}}] {};
	\node (g9) [gaugeSp, above right of =g8,label=above:{\footnotesize{$2(k{-}l)$}}] {};
	\node (g10) [gaugeSp,below right of =g8,label=below:{\footnotesize{$2(k{-}l)$}}] {};
	\node (f1) [flavourSO,right of =g9,label=above:{\footnotesize{$2$}}] {};
	\node (f2) [flavourSO,right of =g10,label=below:{\footnotesize{$2$}}] {};
	\draw  (g1)--(g2) (g2)--(g3) (g3)--(g4) (g4)--(g5) (g5)--(g6) (g6)--(g7) (g7)--(g8) (g8)--(g9) (g8)--(g10) (g9)--(f1) (g10)--(f2);
	\end{tikzpicture}
    }
    \label{eq:magQuiv_5d_E4_finite}
\end{align}
and one verifies that 
\begin{align}
    \dim \Coulomb \eqref{eq:magQuiv_5d_E4_finite} &= 2 \cdot \left[ (2n+4) \cdot k - \left(\dim\sprm(k) - \dim\sprm(l) \right)
    \right] \notag \\
    &= 2 \cdot \dim \Higgs\bigg( \left[\sprm(k) \text{ with } N_f = n{+}2\right] \rightarrow \left[\text{pure } \sprm(l)\right] \bigg)
\end{align}
is consistent with the incomplete Higgsing. The quiver \eqref{eq:magQuiv_5d_E4_finite} yields the finite coupling magnetic quivers for the family $E_{4-2l}\times E_{4-2l}$ and has been classified in \eqref{firstfork} .

\item \ul{Case $n= 2k-2l-2$ for $l\geq 0$.}
To see this, the brane web \eqref{eq:5-brane_ON_and_Op} becomes
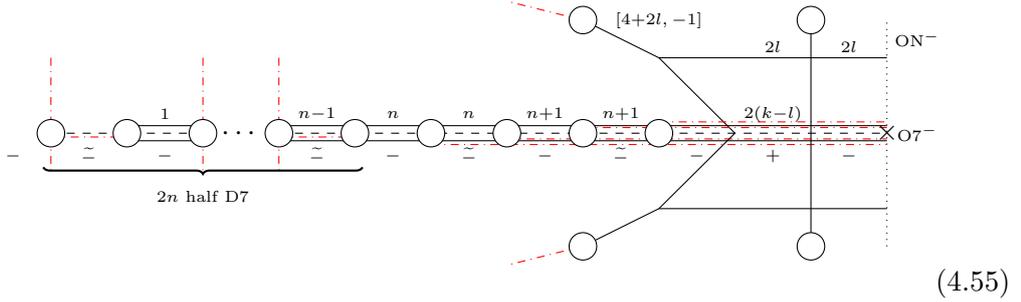
\begin{align}
    \raisebox{-.5\height}{
    \begin{tikzpicture}
    \draw[dashed] (-1,0)--(7,0) (-2,0)--(-4,0);
    \node at (6.5,-0.3) {\tiny{$-$}};
    \node at (5.5,-0.3) {\tiny{$+$}};
    \node at (4.5,-0.3) {\tiny{$-$}};
    \node at (3.5,-0.3) {\tiny{$\widetilde{-}$}};
    \node at (2.5,-0.3) {\tiny{$-$}};
    \node at (1.5,-0.3) {\tiny{$\widetilde{-}$}};
    \node at (0.5,-0.3) {\tiny{$-$}};
    \node at (-0.5,-0.3) {\tiny{$\widetilde{-}$}};
    \node at (-2.5,-0.3) {\tiny{$-$}};
    \node at (-3.5,-0.3) {\tiny{$\widetilde{-}$}};
    \node at (-4.5,-0.3) {\tiny{$-$}};
    \draw[dotted] (7,-1.5)--(7,1.5);
    \node at (7.4,1.25) {\tiny{ON${}^-$}};
    \foreach \i in {-1,1}
    \draw (7,0.1*\i)--(-1,0.1*\i) (-2,0.1*\i)--(-3,0.1*\i);
    \draw (5,0)--(4,1)--(3,1.5);
    \draw (4,1)--(7,1);
    \foreach \i in {-4,-1}
    \monocut{\i,1}{\i,-0.5} 
    \monocut{\i,-0.05}{\i+1,-0.05};
    \monocut{-2,1}{-2,-0.5} 
    \monocut{1,-0.15}{7,-0.15};
    \monocut{2,-0.075}{7,-0.075};
    \monocut{3,0.075}{7,0.075};
    \monocut{4,0.15}{7,0.15};
    \monocut{3,1.5}{2,1.75};
    \monocut{3,-1.5}{2,-1.75};
    \SevenB{4,0}
    \SevenB{3,0}
    \SevenB{2,0}
    \SevenB{1,0}
    \node at (7,0) {$\times$};
    \node at (7.4,0) {\tiny{O7${}^-$}};
    \draw (6,-1.5)--(6,1.5);
    \SevenB{6,1.5}
    \SevenB{6,-1.5}
    \SevenB{0,0}
    \SevenB{-1,0}
    \SevenB{-2,0}
    \SevenB{-3,0}
    \node at (-1.5,0) {$\cdots$};
    \SevenB{-4,0}
    \SevenB{3,1.5}
    \node at (4,1.5) {\tiny{$[4{+}2l,-1]$}};
    \node at (5.5,1.15) {\tiny{$2l$}};
    \node at (6.5,1.15) {\tiny{$2l$}};
    \node at (5.5,0.25) {\tiny{$2(k{-}l)$}};
    \node at (3.5,0.25) {\tiny{$n{+}1$}};
    \node at (2.5,0.25) {\tiny{$n{+}1$}};
    \node at (1.5,0.25) {\tiny{$n$}};
    \node at (0.5,0.25) {\tiny{$n$}};
    \node at (-0.5,0.25) {\tiny{$n{-}1$}};
    \node at (-2.5,0.25) {\tiny{$1$}};
    \draw[decoration={brace,mirror,raise=10pt},decorate,thick]
  (-4.1,-0.1) -- node[below=15pt] {\tiny{$2n$ half D7}} (0.1,-0.1);
      \begin{scope}[yscale=-1,xscale=1]
    \draw (5,0)--(4,1)--(3,1.5);
    \draw (4,1)--(7,1);
      \SevenB{3,1.5}
      \end{scope}
    \end{tikzpicture}
    }
    \label{eq:5-brane_ON_and_Op_Higgs_E3}
\end{align}
where the $[2{+}2k-n,-1]=[4{+}2l,-1]$ 7-brane leads to a  $(4{+}2l,-1)$ 5-brane, which close to the O5 plane is a $(4,-1)$ 5-brane. Due to the presence of four half monodromy lines this 5-brane effectively behaves like a $(2,-1)$ 5-brane. The brane web clearly displays a stack of $2l$ D5 branes in the Coulomb branch phase such that there is a residual $\sprm(l)\times \sprm(l)$ gauge symmetry. The stack of $n+1$ D5 branes closed to the O7 splits as $(k-l-1)+(k-l)$, such that the bifrucation produces two $\sprm$ gauge nodes of rank $(k-l-1)$ and $(k-l)$, respectively. The subweb in the Coulomb branch phase yields a $\sorm(4)$ flavour for the $\sprm(k-l)$ only, due to the double intersection. The magnetic quiver is read off to be
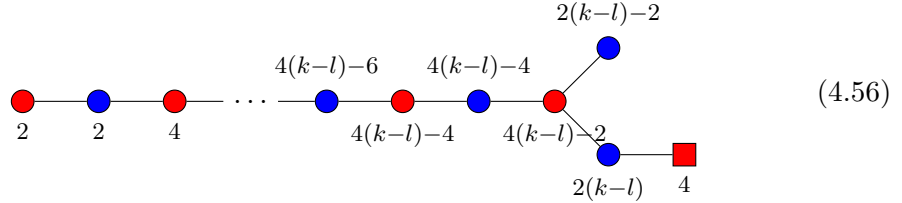
\begin{align}
        \raisebox{-.5\height}{
    \begin{tikzpicture}
	\node (g1) [gaugeSO,label=below:{\footnotesize{$2$}}] {};
	\node (g2) [gaugeSp,right of =g1,label=below:{\footnotesize{$2$}}] {};
	\node (g3) [gaugeSO,right of =g2,label=below:{\footnotesize{$4$}}] {};
	\node (g4) [right of =g3] {$\ldots$};
	\node (g5) [gaugeSp,right of =g4,label=above:{\footnotesize{$4(k{-}l){-}6$}}] {};
	\node (g6) [gaugeSO,right of =g5,label=below:{\footnotesize{$4(k{-}l){-}4$}}] {};
	\node (g7) [gaugeSp,right of =g6,label=above:{\footnotesize{$4(k{-}l){-}4$}}] {};
	\node (g8) [gaugeSO,right of =g7,label=below:{\footnotesize{$4(k{-}l){-}2$}}] {};
	\node (g9) [gaugeSp, above right of =g8,label=above:{\footnotesize{$2(k{-}l){-}2$}}] {};
	\node (g10) [gaugeSp,below right of =g8,label=below:{\footnotesize{$2(k{-}l)$}}] {};
	\node (f2) [flavourSO,right of =g10,label=below:{\footnotesize{$4$}}] {};
	\draw  (g1)--(g2) (g2)--(g3) (g3)--(g4) (g4)--(g5) (g5)--(g6) (g6)--(g7) (g7)--(g8) (g8)--(g9) (g8)--(g10) (g10)--(f2);
	\end{tikzpicture}
    }
    \label{eq:magQuiv_5d_E3_finite}
\end{align}
and one verifies that 
\begin{align}
    \dim \Coulomb \left(\eqref{eq:magQuiv_5d_E3_finite}\right) &= 2 \cdot \left[ (2n+4) \cdot k - \left(\dim\sprm(k) - \dim\sprm(l) \right)
    \right] \notag \\
    &= 2 \cdot \dim \Higgs\bigg( \left[\sprm(k) \text{ with } N_f = n{+}2\right] \rightarrow \left[\text{pure } \sprm(l)\right] \bigg)
\end{align}
is consistent with the incomplete Higgsing. The full moduli space is discussed around \eqref{finalfinitequiver}.
This yields the finite coupling magnetic quivers for the family $E_{3-2l}\times E_{3-2l}$.
\end{itemize}

Moving on to infinite coupling, a case by case analysis is required, as the Higgs branch enhancement depends crucially on $n=n(k)$.
\begin{itemize}
    \item \ul{Maximal case $n=2k+2$.} In the brane web \eqref{eq:5-brane_ON_and_Op_Higgs}, the two NS5 branes become parallel. The infinite coupling point is realised by moving both NS5 branes onto the ON plane. Taking brane creation into account, the web becomes
    \begin{align}
    \raisebox{-.5\height}
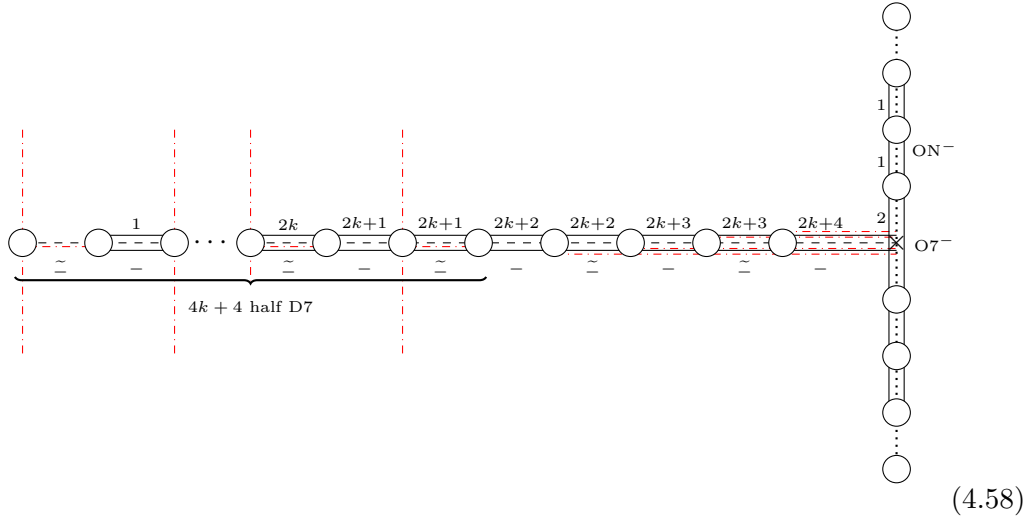
{
    \begin{tikzpicture}
    \draw[dashed] (-3,0)--(5.5,0) (-4,0)--(-6,0);
    \node at (4.5,-0.35) {\tiny{$-$}};
    \node at (3.5,-0.35) {\tiny{$\widetilde{-}$}};
    \node at (2.5,-0.35) {\tiny{$-$}};
    \node at (1.5,-0.35) {\tiny{$\widetilde{-}$}};
    \node at (0.5,-0.35) {\tiny{$-$}};
    \node at (-0.5,-0.35) {\tiny{$\widetilde{-}$}};
    \node at (-1.5,-0.35) {\tiny{$-$}};
    \node at (-2.5,-0.35) {\tiny{$\widetilde{-}$}};
    \node at (-4.5,-0.35) {\tiny{$-$}};
    \node at (-5.5,-0.35) {\tiny{$\widetilde{-}$}};
    \draw[dotted,thick] (5.5,-3)--(5.5,3);
    \node at (6,1.25) {\tiny{ON${}^-$}};
    \foreach \i in {-1,1}
    \draw (5.5,0.1*\i)--(-3,0.1*\i) (-4,0.1*\i)--(-5,0.1*\i);
    \monocut{1,-0.15}{5.5,-0.15};
    \monocut{2,-0.075}{5.5,-0.075};
    \monocut{3,0.075}{5.5,0.075};
    \monocut{4,0.15}{5.5,0.15};
    \SevenB{4,0}
    \SevenB{3,0}
    \SevenB{2,0}
    \SevenB{1,0}
    \node at (5.5,0) {$\times$};
    \node at (6,0) {\tiny{O7${}^-$}};
    \foreach \i in {0,1}
    \draw (5.4+0.2*\i,-2.25)--(5.4+0.2*\i,2.25);
    \monocut{-6,1.5}{-6,-1.5}
    \monocut{-4,1.5}{-4,-1.5}
    \monocut{-3,1.5}{-3,-0.5}
    \monocut{-1,1.5}{-1,-1.5}
    \monocut{-6,-0.05}{-5,-0.05}
    \monocut{-3,-0.05}{-2,-0.05}
    \monocut{-1,-0.05}{0,-0.05}
    \SevenB{0,0}
    \SevenB{-1,0}
    \SevenB{-2,0}
    \SevenB{-3,0}
    \node at (-3.5,0) {$\cdots$};
    \SevenB{-4,0}
    \SevenB{-5,0}
    \SevenB{-6,0}
    \SevenB{5.5,0.75}
    \SevenB{5.5,1.5}
    \SevenB{5.5,2.25}
    \SevenB{5.5,3}
    \node at (4.5,0.25) {\tiny{$2k{+}4$}};
    \node at (3.5,0.25) {\tiny{$2k{+}3$}};
    \node at (2.5,0.25) {\tiny{$2k{+}3$}};
    \node at (1.5,0.25) {\tiny{$2k{+}2$}};
    \node at (0.5,0.25) {\tiny{$2k{+}2$}};
    \node at (-0.5,0.25) {\tiny{$2k{+}1$}};
    \node at (-1.5,0.25) {\tiny{$2k{+}1$}};
    \node at (-2.5,0.25) {\tiny{$2k$}};
    \node at (-4.5,0.25) {\tiny{$1$}};
    \draw[decoration={brace,mirror,raise=10pt},decorate,thick]
  (-6.1,-0.1) -- node[below=15pt] {\tiny{$4k+4$ half D7}}
  (0.1,-0.1);
  \node at (5.3,0.325) {\tiny{$2$}};
  \node at (5.3,0.325+0.75) {\tiny{$1$}};
  \node at (5.3,0.325+1.5) {\tiny{$1$}};
      \begin{scope}[yscale=-1,xscale=1]
    \SevenB{5.5,0.75}
    \SevenB{5.5,1.5}
    \SevenB{5.5,2.25}
    \SevenB{5.5,3}
      \end{scope}
    \end{tikzpicture}
    }
    \label{eq:5-brane_ON_and_Op_infinite_E7}
\end{align}
and now the configuration needs to be decomposed into a maximal subdivision. For the 5-brane away from the ON and O7, only the O5 orientifold is perceived and the magnetic quiver can be straightforwardly derived.  Near the intersection of the three O-planes, a careful consideration is required. Omitting mirror images completely, the relevant part becomes
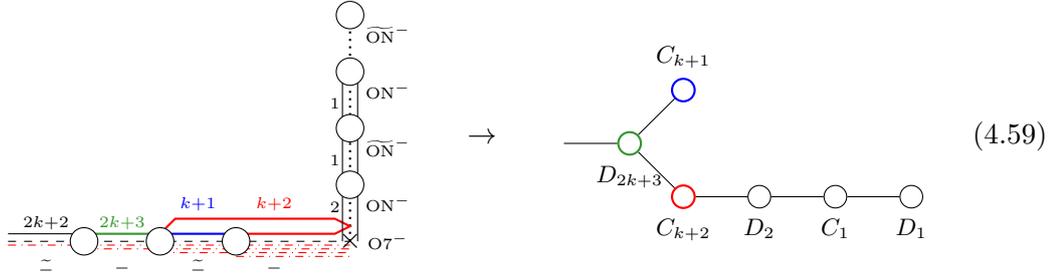
\begin{align}
    \raisebox{-.5\height}{
    \begin{tikzpicture}
    \draw[dashed] (1,0)--(5.5,0);
    \node at (4.5,-0.35) {\tiny{$-$}};
    \node at (3.5,-0.35) {\tiny{$\widetilde{-}$}};
    \node at (2.5,-0.35) {\tiny{$-$}};
    \node at (1.5,-0.35) {\tiny{$\widetilde{-}$}};
    \draw[dotted,thick] (5.5,0)--(5.5,3);
    \node at (6,0.5) {\tiny{ON${}^-$}};
    \node at (6,0.5+0.75) {\tiny{$\widetilde{\mathrm{ON}}^-$}};
    \node at (6,0.5+1.5) {\tiny{ON${}^-$}};
    \node at (6,0.5+2.25) {\tiny{$\widetilde{\mathrm{ON}}^-$}};
    \draw (2,0.1)--(1,0.1);
    \draw[thick,myGreen] (3,0.1)--(2,0.1);
    \draw[thick,blue] (4,0.1)--(3,0.1);
    \draw[thick,red] (3,0.1)--(3.2,0.3)--(5.3,0.3)--(5.5,0.2)--(5.3,0.1)--(4,0.1);
    \monocut{1,-0.05}{5.5,-0.05}
    \monocut{2,-0.1}{5.5,-0.1}
    \monocut{3,-0.15}{5.5,-0.15}
    \monocut{4,-0.2}{5.5,-0.2}
    \SevenB{4,0}
    \SevenB{3,0}
    \SevenB{2,0}
    \node at (5.5,0) {$\times$};
    \node at (6,0) {\tiny{O7${}^-$}};
    \foreach \i in {0,1}
    \draw (5.4+0.2*\i,0)--(5.4+0.2*\i,2.25);
    \SevenB{5.5,0.75}
    \SevenB{5.5,1.5}
    \SevenB{5.5,2.25}
    \SevenB{5.5,3}
    \node[red] at (4.5,0.5) {\tiny{$k{+}2$}};
    \node[blue] at (3.5,0.5) {\tiny{$k{+}1$}};
    \node[myGreen] at (2.5,0.25) {\tiny{$2k{+}3$}};
    \node at (1.5,0.25) {\tiny{$2k{+}2$}};
  \node at (5.3,0.45) {\tiny{$2$}};
  \node at (5.3,0.325+0.75) {\tiny{$1$}};
  \node at (5.3,0.325+1.5) {\tiny{$1$}};
   \end{tikzpicture}
    }
    \quad \rightarrow \quad 
    \raisebox{-.5\height}{
    \begin{tikzpicture}
	\node (g7) {};
	\node[thick,myGreen] (g8) [gauge,right of =g7,label=below:{\footnotesize{$D_{2k{+}3}$}}] {};
	\node[thick,blue] (g9) [gauge, above right of =g8,label=above:{\footnotesize{$C_{k{+}1}$}}] {};
	\node[thick,red] (g10) [gauge,below right of =g8,label=below:{\footnotesize{$C_{k{+}2}$}}] {};
	\node (g11) [gauge,right of =g10,label=below:{\footnotesize{$D_2$}}] {};
	\node (g12) [gauge,right of =g11,label=below:{\footnotesize{$C_1$}}] {};
	\node (g13) [gauge,right of =g12,label=below:{\footnotesize{$D_1$}}] {};
	\draw (g7)--(g8) (g8)--(g9) (g8)--(g10) (g10)--(g11) (g11)--(g12) (g12)--(g13);
	\end{tikzpicture}
    }
\end{align}
which allows to derive the magnetic quiver as follows: The green stack of D5 yields a $D$-type node, while the blue and red subwebs contribute $C$-type algebras. The subwebs on $(0,1)$ 5-branes on top of the ON contribute $D$ and $C$-type magnetic gauge algebras as above. Next, the generalised intersection numbers imply that the green and red node are connected to the green node, but red and blue nodes are not connected to each other. This follows because they share two 7-branes: on one they end on the same side, while on the other they end on opposite sides. The gauge nodes from the $(0,1)$ 5-branes are connected by bifundamental matter in the standard way. Lastly, the red subweb has intersection number one with the central stack of two $(0,1)$ brane because $|\left( \begin{smallmatrix} 1 & 0 \\ 0&1 \end{smallmatrix}
\right)|=1$. The blue subweb has vanishing intersection number with the central $(0,1)$ brane, hence no connections between these nodes.
In summary, the magnetic quiver becomes
    \begin{align}
        \raisebox{-.5\height}{
    \begin{tikzpicture}
	\node (g1) [gaugeSO,label=below:{\footnotesize{$2$}}] {};
	\node (g2) [gaugeSp,right of =g1,label=below:{\footnotesize{$2$}}] {};
	\node (g3) [gaugeSO,right of =g2,label=below:{\footnotesize{$4$}}] {};
	\node (g4) [right of =g3] {$\ldots$};
	\node (g5) [gaugeSp,right of =g4,label=above:{\footnotesize{$4k{+}2$}}] {};
	\node (g6) [gaugeSO,right of =g5,label=below:{\footnotesize{$4k{+}4$}}] {};
	\node (g7) [gaugeSp,right of =g6,label=above:{\footnotesize{$4k{+}4$}}] {};
	\node (g8) [gaugeSO,right of =g7,label=below:{\footnotesize{$4k{+}6$}}] {};
	\node (g9) [gaugeSp, above right of =g8,label=above:{\footnotesize{$2k{+}2$}}] {};
	\node (g10) [gaugeSp,below right of =g8,label=below:{\footnotesize{$2k{+}4$}}] {};
	\node (g11) [gaugeSO,right of =g10,label=below:{\footnotesize{$4$}}] {};
	\node (g12) [gaugeSp,right of =g11,label=below:{\footnotesize{$2$}}] {};
	\node (g13) [gaugeSO,right of =g12,label=below:{\footnotesize{$2$}}] {};
	\draw  (g1)--(g2) (g2)--(g3) (g3)--(g4) (g4)--(g5) (g5)--(g6) (g6)--(g7) (g7)--(g8) (g8)--(g9) (g8)--(g10) (g10)--(g11) (g11)--(g12) (g12)--(g13);
	\end{tikzpicture}
    }
    \label{eq:5d_sp_magQuiv_infinite_E7}
\end{align}
which reproduces the $E_7 \times E_7$ family of Table \ref{EnProdtable2}.
    \item \ul{Case $n=2k+1$.} Starting from \eqref{eq:5-brane_ON_and_Op_Higgs}, the left 5-brane with NS charge is a $(1,-1)$ 5-brane. Moving this brane onto the ON plane together with brane creation leads to
    \begin{align}
    \raisebox{-.5\height}{
    \begin{tikzpicture}
    \draw[dashed] (-3,0)--(5.5,0) (-4,0)--(-6,0);
    \node at (4.5,-0.35) {\tiny{$-$}};
    \node at (3.5,-0.35) {\tiny{$\widetilde{-}$}};
    \node at (2.5,-0.35) {\tiny{$-$}};
    \node at (1.5,-0.35) {\tiny{$\widetilde{-}$}};
    \node at (0.5,-0.35) {\tiny{$-$}};
    \node at (-0.5,-0.35) {\tiny{$\widetilde{-}$}};
    \node at (-1.5,-0.35) {\tiny{$-$}};
    \node at (-2.5,-0.35) {\tiny{$\widetilde{-}$}};
    \node at (-4.5,-0.35) {\tiny{$-$}};
    \node at (-5.5,-0.35) {\tiny{$\widetilde{-}$}};
    \draw[dotted,thick] (5.5,-2)--(5.5,2);
    \node at (6,1.25) {\tiny{ON${}^-$}};
    \foreach \i in {-1,1}
    \draw (5.5,0.1*\i)--(-3,0.1*\i) (-4,0.1*\i)--(-5,0.1*\i);
    \monocut{1,-0.15}{5.5,-0.15};
    \monocut{2,-0.075}{5.5,-0.075};
    \monocut{3,0.075}{5.5,0.075};
    \monocut{4,0.15}{5.5,0.15};
    \SevenB{4,0}
    \SevenB{3,0}
    \SevenB{2,0}
    \SevenB{1,0}
    \node at (5.5,0) {$\times$};
    \node at (6,0) {\tiny{O7${}^-$}};
    \foreach \i in {0,1}
    \draw (5.4+0.2*\i,-0.75)--(5.4+0.2*\i,0.75);
    \draw (5.5,0)--(4.5,1);
    \monocut{-6,1.5}{-6,-1.5}
    \monocut{-4,1.5}{-4,-1.5}
    \monocut{-3,1.5}{-3,-0.5}
    \monocut{-1,1.5}{-1,-1.5}
    \monocut{-6,-0.05}{-5,-0.05}
    \monocut{-3,-0.05}{-2,-0.05}
    \monocut{-1,-0.05}{0,-0.05}
    \monocut{4.5,1}{4.5,2}
    \SevenB{0,0}
    \SevenB{-1,0}
    \SevenB{-2,0}
    \SevenB{-3,0}
    \node at (-3.5,0) {$\cdots$};
    \SevenB{-4,0}
    \SevenB{-5,0}
    \SevenB{-6,0}
    \SevenB{5.5,0.75}
    \SevenB{5.5,1.5}
    \SevenB{4.5,1}
    \node at (4.5,0.25) {\tiny{$2k{+}3$}};
    \node at (3.5,0.25) {\tiny{$2k{+}2$}};
    \node at (2.5,0.25) {\tiny{$2k{+}2$}};
    \node at (1.5,0.25) {\tiny{$2k{+}1$}};
    \node at (0.5,0.25) {\tiny{$2k{+}1$}};
    \node at (-0.5,0.25) {\tiny{$2k$}};
    \node at (-1.5,0.25) {\tiny{$2k$}};
    \node at (-2.5,0.25) {\tiny{$2k{-}1$}};
    \node at (-4.5,0.25) {\tiny{$1$}};
    \node at (3.85,1) {\tiny{$[1,-1]$}};
    \draw[decoration={brace,mirror,raise=10pt},decorate,thick]
  (-6.1,-0.1) -- node[below=15pt] {\tiny{$4k+2$ half D7}}
  (0.1,-0.1);
  \node at (5.7,0.325) {\tiny{$1$}};
    \begin{scope}[yscale=-1,xscale=1]
    \draw (5.5,0)--(4.5,1);
    \SevenB{5.5,0.75}
    \SevenB{5.5,1.5}
    \monocut{4.5,1}{4.5,2}
    \SevenB{4.5,1}
      \end{scope}
    \end{tikzpicture}
    }
    \label{eq:5-brane_ON_and_Op_infinite_E6}
\end{align}
such that one can proceed to analyse the Higgs branch direction via the subweb decomposition. As the configuration away from the O7 and ON can be treated by the results of \cite{Bourget:2020gzi}, one only needs to clarify the decomposition into subwebs around the intersection point
\begin{align}
        \raisebox{-.5\height}{
    \begin{tikzpicture}
    \draw[dashed] (-0.5,0)--(6,0);
    \draw[dotted] (6,2.5)--(6,0);
    \node at (6,0) {$\times$};
    \draw (0,0.1)--(2,0.1);
    \node at (1,0.4) {\tiny{$2k{+}2$}};
    \draw[thick,blue] (2,0.1)--(4,0.1);
    \node[blue] at (2.5,0.5) {\tiny{$k{+}1$}};
    \draw[thick,red] (2,0.1) -- (2.1,0.25) --(5.8,0.25)--(6,0.2)--(5.8,0.15)--(4,0.15);
    \node[red] at (3.5,0.5) {\tiny{$k{+}1$}};
    \draw[myGreen,thick] (4,0.1)--(4.1,0.35)--(5.8,0.35) (6,0.1)--(5.8,0.35)--(5,1);
    \node at (5,1.35) {\tiny{$[1,-1]$}};
    \draw (5.9,0)--(5.9,1);
    \node at (5.75,0.7) {\tiny{$1$}};
    \monocut{-0.5,-0.05}{6,-0.05}
    \monocut{0,-0.1}{6,-0.1}
    \monocut{2,-0.15}{6,-0.15}
    \monocut{4,-0.2}{6,-0.2}
    \SevenB{0,0}
    \SevenB{2,0}
    \SevenB{4,0}
    \SevenB{6,1}
    \SevenB{6,2}
    \SevenB{5,1}
    \node at (1,-0.4) {\tiny{$-$}};
    \node at (3,-0.4) {\tiny{$\widetilde{-}$}};
    \end{tikzpicture}
    }
    \quad \rightarrow \quad
    \raisebox{-.5\height}{
    \begin{tikzpicture}
	\node (g7) {$\cdots$};
	\node (g8) [gauge,right of =g7,label=below:{\tiny{$D_{2k{+}2}$}}] {};
	\node[thick,blue] (g9) [gauge, above right of =g8,label=above:{\tiny{$C_{k{+}1}$}}] {};
	\node[thick,red] (g10) [gauge,below right of =g8,label=below:{\tiny{$C_{k{+}1}$}}] {};
	\node (g11) [gauge,right of =g10,label=below:{\tiny{$U_1$}}] {};
	\node[thick,myGreen] (g12) [gauge,right of =g9,label=above:{\tiny{$U_1$}}] {};
	\draw (g7)--(g8) (g8)--(g9) (g8)--(g10) (g10)--(g11) (g9)--(g12);
	\end{tikzpicture}
    }
\end{align}
where all mirror images have been ignored. First of all, the magnetic gauge nodes are as follows: the black stack of $2k+2$ 5-branes yields a $D_{2k+2}$ gauge algebra, while the red and blue stack yield a $C_{k+1}$ gauge algebra each. The green subweb yields a $\urm(1)$ magnetic gauge group, as the 5-brane does not perceive the presence of the O5 plane. Lastly, the vertical black 5-brane also contributes a $\urm(1)$ node. 
Next, the magnetic gauge nodes are connected by magnetic hypermultiplets in a fashion determined by the generalised stable intersection \cite{Cabrera:2018jxt,Bourget:2020gzi}.
The stack of $2k+2$ D5s on the O5${}^-$ has stable intersection of 1 with both the red and blue stack of D5s, because they end on different sides of the same 7-brane. Hence, the corresponding magnetic gauge nodes are connected by bifundamental magnetic hypermultiplets. Likewise, the red and blue stack have a vanishing generalised intersection number, as they start from the same 7-brane on the same side, but end on another 7-brane from opposite sides. The green subweb has intersection of 1 with the blue subweb, because they end on the same 7-brane from different side. Hence, there is a bifundamental hypermultiplet between the blue $C_{k+1}$ and the green $\urm(1)$ node. 
For the black vertical subweb and the red stack, the intersection number is $|\left(\begin{smallmatrix} 1& 0 \\ 0 & 1 \end{smallmatrix}\right)|=1$, implying a magnetic hypermultiplet in between the nodes.
The green and the black subweb at the intersection point share a vanishing intersection number, but the monodromy cuts needs to be considered carefully to arrived at this result. The two $M_{[1,0]}$ monodromies affect the green $(2,-1)$ 5-brane such that it becomes $M_{[1,0]}^2(2,-1) = (0,1)$. Thus, the stable intersection between the green and black subweb is computed as $|\left(\begin{smallmatrix} 0& 1 \\ 0 & 1 \end{smallmatrix} \right)
|=0$, which is indeed trivial.
Hence, the magnetic quiver is given by
    \begin{align}
        \raisebox{-.5\height}
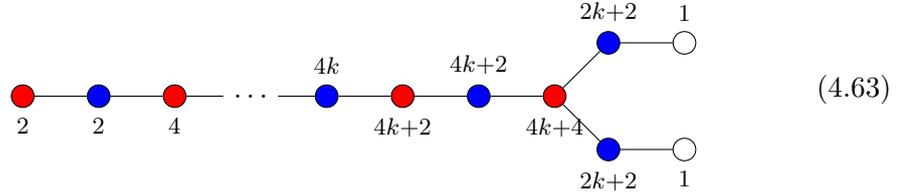
{
    \begin{tikzpicture}
	\node (g1) [gaugeSO,label=below:{\footnotesize{$2$}}] {};
	\node (g2) [gaugeSp,right of =g1,label=below:{\footnotesize{$2$}}] {};
	\node (g3) [gaugeSO,right of =g2,label=below:{\footnotesize{$4$}}] {};
	\node (g4) [right of =g3] {$\ldots$};
	\node (g5) [gaugeSp,right of =g4,label=above:{\footnotesize{$4k$}}] {};
	\node (g6) [gaugeSO,right of =g5,label=below:{\footnotesize{$4k{+}2$}}] {};
	\node (g7) [gaugeSp,right of =g6,label=above:{\footnotesize{$4k{+}2$}}] {};
	\node (g8) [gaugeSO,right of =g7,label=below:{\footnotesize{$4k{+}4$}}] {};
	\node (g9) [gaugeSp, above right of =g8,label=above:{\footnotesize{$2k{+}2$}}] {};
	\node (g10) [gaugeSp,below right of =g8,label=below:{\footnotesize{$2k{+}2$}}] {};
	\node (g11) [gauge,right of =g10,label=below:{\footnotesize{$1$}}] {};
	\node (g12) [gauge,right of =g9,label=above:{\footnotesize{$1$}}] {};
	\draw  (g1)--(g2) (g2)--(g3) (g3)--(g4) (g4)--(g5) (g5)--(g6) (g6)--(g7) (g7)--(g8) (g8)--(g9) (g8)--(g10) (g10)--(g11) (g9)--(g12);
	\end{tikzpicture}
    }
    \label{eq:5d_sp_magQuiv_infinite_E6}
\end{align}
which reproduces the $E_6 \times E_6$ family of Table \ref{EnProdtable2}.
    \item \ul{Case $n=2k$.} The left-hand side brane with NS charge is of type $(2,-1)$, and moving it onto the ON plane, taking brane creation into account, yields the brane web
    \begin{align}
    \raisebox{-.5\height}
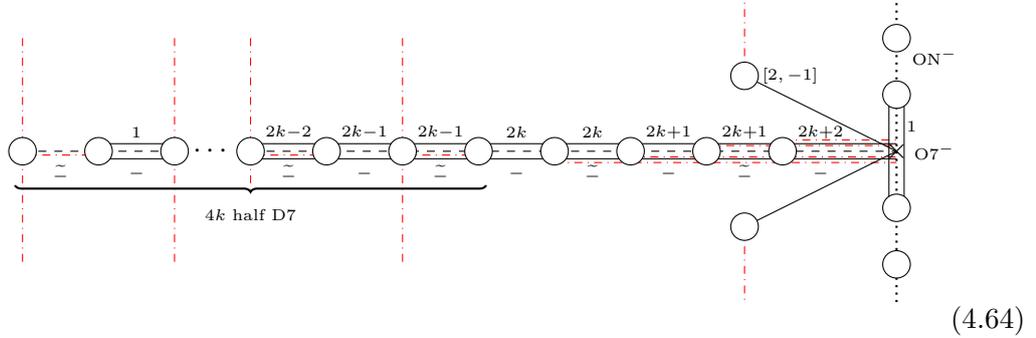
{
    \begin{tikzpicture}
    \draw[dashed] (-3,0)--(5.5,0) (-4,0)--(-6,0);
    \node at (4.5,-0.3) {\tiny{$-$}};
    \node at (3.5,-0.3) {\tiny{$\widetilde{-}$}};
    \node at (2.5,-0.3) {\tiny{$-$}};
    \node at (1.5,-0.3) {\tiny{$\widetilde{-}$}};
    \node at (0.5,-0.3) {\tiny{$-$}};
    \node at (-0.5,-0.3) {\tiny{$\widetilde{-}$}};
    \node at (-1.5,-0.3) {\tiny{$-$}};
    \node at (-2.5,-0.3) {\tiny{$\widetilde{-}$}};
    \node at (-4.5,-0.3) {\tiny{$-$}};
    \node at (-5.5,-0.3) {\tiny{$\widetilde{-}$}};
    \draw[dotted,thick] (5.5,-2)--(5.5,2);
    \node at (6,1.25) {\tiny{ON${}^-$}};
    \foreach \i in {-1,1}
    \draw (5.5,0.1*\i)--(-3,0.1*\i) (-4,0.1*\i)--(-5,0.1*\i);
    \monocut{1,-0.15}{5.5,-0.15};
    \monocut{2,-0.075}{5.5,-0.075};
    \monocut{3,0.075}{5.5,0.075};
    \monocut{4,0.15}{5.5,0.15};
    \SevenB{4,0}
    \SevenB{3,0}
    \SevenB{2,0}
    \SevenB{1,0}
    \node at (5.5,0) {$\times$};
    \node at (6,0) {\tiny{O7${}^-$}};
    \foreach \i in {0,1}
    \draw (5.4+0.2*\i,-0.75)--(5.4+0.2*\i,0.75);
    \draw (5.5,0)--(3.5,1);
    \monocut{-6,1.5}{-6,-1.5}
    \monocut{-4,1.5}{-4,-1.5}
    \monocut{-3,1.5}{-3,-0.5}
    \monocut{-1,1.5}{-1,-1.5}
    \monocut{-6,-0.05}{-5,-0.05}
    \monocut{-3,-0.05}{-2,-0.05}
    \monocut{-1,-0.05}{0,-0.05}
    \monocut{3.5,1}{3.5,2}
    \SevenB{0,0}
    \SevenB{-1,0}
    \SevenB{-2,0}
    \SevenB{-3,0}
    \node at (-3.5,0) {$\cdots$};
    \SevenB{-4,0}
    \SevenB{-5,0}
    \SevenB{-6,0}
    \SevenB{5.5,0.75}
    \SevenB{5.5,1.5}
    \SevenB{3.5,1}
    \node at (4.5,0.25) {\tiny{$2k{+}2$}};
    \node at (3.5,0.25) {\tiny{$2k{+}1$}};
    \node at (2.5,0.25) {\tiny{$2k{+}1$}};
    \node at (1.5,0.25) {\tiny{$2k$}};
    \node at (0.5,0.25) {\tiny{$2k$}};
    \node at (-0.5,0.25) {\tiny{$2k{-}1$}};
    \node at (-1.5,0.25) {\tiny{$2k{-}1$}};
    \node at (-2.5,0.25) {\tiny{$2k{-}2$}};
    \node at (-4.5,0.25) {\tiny{$1$}};
    \node at (4.1,1) {\tiny{$[2,-1]$}};
    \draw[decoration={brace,mirror,raise=10pt},decorate,thick]
  (-6.1,-0.1) -- node[below=15pt] {\tiny{$4k$ half D7}}
  (0.1,-0.1);
  \node at (5.7,0.325) {\tiny{$1$}};
    \begin{scope}[yscale=-1,xscale=1]
    \draw (5.5,0)--(3.5,1);
    \SevenB{5.5,0.75}
    \SevenB{5.5,1.5}
    \monocut{3.5,1}{3.5,2}
    \SevenB{3.5,1}
      \end{scope}
    \end{tikzpicture}
    }
    \label{eq:5-brane_ON_and_Op_infinite_E5}
\end{align}
and the next step consists of the subweb decomposition.
For further clarification, one only needs to detail the decomposition into subwebs around the intersection point
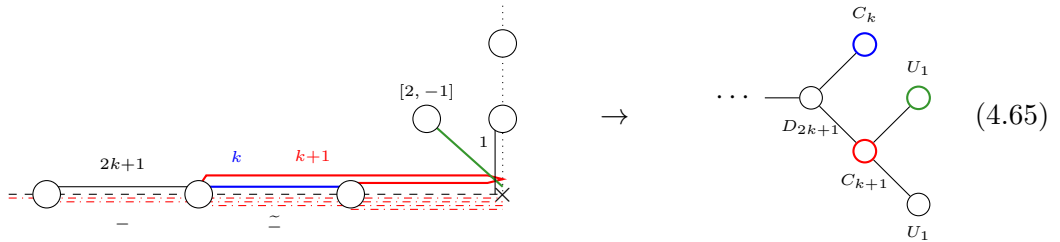
\begin{align}
        \raisebox{-.5\height}{
    \begin{tikzpicture}
    \draw[dashed] (-0.5,0)--(6,0);
    \draw[dotted] (6,2.5)--(6,0);
    \node at (6,0) {$\times$};
    \draw (0,0.1)--(2,0.1);
    \node at (1,0.4) {\tiny{$2k{+}1$}};
    \draw[blue,thick] (2,0.1)--(4,0.1);
    \node[blue] at (2.5,0.5) {\tiny{$k$}};
    \draw[red,thick] (2,0.1) -- (2.1,0.25) --(5.8,0.25)--(6,0.2)--(5.8,0.15)--(4,0.15);
    \node[red] at (3.5,0.5) {\tiny{$k{+}1$}};
    \draw[myGreen,thick] (6,0.1)--(5,1);
    \node at (5,1.35) {\tiny{$[2,-1]$}};
    \draw (5.9,0)--(5.9,1);
    \node at (5.75,0.7) {\tiny{$1$}};
    \monocut{-0.5,-0.05}{6,-0.05}
    \monocut{0,-0.1}{6,-0.1}
    \monocut{2,-0.15}{6,-0.15}
    \monocut{4,-0.2}{6,-0.2}
    \SevenB{0,0}
    \SevenB{2,0}
    \SevenB{4,0}
    \SevenB{6,1}
    \SevenB{6,2}
    \SevenB{5,1}
    \node at (1,-0.4) {\tiny{$-$}};
    \node at (3,-0.4) {\tiny{$\widetilde{-}$}};
    \end{tikzpicture}
    }
    \qquad \rightarrow \qquad
    \raisebox{-.5\height}{
    \begin{tikzpicture}
	\node (g7) {$\cdots$};
	\node (g8) [gauge,right of =g7,label=below:{\tiny{$D_{2k{+}1}$}}] {};
	\node[blue,thick] (g9) [gauge, above right of =g8,label=above:{\tiny{$C_{k}$}}] {};
	\node[red,thick] (g10) [gauge,below right of =g8,label=below:{\tiny{$C_{k{+}1}$}}] {};
	\node (g11) [gauge,below right of =g10,label=below:{\tiny{$U_1$}}] {};
	\node[myGreen,thick] (g12) [gauge,above right of =g10,label=above:{\tiny{$U_1$}}] {};
	\draw (g7)--(g8) (g8)--(g9) (g8)--(g10) (g10)--(g11) (g10)--(g12);
	\end{tikzpicture}
    }
\end{align}
where all mirror images have been ignored.
First of all, the magnetic gauge nodes are as follows \cite{Bourget:2020gzi}: the black stack of $2k+2$ 5-branes yields a $D_{2k+1}$ gauge algebra, while the blue and red stack yield a  $C_k$ and $C_{k+1}$ gauge algebra respectively. The green subweb yields a $\urm(1)$ magnetic gauge group, as the 5-brane does not perceive the presence of the O5 plane. Lastly, the vertical black 5-brane also contributes a $\urm(1)$ node. 
As above, the generalised stable intersection \cite{Cabrera:2018jxt,Bourget:2020gzi} is used to derive the magnetic hypermultiplets.
The stack of $2k+1$ D5s on the O5${}^-$ has stable intersection of 1 with both the red and blue stack of D5s, because they end on different sides of the same 7-brane. Hence, the corresponding magnetic gauge nodes are connected by bifundamental magnetic hypermultiplets. Likewise, the red and blue stack have a vanishing generalised intersection number, as they start from the same 7-brane on the same side, but end on another 7-brane from opposite sides. The green as well as the black subweb at the intersection point have trivial generalised intersection number with the blue subweb. 
The intersection number between the green and red subweb is computed to be $|\left(\begin{smallmatrix} 2& -1 \\ 1 & 0 \end{smallmatrix}\right)|=1$, which implies a magnetic hypermultiplet.
Similarly, for the black vertical subweb and the red stack, the intersection number is $|\left(\begin{smallmatrix} 1& 0 \\ 0 & 1 \end{smallmatrix}\right)|=1$, implying a magnetic hypermultiplet in between the nodes.
The green and the black subweb at the intersection point share a vanishing intersection number, due to the presence of the monodromy cuts. In detail, the two $M_{[1,0]}$ monodromy lines alter the green $(2,-1)$ 5-brane into a $M_{[1,0]}^2 (2,-1)=(0,1)$ 5-brane. Hence, the stable intersection is simply $|\left(\begin{smallmatrix} 0& 1 \\ 0 & 1 \end{smallmatrix} \right)
|=0$. 
The magnetic quiver is given by
    \begin{align}
        \raisebox{-.5\height}
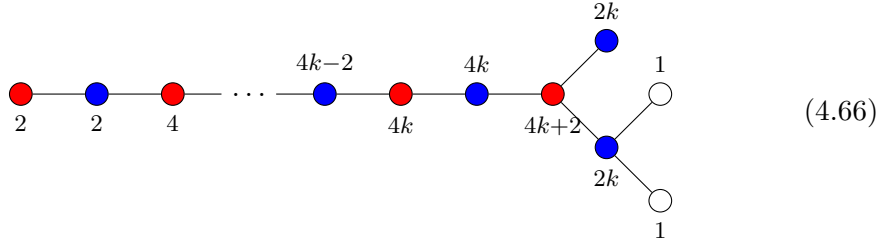
{
    \begin{tikzpicture}
	\node (g1) [gaugeSO,label=below:{\footnotesize{$2$}}] {};
	\node (g2) [gaugeSp,right of =g1,label=below:{\footnotesize{$2$}}] {};
	\node (g3) [gaugeSO,right of =g2,label=below:{\footnotesize{$4$}}] {};
	\node (g4) [right of =g3] {$\ldots$};
	\node (g5) [gaugeSp,right of =g4,label=above:{\footnotesize{$4k{-}2$}}] {};
	\node (g6) [gaugeSO,right of =g5,label=below:{\footnotesize{$4k$}}] {};
	\node (g7) [gaugeSp,right of =g6,label=above:{\footnotesize{$4k$}}] {};
	\node (g8) [gaugeSO,right of =g7,label=below:{\footnotesize{$4k{+}2$}}] {};
	\node (g9) [gaugeSp, above right of =g8,label=above:{\footnotesize{$2k$}}] {};
	\node (g10) [gaugeSp,below right of =g8,label=below:{\footnotesize{$2k$}}] {};
	\node (g11) [gauge,above right of =g10,label=above:{\footnotesize{$1$}}] {};
	\node (g12) [gauge,below right of =g10,label=below:{\footnotesize{$1$}}] {};
	\draw  (g1)--(g2) (g2)--(g3) (g3)--(g4) (g4)--(g5) (g5)--(g6) (g6)--(g7) (g7)--(g8) (g8)--(g9) (g8)--(g10) (g10)--(g11) (g10)--(g12);
	\end{tikzpicture}
    }
    \label{eq:5d_sp_magQuiv_infinite_E5}
\end{align}
which reproduces the $E_5 \times E_5$ family of Table \ref{EnProdtable2}.

\item \ul{Case $n=2k-2l-1$, $k\geq l\geq 0$.}
Starting from \eqref{eq:5-brane_ON_and_Op}, the $(2+2k-n,-1)$ 5-brane becomes a $(2l+3,1)$ brane with $k\geq l \geq 0$. Transitioning to infinite coupling, the 5-brane web becomes
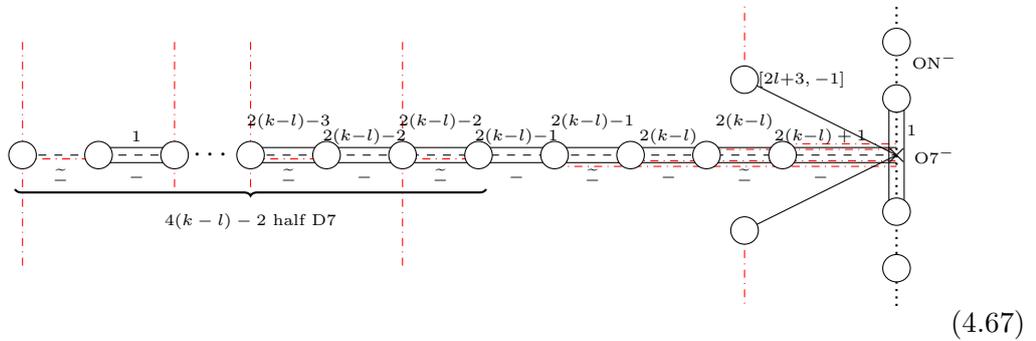
\begin{align}
    \raisebox{-.5\height}{
    \begin{tikzpicture}
    \draw[dashed] (-3,0)--(5.5,0) (-4,0)--(-6,0);
    \node at (4.5,-0.3) {\tiny{$-$}};
    \node at (3.5,-0.3) {\tiny{$\widetilde{-}$}};
    \node at (2.5,-0.3) {\tiny{$-$}};
    \node at (1.5,-0.3) {\tiny{$\widetilde{-}$}};
    \node at (0.5,-0.3) {\tiny{$-$}};
    \node at (-0.5,-0.3) {\tiny{$\widetilde{-}$}};
    \node at (-1.5,-0.3) {\tiny{$-$}};
    \node at (-2.5,-0.3) {\tiny{$\widetilde{-}$}};
    \node at (-4.5,-0.3) {\tiny{$-$}};
    \node at (-5.5,-0.3) {\tiny{$\widetilde{-}$}};
    \draw[dotted,thick] (5.5,-2)--(5.5,2);
    \node at (6,1.25) {\tiny{ON${}^-$}};
    \foreach \i in {-1,1}
    \draw (5.5,0.1*\i)--(-3,0.1*\i) (-4,0.1*\i)--(-5,0.1*\i);
    \monocut{1,-0.15}{5.5,-0.15};
    \monocut{2,-0.075}{5.5,-0.075};
    \monocut{3,0.075}{5.5,0.075};
    \monocut{4,0.15}{5.5,0.15};
    \SevenB{4,0}
    \SevenB{3,0}
    \SevenB{2,0}
    \SevenB{1,0}
    \node at (5.5,0) {$\times$};
    \node at (6,0) {\tiny{O7${}^-$}};
    \foreach \i in {0,1}
    \draw (5.4+0.2*\i,-0.75)--(5.4+0.2*\i,0.75);
    \draw (5.5,0)--(3.5,1);
    \monocut{-6,1.5}{-6,-1.5}
    \monocut{-4,1.5}{-4,-0.5}
    \monocut{-3,1.5}{-3,-0.5}
    \monocut{-1,1.5}{-1,-1.5}
    \monocut{-6,-0.05}{-5,-0.05}
    \monocut{-3,-0.05}{-2,-0.05}
    \monocut{-1,-0.05}{0,-0.05}
    \monocut{3.5,1}{3.5,2}
    \SevenB{0,0}
    \SevenB{-1,0}
    \SevenB{-2,0}
    \SevenB{-3,0}
    \node at (-3.5,0) {$\cdots$};
    \SevenB{-4,0}
    \SevenB{-5,0}
    \SevenB{-6,0}
    \SevenB{5.5,0.75}
    \SevenB{5.5,1.5}
    \SevenB{3.5,1}
    \node at (4.5,0.25) {\tiny{$2(k{-}l)+1$}};
    \node at (3.5,0.45) {\tiny{$2(k{-}l)$}};
    \node at (2.5,0.25) {\tiny{$2(k{-}l)$}};
    \node at (1.5,0.45) {\tiny{$2(k{-}l){-}1$}};
    \node at (0.5,0.25) {\tiny{$2(k{-}l){-}1$}};
    \node at (-0.5,0.45) {\tiny{$2(k{-}l){-}2$}};
    \node at (-1.5,0.25) {\tiny{$2(k{-}l){-}2$}};
    \node at (-2.5,0.45) {\tiny{$2(k{-}l){-}3$}};
    \node at (-4.5,0.25) {\tiny{$1$}};
    \node at (4.25,1) {\tiny{$[2l{+}3,-1]$}};
    \draw[decoration={brace,mirror,raise=10pt},decorate,thick]
  (-6.1,-0.1) -- node[below=15pt] {\tiny{$4(k-l)-2$ half D7}}
  (0.1,-0.1);
  \node at (5.7,0.325) {\tiny{$1$}};
    \begin{scope}[yscale=-1,xscale=1]
    \draw (5.5,0)--(3.5,1);
    \SevenB{5.5,0.75}
    \SevenB{5.5,1.5}
    \monocut{3.5,1}{3.5,2}
    \SevenB{3.5,1}
      \end{scope}
    \end{tikzpicture}
    }
    \label{eq:5-brane_ON_and_Op_infinite_E4}
\end{align}
and the next task is to derive the magnetic quiver. As above, the branes away from the intersection of the O-planes are straightforward to take into account. The novel contributions arise from the vicinity of the intersection. For further clarification, the decomposition into subwebs around the intersection point is given by
\begin{align}
        \raisebox{-.5\height}{
    \begin{tikzpicture}
    \draw[dashed] (-0.5,0)--(6,0);
    \draw[dotted] (6,2.5)--(6,0);
    \node at (6,0) {$\times$};
    \draw (0,0.1)--(2,0.1);
    \node at (1,0.4) {\tiny{$2k{-}2l$}};
    \draw[blue,thick] (2,0.1)--(4,0.1);
    \node[blue] at (2.5,0.5) {\tiny{$k{-}l$}};
    \draw[red,thick] (2,0.1) -- (2.1,0.25) --(5.8,0.25)--(6,0.2)--(5.8,0.15)--(4,0.15);
    \node[red] at (3.5,0.5) {\tiny{$k{-}l$}};
    \draw[myGreen,thick] (4,0.1)--(4.1,0.35)--(5.8,0.35) (6,0.1)--(5.8,0.35)--(5,1);
    \node at (5,1.35) {\tiny{$[2l+3,-1]$}};
    \draw (5.9,0)--(5.9,1);
    \node at (5.75,0.7) {\tiny{$1$}};
    \monocut{-0.5,-0.05}{6,-0.05}
    \monocut{0,-0.1}{6,-0.1}
    \monocut{2,-0.15}{6,-0.15}
    \monocut{4,-0.2}{6,-0.2}
    \SevenB{0,0}
    \SevenB{2,0}
    \SevenB{4,0}
    \SevenB{6,1}
    \SevenB{6,2}
    \SevenB{5,1}
    \node at (1,-0.4) {\tiny{$-$}};
    \node at (3,-0.4) {\tiny{$\widetilde{-}$}};
    \end{tikzpicture}
    }
    \quad \rightarrow \quad
    \raisebox{-.5\height}{
    \begin{tikzpicture}
	\node (g7) {$\cdots$};
	\node (g8) [gauge,right of =g7,label=below:{\tiny{$D_{2k{-}2l}$}}] {};
	\node[blue,thick] (g9) [gauge, above right of =g8,label=above:{\tiny{$C_{k{-}l}$}}] {};
	\node[red,thick] (g10) [gauge,below right of =g8,label=below:{\tiny{$C_{k{-}l}$}}] {};
	\node (g11) [gauge,right of =g10,label=below:{\tiny{$U_1$}}] {};
	\node[myGreen,thick] (g12) [gauge,right of =g9,label=above:{\tiny{$U_1$}}] {};
	\draw (g7)--(g8) (g8)--(g9) (g8)--(g10) (g10)--(g11) (g9)--(g12) (g11)--(g12);
	    \draw[dashed] (g12) to [out=-45,in=45,looseness=1] (g11);
    \node at (2.4,0) {\tiny{$l{+}1$}};
    \node at (3.4,0) {\tiny{$l{+}1$}};
	\end{tikzpicture}
    }
\end{align}
where all mirror images have been ignored. The red and blue subweb lead to the familiar bifurcation into $C$-type magnetic gauge nodes. The green subweb, being away from the O5, leads to a $\urm(1)$ nodes, which has trivial intersection number with the red subweb; while the intersection number with the blue subweb is one.  Similarly, the black vertical $(0,1)$ 5-brane, also giving rise to a $\urm(1)\cong \sorm(2)$, has trivial intersection with the blue subweb; while the intersection number with the red subweb is one. Lastly, the generalised intersection number between the green subweb and the $(0,1)$ 5-brane is evaluated by taking the monodromy cuts into account. The two $M_{[1,0]}$ monodromy cuts alter the green $(2l+4,-1)$ 5-brane into a $M_{[1,0]}^2 (2l+4,-1)=(2l+2,-1)$  such that the intersection number becomes  $| \left( \begin{smallmatrix} 2l+2& -1 \\ 0 & 1 \end{smallmatrix} \right) | = 2l+2$. According to \cite{Akhond:2020vhc}, there are $(l+1)$ hypermultiplets in the $(1,1)$ representation of $\urm(1)\times \urm(1)$ as well as $(l+1)$ hypermultiplets in the $(1,-1)$ representation.
To sum up, the magnetic quiver is proposed to be
    \begin{align}
        \raisebox{-.5\height}{
    \begin{tikzpicture}
	\node (g1) [gaugeSO,label=below:{\footnotesize{$2$}}] {};
	\node (g2) [gaugeSp,right of =g1,label=below:{\footnotesize{$2$}}] {};
	\node (g3) [gaugeSO,right of =g2,label=below:{\footnotesize{$4$}}] {};
	\node (g4) [right of =g3] {$\ldots$};
	\node (g5) [gaugeSp,right of =g4,label=above:{\footnotesize{$4(k{-}l){-}4$}}] {};
	\node (g6) [gaugeSO,right of =g5,label=below:{\footnotesize{$4(k{-}l){-}2$}}] {};
	\node (g7) [gaugeSp,right of =g6,label=above:{\footnotesize{$4(k{-}l){-}2$}}] {};
	\node (g8) [gaugeSO,right of =g7,label=below:{\footnotesize{$4(k{-}l)$}}] {};
	\node (g9) [gaugeSp, above right of =g8,label=above:{\footnotesize{$2(k-l)$}}] {};
	\node (g10) [gaugeSp,below right of =g8,label=below:{\footnotesize{$2(k-l)$}}] {};
	\node (g11) [gauge,right of =g9,label=above:{\footnotesize{$1$}}] {};
	\node (g12) [gauge,right of =g10,label=below:{\footnotesize{$1$}}] {};
	\draw  (g1)--(g2) (g2)--(g3) (g3)--(g4) (g4)--(g5) (g5)--(g6) (g6)--(g7) (g7)--(g8) (g8)--(g9) (g8)--(g10) (g9)--(g11) (g10)--(g12) (g11)--(g12);
    \draw[dashed] (g11) to [out=-45,in=45,looseness=1] (g12);
    \node at (8.4,0) {\tiny{$l{+}1$}};
    \node at (9.4,0) {\tiny{$l{+}1$}};
	\end{tikzpicture}
    }
    \label{eq:5d_sp_magQuiv_infinite_E4}
\end{align}
which reproduces the $E_{4-2l} \times E_{4-2l}$ family of Table \ref{EnProdtable2}.
\item \ul{Case $n=2k-2l-2$, $k\leq l\leq 0$.}
Returning to \eqref{eq:5-brane_ON_and_Op}, the $(2+2k-n,-1)$ 5-brane becomes a $(2l+4,1)$ brane with $k\geq l \geq 0$. At the fixed point, the 5-brane web becomes
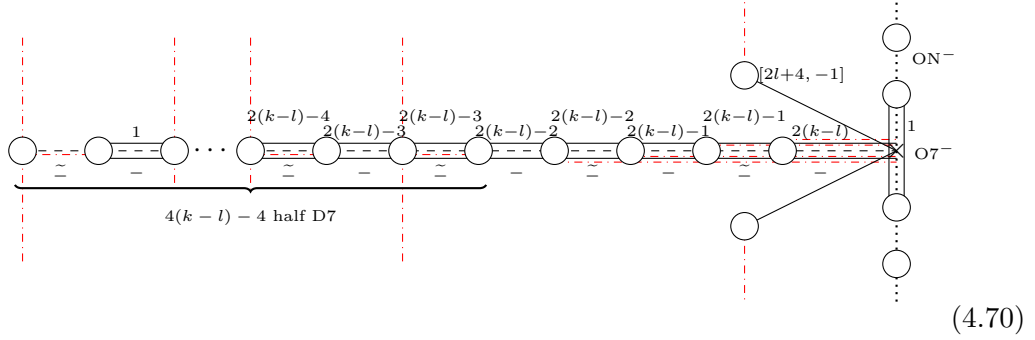
\begin{align}
    \raisebox{-.5\height}{
    \begin{tikzpicture}
    \draw[dashed] (-3,0)--(5.5,0) (-4,0)--(-6,0);
    \node at (4.5,-0.3) {\tiny{$-$}};
    \node at (3.5,-0.3) {\tiny{$\widetilde{-}$}};
    \node at (2.5,-0.3) {\tiny{$-$}};
    \node at (1.5,-0.3) {\tiny{$\widetilde{-}$}};
    \node at (0.5,-0.3) {\tiny{$-$}};
    \node at (-0.5,-0.3) {\tiny{$\widetilde{-}$}};
    \node at (-1.5,-0.3) {\tiny{$-$}};
    \node at (-2.5,-0.3) {\tiny{$\widetilde{-}$}};
    \node at (-4.5,-0.3) {\tiny{$-$}};
    \node at (-5.5,-0.3) {\tiny{$\widetilde{-}$}};
    \draw[dotted,thick] (5.5,-2)--(5.5,2);
    \node at (6,1.25) {\tiny{ON${}^-$}};
    \foreach \i in {-1,1}
    \draw (5.5,0.1*\i)--(-3,0.1*\i) (-4,0.1*\i)--(-5,0.1*\i);
    \monocut{1,-0.15}{5.5,-0.15};
    \monocut{2,-0.075}{5.5,-0.075};
    \monocut{3,0.075}{5.5,0.075};
    \monocut{4,0.15}{5.5,0.15};
    \SevenB{4,0}
    \SevenB{3,0}
    \SevenB{2,0}
    \SevenB{1,0}
    \node at (5.5,0) {$\times$};
    \node at (6,0) {\tiny{O7${}^-$}};
    \foreach \i in {0,1}
    \draw (5.4+0.2*\i,-0.75)--(5.4+0.2*\i,0.75);
    \draw (5.5,0)--(3.5,1);
    \monocut{-6,1.5}{-6,-1.5}
    \monocut{-4,1.5}{-4,-0.5}
    \monocut{-3,1.5}{-3,-0.5}
    \monocut{-1,1.5}{-1,-1.5}
    \monocut{-6,-0.05}{-5,-0.05}
    \monocut{-3,-0.05}{-2,-0.05}
    \monocut{-1,-0.05}{0,-0.05}
    \monocut{3.5,1}{3.5,2}
    \SevenB{0,0}
    \SevenB{-1,0}
    \SevenB{-2,0}
    \SevenB{-3,0}
    \node at (-3.5,0) {$\cdots$};
    \SevenB{-4,0}
    \SevenB{-5,0}
    \SevenB{-6,0}
    \SevenB{5.5,0.75}
    \SevenB{5.5,1.5}
    \SevenB{3.5,1}
    \node at (4.5,0.25) {\tiny{$2(k{-}l)$}};
    \node at (3.5,0.45) {\tiny{$2(k{-}l){-}1$}};
    \node at (2.5,0.25) {\tiny{$2(k{-}l){-}1$}};
    \node at (1.5,0.45) {\tiny{$2(k{-}l){-}2$}};
    \node at (0.5,0.25) {\tiny{$2(k{-}l){-}2$}};
    \node at (-0.5,0.45) {\tiny{$2(k{-}l){-}3$}};
    \node at (-1.5,0.25) {\tiny{$2(k{-}l){-}3$}};
    \node at (-2.5,0.45) {\tiny{$2(k{-}l){-}4$}};
    \node at (-4.5,0.25) {\tiny{$1$}};
    \node at (4.25,1) {\tiny{$[2l{+}4,-1]$}};
    \draw[decoration={brace,mirror,raise=10pt},decorate,thick]
  (-6.1,-0.1) -- node[below=15pt] {\tiny{$4(k-l)-4$ half D7}}
  (0.1,-0.1);
  \node at (5.7,0.325) {\tiny{$1$}};
    \begin{scope}[yscale=-1,xscale=1]
    \draw (5.5,0)--(3.5,1);
    \SevenB{5.5,0.75}
    \SevenB{5.5,1.5}
    \monocut{3.5,1}{3.5,2}
    \SevenB{3.5,1}
      \end{scope}
    \end{tikzpicture}
    }
    \label{eq:5-brane_ON_and_Op_infinite_E3}
\end{align}
and the subweb decomposition provides information on the Higgs branch. 
To clarify the non-trivial features, the decomposition into subwebs around the intersection point is given by
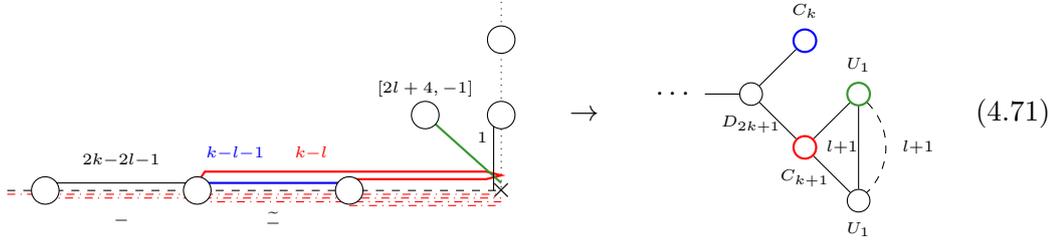
\begin{align}
        \raisebox{-.5\height}{
    \begin{tikzpicture}
    \draw[dashed] (-0.5,0)--(6,0);
    \draw[dotted] (6,2.5)--(6,0);
    \node at (6,0) {$\times$};
    \draw (0,0.1)--(2,0.1);
    \node at (1,0.4) {\tiny{$2k{-}2l{-}1$}};
    \draw[blue,thick] (2,0.1)--(4,0.1);
    \node[blue] at (2.5,0.5) {\tiny{$k{-}l{-}1$}};
    \draw[red,thick] (2,0.1) -- (2.1,0.25) --(5.8,0.25)--(6,0.2)--(5.8,0.15)--(4,0.15);
    \node[red] at (3.5,0.5) {\tiny{$k{-}l$}};
    \draw[myGreen,thick] (6,0.1)--(5,1);
    \node at (5,1.35) {\tiny{$[2l+4,-1]$}};
    \draw (5.9,0)--(5.9,1);
    \node at (5.75,0.7) {\tiny{$1$}};
    \monocut{-0.5,-0.05}{6,-0.05}
    \monocut{0,-0.1}{6,-0.1}
    \monocut{2,-0.15}{6,-0.15}
    \monocut{4,-0.2}{6,-0.2}
    \SevenB{0,0}
    \SevenB{2,0}
    \SevenB{4,0}
    \SevenB{6,1}
    \SevenB{6,2}
    \SevenB{5,1}
    \node at (1,-0.4) {\tiny{$-$}};
    \node at (3,-0.4) {\tiny{$\widetilde{-}$}};
    \end{tikzpicture}
    }
    \quad \rightarrow \quad
    \raisebox{-.5\height}{
    \begin{tikzpicture}
	\node (g7) {$\cdots$};
	\node (g8) [gauge,right of =g7,label=below:{\tiny{$D_{2k{+}1}$}}] {};
	\node[blue,thick] (g9) [gauge, above right of =g8,label=above:{\tiny{$C_{k}$}}] {};
	\node[red,thick] (g10) [gauge,below right of =g8,label=below:{\tiny{$C_{k{+}1}$}}] {};
	\node (g11) [gauge,below right of =g10,label=below:{\tiny{$U_1$}}] {};
	\node[myGreen,thick] (g12) [gauge,above right of =g10,label=above:{\tiny{$U_1$}}] {};
	\draw (g7)--(g8) (g8)--(g9) (g8)--(g10) (g10)--(g11) (g10)--(g12) (g11)--(g12);
	\draw[dashed] (g12) to [out=-45,in=45,looseness=1] (g11);
    \node at (2.2,-0.7) {\tiny{$l{+}1$}};
    \node at (3.2,-0.7) {\tiny{$l{+}1$}};
	\end{tikzpicture}
    }
\end{align}
where all mirror images have been ignored. The red and blue subweb lead to the familiar bifurcation into $C$-type magnetic gauge nodes. The green subweb, being away from the O5, leads to a $\urm(1)$ nodes, which has trivial intersection number with the blue subweb; while the intersection number with the red subweb is one.  Similarly, the black vertical $(0,1)$ 5-brane, also giving rise to a $\urm(1)\cong \sorm(2)$, has trivial intersection with the blue subweb; while the intersection number with the red subweb is one. Moreover, the generalised intersection number between the green subweb and the $(0,1)$ 5-brane is computed by carefully taking action of the two monodromy cuts into account. These affect the green $(2l+4,-1)$ 5-brane, which becomes a $M_{[1,0]}^2(2l+4,-1) = (2l+2,-1)$ 5-brane close to the intersection point of the O-planes. Hence, the intersection number is  $| \left( \begin{smallmatrix} 2l+2& -1 \\ 0 & 1 \end{smallmatrix} \right) | = 2l+2$, implying $(l+1)$ hypermultiplets in the $(1,1)$ representation of $\urm(1)\times \urm(1)$ as well as $(l+1)$ hypermultiplets in the $(1,-1)$ representation.
In conclusion, the magnetic quiver is proposed to be
    \begin{align}
        \raisebox{-.5\height}
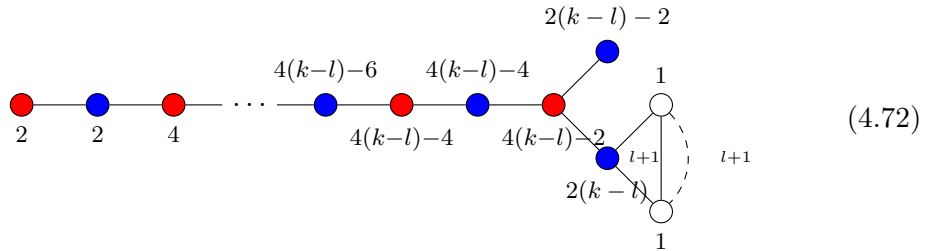
{
    \begin{tikzpicture}
	\node (g1) [gaugeSO,label=below:{\footnotesize{$2$}}] {};
	\node (g2) [gaugeSp,right of =g1,label=below:{\footnotesize{$2$}}] {};
	\node (g3) [gaugeSO,right of =g2,label=below:{\footnotesize{$4$}}] {};
	\node (g4) [right of =g3] {$\ldots$};
	\node (g5) [gaugeSp,right of =g4,label=above:{\footnotesize{$4(k{-}l){-}6$}}] {};
	\node (g6) [gaugeSO,right of =g5,label=below:{\footnotesize{$4(k{-}l){-}4$}}] {};
	\node (g7) [gaugeSp,right of =g6,label=above:{\footnotesize{$4(k{-}l){-}4$}}] {};
	\node (g8) [gaugeSO,right of =g7,label=below:{\footnotesize{$4(k{-}l){-}2$}}] {};
	\node (g9) [gaugeSp, above right of =g8,label=above:{\footnotesize{$2(k-l)-2$}}] {};
	\node (g10) [gaugeSp,below right of =g8,label=below:{\footnotesize{$2(k-l)$}}] {};
	\node (g11) [gauge,above right of =g10,label=above:{\footnotesize{$1$}}] {};
	\node (g12) [gauge,below right of =g10,label=below:{\footnotesize{$1$}}] {};
	\draw  (g1)--(g2) (g2)--(g3) (g3)--(g4) (g4)--(g5) (g5)--(g6) (g6)--(g7) (g7)--(g8) (g8)--(g9) (g8)--(g10) (g10)--(g11) (g10)--(g12) (g11)--(g12);
    \draw[dashed] (g11) to [out=-45,in=45,looseness=1] (g12);
    \node at (8.2,-0.7) {\tiny{$l{+}1$}};
    \node at (9.4,-0.7) {\tiny{$l{+}1$}};
	\end{tikzpicture}
    }
    \label{eq:5d_sp_magQuiv_infinite_E3}
\end{align}
which reproduces the $E_{3-2l} \times E_{3-2l}$ family of Table \ref{EnProdtable2}.
\end{itemize}
As in the 6d setup of Section \ref{sec:branes_6d}, one cannot replace the O7${}^-$ plane with a hypothetical $\widetilde{\text{O7}}^-$, because such orientifolds are inconsistent \cite{Hyakutake:2000mr}.

\section{Conclusions and outlook}
\label{sec:conclusion}
3d $\Ncal=4$ quiver gauge theories with alternating orthogonal and symplectic gauge nodes are less studied and underrepresented. This is mostly due to intrinsic subtleties like absence of FI parameters such that various partition functions can only be evaluated in an unrefined manner. Nevertheless, the aim of this paper has been to demonstrate that orthosymplectic quiver host interesting phenomena that seem unexpected at first glance, see Table \ref{tab:results}.

A central theme of this paper are orthosymplectic quivers that are in fact magnetic quivers for product theories. These theories have been systematically derived based on two guiding principles: (i) from the balancing condition of quivers and (ii) from brane constructions with ON${}^0$ planes. 

Firstly, all orthosymplectic quivers that have balanced $D$-type bifurcation have been classified solely based on the balance condition. For framed quivers, this has been achieved in Section \ref{forkingit}. The theories of Group \ref{rule:framed_D} are generically understood as mirrors of products of $\sprm$ SQCD theories. For unframed quivers, Section \ref{forkingunframed} provides a classification of balanced $D$-type Dynkin quivers. The classification comprises several families: a large class of theories corresponds to products of the exceptional $E_n$ theories, see Tables \ref{EnProdtable} and \ref{EnProdtable2}.  Other product types include families of the type $E_6 \times \sorm(10)$, $E_8 \times \sorm(16) $, and $F_4\times F_4$.  Moreover, a family that is the product of two $B$-type nilpotent orbit closures has been uncovered as well.
Having classified balanced $D$-type orthosymplectic quivers, folding has been used in Section \ref{foldafterfork} to derive balanced $B$-type orthosymplectic quivers that are, in fact, product theories as well. As a first consistency check, the Coulomb branch Hilbert series have been computed perturbatively and compared against the proposed product theories. Agreement is found in all cases.

Secondly, the balanced Dynkin-type orthosymplectic quivers have been derived from brane configurations in Section \ref{branes}. For this, the reasoning is straightforward: On the one hand, $D$-type Dynkin quivers originate from ON planes. On the other hand, the relation between balance and product global symmetry is immanent in the brane system: by construction, the Higgs branch side has a product non-abelian flavour symmetry; thus, all gauge nodes in the Dynkin-type subdiagram of the magnetic quiver need to be balanced. In addition, the condition that the long tail of the orthosymplectic quiver has to end on an $\sorm(2)$ node can be understood from the branes as well. It translates to the condition that the electric side is a degenerate $D$-type quiver, i.e.\ a simple product, as discussed at the end of Section \ref{sec:branes_3d}. With this ingredients at hand, the embeddings into brane configurations in the context of magnetic quivers allowed to derive all balanced $D$-type orthosymplectic quivers that are products.  As a byproduct, the magnetic quiver approach has been extended to included ON planes. The brane configurations serve as a second consistency check for the results.

\paragraph{Class $\mathcal{S}$ product theories.}
Consider class $\mathcal{S}$ theories with untwisted $D$-type punctures that correspond to \emph{good} star-shaped orthosymplectic quivers. If one of the punctures is a maximal puncture, the results of this paper classify all cases where the theory is a product. For class $\mathcal{S}$ theories constructed from a three-punctured spheres, the results are as follows:
\begin{compactitem}
\item The $E_8\times E_8$ family of Table \ref{EnProdtable2} is primarily defined as set of class $\mathcal{S}$ theories for a three-puncture sphere with $D_{2k+5}$ punctures. These theories factorise into two copies of the $E_8$ family that is understood as three-punctured sphere with $D_{k+3}$ punctures: 
\begin{equation}
   \mathcal{H}^{4d} \left( \raisebox{-.5\height}{\scalebox{.6}{\begin{tikzpicture}
        \filldraw[color=blue!60, fill=blue!5, very thick](0,0) circle (2);
        \filldraw[fill=black!100] (1.4,0) circle (.1);
        \filldraw[fill=black!100] (-.7,1) circle (.1);
        \filldraw[fill=black!100] (-.7,-1) circle (.1);
        \node at (1.4,.4) {\large $(1^{4k+10})$};
        \node at (-.7,1.4) {\large$(2k+5,2k+5)$};
        \node at (-.7,-.6) {\large$(2k+3,2k+3,3,1)$};
    \end{tikzpicture}}} \right) =  \mathcal{H}^{4d} \left( \raisebox{-.5\height}{\scalebox{.6}{\begin{tikzpicture}
        \filldraw[color=blue!60, fill=orange!5, very thick](0,0) circle (2);
        \filldraw[fill=black!100] (1.4,0) circle (.1);
        \filldraw[fill=black!100] (-.7,1) circle (.1);
        \filldraw[fill=black!100] (-.7,-1) circle (.1);
        \node at (1.4,.4) {\large $(1^{2k+6})$};
        \node at (-.7,1.4) {\large$(1^{2k+6})$};
        \node at (-.7,-.6) {\large$(2k+3,3)$};
    \end{tikzpicture}}} \right)^2 \,,
    \label{eq:class-S_E8xE8}
\end{equation}
where the partitions are the Nahm partitions of the nilpotent orbits, following the same convention as \cite{Distler:2017xba,Distler:2018gbc}. In particular,  partition $(1^{2n})$ of $D_n$ corresponds to the maximal nilpotent orbit.
\item The $E_7\times E_7$ family, see Table \ref{EnProdtable2}, is firstly defined by a three-punctured sphere with $D_{2k+3}$ punctures. The latter factorises into two copies of the so-called $E_7$ family that are defined by $D_{k+2}$ punctures:
\begin{equation}
   \mathcal{H}^{4d} \left( \raisebox{-.5\height}{\scalebox{.6}{\begin{tikzpicture}
        \filldraw[color=blue!60, fill=blue!5, very thick](0,0) circle (2);
        \filldraw[fill=black!100] (1.4,0) circle (.1);
        \filldraw[fill=black!100] (-.7,1) circle (.1);
        \filldraw[fill=black!100] (-.7,-1) circle (.1);
        \node at (1.4,.4) {\large $(1^{4k+6})$};
        \node at (-.7,1.4) {\large$(2k+1,2k+1,1^4)$};
        \node at (-.7,-.6) {\large$(2k+3,2k+3)$};
    \end{tikzpicture}}} \right) =  \mathcal{H}^{4d} \left( \raisebox{-.5\height}{\scalebox{.6}{\begin{tikzpicture}
        \filldraw[color=blue!60, fill=orange!5, very thick](0,0) circle (2);
        \filldraw[fill=black!100] (1.4,0) circle (.1);
        \filldraw[fill=black!100] (-.7,1) circle (.1);
        \filldraw[fill=black!100] (-.7,-1) circle (.1);
        \node at (1.4,.4) {\large $(1^{2k+4})$};
        \node at (-.7,1.4) {\large$(1^{2k+4})$};
        \node at (-.7,-.6) {\large$(2k+1,1^3)$};
    \end{tikzpicture}}} \right)^2 \, .
    \label{eq:class-S_E7xE7}
\end{equation}
\item The $E_6\times E_6$ family of Table \ref{EnProdtable2} can be understood as the class $\mathcal{S}$ theory of a three-punctured sphere with $D_{2k+2}$ punctures. Again, it factorises into two copies of the $E_6$ family, defined by $A_{2k+1}$ punctures: 
\begin{equation}
   \mathcal{H}^{4d} \left( \raisebox{-.5\height}{\scalebox{.6}{\begin{tikzpicture}
        \filldraw[color=blue!60, fill=blue!5, very thick](0,0) circle (2);
        \filldraw[fill=black!100] (1.4,0) circle (.1);
        \filldraw[fill=black!100] (-.7,1) circle (.1);
        \filldraw[fill=black!100] (-.7,-1) circle (.1);
        \node at (1.4,.4) {\large $(1^{4k+4})$};
        \node at (-.7,1.4) {\large$(2k+1,2k+1,1^2)$};
        \node at (-.7,-.6) {\large$(2k+1,2k+1,1^2)$};
    \end{tikzpicture}}} \right) =  \mathcal{H}^{4d} \left( \raisebox{-.5\height}{\scalebox{.6}{\begin{tikzpicture}
        \filldraw[color=blue!60, fill=orange!5, very thick](0,0) circle (2);
        \filldraw[fill=black!100] (1.4,0) circle (.1);
        \filldraw[fill=black!100] (-.7,1) circle (.1);
        \filldraw[fill=black!100] (-.7,-1) circle (.1);
        \node at (1.4,.4) {\large $(2k+1,1)$};
        \node at (-.7,1.4) {\textcolor{red}{\large$(1^{2k+3})$}};
        \node at (-.7,-.6) {\textcolor{red}{\large$(1^{2k+3})$}};
    \end{tikzpicture}}} \right)^2 \,,
    \label{eq:class-S_E6xE6}
\end{equation}
where the red partition denotes the B partitions of the twisted $A_{2n+1}$ punctures.
\end{compactitem}
 The three families \eqref{eq:class-S_E8xE8}--\eqref{eq:class-S_E6xE6} have been detailed in Section \ref{sec:En_EN_5d} and Sections  \ref{sec:6d_ON_and_Op}, \ref{sec:5d_ON_and_Op}. The rarity of such decomposable three-punctured spheres has been noted in \cite{Distler:2017xba}, where examples for $D_N$ theories with $N=4$ have been presented. In response to one of the observations in \cite{Distler:2017xba}, the results of this paper show that only when $k=1$ does the Coulomb branch carry exceptional global symmetry; whereas for $k>1$, the global symmetry is composed of classical groups.

For a four-punctured sphere with one maximal puncture, the only family that factorises is the $l=1$ limiting case of the $E_8 \times \mathrm{SO}(16)$ family in Section \ref{finalfamily} given by $D_{2k+1}$ punctures. The moduli space is the product of the Higgs branch of a class $\mathcal{S}$ theory with $D_{k+1}$ punctures and the nilpotent orbit closure $\overline{\mathcal{O}}^{\mathfrak{so}(4k+4l)}_{(2^{2k},1^{4l})}$. 
\begin{equation}
   \mathcal{H}^{4d} \left( \raisebox{-.5\height}{\scalebox{.6}{\begin{tikzpicture}
        \filldraw[color=blue!60, fill=blue!5, very thick](0,0) circle (2);
        \filldraw[fill=black!100] (1.4,-1) circle (.1);
        \filldraw[fill=black!100] (-.7,1) circle (.1);
        \filldraw[fill=black!100] (-.7,-0.5) circle (.1);
          \filldraw[fill=black!100] (1.4,1) circle (.1);
        \node at (1.4,-.6) {\large$(2k+1,2k+1)$};
        \node at (-1.1,1.4) {\large $(1^{4k+2})$};
        \node at (-1.1,-.2) {\large$(2k+1,2k+1)$};
            \node at (1.4,1.4) {\large $(4k-1,3)$};
    \end{tikzpicture}}} \right) =  \mathcal{H}^{4d} \left( \raisebox{-.5\height}{\scalebox{.6}{\begin{tikzpicture}
        \filldraw[color=blue!60, fill=orange!5, very thick](0,0) circle (2);
        \filldraw[fill=black!100] (1.4,0) circle (.1);
        \filldraw[fill=black!100] (-.7,1) circle (.1);
        \filldraw[fill=black!100] (-.7,-1) circle (.1);
        \node at (1.4,.4) {\large $(1^{2k+2})$};
        \node at (-.7,1.4) {\large $(1^{2k+2})$};
        \node at (-.7,-.6) {\large $(2k-1,3)$};
    \end{tikzpicture}}} \right)\times \overline{\mathcal{O}}^{\mathfrak{so}(4k+4)}_{(2^{2k},1^{4})}. 
\end{equation}
For higher punctures, it can be shown that the fork cannot be balanced and, therefore, the product structure does not appear. For class $\mathcal{S}$ theories without maximal punctures, one yet has to find an example where the theory factorises. For  \emph{bad} star-shaped quivers, no further analysis has been attempted in this work. Nonetheless, there are conjectures where product theories can arise from bad theories; for examples, a different realisation of the $E_8 \times E_8$ theory as a bad quiver has been given in \cite{Akhond:2021knl}.

\paragraph{Outlook.}
Along with \cite{Bourget:2020xdz}, orthosymplectic quivers of balanced $ABCD$-types have been studied in detail. The next step is to extend this to exceptional $EFG$-type orthosymplectic quivers as well. 

Furthermore, due to the balancing conditions of orthosymplectic quivers, it is possible to construct balanced quivers beyond Dynkin types. So far, all known examples are either bad quivers or free theories. 
For example, take the $k=0$ members of the $E_n$ families where $4\leq n \leq 8$ as shown in \cite[Fig.\ 18]{Bourget:2020xdz}. These are free theories of $2^{n-4}$ free hypermultiplets; thus, the Coulomb branch is $\mathbb{H}^{2^{n-4}}$. Similarly, for the $E_n\times E_n$ family in Table \ref{EnProdtable2}, the theories are $2^{n-3}$ free hypermultiplets. The results are summarised in Table \ref{EnProdtable4}. Notice that all the nodes are balanced, and for $n=6,7,8$ the balanced set of nodes is not a Dynkin diagram of any finite algebra. For $n=8$ case, the Dynkin diagram of $E_{12}$ seems to arise. The discussion of balanced Dynkin diagrams beyond finite type is left for future work. 
\begin{table}[t]
    \centering
    \scalebox{0.9}{
    \begin{tabular}{ccc} \toprule
   Family & Orthosymplectic quiver & Coulomb branch \\ \midrule 
      $E_8$ & 
         \scalebox{0.8}{     \raisebox{-.5\height}{\begin{tikzpicture}
	\begin{pgfonlayer}{nodelayer}
		\node [style=redgauge] (0) at (0, 0) {};
		\node [style=bluegauge] (1) at (1, 0) {};
		\node [style=redgauge] (2) at (2, 0) {};
		\node [style=bluegauge] (3) at (3, 0) {};
		\node [style=redgauge] (4) at (4, 0) {};
		\node [style=bluegauge] (5) at (5, 0) {};
		\node [style=redgauge] (6) at (6, 0) {};
		\node [style=bluegauge] (7) at (7, 0) {};
		\node [style=redgauge] (8) at (8, 0) {};
		\node [style=bluegauge] (9) at (9, 0) {};
		\node [style=redgauge] (10) at (10, 0) {};
		\node [style=none] (15) at (2, -0.5) {10};
		\node [style=none] (16) at (3, -0.5) {8};
		\node [style=none] (17) at (4, -0.5) {8};
		\node [style=none] (18) at (5, -0.5) {6};
		\node [style=none] (19) at (6, -0.5) {6};
		\node [style=none] (20) at (7, -0.5) {4};
		\node [style=none] (21) at (8, -0.5) {4};
		\node [style=none] (22) at (9, -0.5) {2};
		\node [style=none] (23) at (10, -0.5) {2};
		\node [style=bluegauge] (30) at (2, 1.25) {};
		\node [style=none] (31) at (2, 1.75) {4};
		\node [style=none] (32) at (1, -0.5) {6};
		\node [style=none] (33) at (0, -0.5) {4};
	\end{pgfonlayer}
	\begin{pgfonlayer}{edgelayer}
		\draw (30) to (2);
		\draw (0) to (1);
		\draw (1) to (2);
		\draw (7) to (2);
		\draw (7) to (10);
	\end{pgfonlayer}
\end{tikzpicture}}
} &\Large{ $\mathbb{H}^{32}$ }  \\
$E_7$ & 
\scalebox{0.8}{    \raisebox{-.5\height}{\begin{tikzpicture}
	\begin{pgfonlayer}{nodelayer}
		\node [style=redgauge] (0) at (4, 0) {};
		\node [style=bluegauge] (1) at (5, 0) {};
		\node [style=redgauge] (6) at (6, 0) {};
		\node [style=bluegauge] (7) at (7, 0) {};
		\node [style=redgauge] (8) at (8, 0) {};
		\node [style=bluegauge] (9) at (9, 0) {};
		\node [style=redgauge] (10) at (10, 0) {};
		\node [style=none] (15) at (6, -0.5) {6};
		\node [style=none] (20) at (7, -0.5) {4};
		\node [style=none] (21) at (8, -0.5) {4};
		\node [style=none] (22) at (9, -0.5) {2};
		\node [style=none] (23) at (10, -0.5) {2};
		\node [style=bluegauge] (30) at (6, 1.5) {};
		\node [style=none] (31) at (6, 2) {2};
		\node [style=none] (32) at (5, -0.5) {4};
		\node [style=none] (33) at (4, -0.5) {4};
		\node [style=bluegauge] (34) at (3, 0) {};
		\node [style=redgauge] (35) at (2, 0) {};
		\node [style=none] (36) at (3, -0.5) {2};
		\node [style=none] (37) at (2, -0.5) {2};
	\end{pgfonlayer}
	\begin{pgfonlayer}{edgelayer}
		\draw (0) to (1);
		\draw (30) to (6);
		\draw (6) to (1);
		\draw (7) to (6);
		\draw (0) to (35);
		\draw (7) to (10);
	\end{pgfonlayer}
\end{tikzpicture}}
}
&\Large{ $\mathbb{H}^{16}$}     \\
$E_6$ & 
\scalebox{0.8}{    \raisebox{-.5\height}{\begin{tikzpicture}
	\begin{pgfonlayer}{nodelayer}
		\node [style=redgauge] (0) at (6, 0) {};
		\node [style=bluegauge] (1) at (7, 0) {};
		\node [style=redgauge] (8) at (8, 0) {};
		\node [style=bluegauge] (9) at (9, 0) {};
		\node [style=redgauge] (10) at (10, 0) {};
		\node [style=none] (19) at (8, -0.5) {4};
		\node [style=none] (20) at (9, -0.5) {2};
		\node [style=none] (21) at (10, -0.5) {2};
		\node [style=bluegauge] (30) at (8, 1) {};
		\node [style=none] (31) at (8.5, 1) {2};
		\node [style=none] (32) at (7, -0.5) {2};
		\node [style=none] (33) at (6, -0.5) {1};
		\node [style=redgauge] (34) at (8, 2) {};
		\node [style=none] (35) at (8.5, 2) {1};
		\node [style=gauge3] (36) at (8, 2) {};
		\node [style=gauge3] (37) at (6, 0) {};
	\end{pgfonlayer}
	\begin{pgfonlayer}{edgelayer}
		\draw (0) to (1);
		\draw (8) to (1);
		\draw (8) to (30);
		\draw (30) to (34);
		\draw (8) to (10);
	\end{pgfonlayer}
\end{tikzpicture}}} &\Large{  $\mathbb{H}^8$}  
\\
$E_5$ & 
\scalebox{0.8}{    \raisebox{-.5\height}{\begin{tikzpicture}
	\begin{pgfonlayer}{nodelayer}
		\node [style=redgauge] (0) at (7.5, -0.5) {};
		\node [style=bluegauge] (1) at (8.5, -1.25) {};
		\node [style=redgauge] (10) at (9.75, -1.25) {};
		\node [style=none] (19) at (9.75, -1.75) {2};
		\node [style=none] (32) at (8.5, -1.75) {2};
		\node [style=none] (33) at (7.5, 0) {1};
		\node [style=redgauge] (34) at (7.5, -2) {};
		\node [style=none] (35) at (7.5, -2.5) {1};
		\node [style=gauge3] (36) at (7.5, -0.5) {};
		\node [style=gauge3] (37) at (7.5, -2) {};
	\end{pgfonlayer}
	\begin{pgfonlayer}{edgelayer}
		\draw (0) to (1);
		\draw (10) to (1);
		\draw (1) to (34);
	\end{pgfonlayer}
\end{tikzpicture}}
} & \Large{  $\mathbb{H}^4$} \\
$E_4$ & 
 \scalebox{0.8}{
 \raisebox{-.5\height}{
\begin{tikzpicture}
	\begin{pgfonlayer}{nodelayer}
		\node [style=gauge3] (33) at (9.5, 1) {};
		\node [style=gauge3] (34) at (9.5, -0.55) {};
		\node [style=none] (35) at (9.5, 1.25) {1};
		\node [style=none] (36) at (9.5, -1.05) {1};
		\node [style=none] (37) at (9.075, 0.1) {};
	\end{pgfonlayer}
	\begin{pgfonlayer}{edgelayer}
		\draw (33) to (34);
		\draw [style=new edge style 1, bend right, looseness=0.75] (33) to (37.center);
		\draw [style=new edge style 1, bend right=45, looseness=0.75] (37.center) to (34);
	\end{pgfonlayer}
\end{tikzpicture}
}}  &\Large{ $\mathbb{H}^2$} \\  
\bottomrule
    \end{tabular}}
    \caption{The $k=0$ members of the $E_n\times E_n$ family of Table \ref{EnProdtable2} for $4 \leq n \leq 8$. These quivers are magnetic quivers for free hypermultiplets such that Coulomb branches are flat spaces. The Coulomb branch Hilbert series are given by $\mathrm{PE}\left[ 2(n-3)\ t \right]$.}
    \label{EnProdtable4}
\end{table}

\paragraph{Acknowledgements.}
We are indebted to Amihay Hanany for countless insightful discussions and collaborations on related projects.
We would like to acknowledge helpful discussions with Mohammad Akhond, Jacques Distler, Sung-Soo Kim, Satoshi Nawata, Futoshi Yagi, and Ruidong Zhu.
M.S. was further supported by the National Natural Science Foundation of China (grant no.\ 11950410497), and the China Postdoctoral Science Foundation (grant no.\ 2019M650616).
M.S. is grateful for the warm hospitality of Fudan University, Department of Physics during the intermediate stage of this work. 
Z.Z. is thankful for the kind hospitality of YMSC and Tsinghua University where part of this work is completed.
\appendix
\section{Brane configurations}
\label{app:branes}
In this appendix, some facts on branes with orientifolds are recalled, cf.\ \cite{Evans:1997hk,Hanany:1999sj,Hanany:2000fq}. 
\begin{compactitem}
\item An Op${}^\pm$ becomes an Op${}^\mp$ when passing through a half NS5; similarly, an $\widetilde{\mathrm{Op}}^{\pm}$ turns into an $\widetilde{\mathrm{Op}}^{\mp}$ plane.
\item An Op${}^{\pm}$ plane becomes a $\widetilde{\mathrm{Op}}^{\pm}$ plane when passing a half D(p+2), and vice versa.
\end{compactitem}

For O3 planes, S-duality acts as follows
\begin{align}
\mathrm{O3}^- \xleftrightarrow{\; S \;} \mathrm{O3}^- 
\, , \qquad 
\mathrm{O3}^+ \xleftrightarrow{\; S \;} \widetilde{\mathrm{O3}}^-
\, , \qquad
\widetilde{\mathrm{O3}}^+ \xleftrightarrow{\; S \;} \widetilde{\mathrm{O3}}^+
\end{align}
i.e.\ transforms the O3 planes among each other in the pattern of GNO duality.
For O5 planes, S-duality is argued to introduce a different type of orientifold, called ON planes. These are defined by the S-duality properties
\begin{align}
\mathrm{O5}^\pm \xleftrightarrow{\; S \;} \mathrm{ON}^\pm    
\, ,\qquad 
\widetilde{\mathrm{O5}}^\pm \xleftrightarrow{\; S \;} \widetilde{\mathrm{ON}}^\pm  
\end{align}
The various orientifolds may couple to the RR or NS two-forms and carry charges (in units of physical branes) as indicated in Table \ref{tab:O-plane_charges}, which follow \cite{Hanany:1999sj,Feng:2000eq}.

\begin{table}[t]
\centering
\begin{tabular}{c|cc||c|cc}
\toprule 
Op &     RR  &  NS   &  ON &   RR  &  NS  \\ \midrule
O$p\pm$  & $\pm2^{p-5}$   & 0  & ON${}^\pm$  & $0$   & $\pm 1$\\
$\widetilde{\text{O$p$}}^-$  & $\frac{1}{2}-2^{p-5}$   & 0 & $\widetilde{\mathrm{ON}}^-$  & $0$   & $\frac{1}{2}$\\
$\widetilde{\text{O$p$}}^+$  & $2^{p-5}$   & 0 & $\widetilde{\mathrm{ON}}^+$  & $0$   & -\\ \bottomrule
\end{tabular}
\caption{The RR and NS charges of the various O-planes.}
\label{tab:O-plane_charges}
\end{table}

\begin{figure}[t]
    \centering
    \begin{subfigure}{0.3\textwidth}
    \centering
    \begin{tikzpicture}
        \draw (0,1)--(0,-1) (1,1)--(1,-1) (2,1)--(2,-1);
        \draw[dashed] (3,1)--(3,-1);
        \draw [line join=round,decorate, decoration={zigzag, segment length=4,amplitude=.9,post=lineto,post length=2pt}] (0,0)--(1,0) (1,-0.5)--(2,-0.5);
         \draw [line join=round,decorate, decoration={zigzag, segment length=4,amplitude=.9,post=lineto,post length=2pt}] (1,0.5) .. controls (3.35,0.25) .. (2,0);
         \node at (3.35,0.75) {\tiny{O3${}^-$}};
         \node at (0.25,0.75) {\tiny{D3}};
         \node at (1.5,-0.25) {\tiny{F1}};
    \end{tikzpicture}
    \caption{}
    \label{subfig:F1_O3-}
    \end{subfigure}
    \hfill
    \begin{subfigure}{0.3\textwidth}
    \centering
    \begin{tikzpicture}
        \draw (0,1)--(0,-1) (1,1)--(1,-1) (2,1)--(2,-1);
        \draw[dashed] (3,1)--(3,-1);
        \draw [line join=round,decorate, decoration={zigzag, segment length=4,amplitude=.9,post=lineto,post length=2pt}] (0,0)--(1,0) (1,-0.5)--(2,-0.5) (2,0.5)--(3,0.5);
         \node at (3.35,0.75) {\tiny{${\widetilde{\mathrm{O3}}}^-$}};
         \node at (0.25,0.75) {\tiny{D3}};
         \node at (1.5,-0.25) {\tiny{F1}};
    \end{tikzpicture}
    \caption{}
    \label{subfig:F1_O3t-}
    \end{subfigure}
    \hfill
    \begin{subfigure}{0.3\textwidth}
    \centering
    \begin{tikzpicture}
        \draw (0,1)--(0,-1) (1,1)--(1,-1) (2,1)--(2,-1);
        \draw[dashed] (3,1)--(3,-1);
        \draw [line join=round,decorate, decoration={zigzag, segment length=4,amplitude=.9,post=lineto,post length=2pt}] (0,0)--(1,0) (1,-0.5)--(2,-0.5);
         \draw [line join=round,decorate, decoration={zigzag, segment length=4,amplitude=.9,post=lineto,post length=2pt}] (2,0.5) .. controls (3.35,0.25) .. (2,0);
         \node at (3.35,0.75) {\tiny{O3${}^+$}};
         \node at (0.25,0.75) {\tiny{D3}};
         \node at (1.5,-0.25) {\tiny{F1}};
    \end{tikzpicture}
    \caption{}
    \label{subfig:F1_O3+}
    \end{subfigure}
    \\
    \begin{subfigure}{0.3\textwidth}
    \centering
    \begin{tikzpicture}
        \draw (0,1)--(0,-1) (1,1)--(1,-1) (2,1)--(2,-1);
        \draw[dashed] (3,1)--(3,-1);
        \draw (0,0)--(1,0) (1,-0.5)--(2,-0.5);
        \draw (1,0.5) .. controls (3.35,0.25) .. (2,0);
         \node at (3.35,0.75) {\tiny{O3${}^-$}};
         \node at (0.25,0.75) {\tiny{D3}};
         \node at (1.5,-0.25) {\tiny{D1}};
    \end{tikzpicture}
    \caption{}
    \label{subfig:D1_O3-}
    \end{subfigure}
    \hfill
    \begin{subfigure}{0.3\textwidth}
    \centering
    \begin{tikzpicture}
        \draw (0,1)--(0,-1) (1,1)--(1,-1) (2,1)--(2,-1);
        \draw[dashed] (3,1)--(3,-1);
        \draw (0,0)--(1,0) (1,-0.5)--(2,-0.5) (2,0.5)--(3,0.5);
         \node at (3.35,0.75) {\tiny{O3${}^+$}};
         \node at (0.25,0.75) {\tiny{D3}};
         \node at (1.5,-0.25) {\tiny{D1}};
    \end{tikzpicture}
    \caption{}
    \label{subfig:D1_O3t-}
    \end{subfigure}
    \hfill
    \begin{subfigure}{0.3\textwidth}
    \centering
    \begin{tikzpicture}
        \draw (0,1)--(0,-1) (1,1)--(1,-1) (2,1)--(2,-1);
        \draw[dashed] (3,1)--(3,-1);
        \draw (0,0)--(1,0) (1,-0.5)--(2,-0.5);
         \draw (2,0.5) .. controls (3.35,0.25) .. (2,0);
         \node at (3.35,0.75) {\tiny{${\widetilde{\mathrm{O3}}}^-$}};
         \node at (0.25,0.75) {\tiny{D3}};
         \node at (1.5,-0.25) {\tiny{D1}};
    \end{tikzpicture}
    \caption{}
    \label{subfig:D1_O3+}
    \end{subfigure}
    \caption{F1 and D1 pattern on a stack of D3s on top of O3 planes.}
    \label{fig:D1_and_F1_on_O3}
\end{figure}
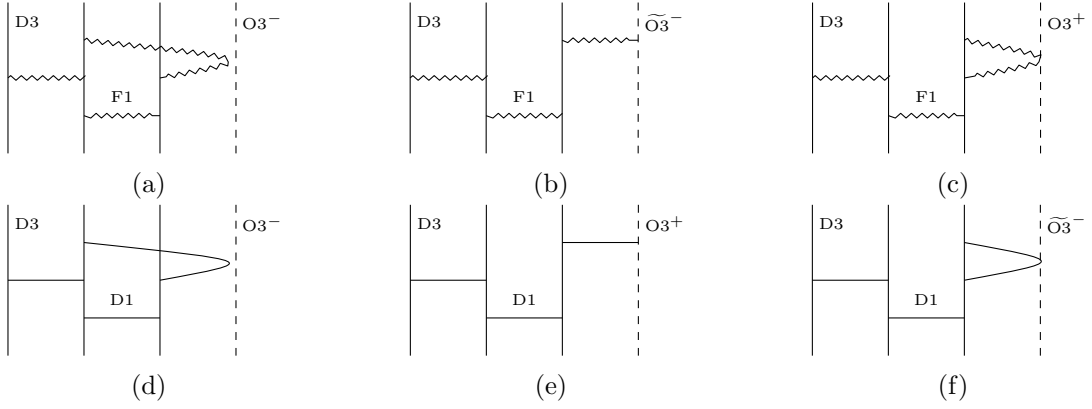

\paragraph{Branes end on O-planes.}
Recall from \cite{Hanany:2001iy}, the pattern how F1 and D1 behave near O3 planes, see Figure \ref{fig:D1_and_F1_on_O3}. The F1 cannot end on O3${}^-$ and cannot end on the mirror D3. The F1 can end on $\widetilde{\mathrm{O3}}^-$, due to the stuck half D3. The F1 cannot end on O3$^+$, but can end on the mirror D3. As a result, the O-planes produce $D$-type, $B$-type, and $C$-type gauge algebras, respectively. Upon S-duality, one obtains the pattern of D1 strings.

Assuming that T-duality can be applied to the systems of D3-D1-O3 in Figure \ref{subfig:D1_O3-}--\ref{subfig:D1_O3+}, one obtains the identical suspension patterns of D$p$ branes on a stack of D$(p+2)$ branes near an O$(p+2)$ planes. 
\paragraph{Monodromy of 7-branes.} For a $[p,q]$ 7-brane, the associated monodromy matrix is
\begin{align}
M_{[p,q]} = \begin{pmatrix} 1-p q & p^2 \\ - q^2 & 1+ pq
\end{pmatrix}
\end{align}
and the action is clockwise. The treatment of $\widetilde{\mathrm{O5}}^\pm$ planes and (half-) monodromy cuts follows the arguments \cite{Zafrir:2015ftn}, similarly to \cite[App.\ A]{Bourget:2020gzi}.
\paragraph{Rules for magnetic quivers with O-planes.}
For the 6d brane systems, one needs to take care of the three types of O-planes. Magnetic quivers in presence of O8${}^-$ have been derived in \cite{Cabrera:2019izd} and shown to result in $D$-type quivers. Likewise, magnetic quivers on the presence of O6 planes are detailed in \cite{Cabrera:2019dob}, which yield orthosymplectic quivers.

As a first step, the inclusion of ON is proposed to yield the following magnetic quiver rule
\begin{subequations}
\begin{align}
    \raisebox{-.5\height}{
    \begin{tikzpicture}
    \draw (-1,0)--(2,0);
    \ns{0,0}
    \onz{2,0}
    \node at (1,0.25) {\tiny{$2k$}};
    \node at (-0.5,0.25) {\tiny{$2k$}};
%
	\node (g1) [gauge] at (5,0) {};
    \node (f1) [gauge,above right of=g1, label=right:{\footnotesize{$\surm(k)$}}] {};
    \node (f2) [gauge,below right of=g1, label=right:{\footnotesize{$\surm(k)$}}] {};
    \node (b1) [left of =g1] {$\cdots$};
    \draw (g1)--(f1) (g1)--(f2) (g1)--(b1);
	\end{tikzpicture}
    }
    \\
\raisebox{-.5\height}{
    \begin{tikzpicture}
    \draw (-1,0.5)--(0,0.5) (-1,0.3)--(0,0.3) (-1,-0.5)--(0,-0.5);
    \node at (-0.5,-0.1) {$\vdots$};
    \draw (0,0.75)--(2,0.75) (0,-0.75)--(2,-0.75);
    \draw (0,1)--(0,-1) (-1,1)--(-1,-1);
    \node at (-1.5,0) {$\cdots$};
    \ns{0.5,0}
    \ns{1.5,0}
    \on{2,0}
    \node at (1,0.95) {\tiny{$2k$}};
    \node at (-0.5,0.75) {\tiny{$2k{-}1$}};
	\node (g1) [gauge,label=below:{\footnotesize{$\surm(2k{-}1)$}}] at (5,0) {};
	\node (g2) [gaugeSp,right of=g1, label=below right:{\footnotesize{$\usprm(2k)$}}] {};
    \node (f1) [gaugeSO,above right of=g2, label=right:{\footnotesize{$\sorm(2)$}}] {};
    \node (f2) [gaugeSO,above left of=g2, label=left:{\footnotesize{$\sorm(2)$}}] {};
    \node (b1) [left of =g1] {$\cdots$};
    \draw (g1)--(g2) (g2)--(f1) (g2)--(f2) (g1)--(b1);
    \end{tikzpicture}
    }
    \\
\raisebox{-.5\height}{
    \begin{tikzpicture}
    \draw (-1,0.5)--(0,0.5) (-1,0.3)--(0,0.3) (-1,-0.5)--(0,-0.5);
    \node at (-0.5,-0.1) {$\vdots$};
    \draw (0,0.75)--(2,0.75) (0,-0.75)--(2,-0.75);
    \draw (0,1)--(0,-1) (-1,1)--(-1,-1);
    \node at (-1.5,0) {$\cdots$};
    \ns{2,0}
    \node at (2.3,0.2) {\footnotesize{$\times 4$}};
    \node at (1,0.95) {\tiny{$2k$}};
    \node at (-0.5,0.75) {\tiny{$2k{-}1$}};
	\node (g1) [gauge,label=below:{\footnotesize{$\surm(2k{-}1)$}}] at (5,0) {};
	\node (g2) [gaugeSp,right of=g1, label=below right:{\footnotesize{$\usprm(2k)$}}] {};
    \node (f1) [gaugeSO,above of=g2, label=right:{\footnotesize{$\sorm(4)$}}] {};
    \node (b1) [left of =g1] {$\cdots$};
    \draw (f1) to [out=45,in=135,looseness=10] (f1);
    \draw (g1)--(g2) (g2)--(f1) (g1)--(b1);
    \end{tikzpicture}
    }
\end{align}
\end{subequations}
i.e.\ the ON${}^-$ planes acts via a projection magnetic degrees of freedom which reduce the magnetic vector multiplet to $\sprm(k)$. Similarly, the magnetic degrees of freedom from the NS5 branes away from the ON are encoded in two $\sorm(2)$ magnetic vector multiplets.
Once the 4 half NS5s are coincident with the ON, the magnetic degrees of freedom are argued to give rise to $\sorm(4)$ SYM theory, i.e.\ an $\sorm(4)$ magnetic vector multiplet plus an adjoint hypermultiplet.

Having O6, O8, and ON present, the magnetic quiver rules become a combination of the above.

\section{Global symmetry enhancement beyond balanced nodes}
\label{u1guy} 
As described in Section \ref{forkingit}, the following quiver has an additional $\sorm(2)$ subgroup in the global symmetry that cannot be read off from balancing conditions:
\begin{equation}
  \scalebox{0.9}{      \raisebox{-.5\height}{\begin{tikzpicture}
	\begin{pgfonlayer}{nodelayer}
		\node [style=miniU] (0) at (-0.5, 0) {};
		\node [style=miniBlue] (1) at (0.5, 0) {};
		\node [style=miniU] (2) at (1.5, 0) {};
		\node [style=none] (3) at (2.25, 0) {$\dots$};
		\node [style=miniU] (5) at (3, 0) {};
		\node [style=miniBlue] (6) at (4, 0.75) {};
		\node [style=miniBlue] (7) at (4, -0.75) {};
		\node [style=none] (8) at (-0.5, -0.75) {};
		\node [style=none] (9) at (3, -0.75) {};
		\node [style=none] (10) at (1.25, -1.25) {$n-2$};
		\node [style=none] (11) at (1.25, -2) {$G_{\mathrm{global}}= \sorm(n+1)\times \sorm(n+1)$};
		\node [style=none] (12) at (-0.5, -0.5) {2};
		\node [style=miniBlue] (13) at (9, 0) {};
		\node [style=miniU] (14) at (10, 0) {};
		\node [style=none] (15) at (10.75, 0) {$\dots$};
		\node [style=miniU] (16) at (11.5, 0) {};
		\node [style=miniBlue] (17) at (12.5, 0.75) {};
		\node [style=miniBlue] (18) at (12.5, -0.75) {};
		\node [style=none] (19) at (9, -0.75) {};
		\node [style=none] (20) at (11.5, -0.75) {};
		\node [style=none] (21) at (10, -1.3) {$n-3$};
		\node [style=none] (22) at (10, -2) {$G_{\mathrm{global}}= \sorm(n-1)\times \sorm(n-1)\times \sorm(2)$};
		\node [style=none] (23) at (8, -0.5) {2};
		\node [style=flavourRed] (24) at (8, 0) {};
		\node [style=none] (25) at (9, -0.5) {2};
		\node [style=none] (26) at (4.75, 0) {};
		\node [style=none] (27) at (7, 0) {};
		\node [style=none] (28) at (5.75, 0.5) {Ungauge SO(2)};
	\end{pgfonlayer}
	\begin{pgfonlayer}{edgelayer}
		\draw (0) to (2);
		\draw (5) to (6);
		\draw (5) to (7);
		\draw [style=brace] (9.center) to (8.center);
		\draw (16) to (17);
		\draw (16) to (18);
		\draw [style=brace] (20.center) to (19.center);
		\draw (24) to (13);
		\draw (13) to (14);
		\draw [style=->] (26.center) to (27.center);
	\end{pgfonlayer}
\end{tikzpicture}}}
\label{anomalous}
\end{equation}
In this section, an explanation to this additional $\sorm(2)$ is proposed. 
From the brane set up, one can shown that when $n=5$, the following mirror pairs are established:
\begin{equation}
  \scalebox{0.9}{     \raisebox{-.5\height}{ \begin{tikzpicture}
	\begin{pgfonlayer}{nodelayer}
		\node [style=miniU] (0) at (-1.5, 0) {};
		\node [style=miniBlue] (1) at (-0.5, 0) {};
		\node [style=miniU] (5) at (0.5, 0) {};
		\node [style=miniBlue] (6) at (1.5, 0.75) {};
		\node [style=miniBlue] (7) at (1.5, -0.75) {};
		\node [style=none] (11) at (1.25, 2) {$G_{\mathrm{global}}= \sorm(6)\times \sorm(6)$};
		\node [style=none] (12) at (-1.5, -0.5) {2};
		\node [style=none] (22) at (9, 2) {$G_{\mathrm{global}}= \sorm(4)\times \sorm(4)\times \sorm(2)$};
		\node [style=none] (26) at (4, 0) {};
		\node [style=none] (27) at (6.25, 0) {};
		\node [style=none] (28) at (5, 0.5) {Ungauge SO(2)};
		\node [style=flavorRed] (29) at (2.5, 0.75) {};
		\node [style=flavorRed] (30) at (2.5, -0.75) {};
		\node [style=miniBlue] (32) at (8, 0) {};
		\node [style=miniU] (33) at (9, 0) {};
		\node [style=miniBlue] (34) at (10, 0.75) {};
		\node [style=miniBlue] (35) at (10, -0.75) {};
		\node [style=flavorRed] (36) at (11, 0.75) {};
		\node [style=flavorRed] (37) at (11, -0.75) {};
		\node [style=none] (38) at (11.5, -0.75) {2};
		\node [style=none] (39) at (11.5, 0.75) {2};
		\node [style=none] (40) at (7, -0.5) {2};
		\node [style=none] (41) at (3, 0.75) {2};
		\node [style=none] (42) at (3, -0.75) {2};
		\node [style=none] (43) at (10, -1.25) {2};
		\node [style=none] (44) at (10, 1.25) {2};
		\node [style=none] (45) at (8, -0.5) {2};
		\node [style=none] (46) at (1.5, 1.25) {2};
		\node [style=none] (47) at (1.5, -1.25) {2};
		\node [style=none] (48) at (-0.5, -0.5) {2};
		\node [style=none] (49) at (9, -0.5) {4};
		\node [style=none] (50) at (0.5, -0.5) {4};
		\node [style=none] (51) at (0, -2.5) {};
		\node [style=none] (52) at (0, -3.5) {};
		\node [style=none] (53) at (10, -2.5) {};
		\node [style=none] (54) at (10, -3.5) {};
		\node [style=none] (55) at (0.85, -3) {3d mirror};
		\node [style=none] (56) at (11, -3) {3d mirror};
		\node [style=flavourRed] (57) at (7, 0) {};
		\node [style=miniBlue] (58) at (1.5, -4.75) {};
		\node [style=miniBlue] (59) at (1.5, -6.25) {};
		\node [style=flavorRed] (60) at (0.75, -5.5) {};
		\node [style=flavorRed] (61) at (0.75, -5.5) {};
		\node [style=none] (62) at (0.25, -5.5) {6};
		\node [style=none] (64) at (1.5, -4.25) {2};
		\node [style=none] (65) at (1.5, -6.75) {2};
		\node [style=miniBlue] (66) at (10, -4.75) {};
		\node [style=miniBlue] (67) at (10, -6.25) {};
		\node [style=flavorRed] (68) at (9.25, -5.5) {};
		\node [style=flavorRed] (69) at (9.25, -5.5) {};
		\node [style=none] (70) at (8.75, -5.5) {4};
		\node [style=none] (71) at (10, -4.25) {2};
		\node [style=none] (72) at (10, -6.75) {2};
		\node [style=miniU] (73) at (10.75, -5.5) {};
		\node [style=none] (74) at (11.25, -5.5) {2};
		\node [style=none] (75) at (4.25, -5.5) {};
		\node [style=none] (76) at (6.5, -5.5) {};
		\node [style=none] (77) at (5.25, -5) {Gauge SO(2)};
	\end{pgfonlayer}
	\begin{pgfonlayer}{edgelayer}
		\draw (5) to (6);
		\draw (5) to (7);
		\draw [style=->] (26.center) to (27.center);
		\draw (0) to (5);
		\draw (6) to (29);
		\draw (7) to (30);
		\draw (33) to (34);
		\draw (33) to (35);
		\draw (34) to (36);
		\draw (35) to (37);
		\draw [style=->] (51.center) to (52.center);
		\draw [style=->] (52.center) to (51.center);
		\draw [style=->] (53.center) to (54.center);
		\draw [style=->] (54.center) to (53.center);
		\draw (57) to (33);
		\draw (58) to (60);
		\draw (59) to (61);
		\draw (66) to (68);
		\draw (67) to (69);
		\draw (66) to (73);
		\draw (73) to (67);
		\draw [style=->] (75.center) to (76.center);
	\end{pgfonlayer}
\end{tikzpicture}}}
\end{equation}
The bottom left quiver is obtained by studying the brane system in \eqref{eq:ex_product_OSp}. The two $\mathrm{USp}(2)$ gauge nodes shares the same D5 branes in the brane system (hence, the same flavour node). This, of course, is equivalent to the product of two copies of $\mathrm{USp}(2)$ gauge theory with $\sorm(6)$ flavour symmetry. Now, since the action of ungauging an $\mathrm{SO}(2)$ is performed on the top quiver, the equivalent action on the mirror would be the gauging of $\mathrm{SO}(2)$. As the flavour node is shared between the two gauge nodes in the mirror quiver, the gauged $\mathrm{SO}(2)$ will be shared as well, giving the quiver on the right. To summarise the effect, in the mirror quiver the  original $\mathrm{SO}(6)\times \mathrm{SO}(6) $ flavour symmetry is first broken into $\mathrm{SO}(4)\times \mathrm{SO}(4) \times \mathrm{SO}(2) \times \mathrm{SO}(2) $. The diagonal $ \mathrm{SO}(2)$ is then gauged, but a  $ \mathrm{SO}(2)$ factor remains. This explains the additional abelian factor in the Higgs branch global symmetry of the mirror, and hence the Coulomb branch global symmetry in the original quiver.  
The Coulomb branch Hilbert series of the top right quiver (and equivalently, the Higgs branch Hilbert series of the bottom right quiver) is:
\begin{align}
\mathrm{HS} = \frac{1{+}8 t^2 {+}59 t^4{+}176 t^6{+}364 t^8{+}432 t^{10}{+}364 t^{12}{+}176 t^{14}{+}59 t^{16}{+}8 t^{18}{+}t^{20}}{(1-t^2)^{5}
   \left(1-t^4\right)^5} \,.
\end{align}
A similar argument can be applied to the additional $\sorm(2)$ factor in the Coulomb branch global symmetry of the $B$-type orthosymplectic quiver \eqref{anomalousfold}.  \eqref{anomalous} and  \eqref{anomalousfold} remain the only framed $D$-type and $B$-type orthosymplectic quiver we know of whose Coulomb branch global symmetry cannot be naively read off from the balance condition. 

\section{Hilbert series results}
\label{app:HS}
In this appendix, the Coulomb branch Hilbert series calculations of some of the quivers in this paper are presented. The computations use the monopole formula introduced in \cite{Cremonesi:2013lqa}. In cases of unframed orthosymplectic quivers with $\sorm(2n)$, $\mathrm{USp}(2n)$ and $\urm(n)$ quivers, there is an overall $\mathbb{Z}_2$ which acts trivial on the matter content. For the quivers relevant here, the $\mathbb{Z}_2$ is removed from the gauge groups such that the magnetic lattices becomes as discussed in \cite{Bourget:2020xdz}. For smaller quivers results are provided up to order $t^{20}$, while fewer orders are shown for larger quivers, due to computational complexity.

\begin{table}[h]
\ra{2}
    \centering
\begin{tabular}{ll}
    \toprule
         & Coulomb branch Hilbert series  \\  \midrule
         $\substack{n=3,\\k=1 \\l=0}$     & \scriptsize{\parbox{13cm}{$1+30 t^2+393 t^4+3120 t^6+17706 t^8+78972 t^{10}+293770 t^{12}+948336 t^{14}+2731755
   t^{16}+7166042 t^{18}+17382963 t^{20}+\dots \\ = (1+15 t^2+84 t^4+300 t^6+825 t^8+1911 t^{10}+3920 t^{12}+7344 t^{14}+12825 t^{16}+21175
   t^{18}+33396 t^{20}\dots)^2$} }\\ \midrule
         $\substack{n=4,\\k=1 \\l=0}$     &\scriptsize{\parbox{13cm}{$ 1+56 t^2+1384 t^4+20650 t^6+215636 t^8+1720264 t^{10}+11106025 t^{12}+60413096 t^{14}            +\dots \\
    = (                  1+28 t^2+300 t^4+1925 t^6+8918 t^8+32928 t^{10}+102816 t^{12}+282150 t^{14}               +\dots )^2$} }  \\ \midrule
    $\substack{n=5,\\k=1 \\l=0}$ &\scriptsize{\parbox{13cm}{$ 1+90 t^2+3565 t^4+84588 t^6+1386700 t^8+17100048 t^{10}+167857416 t^{12}+1365541740
   t^{14}               +\dots \\
    = ( 1+45 t^2+770 t^4+7644 t^6+52920 t^8+282744 t^{10}+1241460 t^{12}+4671810 t^{14}                                +\dots )^2$} } \\ \midrule 
         $\substack{n=5,\\k=2 \\l=0}$ &\scriptsize{\parbox{13cm}{$1+90 t^2+3985 t^4+115368 t^6+2454040 t^8+40922106 t^{10}+557904916 t^{12}+6407626218
   t^{14}+63421757476 t^{16}+550795731040 t^{18}+4259031312565 t^{20}         +\dots \\
    = (1+45 t^2+980 t^4+13584 t^6+135540 t^8+1049433 t^{10}                             +\dots )^2$} } \\ \midrule
    $\substack{n=6,\\k=1 \\l=0}$ &\scriptsize{\parbox{13cm}{$ 1+132 t^2+8622 t^4+369798 t^6+11686995 t^8+289989018 t^{10}+5884262173 t^{12}+100491980996
   t^{14}               +\dots \\
    = (1+66 t^2+2133 t^4+44121 t^6+656667 t^8+7544394 t^{10}+70199051 t^{12}+547856023 t^{14}                                 +\dots )^2$} } \\
\bottomrule
    \end{tabular}
    \caption{Hilbert series results for \eqref{weirdcase}. The first line displays the Coulomb branch Hilbert series for \eqref{weirdcase}, while the second line displays the factorisation \eqref{productfiniteX}. Each factor matches the Higgs branch Hilbert series of $\sprm(k)$ or $\sprm(k{+}l)$ SQCD with $N_f{=}n$.}
    \label{weirdcase1}
\end{table}

\begin{table}[]
\ra{1.5}
    \centering
    \begin{tabular}{ll}
    \toprule
         & Coulomb branch Hilbert series  \\  \midrule
    $E_{3-2l}^{k=2,l=0}$     &\scriptsize{\parbox{13cm}{$         1+58 t^2+32 t^3+1569 t^4+1600 t^5+27220 t^6+37856 t^7+348124 t^8+577920
   t^9+3540936 t^{10}+6502720 t^{11}+29981572 t^{12}+58099072 t^{13}+217747736
   t^{14}+432151456 t^{15}+1385881186 t^{16}+2764473568 t^{17}+7858110900
   t^{18}+15572468640 t^{19}+40224531398 t^{20}       +\dots \\
    = (   1 + 29 t^2 + 16 t^3 + 364 t^4 + 336 t^5 + 2926 t^6 + 3360 t^7 + 
 17584 t^8 + 22400 t^9 + 85260 t^{10} + 114240 t^{11} + 349572 t^{12} + 
 479232 t^{13} + 1251816 t^{14} + 1729488 t^{15} + 4008081 t^{16} + 
 5534496 t^{17} + 11680405 t^{18} + 16045920 t^{19} + 31415582 t^{20}                              +\dots )^2$} }  \\ \midrule
    $E_{3-2l}^{k=3,l=1}$  &\scriptsize{\parbox{13cm}{$   1+58 t^2+1601 t^4+28424 t^6+369156 t^8+3771348 t^{10}+31759488
   t^{12}+227801304 t^{14}+1425775758 t^{16}+7933063516 t^{18}+39822908626
   t^{20}             +\dots \\
    = (        1 + 29 t^2 + 380 t^4 + 3192 t^6 + 19810 t^8 + 98224 t^{10} + 
 409016 t^{12} + 1480548 t^{14} + 4776849 t^{16} + 13998385 t^{18} + 
 37805264 t^{20}                         +\dots )^2$} } \\ \midrule
    $E_{3-2l}^{k=3,l=0}$ &\scriptsize{\parbox{13cm}{$       1+134 t^2+9017 t^4+403982 t^6+13508026 t^8+358747158 t^{10} +\dots \\
    = ( 1+67 t^2+2264 t^4+50303 t^6+820864 t^8+10489699 t^{10}+\dots )^2$} } \\ \hline
    
      $E_{3=2l}^{k=4,l=1}$ &\scriptsize{\parbox{13cm}{$      1+134 t^2+8889 t^4+128 t^5+388366 t^6+15616 t^7+12558722 t^8+943360
   t^9+320512966 t^{10}       +\dots \\
    = (     1 + 67 t^2 + 2200 t^4 + 64 t^5 + 46783 t^6 + 3520 t^7 + 724900 t^8 + 
 95040 t^9 + 8763535 t^{10}                          +\dots )^2$} } \\
\bottomrule
    \end{tabular}
    \caption{Hilbert series results for theories of Tables \ref{EnProdtable} and \ref{EnProdtable2}. The first line displays the Coulomb branch Hilbert series for the proposed quiver, while the second line displays the factorisation into a product. The known Hilbert series for the factors match these findings.}
    \label{EnProdtableT}
\end{table}

\begin{table}[]
\ra{2}
    \centering
    \begin{tabular}{ll}
    \toprule
         & Coulomb branch Hilbert series  \\ \midrule
    $\substack{n=5,\\ k=1,\\ l=1}$     & \scriptsize{\parbox{13cm}{$1+123 t^2+6710 t^4+220812 t^6+5027328 t^8+85856661 t^{10}+1160906385 t^{12}+12919480684
   t^{14}+121836107745 t^{16}           +\dots \\
    = (    1+78 t^2+2430 t^4+43758 t^6+537966 t^8+4969107 t^{10}+36685506 t^{12}+225961450 t^{14}+1198006524
   t^{16}                             +\dots )\\ 
   \times( 1+45 t^2+770 t^4+7644 t^6+52920 t^8+282744 t^{10}+1241460 t^{12}+4671810 t^{14}+15520791 t^{16}                          +\dots )$} } \\ \midrule
    $\substack{n=7,\\k=1,\\l=2}$ & \scriptsize{\parbox{13cm}{$ 1+183 t^2+128 t^3+15625 t^4+21632 t^5+849367 t^6+1687168 t^7+33584330 t^8+83001984 t^9+1039787073
   t^{10}+2940462720 t^{11}+26377189292 t^{12}+80780840064 t^{13}+564910951617
   t^{14}+1804412218752 t^{15}+10430788358641 t^{16}+33865771104384 t^{17}+168680900002512
   t^{18}+547066306438912 t^{19}+2419061167200157 t^{20}           +\dots \\
    = (    1+91 t^2+3080 t^4+58344 t^6+741741 t^8+7014007 t^{10}+52676624 t^{12}+328671200 t^{14}+1760034276
   t^{16}+8288708428 t^{18}+34981096384 t^{20}                             +\dots ) \\ 
   \times (1+92 t^2+128 t^3+4173 t^4+9984 t^5+127920 t^6+384384 t^7+2981381 t^8+9804288 t^9+55764111
   t^{10}+186920448 t^{11}+863386770 t^{12}+2844037120 t^{13}+11315720064 t^{14}+35984655616
   t^{15}+127863652644 t^{16}+389538715904 t^{17}+1265346923490 t^{18}+3684348495360
   t^{19}+11116073388432 t^{20} +\dots)$} }\\  \midrule   $\substack{n=7,\\ k=2,\\ l=1}$     & \scriptsize{\parbox{13cm}{$1 + 183 t^2 + 128 t^3 + 16626 t^4 + 21632 t^5 + 1000427 t^6 + 1815296 t^7 + 44866419 t^8 + 100543872 t^9 + 1599857948 t^{10}      +\dots \\
    = (  1+92 t^2+128 t^3+4173 t^4+9984 t^5+127920 t^6+384384 t^7+2981381 t^8+9804288 t^9+55764111 t^{10}     +\dots )\\ 
   \times(1+91 t^2+4081 t^4+117312 t^6+2421601 t^8+38416378 t^{10}                     +\dots )$} } \\
\bottomrule
    \end{tabular}
    \caption{Hilbert series results for \eqref{weirder}. The first line displays the Coulomb branch Hilbert series for \eqref{weirder}, while the second line displays the factorisation into a product. Each factor is consistent with the known Hilbert series for the individual theories in \eqref{weirder_product}.}
    \label{weirder1}
\end{table}

\begin{table}[]
\ra{2}
    \centering
    \begin{tabular}{ll}
    \toprule
         & Coulomb branch Hilbert series  \\ \midrule
      $\substack{n=6\\k=2,\\l=1}$    & \scriptsize{\parbox{13cm}{$   1+32 t+594 t^2+8096 t^3+87208 t^4+771936 t^5+5780544 t^6+37501728 t^7+214959492 t^8+1106375776
   t^9+5181666776 t^{10}+22327527456 t^{11}+89331239192 t^{12}         +\dots \\
    = (  1+32 t+528 t^2+5984 t^3+52360 t^4+376992 t^5+2324784 t^6+12620256 t^7+61523748 t^8+273438880
   t^9+1121099408 t^{10}   +\dots ) \\ 
   \times (1 + 66 t^2 + 2133 t^4 + 44121 t^6 + 656667 t^8 + 7544394 t^{10} + 
 70199051 t^{12} +\dots$} } \\ \midrule
   $\substack{n=8\\k=3,\\l=1}$ & \scriptsize{\parbox{13cm}{$ 1+368 t^2+63884 t^4+7048932 t^6+560375300 t^8             +\dots \\
    = (  1+248 t^2+27000 t^4+1763125 t^6+79143000 t^8                           +\dots )\times(1+120 t^2+7124 t^4+279055 t^6+8103660 t^8 +\dots)$} }\\  \midrule
   $\substack{n=6\\k=1,\\l=2}$ & \scriptsize{\parbox{13cm}{$ 1 + 32 t + 594 t^2 + 8096 t^3 + 88846 t^4 + 824352 t^5 + 6668508 t^6 + 48042720 t^7 + 313144128 t^8 + 1869228064 t^9 + 10318091234 t^{10}         +\dots \\
    = (                     1+32 t+528 t^2+5984 t^3+52360 t^4+376992 t^5+2324784 t^6+12620256 t^7+61523748 t^8+273438880
   t^9+1121099408 t^{10}    +\dots )\\ \times (1+66 t^2+1638 t^4+23100 t^6+222156 t^8+1613898 t^{10}+\dots)$} }\\ \midrule
     $\substack{n=8\\k=2,\\l=2}$ & \scriptsize{\parbox{13cm}{$ 1 + 368 t^2 + 63884 t^4 + 7040924 t^6 + 557673276 t^8 + 34004993996 t^{10}     +\dots \\
    = (                   1+248 t^2+27000 t^4+1763125 t^6+79143000 t^8+2642777280 t^{10}    +\dots )\\ \times (1+120 t^2+7124 t^4+271047 t^6+7387620 t^8+154155456 t^{10}+\dots)$} }\\\bottomrule
    \end{tabular}
    \caption{Hilbert series results for \eqref{betterbelastguy}. The first line displays the Coulomb branch Hilbert series for \eqref{betterbelastguy}, while the second line displays the factorisation into a product. Each factor is consistent with the known Hilbert series for the individual theories in \eqref{finallygotitright}.}
    \label{betterbelastguy1}
\end{table}

\begin{table}[]
\ra{2}
    \centering
    \begin{tabular}{ll}
    \toprule
         & Coulomb branch Hilbert series  \\ \midrule
     $\substack{k=1}$    & \scriptsize{\parbox{13cm}{$  1+104 t^2+4810 t^4+134264 t^6+2597465 t^8+37800384 t^{10}+437076562 t^{12}+4175738528
   t^{14}+33943481936 t^{16}+240150358656 t^{18}+1505698109306 t^{20}              +\dots \\
    = (    1+52 t^2+1053 t^4+12376 t^6+100776 t^8+627912 t^{10}+3187041 t^{12}+13748020 t^{14}+51949755
   t^{16}+175847880 t^{18}+542393670 t^{20}                             +\dots )^2$} } \\ \midrule
     $\substack{k=2}$ & \scriptsize{\parbox{13cm}{$  1+156 t^2+128 t^3+12064 t^4+18304 t^5+620840 t^6+1297792 t^7+23982530 t^8+60845696 t^9+741714140
   t^{10}+2122578432 t^{11}+19100059850 t^{12}+58780377472 t^{13}+420308771752
   t^{14}+1346411945088 t^{15}+8051191269563 t^{16}              +\dots \\
    = (                       1+78 t^2+64 t^3+2990 t^4+4160 t^5+75152 t^6+133056 t^7+1393119 t^8+2796352 t^9+20320924
   t^{10}+43544384 t^{11}+243181003 t^{12}+537295616 t^{13}+2459411330 t^{14}+5486291136
   t^{15}+21499028967 t^{16}          +\dots )^2$} }\\ \bottomrule 
    \end{tabular}
\caption{Hilbert series results for \eqref{f4f4}. The first line displays the Coulomb branch Hilbert series for \eqref{f4f4}, while the second line displays the factorisation into a product. Each factor agrees with the known Hilbert series.}
    \label{f4f41}
\end{table}

\begin{table}[]
\ra{2}
    \centering
    \begin{tabular}{ll}
    \toprule
         & Coulomb branch Hilbert series  \\ \midrule
    $k=1$     & \scriptsize{\parbox{13cm}{$ 1+42 t^2+777 t^4+8706 t^6+68880 t^8+421162 t^{10}+2109885 t^{12}+9015666 t^{14}+33826779
   t^{16}+113879276 t^{18}+349750894 t^{20}           +\dots \\
    = (1+21 t^2+168 t^4+825 t^6+3003 t^8+8918 t^{10}+22848 t^{12}+52326 t^{14}+109725 t^{16}+214291
   t^{18}+394680 t^{20}                                 +\dots )^2$} } \\ \midrule
    $k=2$ &\scriptsize{\parbox{13cm}{$  1+110 t^2+5973 t^4+212606 t^6+5570246 t^8+114525004 t^{10}+1925660177 t^{12}+27266117052
   t^{14}+332379773462 t^{16}+3549706986122 t^{18}+33687210855500 t^{20}              +\dots \\
    = (           1+55 t^2+1474 t^4+25233 t^6+310970 t^8+2965710 t^{10}+22994114 t^{12}+150219706 t^{14}+849547985
   t^{16}+4247189980 t^{18}+19084874832 t^{20}                      +\dots )^2$} } \\ \midrule
        $k=3$ &\scriptsize{\parbox{13cm}{$ 1+210 t^2+21915 t^4+1515458 t^6               +\dots \\
    = (      1+105 t^2+5445 t^4+186004 t^6                           +\dots )^2$} } \\ \bottomrule 
    \end{tabular}
        \caption{Hilbert series results for \eqref{soddxsodd}. The first line displays the Coulomb branch Hilbert series for \eqref{soddxsodd}, while the second line displays the factorisation into a product. Each factor agrees with the known Hilbert series.}
    \label{soddxsodd1}
\end{table}


\begin{table}[]
\ra{2}
    \centering
    \begin{tabular}{ll}
    \toprule
         & Coulomb branch Hilbert series  \\ \midrule
    $k=1$     & \scriptsize{\parbox{13cm}{$ 1+25 t^2+269 t^4+1749 t^6+8190 t^8+30478 t^{10}+95710 t^{12}+263758 t^{14}+655303 t^{16}+1496495
   t^{18}+3186755 t^{20}               +\dots \\
    = ( 1+10 t^2+35 t^4+84 t^6+165 t^8+286 t^{10}+455 t^{12}+680 t^{14}+969 t^{16}+1330 t^{18}+1771
   t^{20}                                +\dots )\\ \times (  1+15 t^2+84 t^4+300 t^6+825 t^8+1911 t^{10}+3920 t^{12}+7344 t^{14}+12825 t^{16}+21175
   t^{18}+33396 t^{20}                               +\dots ) $} } \\ \midrule
    $k=2$ &\scriptsize{\parbox{13cm}{$ 1+81 t^2+3221 t^4+83585 t^6+1591224 t^8+23721865 t^{10}+288966230 t^{12}+2965143530
   t^{14}+26230232434 t^{16}+203743162413 t^{18}+1410507872500 t^{20}               +\dots \\
    = ( 1+45 t^2+980 t^4+13584 t^6+135540 t^8+1049433 t^{10}+6636245 t^{12}+35562384 t^{14}+166007686
   t^{16}+689482430 t^{18}+2590532352 t^{20}                                +\dots )\\ \times (  1+36 t^2+621 t^4+6776 t^6+53160 t^8+324648 t^{10}+1628913 t^{12}+6977828 t^{14}+26250939
   t^{16}+88600616 t^{18}+272745382 t^{20}                               +\dots ) $} }  \\ \midrule
    $k=3$ & \scriptsize{\parbox{13cm}{$ 1+169 t^2+14169 t^4+785875 t^6+32440512 t^8+1063059375 t^{10}               +\dots \\
    = (  1+91 t^2+4081 t^4+120315 t^6+2621801 t^8+45026982 t^{10}+634993345 t^{12}                               +\dots )\\ \times (        1+78 t^2+2990 t^4+75152 t^6+1393119 t^8+20320924 t^{10}                         +\dots ) $} } \\ \bottomrule
    \end{tabular}
    \caption{Hilbert series results for \eqref{foldedBtypefirst}. The first line displays the Coulomb branch Hilbert series for \eqref{foldedBtypefirst}, while the second line displays the factorisation into a product. Each factor agrees with the known Hilbert series for $\overline{\mathcal{O}}^{\mathfrak{so}(4k+1)}_{(2^{2k},1^{1})}$ and $\overline{\mathcal{O}}^{\mathfrak{so}(4k+2)}_{(2^{2k},1^{2})}$, respectively.}
    \label{foldedBtypefirst1}
\end{table}

\begin{table}[]
\ra{2}
    \centering
    \begin{tabular}{ll}
    \toprule
         & Coulomb branch Hilbert series  \\  \midrule
         $\substack{n=4,\\ k=1,\\l=0}$     & \scriptsize{\parbox{13cm}{$ 1+49 t^2+1056 t^4+13754 t^6+125846 t^8+884108 t^{10}+5054105 t^{12}+24478785 t^{14}               +\dots \\
    = (  1+28 t^2+300 t^4+1925 t^6+8918 t^8+32928 t^{10}+102816 t^{12}+282150 t^{14}                               +\dots )\\ \times (1+21 t^2+168 t^4+825 t^6+3003 t^8+8918 t^{10}+22848 t^{12}+52326 t^{14}                                 +\dots ) $} }  \\ \midrule
    $\substack{n=5,\\ k=1,\\l=0}$ & \scriptsize{\parbox{13cm}{$1+81 t^2+2885 t^4+61643 t^6+912366 t^8+10189480 t^{10}+90890304 t^{12}+674183898 t^{14}                +\dots \\
    = (   1+36 t^2+495 t^4+4004 t^6+22932 t^8+102816 t^{10}+383724 t^{12}+1241460 t^{14}                              +\dots )\\ \times (1+45 t^2+770 t^4+7644 t^6+52920 t^8+282744 t^{10}+1241460 t^{12}+4671810 t^{14}                                 +\dots ) $} } \\ \midrule
    $\substack{n=6,\\ k=2,\\l=0}$ & \scriptsize{\parbox{13cm}{$ 1+121 t^2+7237 t^4+283953 t^6+8203712 t^8+186007152 t^{10}+3448403056 t^{12}+53812917025 t^{14}  +\dots \\
    = (     1+55 t^2+1474 t^4+25233 t^6+310970 t^8+2965710 t^{10}+22994114 t^{12}+150219706 t^{14}                   +\dots )\\ \times ( 1+66 t^2+2133 t^4+44121 t^6+656667 t^8+7544394 t^{10}+70199051 t^{12}+547856023 t^{14}                  +\dots ) $} } \\\midrule
    $\substack{n=6, \\ k=1,\\ l=1}$     & \scriptsize{\parbox{13cm}{$  1+121 t^2+6907 t^4+250590 t^6+6536574 t^8+131359647 t^{10}+2127959063 t^{12}+28706249847
   t^{14}+330553477948 t^{16}                           +\dots \\
    = (   1+66 t^2+2133 t^4+44121 t^6+656667 t^8+7544394 t^{10}+70199051 t^{12}+547856023 t^{14}+3682768464
   t^{16}                                       +\dots )\\ 
   \times (  1+55 t^2+1144 t^4+13650 t^6+112200 t^8+703494 t^{10}+3586440 t^{12}+15520791 t^{14}+58790875
   t^{16}                                             +\dots ) $} }  \\
\bottomrule
    \end{tabular}
    \caption{Hilbert series results for \eqref{missingguy}. The first line displays the Coulomb branch Hilbert series for \eqref{missingguy}, while the second line displays the factorisation. Each factor agrees with the known Hilbert series for $\overline{\mathcal{O}}^{\mathfrak{so}(2n-1)}_{(2^{2k},1^{2n-4k-1})}$ and $\overline{\mathcal{O}}^{\mathfrak{so}(2n)}_{(2^{2k+2l},1^{2n-4k-4l})}$, respectively.}
    \label{missingguy1}
\end{table}

\begin{table}[]
\ra{2}
    \centering
    \begin{tabular}{ll}
    \toprule
         & Coulomb branch Hilbert series  \\  \midrule
    $k=1$     &  \scriptsize{\parbox{13cm}{$ 1+130 t^2+7539 t^4+264628 t^6+6438276 t^8+117582633 t^{10}+1700156133 t^{12}+20221864235 t^{14}+203634547831 t^{16}+1775507329154 t^{18}+13646329994778 t^{20}               +\dots \\
    = (   1+78 t^2+2430 t^4+43758 t^6+537966 t^8+4969107 t^{10}+36685506
   t^{12}+225961450 t^{14}+1198006524 t^{16}+5597569328 t^{18}+23474156784
   t^{20}                              +\dots )\\ \times (                       1+52 t^2+1053 t^4+12376 t^6+100776 t^8+627912 t^{10}+3187041 t^{12}+13748020 t^{14}+51949755 t^{16}+175847880 t^{18}+542393670 t^{20}          +\dots ) $} } \\ \midrule
    $k=2$ & \scriptsize{\parbox{13cm}{$    1+170 t^2+192 t^3+14339 t^4+30016 t^5+811838 t^6+2328704 t^7+34914970 t^8+119841920 t^9+1216054981 t^{10}+4608606912 t^{11}+35579828088 t^{12}+141340385344
   t^{13}+894676190640 t^{14}+3601291105024 t^{15}+19640751761433 t^{16}+78397263587712 t^{17}+380968232656361 t^{18}+1488093619535168 t^{19}+6594271153605775
   t^{20}            +\dots \\
    = (   1+92 t^2+128 t^3+4173 t^4+9984 t^5+127920 t^6+384384 t^7+2981381 t^8+9804288 t^9+55764111 t^{10}+186920448 t^{11}+863386770 t^{12}+2844037120 t^{13}+11315720064
   t^{14}+35984655616 t^{15}+127863652644 t^{16}+389538715904 t^{17}+1265346923490 t^{18}+3684348495360 t^{19}+11116073388432 t^{20}                              +\dots )\\ \times(   1+78 t^2+64 t^3+2990 t^4+4160 t^5+75152 t^6+133056 t^7+1393119 t^8+2796352 t^9+20320924 t^{10}+43544384 t^{11}+243181003 t^{12}+537295616 t^{13}+2459411330
   t^{14}+5486291136 t^{15}+21499028967 t^{16}+47797193600 t^{17}+165358919076 t^{18}+363477599744 t^{19}+1135433776023 t^{20}                              +\dots ) $} } \\ \bottomrule
    \end{tabular}
    \caption{Hilbert series results for \eqref{E6F4family}. The first line displays the Coulomb branch Hilbert series for \eqref{E6F4family}, while the second line displays the factorisation into a product. Each factor agrees with the known Hilbert series.}
    \label{E6F4family1}
\end{table}

\begin{table}[]
\ra{2}
    \centering
    \begin{tabular}{ll}
    \toprule
         & Coulomb branch Hilbert series  \\ \midrule
    $\substack{ k=1,\\ l=1}$      &  \scriptsize{\parbox{13cm}{$1+114 t^2+5733 t^4+173852 t^6+3651348 t^8+57617325 t^{10}+721201692 t^{12}+7444836763
   t^{14}+65256772383 t^{16}+496825085316 t^{18}+3345139658705 t^{20}              +\dots \\
    = (       1+36 t^2+495 t^4+4004 t^6+22932 t^8+102816 t^{10}+383724 t^{12}+1241460 t^{14}+3581721
   t^{16}+9406540 t^{18}+22837815 t^{20}          +\dots )\\ \times ( 1+78 t^2+2430 t^4+43758 t^6+537966 t^8+4969107 t^{10}+36685506 t^{12}+225961450 t^{14}+1198006524
   t^{16}+5597569328 t^{18}+23474156784 t^{20}      +\dots ) $} } \\ \midrule
      $\substack{ k=2,\\ l=1}$   &\scriptsize{\parbox{13cm}{$1+170 t^2+128 t^3+13624 t^4+19968 t^5+700114 t^6+1454336 t^7+26309217 t^8+67287040 t^9+777105979
   t^{10}                +\dots \\
    = (           1+92 t^2+128 t^3+4173 t^4+9984 t^5+127920 t^6+384384 t^7+2981381 t^8+9804288 t^9+55764111 t^{10}                      +\dots )\\ \times  (  1+78 t^2+2275 t^4+37400 t^6+415701 t^8+3461458 t^{10}                               +\dots ) $} }  \\ \bottomrule
    \end{tabular}
    \caption{Hilbert series results for \eqref{foldedannoyingquiver}. The first line displays the Coulomb branch Hilbert series for \eqref{foldedannoyingquiver}, while the second line displays the factorisation into a product. Each factor agrees with the known Hilbert series.}
    \label{foldedannoyingquiver1}
\end{table}

\begin{table}[]
\ra{2}
    \centering
    \begin{tabular}{ll}
    \toprule
         & Coulomb branch Hilbert series  \\ \midrule
      $\substack{k=2,\\ l=1}$      & \scriptsize{\parbox{13cm}{$ 1+32 t+583 t^2+7744 t^3+82874 t^4+753280 t^5+6008089 t^6+42982688 t^7+280199502 t^8               +\dots \\
    = (1+55 t^2+1474 t^4+25233 t^6+310970 t^8                                 +\dots )\\ \times (1+32 t+528 t^2+5984 t^3+52360 t^4+376992 t^5+2324784 t^6+12620256 t^7+61523748 t^8                                 +\dots ) $} }  \\ \midrule
    $\substack{k=3,\\ l=1}$ & \scriptsize{\parbox{13cm}{$ 1+353 t^2+58485 t^4+6134489 t^6+462119507 t^8               +\dots \\
    = (       1+105 t^2+5445 t^4+186004 t^6+4704390 t^8                          +\dots )\\ \times (              1+248 t^2+27000 t^4+1763125 t^6+79143000 t^8                   +\dots ) $} }\\ \bottomrule
    \end{tabular}
    \caption{Hilbert series results for \eqref{foldedfinaleguy}. The first line displays the Coulomb branch Hilbert series for \eqref{foldedfinaleguy}, while the second line displays the factorisation into a product. Each factor agrees with the known Hilbert series.}
    \label{foldedfinaleguy1}
\end{table}

\begin{table}[]
\ra{2}
    \centering
    \begin{tabular}{ll}
    \toprule
         & Coulomb branch Hilbert series  \\ \midrule
    $k=1$     & \scriptsize{\parbox{13cm}{$1+88 t^2+3420 t^4+80028 t^6+1298687 t^8+15893460 t^{10}+155107049 t^{12}+1256071828
   t^{14}+8694268479 t^{16}+52624765960 t^{18}+283597751451 t^{20}                +\dots \\
    = ( 1+52 t^2+1053 t^4+12376 t^6+100776 t^8+627912 t^{10}+3187041 t^{12}+13748020 t^{14}+51949755
   t^{16}+175847880 t^{18}+542393670 t^{20}                                +\dots )\\
   \times(   1+36 t^2+495 t^4+4004 t^6+22932 t^8+102816 t^{10}+383724 t^{12}+1241460 t^{14}+3581721
   t^{16}+9406540 t^{18}+22837815 t^{20}                              +\dots ) $} }  \\ \midrule
    $k=2$ &\scriptsize{\parbox{13cm}{$ 1+144 t^2+96 t^3+10271 t^4+12608 t^5+485497 t^6+820352 t^7+17130124 t^8+35239616 t^9+481174913
   t^{10}               +\dots \\
    = (     1+78 t^2+64 t^3+2990 t^4+4160 t^5+75152 t^6+133056 t^7+1393119 t^8+2796352 t^9+20320924 t^{10}                            +\dots )\\ 
    \times ( 1+66 t^2+32 t^3+2133 t^4+1728 t^5+44583 t^6+45760 t^7+678117 t^8+794112 t^9+8037711 t^{10}                                +\dots ) $} }  \\ \bottomrule
    \end{tabular}
    \caption{Hilbert series results for \eqref{F4SO9family}. The first line displays the Coulomb branch Hilbert series for \eqref{F4SO9family}, while the second line displays the factorisation into a product. Each factor agrees with the known Hilbert series.}
    \label{F4SO9family1}
\end{table}

\begin{table}[]
\ra{2}
    \centering
    \begin{tabular}{ll}
    \toprule
         & Coulomb branch Hilbert series  \\ \midrule
    $k=1$     & \scriptsize{\parbox{13cm}{$1+36 t^2+567 t^4+5409 t^6+36615 t^8+192899 t^{10}+839021 t^{12}+3136395 t^{14}+10368906
   t^{16}+30964325 t^{18}+84878874 t^{20}      +\dots \\
    = (1+15 t^2+84 t^4+300 t^6+825 t^8+1911 t^{10}+3920 t^{12}+7344 t^{14}+12825 t^{16}+21175
   t^{18}+33396 t^{20}                          +\dots )\\ 
   \times ( 1+21 t^2+168 t^4+825 t^6+3003 t^8+8918 t^{10}+22848 t^{12}+52326 t^{14}+109725 t^{16}+214291
   t^{18}+394680 t^{20} +\dots ) $} }  \\ \midrule
    $k=2$ &\scriptsize{\parbox{13cm}{$1+100 t^2+4929 t^4+159047 t^6+3773635 t^8+70214649 t^{10}+1068107756 t^{12}+13683068037
   t^{14}+150962851740 t^{16}+1460049948432 t^{18}+12558512227814 t^{20}     +\dots \\
    = (    1+45 t^2+980 t^4+13584 t^6+135540 t^8+1049433 t^{10}+6636245 t^{12}+35562384 t^{14}+166007686
   t^{16}+689482430 t^{18}+2590532352 t^{20}                      +\dots )\\ 
   \times (1+55 t^2+1474 t^4+25233 t^6+310970 t^8+2965710 t^{10}+22994114 t^{12}+150219706 t^{14}+849547985
   t^{16}+4247189980 t^{18}+19084874832 t^{20}    +\dots ) $} }  \\ \midrule
   $k=3$ & \scriptsize{\parbox{13cm}{$ 1+196 t^2+19081 t^4+1230319 t^6+59106675 t^8+2256496369 t^{10}+71299059243 t^{12}+1917659317006
   t^{14}+44815111648324 t^{16}+924438389586125 t^{18}+17043068970722031 t^{20} +\dots \\   = (1+91 t^2+4081 t^4+120315 t^6+2621801 t^8+45026982 t^{10}+634993345 t^{12}+7569785080
   t^{14}+77969074288 t^{16}+705979238874 t^{18}+5699104231898 t^{20} +\dots )\\ 
   \times(1+105 t^2+5445 t^4+186004 t^6+4704390 t^8+93883293 t^{10}+1539459830 t^{12}+21340594394
   t^{14}+255505157414 t^{16}+2687035828320 t^{18}+25164702542013 t^{20} +\dots ) $} }  \\ \bottomrule
    \end{tabular}
    \caption{Hilbert series results for \eqref{soddfold}. The first line displays the Coulomb branch Hilbert series for \eqref{soddfold}, while the second line displays the factorisation. Each factor agrees with the known Hilbert series for $\overline{\mathcal{O}}^{\mathfrak{so}_{4k+2}}_{[2^{2k-2},1^2]}$ and $\overline{\mathcal{O}}^{\mathfrak{so}_{4k+3}}_{[2^{2k-2},1^3]}$, respectively.}
    \label{soddfold1}
\end{table}

\begin{table}[]
\ra{2}
    \centering
    \begin{tabular}{ll}
    \toprule
         & Coulomb branch Hilbert series  \\ \midrule
    $\substack{p=2,\\ q=1,\\ k=1}$     & \scriptsize{\parbox{13cm}{$1+30 t^2+413 t^4+3595 t^6+22845 t^8+115073 t^{10}+484367 t^{12}+1766649 t^{14}+5731324
   t^{16}+16864225 t^{18}+45687006 t^{20} +\dots \\
    = (1+15 t^2+84 t^4+300 t^6+825 t^8+1911 t^{10}+3920 t^{12}+7344 t^{14}+12825 t^{16}+21175
   t^{18}+33396 t^{20}  +\dots )\\ \times (1+15 t^2+104 t^4+475 t^6+1659 t^8+4802 t^{10}+12096 t^{12}+27378 t^{14}+56925 t^{16}+110473
   t^{18}+202488 t^{20} +\dots) $} }  \\ \midrule
    $\substack{p=2,\\ q=1,\\ k=2}$ &\scriptsize{\parbox{13cm}{$1+70 t^2+40 t^3+2415 t^4+2520 t^5+55560 t^6+78624 t^7+965011 t^8+1631240 t^9+13512030 t^{10}   +\dots \\
    = ( 1+24 t^2+200 t^4+1000 t^6+3675 t^8+10976 t^{10}+28224 t^{12}+64800 t^{14}+136125
   t^{16}+266200 t^{18}+490776 t^{20}+\dots )\\ \times ( 1+24 t^2+275 t^4+2024 t^6+10976 t^8+47552 t^{10}+173448 t^{12}+552200 t^{14}+1574573
   t^{16}+4098952 t^{18}+9884303 t^{20}+\dots $} } 
   \\ \midrule
    $\substack{p=3,\\ q=2,\\ k=1}$ &\scriptsize{\parbox{13cm}{$1+48 t^2+1051 t^4+14424 t^6+142227 t^8+1089952 t^{10}+6836169 t^{12}+36400128 t^{14}+169074122 t^{16}+699481456 t^{18}+2619981678
   t^{20} +\dots \\
    = (1+35 t^2+594 t^4+6370 t^6+49160 t^8+295974 t^{10}+\dots )\\ \times ( 1+35 t^2+594 t^4+6545 t^6+53129 t^8+341810 t^{10}+\dots $} } 
   \\ \bottomrule
    \end{tabular}
    \caption{Hilbert series results for \eqref{eq:ex_product_different_mirror}. The first line displays the Coulomb branch Hilbert series for \eqref{eq:ex_product_different_mirror}, while the second line displays the factorisation into a product of two identical factors. Each factor agrees with the known Hilbert series.}
    \label{marcussuperquiver}
\end{table}

\begin{table}[]
\ra{2}
    \centering
    \begin{tabular}{ll}
    \toprule
         & Coulomb branch Hilbert series  \\ \midrule
    $k=2$     & \scriptsize{\parbox{13cm}{$1+56 t^2+1384 t^4+20650 t^6+215636 t^8+1720264 t^{10}+11106025 t^{12}+60413096 t^{14}+285191074
   t^{16}+1194750560 t^{18}+4519533676 t^{20}      +\dots \\
    = (1+28 t^2+300 t^4+1925 t^6+8918 t^8+32928 t^{10}+102816 t^{12}+282150 t^{14}+698775 t^{16}+1591876
   t^{18}+3383380 t^{20}    +\dots )^2 $} }  \\ \midrule
    $k=3$ &\scriptsize{\parbox{13cm}{$1+72 t^2+80 t^3+2556 t^4+5120 t^5+62010 t^6+162288 t^7+1165484 t^8+3433728 t^9+17928592 t^{10}   +\dots \\
    = (   1+36 t^2+40 t^3+630 t^4+1120 t^5+7525 t^6+15624 t^7+68592 t^8+147800 t^9+502074 t^{10}+\dots )^2 $} }  \\  \bottomrule
    \end{tabular}
    \caption{Hilbert series results for \eqref{eq:6d_magQuiv_Unitary_finite}. The first line displays the Coulomb branch Hilbert series for \eqref{eq:6d_magQuiv_Unitary_finite}, while the second line displays the factorisation into a product of two identical factors. Each factor agrees with the known Hilbert series.}
    \label{eq:6d_magQuiv_Unitary_finite1}
\end{table}

\begin{table}[]
\ra{2}
    \centering
    \begin{tabular}{ll}
    \toprule
         & Coulomb branch Hilbert series  \\ \midrule
    $k=2$     & \scriptsize{\parbox{13cm}{$1+42 t^2+831 t^4+10500 t^6+96727 t^8+700938 t^{10}+4198425 t^{12}+21520704 t^{14}+96852826 t^{16}+390254420
   t^{18}+1429637622 t^{20}  +\dots \\
    = (1+21 t^2+195 t^4+1155 t^6+5096 t^8+18228 t^{10}+55692 t^{12}+150480 t^{14}+368445 t^{16}+832117 t^{18}+1756755
   t^{20}  +\dots )^2 $} }  \\ \midrule
    $k=3$ &\scriptsize{\parbox{13cm}{$1+70 t^2+40 t^3+2415 t^4+2520 t^5+55560 t^6+78624 t^7+965011 t^8+1631240 t^9+13512030 t^{10}   +\dots \\
    = (  1+35 t^2+20 t^3+595 t^4+560 t^5+6755 t^6+7812 t^7+57868 t^8+73900 t^9+398370 t^{10}+\dots )^2 $} }  \\ \bottomrule
    \end{tabular}
    \caption{Hilbert series results for \eqref{eq:6d_magQuiv_Unitary_infinite}. The first line displays the Coulomb branch Hilbert series for \eqref{eq:6d_magQuiv_Unitary_infinite}, while the second line displays the factorisation into a product of two identical factors. Each factor agrees with the known Hilbert series.}
    \label{eq:6d_magQuiv_Unitary_infinite1}
\end{table}

\FloatBarrier
\bibliographystyle{JHEP}
\bibliography{bibli.bib}

\end{document}